%% file: ALERT-TagDVCS.tex
\documentclass[letter,12pt]{report}

\input{formatAndDefs}

\input{preamble}

\begin{document}

\input{TitlePage}
\input{TitlePage_RG}

\renewcommand{\baselinestretch}{1.10}

\input{Abstract}

\tableofcontents

\input{Preface}

\input{Introduction}

\input{Motivation}

\input{Formalism}

\input{Detectors}

\input{Experiment}

\input{Conclusion}

\input{appendix}

\bibliographystyle{ieeetr}
\bibliography{biblio,tagged_dvcs}

\end{document}

%% file: formatAndDefs.tex
\usepackage{amsmath,amssymb}             
\usepackage[latin1]{inputenc}
\usepackage[OT1]{fontenc}
\usepackage[left=2.5cm,right=2.5cm,top=2cm,bottom=2cm,includefoot,includehead,headheight=13.6pt]{geometry}
\usepackage{setspace}
\usepackage{lineno}
\usepackage{footmisc}
\usepackage{indentfirst}
\usepackage{siunitx}
\usepackage{lmodern}
\usepackage{bm}
\usepackage{float}
\usepackage{tabu}
\usepackage{multirow}

\usepackage{graphicx,type1cm,eso-pic,color}

\usepackage{color}
\definecolor{linkcol}{rgb}{0,0,0.4} 
\definecolor{citecol}{rgb}{0.5,0,0} 

\usepackage{array}
\newcolumntype{L}[1]{>{\raggedright\let\newline\\\arraybackslash\hspace{0pt}}m{#1}}
\newcolumntype{C}[1]{>{\centering\let\newline\\\arraybackslash\hspace{0pt}}m{#1}}
\newcolumntype{R}[1]{>{\raggedleft\let\newline\\\arraybackslash\hspace{0pt}}m{#1}}

\usepackage{rotating}                    
\usepackage{fancyhdr}                    

\let\headruleORIG\headrule
\renewcommand{\headrule}{\color{black} \headruleORIG}

\usepackage{colortbl}
\arrayrulecolor{black}

\fancypagestyle{plain}{
  \fancyhead{}
  \fancyfoot{}
  
}

\makeatletter

\def\cleardoublepage{\clearpage\if@twoside \ifodd\c@page\else%
  \hbox{}%
  \thispagestyle{empty}
  \newpage%
  \if@twocolumn\hbox{}\newpage\fi\fi\fi}

\makeatother
 
{%

\hrulefill
\vspace*{0.5cm}%
\end{minipage}
}

{ \begin{list}%
	{$\bullet$}%
	{\setlength{\labelwidth}{25pt}%
	 \setlength{\leftmargin}{30pt}%
	 \setlength{\itemsep}{\parsep}}}%
{ \end{list} }

\renewcommand{\epsilon}{\varepsilon}


%% file: preamble.tex
\usepackage[most]{tcolorbox}
\usepackage{lmodern}

\usepackage{xparse}

\newcommand\numberthis{\addtocounter{equation}{1}\tag{\theequation}}

\def\derivationtext{Derivation} 

\NewDocumentEnvironment{derivation}{ O{} }
{
\colorlet{colexam}{red!55!black} 
\newtcolorbox[use counter=derivationcounter]{derivationbox}{%
    empty,
    title={\derivationtext\ \thetcbcounter: #1},
    attach boxed title to top left,
       minipage boxed title,
    boxed title style={empty,size=minimal,toprule=0pt,top=4pt,left=3mm,overlay={}},
    coltitle=colexam,fonttitle=\bfseries,
    before=\par\medskip\noindent,parbox=false,boxsep=0pt,left=3mm,right=0mm,top=2pt,breakable,pad at break=0mm,
       before upper=\csname @totalleftmargin\endcsname0pt, 
    overlay unbroken={\draw[colexam,line width=.5pt] ([xshift=-0pt]title.north west) -- ([xshift=-0pt]frame.south west); },
    overlay first={\draw[colexam,line width=.5pt] ([xshift=-0pt]title.north west) -- ([xshift=-0pt]frame.south west); },
    overlay middle={\draw[colexam,line width=.5pt] ([xshift=-0pt]frame.north west) -- ([xshift=-0pt]frame.south west); },
    overlay last={\draw[colexam,line width=.5pt] ([xshift=-0pt]frame.north west) -- ([xshift=-0pt]frame.south west); },%
    }
\begin{derivationbox}}
{\end{derivationbox}\endlist}

\usepackage{hyperref}
\hypersetup
{
bookmarksopen=true,
pdftitle="Tagged DVCS - ALERT Run Group Proposal",
pdfauthor="W. R. Armstrong", 
pdfsubject="Tagged DVCS - ALERT Run Group Proposal", 
pdfmenubar=true, 
pdfhighlight=/O, 
colorlinks=true, 
pdfpagemode=UseNone, 
pdfpagelayout=SinglePage, 
pdffitwindow=true, 
linkcolor=linkcol, 
citecolor=citecol, 
urlcolor=linkcol 
}

%% file: TitlePage.tex
\begin{titlepage}
     \begin{center}
       \vspace*{-1.8cm}
       \noindent \Huge \textbf{Spectator-Tagged Deeply Virtual Compton Scattering on Light Nuclei} \\
     \end{center}
  
\renewcommand{\thefootnote}{\fnsymbol{footnote}}
     \begin{center}
       \vspace*{1.0cm}
       \noindent {W.~R.~Armstrong$^\dagger$\footnote[3]{Contact person}, J.~Arrington, I.~Clo\"{e}t, A.~Freese, K.~Hafidi\footnote[2]{Spokesperson}, M.~Hattawy, S.~Johnston, D.~Potterveld, P.~Reimer, S.~Riordan, Z.~Ye} \\
       \vspace*{0.2cm}
       \noindent \emph{Argonne National Laboratory, Lemont, IL 60439, USA} \\
       \vspace*{0.7cm}
       \noindent {J. Ball, M. Defurne, M. Gar\c{c}on, H. Moutarde, S. Procureur, F. Sabati\'e} \\
       \vspace*{0.2cm}
       \noindent \emph{CEA, Centre de Saclay, Irfu/Service de Physique Nucl\'eaire, 91191 Gif-sur-Yvette, France} \\
       \vspace*{0.7cm}
       \noindent {W. Cosyn} \\
       \vspace*{0.2cm}
       \noindent \emph{Department of Physics and Astronomy, Proeftuinstraat 86, Ghent University, 9000 Ghent, Belgium} \\
       \vspace*{0.7cm}
       \noindent {M. Mazouz} \\
       \vspace*{0.2cm}
       \noindent \emph{Facult\'e des Sciences de Monastir, 5000 Tunisia} \\
       \vspace*{0.7cm}
       \noindent {A. Accardi} \\
       \vspace*{0.2cm}
       \noindent \emph{Hampton University, Hampton, VA 23668, USA} \\
       \vspace*{0.7cm}
       \noindent {J.~Bettane, R.~Dupr\'{e}$^\dagger$, 
                  M.~Guidal, D.~Marchand, C.~Mu\~noz, S.~Niccolai, E.~Voutier} \\
       \vspace*{0.2cm}
       \noindent \emph{Institut de Physique Nucl\'eaire, CNRS-IN2P3, Univ. Paris-Sud, Universit\'e Paris-Saclay, 91406 Orsay Cedex, France} \\
       \vspace*{0.7cm}
       \noindent {K. P. Adhikari, J. A. Dunne, D. Dutta,  M. L. Kabir, L. El Fassi,  L. Ye} \\
       \vspace*{0.2cm}
       \noindent \emph{Mississippi State University, Mississippi State, MS 39762, USA} \\
       \vspace*{0.7cm}
       \noindent {M. Amaryan, G.~Charles, G. Dodge} \\
       \vspace*{0.2cm}
       \noindent \emph{Old Dominion University, Norfolk, VA 23529, USA} \\
       \vspace*{0.7cm}
       \noindent {V. Guzey} \\
       \vspace*{0.2cm}
       \noindent \emph{Petersburg Nuclear Physics Institute, National Research Center "Kurchatov Institute", 
                                  Gatchina, 188300, Russia} \\
       \vspace*{0.7cm}
       \noindent {N. Baltzell, F.~X.~Girod, V. Kubarovsky, K. Park, S. Stepanyan} \\
       \vspace*{0.2cm}
       \noindent \emph{Thomas Jefferson National Accelerator Facility, Newport News, VA 23606, USA} \\
       \vspace*{0.7cm}
       \noindent {B.~Duran, S.~Joosten, Z.-E.~Meziani$^\dagger$, M.~Paolone, M.~Rehfuss, N.~Sparveris} \\
       \vspace*{0.2cm}
       \noindent \emph{Temple University, Philadelphia, PA 19122, USA} \\
       \vspace*{0.7cm}
       \noindent {F. Cao, K. Joo, A. Kim, N. Markov} \\
       \vspace*{0.2cm}
       \noindent \emph{University of Connecticut, Storrs, CT 06269, USA} \\
       \noindent {S. Scopetta} \\
       \vspace*{0.2cm}
       \noindent \emph{Universit\`a di Perugia, INFN, Italy} \\
       \vspace*{0.7cm}
       \noindent {B. McKinnon}\\
       \vspace*{0.2cm}
       \noindent \emph{University of Glasgow, Glasgow G12 8QQ, United Kingdom}\\
       \vspace*{0.7cm}
       \noindent {W. Brooks, A. El-Alaoui} \\
       \vspace*{0.2cm}
       \noindent \emph{Universidad T\'ecnica Federico Santa Mar\'ia, Valpara\'iso, Chile} \\
       \vspace*{0.7cm}
       \noindent {S. Liuti} \\
       \vspace*{0.2cm}
       \noindent \emph{University of Virginia, Charlottesville, VA 22903, USA} \\
       \vspace*{0.7cm}
       \noindent {D. Jenkins} \\
       \vspace*{0.2cm}
       \noindent \emph{Virginia Polytechnic Institute and State University, Blacksburg, VA 24061, USA } \\
       \vspace*{1.1cm}
       \noindent {\Large \textbf{a CLAS Collaboration Proposal} } \\
      \end{center}
\renewcommand*{\thefootnote}{\arabic{footnote}}

\date{\today}

\end{titlepage}
\sloppy

\titlepage

%% file: TitlePage_RG.tex
\setcounter{page}{3}
      \renewcommand{\thefootnote}{\fnsymbol{footnote}}  
     \begin{center}
       \vspace*{-1.0cm}
      \noindent {\Large \textbf{Jefferson Lab PAC 45}} \\
      \vspace*{0.8cm}
       \noindent \Huge \textbf{Nuclear Exclusive and Semi-inclusive Measurements with a New CLAS12 Low Energy Recoil Tracker} \\
       \vspace*{0.8cm}
       \noindent \Large \textbf{ALERT Run Group\footnote[2]{Contact Person: Kawtar Hafidi (kawtar@anl.gov)} } \\      
       \vspace*{2.0cm}
       {\large\textbf{EXECUTIVE SUMMARY}}
     \end{center}
 
 \vspace*{0.4cm}

In this run group, we propose a comprehensive physics program to investigate 
the fundamental structure of the $^4$He nucleus. 
An important focus of this program is on 
the coherent exclusive Deep Virtual Compton Scattering (DVCS) and Deep 
Virtual Meson Production (DVMP) with emphasis on $\phi$ meson production. These are 
particularly powerful tools enabling model-independent nuclear 3D tomography 
through the access of partons' position in the transverse plane. These 
exclusive measurements will give the chance to compare directly 
the quark and gluon radii of the helium nucleus. 
Another important measurement proposed in this program is the study of the 
partonic structure of bound nucleons. To this end, we propose next generation 
nuclear measurements in which low energy recoil nuclei are detected. The 
tagging of recoil nuclei in deep inelastic reactions is a powerful technique, 
which will provide unique information about the nature of medium modifications 
through the measurement of the EMC ratio and its dependence on the nucleon 
off-shellness. 
Finally, we propose to measure incoherent spectator-tagged DVCS 
on light nuclei (d, $^4$He) where the observables are sensitive to the
Generalized Parton Distributions (GPDs) of a quasi-free neutron for the case of the deuteron, and bound proton and neutron for the case of $^4$He. The objective is to 
study and separate nuclear effects and their manifestation in GPDs.
The fully exclusive kinematics provide a novel approach for studying
final state interactions in the measurements of the 
beam spin asymmetries and the off-forward EMC ratio.\\

At the heart of this program is the Low Energy Recoil Tracker (ALERT) 
combined with the CLAS12 detector. The ALERT detector is composed of a stereo 
drift chamber for track reconstruction and an array of scintillators for 
particle identification. Coupling these two types of fast detectors will allow 
ALERT to be included in the trigger for efficient background rejection, while 
keeping the material budget as low as possible for low energy particle 
detection. ALERT will be installed inside the solenoid magnet instead of the 
CLAS12 Silicon Vertex Tracker and Micromegas tracker. We will use an 11 GeV 
longitudinally polarized electron beam (80\% polarization) of up to 1000~nA on a gas target 
straw filled with deuterium or $^4$He at 3 atm to obtain a luminosity up to
$6\times10^{34}$~nucleon~cm$^{-2}$s$^{-1}$. In addition we will need to run 
hydrogen and $^4$He targets at different beam energies for detector 
calibration. The following table summarizes our beam time request: \\

\newcommand{\minitab}[2][l]{\begin{tabular}{#1}#2\end{tabular}}
\begin{table}[ht!]
\label{tab:beamTimeRequest}
\center
\bgroup
\def\arraystretch{1.2}%
\tabulinesep=1.5mm
\begin{tabu}{C{3.1cm}C{2.8cm}C{1.6cm}C{2.3cm}C{1.6cm}C{2.5cm}}
\tabucline[2pt]{-}
\bf Configurations  & \bf Proposals & \bf Targets   & \bf Beam time request  & \bf Beam current & \bf Luminosity$^*$ \\
                    &                    &               & days    & nA       & n/cm$^{2}$\!/s     \\
\tabucline[1pt]{-}                                                   
{Commissioning}     & All$^\dagger$      & $^1$H, $^4$He & 5       & Various  & Various            \\
A                   & Nuclear GPDs       & $^4$He        & 10      & 1000      & $6\times10^{34}$   \\
B                   & Tagged EMC \& DVCS & $^2$H         & 20      & 500      & $3\times10^{34}$   \\
C                   & All$^\dagger$      & $^4$He        & 20      & 500      & $3\times10^{34}$   \\
\tabucline[1pt]{-}                                                   
{\bf TOTAL}         &                    & \,            & \bf 55  & \,       & \,                 \\
\tabucline[2pt]{-}
\end{tabu}
\egroup
\end{table}

\footnotetext[1]{This luminosity value is 
   based on the effective part of the target. When accounting for the target's 
   windows, which are outside of the ALERT detector, it is increased by 60\%.}
   
\footnotetext[2]{``All'' includes the four proposals of the run group: Nuclear GPDs, Tagged EMC, Tagged DVCS and Extra Topics. Note that the beam time request is only driven by the three first proposals.}
   
\renewcommand*{\thefootnote}{\arabic{footnote}}

\date{\today}

\sloppy

\titlepage

%% file: Abstract.tex
\setcounter{page}{5}
\addcontentsline{toc}{chapter}{Abstract}

     \begin{center}
{\large\textbf{Abstract}}
    \end{center}
\vspace*{0.4cm}

\input{abstract/tagged_dvcs_abstract}

\newpage

%% file: abstract/tagged_dvcs_abstract.tex
The three-dimensional picture of quarks and gluons in the 
proton is set to be revealed through Deeply virtual Compton scattering  
while a critically important puzzle in the one-dimensional picture remains, 
namely, the origins of the EMC effect. Incoherent nuclear DVCS, i.e. DVCS on 
a nucleon inside a nucleus, can reveal the 3D partonic structure of 
the \emph{bound nucleon} and shed a new light on the EMC effect.
However, the Fermi motion of the struck nucleon, off-shell effects and 
final-state interactions (FSIs) complicate this parton level interpretation. 
We propose here a measurement of incoherent DVCS with a tagging of the 
recoiling spectator system (nucleus A-1) to systematically control nuclear effects.
Through spectator-tagged DVCS, a fully detected final state presents a 
\emph{unique opportunity}
to systematically study these nuclear effects and cleanly observe possible 
modification of the nucleon's quark distributions.\\


We propose to measure the DVCS beam-spin asymmetries (BSAs) on $^4$He and deuterium 
targets. The reaction $^4$He$(e,e^{\prime}\gamma\,p\,^3$H$)$ with a fully detected final state 
has the rare ability to simultaneously quantify FSIs, measure initial nucleon momentum,
and provide a sensitive probe to other nuclear effects at the parton level.
The DVCS BSA on a (quasi-free) neutron will be measured by tagging a spectator proton with a 
deuteron target. Similarly, a bound neutron measurement detects a spectator $^3$He off a $^4$He target.
These two observables will allow for a self-contained measurement of the neutron \emph{off-forward EMC Effect}.\\

We will also measure the impact of final state interactions on incoherent DVCS 
when the scattered electron, the real photon, and the struck proton are 
detected in the final state. This will help understand the measurements 
performed on helium during the previous CLAS E-08-024 experiment and will allow 
better measurements of the same channel where both statistics and kinematic
coverage are extended. The measurement of neutron DVCS by tagging the recoil 
proton from a deuterium target is highly complementary to the approved 
CLAS12 experiment E12-11-003 which will also measure quasi-free neutron DVCS by 
detecting the scattered neutron.

%% file: Preface.tex
\setlength\parskip{\baselineskip}
\chapter*{Preface to the PAC 45 Edition\markboth{\bf Preface to the PAC 45 Edition}{Preface to the PAC 45 Edition}}

\addcontentsline{toc}{chapter}{Preface to the PAC45 Edition}

This proposal was submitted to the PAC44 and was deferred as part of the run 
group proposal. PAC44 raised a few issues that we specifically address in 
section \ref{sec:PACAnswers}. However, here we take a moment to comment on 
their concern regarding the final state interactions because it is directed at 
this proposal and not the whole run group.

The PAC44 was skeptical about the kinematic cuts which identify FSIs and our 
claim to unambiguously identify FSIs. Paraphrased, our initial statement, ``we 
can unambiguously identify \emph{FSI-free} kinematics in a model independent 
way'', was revised to be: ``we can unambiguously identify kinematics with
\emph{significant FSIs} in a model independent way.''
The main distinction is FSI-free kinematics will always require a model
while the inverse is not necessarily true. The nuance being that kinematics 
which are identical to the plane wave impulse approximation result could have FSIs that 
must be modeled at the amplitude level.

In order to make clear our approach, this updated version of the 
proposal dedicates a new appendix (\ref{chap:appendixKine}) to the unique 
kinematic leverage that spectator-tagged DVCS affords in understanding FSIs.  
The reader is urged to read this section before Chapter 1 as it should clarify 
concerns in regard to the FSIs.

%% file: Introduction.tex
\setlength\parskip{\baselineskip}%
\chapter*{Introduction\markboth{\bf Introduction}{Introduction}}
\label{chap:intro}
\addcontentsline{toc}{chapter}{Introduction}

\begin{figure}[!hb]
   \centering
   \includegraphics[width=0.40\textwidth]{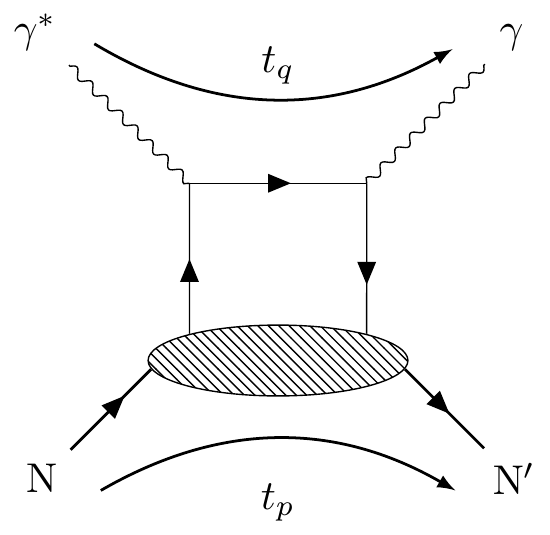}
   \caption{\label{fig:Handbagintro}The DVCS handbag diagram showing the two 
   ways the momentum transfer can be calculated. The hatched vertex represents   
   the non-perturbative GPD.
 }
\end{figure}

Deeply virtual Compton scattering is widely used to extract information about 
the generalized parton distributions of the nucleon. Its usefulness 
comes from the fact that the final state photon does not interact strongly (at 
leading order), requiring no additional non-perturbative formation mechanism.  
That is, the process in which the active quark radiates a final state photon is 
well understood, therefore, it is very useful for extracting information about 
the unknown non-perturbative vertex shown in Figure~\ref{fig:Handbagintro}.

The extracted GPDs offer a three dimensional picture of how quarks and gluons 
are distributed in the nucleon. DVCS measurements on the 
proton~\cite{Chekanov:2003ya,Aktas:2005ty,Camacho:2006qlk,Girod:2007aa} and 
neutron~\cite{Mazouz:2007aa} have already begun to provide insight into this 
slowly developing picture of the nucleon, however, without a free neutron 
target a flavor separation will always require using a quasi-free neutron 
target bound in light nuclei such as deuterium or $^3$He. Such an extraction 
requires control of numerous nuclear effects: Fermi motion, off-shellness of 
the nucleons, mean field modified nucleons, short-range correlations (SRC), and 
final-state interactions. 

Most observables involving nuclear targets ({\it e.g.}, inclusive deep-inelastic 
scattering (DIS), tagged DIS, inclusive quasi-elastic, semi-inclusive nucleon knockout, and polarization 
transfer in quasi-elastic scattering) are sensitive to many of these nuclear 
effects.  Some experiments have been conducted in such a way as to mitigate or 
provide some systematic control over the size of these 
effects~\cite{Rvachev:2004yr,Paolone:2010qc,Malace:2010ft}.  However, as 
discussed in the next chapter, the very nature of each experiment often 
precludes control of one or more nuclear effect mentioned above.  Therefore, it 
is difficult to unambiguously draw conclusions from these measurements as to 
whether a nucleon is modified in a nuclear environment.

Much like the DVCS observables' ability to cleanly access information about the 
GPDs, tagged incoherent DVCS analogously provides a method for cleanly 
extracting nuclear effects from the observables. In a fully exclusive reaction, 
the over-determined kinematics yield two measurements of the same momentum 
transfer (see Figure~\ref{fig:Handbagintro}). Within the plane wave Born 
approximation (PWBA), the momentum transfer between the virtual and real photon 
is completely insensitive to FSIs. On the other side of the diagram, the 
momentum transfer calculated between the initial and final nucleon is quite 
sensitive to FSIs under the assumption that the plane wave impulse approximation 
(PWIA) holds\footnote{See appendix~\ref{chap:appendixKine} for a detailed 
discussion of kinematics and the PWIA.}. That is to say, any relative deviation 
between the two momentum transfers can be attributed to the breakdown of the 
PWIA. In this way we can identify the kinematics where FSIs are significant and 
where they are minimal.  Unlike, fully exclusive quasi elastic knockout 
reactions, tagged DVCS has a unique opportunity to simultaneously probe nuclear 
effects at the parton level.

Therefore, we propose to measure tagged DVCS beam spin asymmetries on two 
nuclear targets ($^2$H and $^4$He) to unambiguously determine if the nucleon is 
modified in a nuclear environment. We will measure three beam spin asymmetries 
through tagged incoherent DVCS using two gas targets. The experiment requires 
the measurement of the three main processes:
\begin{enumerate}
   \item $^4$He$\,+\,\gamma^{*} \longrightarrow \gamma\,+\,p\,+\,^3$H
   \item $^4$He$\,+\,\gamma^{*} \longrightarrow \gamma\,+\,(n)\,+\,^3$He
   \item $^2$H$\,+\,\gamma^{*}  \longrightarrow \gamma\,+\,(n)\,+\,p$
\end{enumerate}
The first process provides a measurement of bound proton DVCS, but more 
importantly, by also detecting the final state proton, provides the 
over-determined kinematics needed to systematically probe the size of FSIs. This 
measurement is of critical importance to unwinding the nuclear effects when 
analyzing the last two processes, bound and quasi-free neutron DVCS, where the 
active nucleon (a neutron) goes undetected. A self-contained analysis of the 
nuclear effects on a neutron will be compared to a similar analysis of the 
proton. In addition to the proposed measured bound proton results, the latter 
will make use of previously measured results and already approved free proton 
DVCS~\cite{E1206119,E1206114} experiments.
In this way we will extract for both the proton and neutron an ``off-forward 
EMC effect'', {\it i.e.}, the ratio of a bound nucleon's off-forward structure 
function in $^4$He to a free nucleon's off-forward structure function.

A large acceptance detector system capable of running at high luminosity along 
with the ability to detect or ``tag'' the low energy recoil spectator system is 
needed to perform such measurements.
The ideal choice for this experiment is clearly CLAS12 augmented with \emph{a 
low energy recoil tracker} (ALERT) that cleanly identifies the recoiling 
spectator system down to the lowest possible momentum.  The ALERT detector 
consists of a small drift chamber that is insensitive to minimum ionizing 
particles, and a surrounding scintillator hodoscope that principally provides 
TOF information.

The outline of this proposal is as follows.  Chapter one will provide a 
detailed motivation for the experiment.  Chapter two will present the formalism 
and the observables we aim to measure.  Chapter three presents  a discussion of 
detector requirements, specifically, the detectors of CLAS12 and the need for a 
new low energy recoil recoil detector.  Chapter four will discuss the 
experimental outputs, kinematic coverage, projected results, and required 
beam time for the proposed experiment.
The first time reader is encouraged to first read the 
Appendix~\ref{chap:appendixKine} which covers in detail the kinematics, PWIA, 
and FSIs.


%% file: Motivation.tex
\setlength\parskip{\baselineskip}%
\chapter{Physics Motivations}
\label{chap:physics}

Before diving into the details of tagged incoherent DVCS, we will first explain 
why an understanding of the nature and origins of nuclear effects is important 
for determining the nucleon structure at the parton level.  We will begin this 
chapter with a discussion of nuclear effects and the  challenges they present 
to experiment. This is followed by a quick overview of GPDs and their 
importance to understand the partonic structure of nuclear matter.  The 
kinematics of incoherent DVCS are discussed, highlighting the critical 
importance of the spectator tagging method. We will emphasize the unique 
opportunity tagged DVCS has to finally settle the more than three decades old 
question: \emph{is the partonic structure of the nucleon modified in presence 
of a nuclear medium?}
\section{Nuclear Effects} \label{sec:NuclearEffects}
\subsection{The EMC Legacy}\label{sec:EMCLegacy}
Measurements of the longitudinal parton distribution functions (PDFs) with 
polarized beams and targets have provided a detailed one dimensional mapping of 
the quark distributions in the nucleon. QCD has been successful in describing 
the evolution of these distributions across scales differing by many orders of 
magnitude. The European Muon Collaboration not only observed the 
so-called ``EMC effect'', but they also created the poorly named ``spin 
crisis''. The EMC effect originates from their observation that the naive 
expectation of the quark distributions in nuclei, {\it i.e.}, they are the sum 
of the quark distributions for $A$ free nucleons, was not 
observed~\cite{Aubert:1983xm}.  However, consensus has yet to be reached in how 
to explain this effect. The spin crisis began when it was discovered that the 
spin of the quarks only carry a small fraction of the nucleon's total spin.  It 
was soon understood that, through the non-perturbative dynamics of QCD, the 
remaining angular momentum will likely come in the form of quark orbital 
angular momentum (OAM) and gluon angular momentum.  The EMC experiments gave us 
a ``crisis'' that was quickly understood, and an empirical ``effect'' whose 
origins remain ambiguous more than 3 decades later.\footnote{In hindsight, 
   perhaps the ``EMC effect'' should have been called the ``EMC Crisis'', and 
the ``spin crisis'' called the ``EMC spin effect''.}
\subsection{Measuring Medium Modified Nucleons}\label{sec:MediumModified}
The EMC effect  showed the possibility of medium modifications to the nucleon 
may be significant in deep inelastic scattering\footnote{Perhaps the earliest 
known medium modification of the nucleon is the free neutron lifetime compared 
to the significantly longer lifetime when bound in a nucleus.}.  However, the 
degree to which these modifications can be cleanly studied in inclusive or 
semi-inclusive processes is made difficult by the possible presence of final 
state interactions.  Furthermore, when considering the Fermi motion of a 
bound nucleon, there is a probability of finding a nucleon moving with 
large relative momenta which corresponds to a configuration where the two 
nucleons are separated by a small distance. By selecting these dense 
configurations through spectator tagging in hard processes, it is not 
unreasonable to expect sizable modifications relative to the mean field 
nucleons~\cite{Weinstein:2010rt}.
Therefore, knowing precisely the initial momentum of the struck nucleon is key 
for understanding the short and long range nuclear effects.  Isolating the 
configuration space effects from the FSIs presents a significant obstacle to 
drawing a definitive conclusion about medium modifications.

Similar medium modifications are expected to manifest themselves in the elastic 
form factors of bound nucleons. Observation of saturation of the Coulomb sum 
rule (CSR), which is accessible through measurements of the longitudinal 
response function in quasi-elastic scattering off nuclei, has been debated for 
some time~\cite{Meziani:1984is,Morgenstern:2001jt}. An observed quenching of 
the sum rule would indicate that nucleons are modified in such a way that the 
net charge response of the parent nucleus is much more complicated than a 
simple sum of nucleons.  Recent, QCD inspired theoretical work predicts a 
dramatic quenching of the sum rule~\cite{Cloet:2015tha}. Furthermore, 
observations of short range correlated nucleon pairs in knockout reactions have 
challenged us to confront our ignorance of the short-range part of the N-N 
potential.  

While a consensus has yet to be reached in explaining the origins of and the 
connections between the EMC effect, short range correlations, and quenching of 
the Coulomb sum rule, medium modifications of the nucleon is expected to play 
an important role in these phenomena.
\subsection{Why \emph{Tagged} DVCS?}
DVCS is poised to provide some much needed contact between the EMC effect and 
short-range correlations. The two phenomena are observed through notably  
different processes but the connection between the inelastic and elastic 
observables is due to the properties of GPDs (see~\ref{sec:GPDs}). In the 
forward limit the GPDs reduce to the longitudinal parton distributions whose 
modification may explain the EMC effect, and in the off-forward case they 
reduce to the form factors, thus, describing elastic scattering off of the 
nucleon.  Therefore, measurements of nucleon GPDs in nuclei will bridge the gap 
between these two processes and will shed light on the connections between the 
EMC effect and short range correlations.  The sensitivity of the GPDs to medium 
modifications is a significant motivation, however, \emph{tagged} incoherent 
DVCS provides an unprecedented handle on quantifying and systematically 
controlling the various nuclear effects.

First, let us consider the inclusive DIS measurements where only the scattered 
electron is detected and the exchanged virtual photon interacts at the parton 
level. The nucleon containing the struck quark may potentially be in a 
short-range correlated N-N pair, therefore, tagging the spectator system in the 
PWIA provides the experimental handle needed to compare the contributions to 
the EMC effect from SRC nucleons versus the nucleons in the mean field. This 
measurement is part of the ``Tagged EMC'' proposal found in the current 
proposal's run group. Here, FSIs are the principle challenge for this method 
which become amplified at larger initial nucleon momentum. The re-interaction 
of the spectator system (A-1) with hadronizing fragments (X) can alter the 
detected momentum of the spectator system.  Therefore model calculations have 
to be used to explore kinematics where FSIs can 
minimized~\cite{CiofidegliAtti:2002as}.

Similarly, for inclusive quasi-elastic scattering we would like to measure the 
nucleon elastic form factor modifications associated with the SRC and the 
mean-field nucleons. Therefore, by detecting the knockout nucleon or a 
spectator recoil, the initial nucleon's momentum can be determined (within the 
PWIA).  If both are detected, the over-determined kinematics allow for a second 
calculation of the momentum transfer.  However, for large nuclei the 
possibility of detecting the full (A-1) recoil system  becomes nearly 
impossible. Furthermore, FSIs in the form of meson exchange currents (MEC) can 
become rather troublesome even for measurements of induced polarization in 
quasi-elastic knock-out reactions. Therefore, again, we find an explicit model 
dependence spoiling the interpretation of medium modifications.  

Finally, incoherent DVCS has a unique combination characteristics found in DIS 
and quasi-elastic scattering that make it the ideal process for exploring 
nuclear effects. Like both processes, tagging the recoil spectator system  
serves to identify and separate the mean field (low momentum) nucleons from the 
SRC (high momentum) nucleons. Similar to DIS, DVCS has a parton level 
interpretation, and like elastic scattering the process is exclusive. The 
latter property allows for systematic control of the FSIs through the redundant 
measurement of the momentum transfer, $t$. Therefore, tagged incoherent DVCS 
provides a \emph{model independent} method for studying and accounting for 
final state interactions while providing an observable that is uniquely 
sensitive to the medium modifications. 


%
\section{Generalized Parton Distributions}
\subsection{Spin Sum Rule}
By studying the ``off-forward parton distributions'', Ji derived a sum 
rule~\cite{Ji:1996ek} which is a gauge invariant decomposition of the nucleon 
spin. It relates integrals of the GPDs to the quark total angular momentum and 
is written as
\begin{equation}\label{eq:JisSumRule}
   J_q = \frac{1}{2}\int dx\,x\left[ H_q(x,\xi,t=0;Q^2) + 
   E_q(x,\xi,t=0;Q^2)\right]
\end{equation}
where the $H_q$ and $E_q$ are the leading-twist chiral-even quark GPDs. An 
identical  expression for the total gluon orbital angular momentum is obtained 
using the two gluon GPDs, $H_g$ and $E_g$. The total nucleon spin is simply 
written as
\begin{equation}
   \frac{1}{2} = \sum_q J_q + J_g ,
\end{equation}
where the sum is over light quarks and anti-quarks. A topic of heavy discussion 
over the past five years or so has been about the decomposition of the nucleon 
spin. Ji showed that the gluon angular momentum cannot be broken into spin and 
orbital in a gauge invariant way, however, the quark can. In fact, the 
polarized PDFs provide the quark spin contribution in the forward limit
\begin{equation}\label{eq:quarkspin}
   \Delta \Sigma_q (Q^2) = \int dx\, x\,\tilde{H}_q(x,\xi=0,t=0;Q^2)
\end{equation}
where $\Delta\Sigma_q$ is the integral of the polarized PDF $\Delta q$. Therefore, we arrive at an expression for the quark OAM
\begin{equation}
L_q = J_q - \frac{1}{2}\Delta\Sigma_q
\end{equation}
which, through equations~\ref{eq:JisSumRule} and~\ref{eq:quarkspin}, is a 
function of the quark GPDs $E$, $H$, and $\tilde{H}$.

The GPDs of the up and down quarks can be extracted from measurements on the 
neutron and proton using  isospin symmetry
\begin{align}
   F_u &= \frac{3}{5} \left( 4 F^p - F^n\right)\\
   F_d &= \frac{3}{5}\left( 4 F^n - F^p \right)
\end{align}
where $F \in [H,E,\tilde{H},\tilde{E}]$. The flavor separation is straight 
forward if equal data on the proton and neutron GPDs are available.  However, 
clean neutron data are not available due to a non-existent free-neutron target, 
and when neutron measurements are made using nuclear targets they suffer from a 
variety of nuclear effects previously discussed. Part of the proposed 
experiment would provide a precise neutron measurement from deuteron with a 
technique similar to the one used by the BoNuS experiment\footnote{BoNuS 
stands for ``Barely off-shell Nucleon Structure''.} to extract the neutron 
structure function at high x~\cite{bonus6,bonus12}.

\begin{figure}
   \centering
   \includegraphics[width=0.60\textwidth]{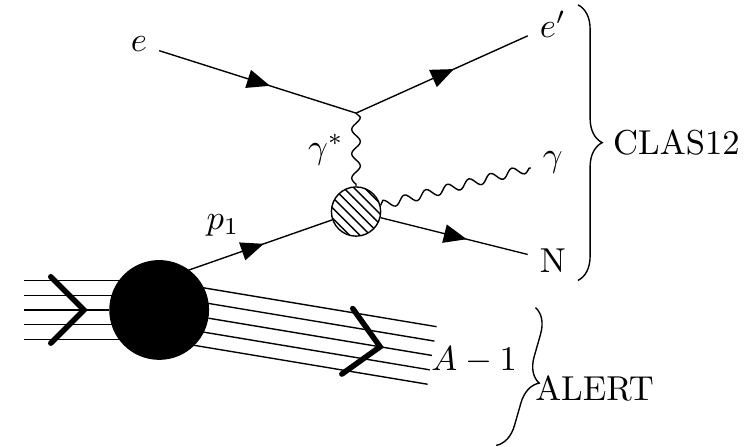}
   \caption{\label{fig:taggedDVCS}Tagged DVCS diagram showing the detection of 
   the forward DVCS final state particles in CLAS12 and the detection of the 
recoiling spectator system (A-1) in ALERT. The hatched circle represents the 
hard DVCS process.}
\end{figure}

\subsection{Polarized EMC Effect}\label{sec:polEMCeffect}
Using polarized nuclear targets, the polarized EMC effect can be measured and 
is predicted to be larger than the ``unpolarized'' EMC 
effect~\cite{Cloet:2005rt}.  The polarized EMC effect is measured through the 
ratio of spin structure functions $g_1$ of a bound nucleon to that of a free 
nucleon. DIS measurements typically require both a longitudinally polarized 
target and longitudinally polarized beam to measure $g_1$. The $\sin\phi$ 
harmonic of the neutron DVCS beam spin asymmetry is
\begin{equation}\label{eq:ALUneutronEMC}
   A_{LU,n}^{\sin\phi} \propto \operatorname{Im}( F_1^n \mathcal{H}^n- 
   \frac{t}{4M^2} F_2^n \mathcal{E}^n+ 
   \frac{x_B}{2}(F_1+F_2)^n\tilde{\mathcal{H}}^n)~.
\end{equation}
The first term is suppressed by $F_1^n$ and if for the moment we neglect the 
$\mathcal{E}$ term, the ratio of this asymmetry for a bound neutron to a free 
neutron is
\begin{equation}\label{eq:PolEMCRatio}
   R_{AL,n}^{\sin\phi} = \frac{A_{LU,n^{*}}^{\sin\phi}}{A_{LU,n}^{\sin\phi}} 
   \simeq 
   \frac{G_M^{n^*}(t)}{G_M^n(t)}\frac{\operatorname{Im}(\tilde{\mathcal{H}}^{n^*}(\xi,\xi,t))}{\operatorname{Im}(\tilde{\mathcal{H}}^{n}(\xi,\xi,t))}
\end{equation}
which in the forward limit becomes
\begin{equation}\label{eq:FWDPolEMCRatio}
   R_{AL,n}^{\sin\phi} \longrightarrow 
   \frac{\mu_{n^*}}{\mu_{n}}\frac{g_1^{n^*}(x)}{g_1^{n}(x)}~,
\end{equation}
where $\mu_{n^*}/\mu_{n}$ is the ratio of the bound neutron magnetic moment to the 
free neutron magnetic moment, and $g_1^{n^*}/g_1^n$ is similarly the ratio of 
the bound to free neutron spin structure functions.

Equations~\ref{eq:PolEMCRatio} and~\ref{eq:FWDPolEMCRatio} are rather 
interesting for a few reasons. First, they can be used to draw conclusions 
about the behavior of the \emph{polarized} quark distributions in unpolarized nuclei 
without using a polarized target. But we must note the unjustified neglect of the 
$\mathcal{E}^{n}$ term which complicates subsequent analysis.
We point out this term in the ratio of Equation~\ref{eq:PolEMCRatio} because it 
highlights the observable's sensitivity to medium modifications. Specifically
noting that a modification to the nucleon's static properties, such as, 
anomalous magnetic moment or polarization-dependent transverse-charge 
distribution (see~\cite{Burkardt:2002hr}), would also manifest themselves 
through the magnetic form factor whose ratio also appears in this observable.

Also, a measurement of the BSA in Equation~\ref{eq:ALUneutronEMC} will provide 
important model constraints on the GPD $E^n$ and measurements of the ratio with 
$^4$He would further constrain nuclear GPD models. This is particularly 
motivating in the context of Equation~\ref{eq:JisSumRule}, where $E^n$ is 
clearly an important quantity for understanding the quark orbital angular 
momentum.


%
\subsection{Models of Nuclear Effects}
To understand the potential sources of observable nuclear effects, we take the 
ratio of beam spin asymmetries for a bound nucleon to that on the free nucleon 
target.  Here we discuss just two models that make very different predictions 
for similar ratios based on the presumed sources of the nuclear effects.

\begin{figure}
   \centering
   \includegraphics{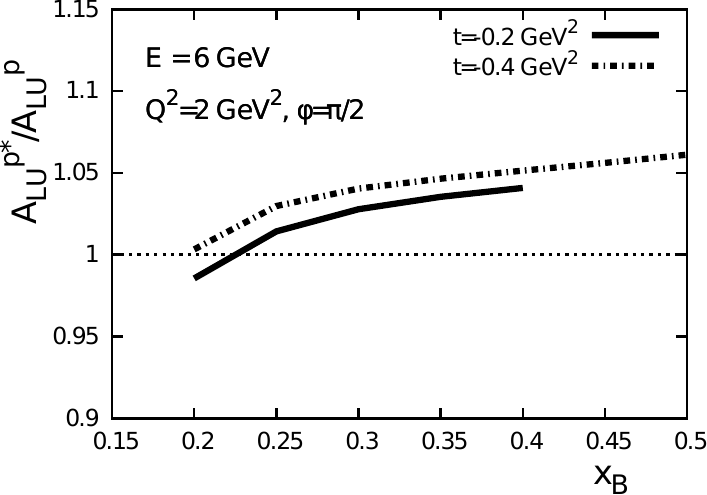}
   \caption{\label{fig:ALUratio}Beam spin asymmetry ratio of a bound proton to 
   a free proton.  Reproduced from~\cite{Guzey:2008fe}.}
\end{figure}

\begin{figure}
   \centering
   \includegraphics{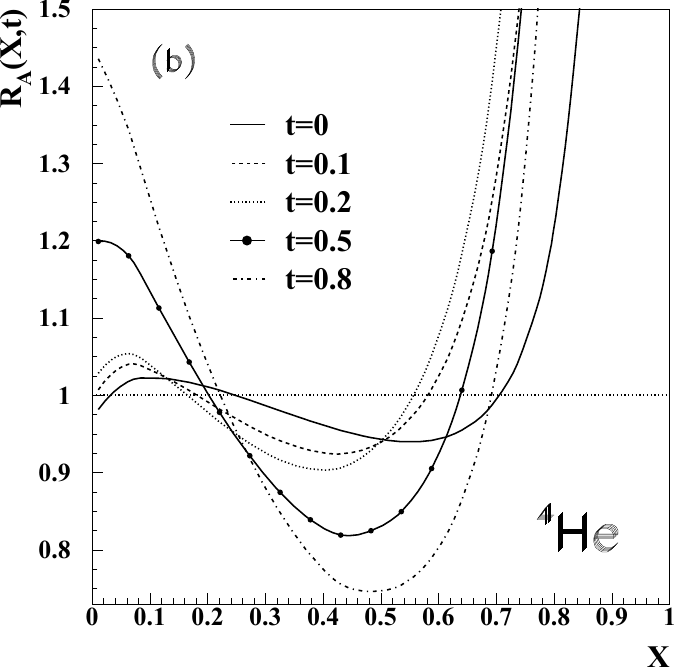}
   \caption{\label{fig:RAratio}Predictions for the ratio given in equation 
   \ref{eq:RA}. Reproduced from~\cite{Liuti:2005qj}.}
\end{figure}

First, predictions for the ratio of beam spin asymmetry at 6 GeV are shown in 
Figure~\ref{fig:ALUratio}, which shows the bound proton beam spin asymmetry, 
$A_{LU}^{p^{*}}$, to the free proton $A_{LU}^{p}$~\cite{Guzey:2008fe}.  These 
calculations use the medium modified GPDs as calculated from the quark-Meson 
coupling model. However, they do not include FSIs and predict their 
contribution is at most a few percent. In another calculation, Liuti et 
al.~\cite{Liuti:2005qj,Liuti:2006dx} use a realistic spectral function and 
consider off-shell effects. This is a more traditional approach to explaining 
differences in a bound nucleon. They make predictions for the ratio
\begin{equation}\label{eq:RA}
   R_A(x,\xi=0,t) = \frac{H_A(x,\xi=0,t) F_N(t)}{H_N(x,\xi=0,t) F_A(t)}
\end{equation}
which is shown in Figure~\ref{fig:RAratio}. For more discussions of modeling 
nuclear effects see~\cite{Dupre:2015jha,Atti:2015eda}.

It is clear that the role of off-shellness, and final state interactions in 
nuclei needs to be better understood if we are to conclude that the nucleon 
structure is modified by the nuclear medium. With spectator tagging, we will be 
able to test these models over a broad range of spectator momentum and angles.  
This tagging technique can be used as a knob to tune the effect of final state 
interactions and either maximized or minimized it.%
%
\begin{figure}
   \centering
   \includegraphics[width=0.49\textwidth]{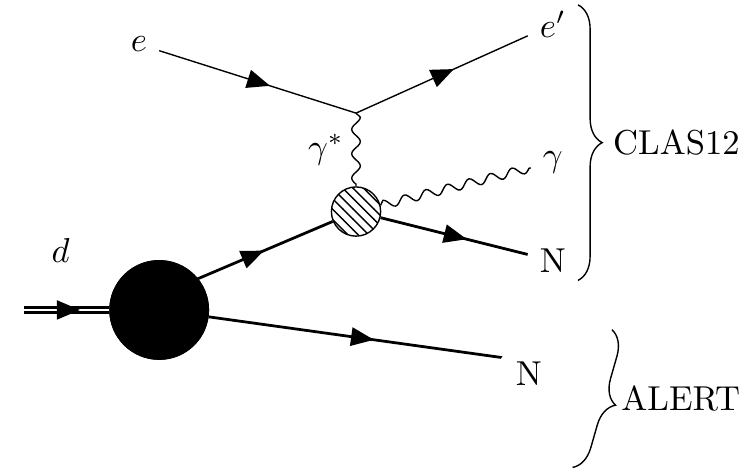}
   \includegraphics[width=0.49\textwidth]{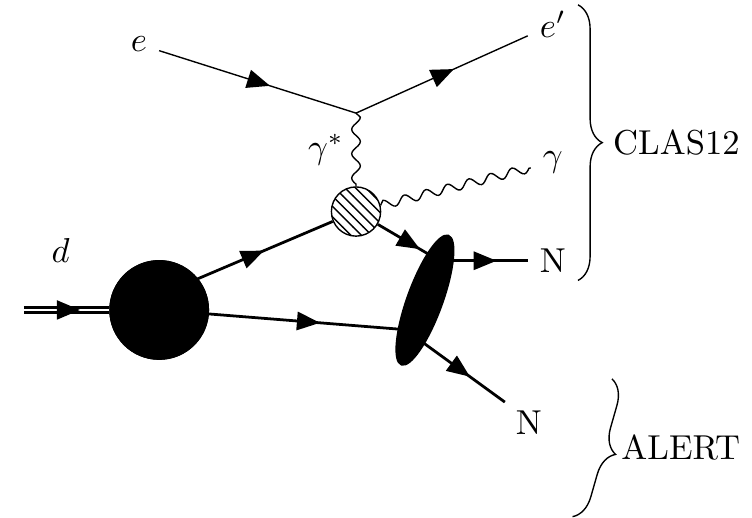}
   \caption{\label{fig:deuteronFSIdiagram}PWIA diagram for incoherent DVCS on 
   the deuteron (left) and with the inclusion of final state interactions 
(right).}
\end{figure}

\begin{figure}
   \centering
   \includegraphics{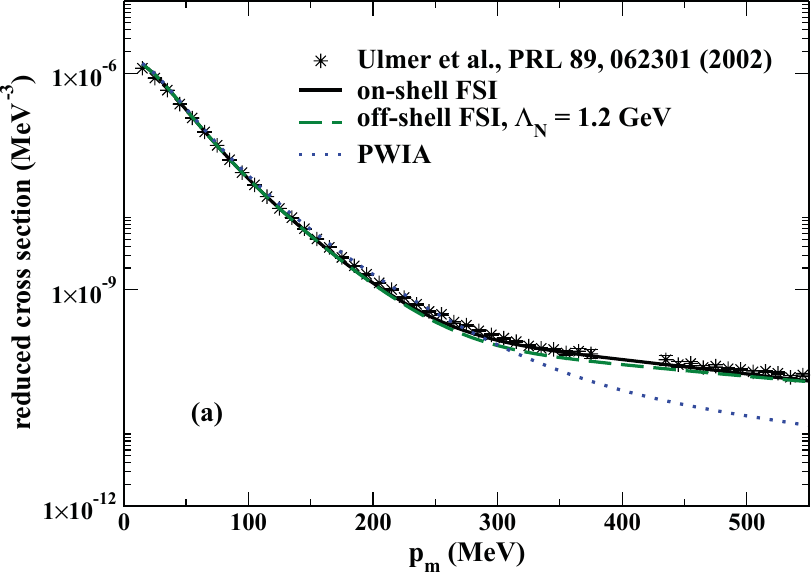}
   \caption{\label{fig:quasiElasticDeuteronFSI}Reduced quasi-elastic scattering 
   cross section  on deuteron. Reproduced from ~\cite{Jeschonnek:2008zg}.  The 
   reduced cross section includes a term multiplying the cross section by $M_d 
 f_{\text{rec}}/(\sigma_{\text{Mott}} m_p m_n p_p)$, where the $f_{\text{rec}}$ 
 is the recoil factor, and $\sigma_{\text{Mott}}$ is the Mott cross section.  
 See reference~\cite{Jeschonnek:2008zg} for more details.}
\end{figure}

\begin{figure}
   \centering
   \includegraphics{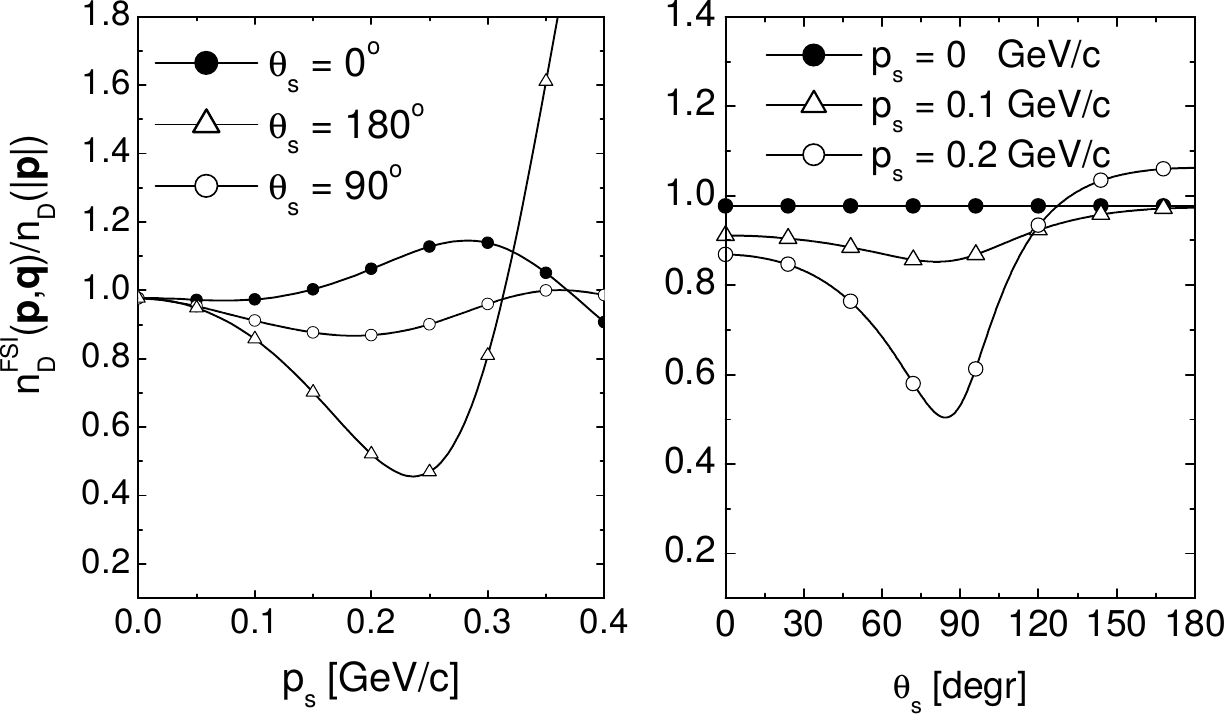}
   \caption{\label{fig:deuteronFSI} Ratio of cross sections for the FSI model 
   from~\cite{CiofidegliAtti:2003pb} to PWIA calculation as a function of
   the spectator momentum (left) and spectator angle (right).}
\end{figure}

To understand the regions where FSIs are expected to be significant, we first 
look at the deuteron. Consider the quasi-elastic scattering on a quasi-free 
nucleon as shown in Figure\,\ref{fig:deuteronFSIdiagram}. Measurements of the  
cross section as a function of missing momentum are shown in 
Figure\,\ref{fig:quasiElasticDeuteronFSI} along with model calculations in PWIA 
with different final state interactions. From model calculations it was found 
that the PWIA was insufficient for describing the data at missing momenta above 
300 MeV/c.
Similarly, the size of the FSI strength as a function of spectator momentum 
(left) and angle relative to the momentum transfer, $\theta_s$, (right) is 
shown in 
Figure~\ref{fig:deuteronFSI}~\cite{CiofidegliAtti:2003pb,CiofidegliAtti:2002as}.  
At low recoil momentum and backwards spectator angle, the FSIs are negligible, 
where at high momenta perpendicular to the momentum transfer, the FSIs are 
maximized.

\section{Motivation Summary}
In summary, we propose to perform the following key measurements using CLAS12 and ALERT for the low energy spectator recoil tagging:
\begin{itemize}
\item \emph{Bound proton DVCS} with a $^4$He target where the final state is 
   fully detected by tagging a spectator $^3$H and the struck proton is 
   detected in CLAS12.  The PWIA will be tested by the redundant measurement of 
   the momentum transfer as explained above and in great detail in appendix 
   \ref{chap:appendixKine}. Thus, kinematics with significant FSIs are 
   identified in a completely \emph{model independent} way. 
\item \emph{Bound neutron DVCS} with a $^4$He target where the neutron goes 
  undetected and the spectator $^3$He is detected in ALERT. Using the same 
  kinematics identified in the previous measurement, and using iso-spin 
  (charge) symmetry, we can conclude that the struck neutron feels the same  
  final state interactions as the struck proton.
\item \emph{Quasi-free neutron DVCS} with a $^2$H target where the recoil 
  proton is tagged and the struck neutron goes undetected.
\end{itemize}

%% file: Formalism.tex
\setlength\parskip{\baselineskip}%
\chapter{Formalism and Experimental Observables}
\label{chap:formalism}
\section{Deeply Virtual Compton Scattering}
The cross section for DVCS on a spin-1/2 target can be parameterized in terms 
of four helicity conserving GPDs: $H^q$, $E^q$, $\tilde{H}^q$, and 
$\tilde{E}^q$. For spin-0 targets, such as $^{4}$He, the cross section is 
parameterized with just one helicity conserving GPD~\cite{Guzey:2003jh}. For 
spin-1 targets like the deuteron, the cross section is parameterized with nine 
GPDs~\cite{Cano:2003ju,Kirchner:2003wt}.

\begin{figure}
   \centering
   \includegraphics[width=0.60\textwidth]{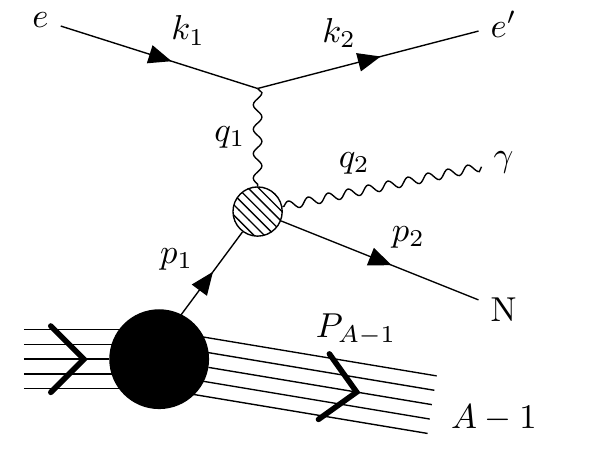}
   \caption{\label{fig:dvcsMomenta}Incoherent DVCS process with the momentum 
   definitions labeled.}
\end{figure}

The DVCS cross section is written as
\begin{equation}
\frac{d\sigma}{dx_A\,dy\,dt\,d\phi\,d\varphi} = \frac{\alpha^3 x_A y}{16 \pi^2 
Q^2 \sqrt{1+\epsilon^2}} \left| \frac{\mathcal{T}}{e^3} \right|^2
\end{equation}
where
\begin{equation}
   \epsilon \equiv 2x_A \frac{M_A}{Q},
\end{equation}
$x_A=Q^2/(2p_1\cdot q_1$) is the scaling variable, $y= (p_1\cdot q_1)/(p_1\cdot 
k_1)$ is the photon energy fraction, $\phi$ is the angle between the leptonic 
and hadronic planes, $\varphi$ is the scattered electron's azimuthal angle, $Q^2= 
-q_1^2$, and $q_1=k_1-k_2$. The particle momentum definitions are shown in 
Figure~\ref{fig:dvcsMomenta}. We use the BMJ\footnote{The Belitsky, M\"{u}ller, and
Ji reference frame. See \cite{Braun:2014paa} for a nice discussion of the 
various reference frames.} 
convention~\cite{Braun:2014paa,Belitsky:2001ns,Belitsky:2010jw,Belitsky:2012ch} 
for defining the momentum transfer where the target nucleus is initially at 
rest, $\Delta = p_1-p_2$ and $t=\Delta^2$. The Bjorken variable  is related to 
the scaling variable by
\begin{equation}
   x_{\text{B}} = \frac{Q^2}{2 M_N E\,y} \simeq A x_A
\end{equation}
where $M_N$ is the nucleon mass and $E$ is the beam energy. Another scaling variable called skewedness is
\begin{equation}
\xi = \frac{x_A}{2 - x_A} + \mathcal{O}(1/Q^2)
\end{equation}
where the power suppressed contributions originate with the selection of the 
BMJ frame convention needed to unambiguously define the leading-twist 
approximation used in this proposal~\cite{Braun:2014paa}.

\begin{figure}[!hbt]
   \centering
   \includegraphics[width=0.20\textwidth]{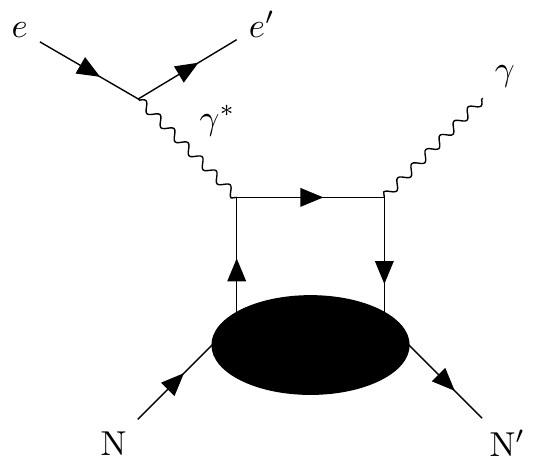}
   \includegraphics[width=0.20\textwidth]{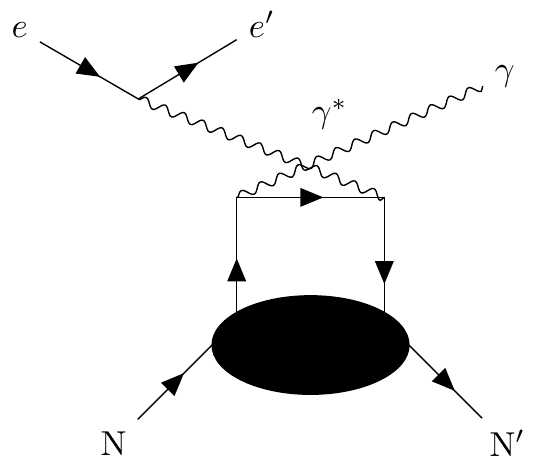}
   \includegraphics[width=0.20\textwidth]{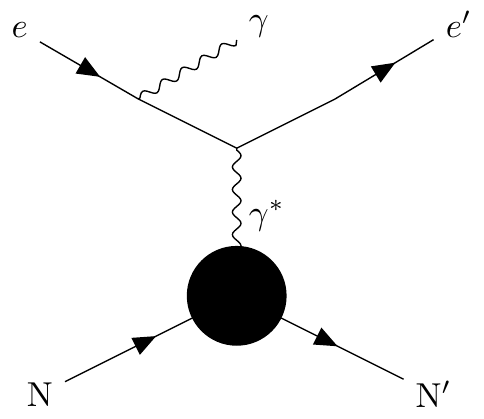}
   \includegraphics[width=0.20\textwidth]{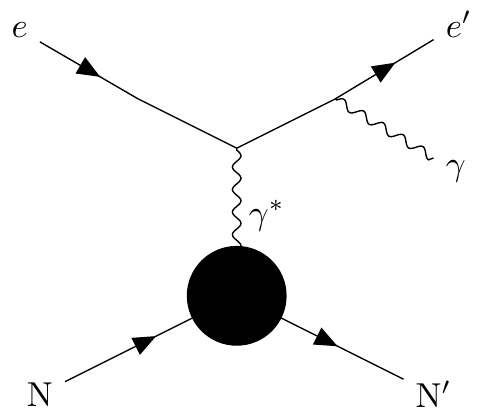}
   \caption{\label{fig:DVCShandbag}DVCS handbag diagram and BH contributions 
   used for calculating DVCS amplitudes.}
\end{figure}

The amplitude is the sum of the DVCS and Bethe-Heitler (BH) amplitudes, and 
when squared has terms
\begin{equation}
   \mathcal{T}^2 = \left|\mathcal{T}_{\text{BH}}\right|^2 + 
   \left|\mathcal{T}_{\text{DVCS}}\right|^2 + \mathcal{I}
\end{equation}
where the first is the BH contribution, the second is the DVCS part, and the last 
term is the interference part,
\begin{equation}
   \mathcal{I} = \mathcal{T}_{\text{DVCS}}\mathcal{T}_{\text{BH}}^{*} + 
   \mathcal{T}_{\text{DVCS}}^{*}\mathcal{T}_{\text{BH}}.
\end{equation}
The corresponding amplitudes are calculated with the diagrams shown in 
Figure~\ref{fig:DVCShandbag}. The details of contracting the DVCS tensor with 
various currents and tensors can be found in~\cite{Kirchner:2003wt}.
The resulting expressions for the amplitudes are
\begin{align}
   \left|\mathcal{T}_{\text{BH}}\right|^2 &= 
   \frac{e^6(1+\epsilon^2)^{-2}}{x_A^2\,y^2\,t\,
   \mathcal{P}_1(\phi)\mathcal{P}_2(\phi)} \left\{ c_0^{\text{BH}} + 
   \sum_{n=1}^{2}\left[ c_n^{\text{BH}}\cos(n\phi) +s_n^{\text{BH}}\cos(n\phi) 
\right] \right\} \\
\left|\mathcal{T}_{\text{DVCS}}\right|^2 &= \frac{e^6}{y^2\,Q^2}\left\{ 
c_0^{\text{DVCS}} + \sum_{n=1}^{2}\left[ c_n^{\text{DVCS}}\cos(n\phi) 
+s_n^{\text{DVCS}}\cos(n\phi) \right] \right\}\\
   \mathcal{I} &= \frac{e^6(1+\epsilon^2)^{-2}}{x_A\,y^3\,t\,
   \mathcal{P}_1(\phi)\mathcal{P}_2(\phi)}\left\{ c_0^{\mathcal{I}} + 
   \sum_{n=1}^{3}\left[ c_n^{\mathcal{I}}\cos(n\phi) 
+s_n^{\mathcal{I}}\cos(n\phi) \right] \right\}
\end{align}
The functions $c_0$, $c_n$, and $s_n$ are called \emph{Fourier coefficients} 
and they depend on the kinematic variables and the operator decomposition of 
the DVCS tensor for a target with a given spin. At leading twist there is a 
straightforward form factor decomposition which relates the vector and 
axial-vector operators with the so-called Compton form factors 
(CFFs)~\cite{Belitsky:2000gz}. The Compton form factors appearing in the DVCS 
amplitudes are integrals of the type
\begin{equation}
   \mathcal{F} = \int_{-1}^{1} dx F(\mp x,\xi,t) C^{\pm}(x,\xi)
\end{equation}
where the coefficient functions at leading order take the form
\begin{equation}
   C^{\pm}(x,\xi) = \frac{1}{x-\xi + i\epsilon} \pm \frac{1}{x+\xi - 
   i\epsilon}.
\end{equation}
We plan on measuring the beam spin asymmetry as a function of $\phi$
\begin{equation}
   A_{LU}(\phi) = \frac{d\sigma^{\uparrow}(\phi) - 
   d\sigma^{\downarrow}(\phi)}{d\sigma^{\uparrow}(\phi) + 
   d\sigma^{\downarrow}(\phi)}
\end{equation}
where the arrows indicate the electron beam helicity. 

%
\subsection{DVCS Beam Spin Asymmetry}
%

Through the Bethe-Heitler dominance of the first sine harmonic of the beam spin 
asymmetry
\begin{equation}
   A_{LU}^{\sin\phi} = \frac{1}{\pi} \int_{\pi}^{\pi} d\phi \sin\phi 
   A_{LU}(\phi)
\end{equation}
is proportional to the following combination of Compton form 
factors~\cite{Guidal:2013rya}
\begin{equation}
   A_{LU}^{\sin\phi} \propto \operatorname{Im}( F_1 \mathcal{H}- \frac{t}{4M^2} 
   F_2 \mathcal{E}+ \frac{x_B}{2}(F_1+F_2)\tilde{\mathcal{H}})
\end{equation}
which is dominated by $\operatorname{Im}(\mathcal{H})$ for the proton, and
dominantly sensitive to $\operatorname{Im}(\mathcal{E})$ and 
$\operatorname{Im}(\tilde{\mathcal{H}})$ for the neutron.

\begin{figure}
   \centering
   \includegraphics[width=0.5\textwidth,trim=0mm 10mm 0mm 5mm, 
   clip]{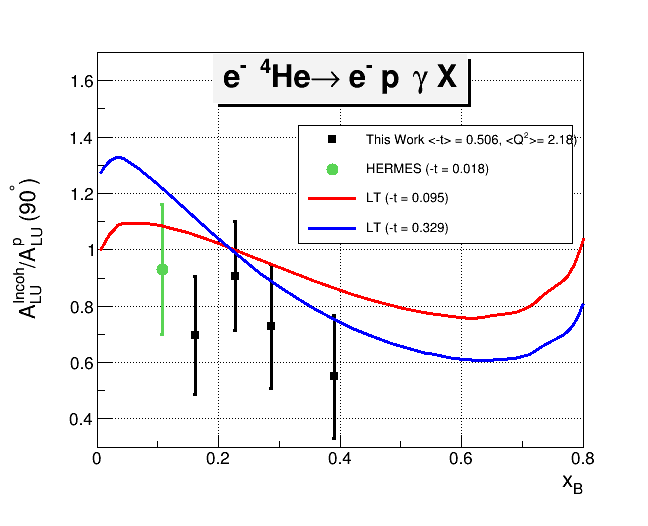}
   \caption{\label{fig:eg6Result}The beam spin asymmetry from 
      eg6~\cite{eg6_note} and HERMES along with models from Liuti and Taneja~\cite{Liuti:2005gi}.}
\end{figure}

Recent measurement~\cite{eg6_note} of incoherent DVCS by the CLAS collaboration 
conducted during the 6 GeV era (E08-024) have indeed shown significant 
modification of the proton beam spin asymmetry in $^4$He without the 
possibility to decipher between the nuclear effects presented above. These 
results are shown in Figure~\ref{fig:eg6Result}. In these measurements the SRC 
and mean field nucleons are not separated and the FSIs remain unchecked.

\section{Tagged DVCS Reactions}

The ALERT detector combined with CLAS12 provides a unique opportunity to 
measure incoherent exclusive processes on light nuclei.
As mentioned in the previous chapters, tagging low momentum spectator recoils 
in exclusive knockout reactions provides the experimental leverage needed to 
separate and cleanly study a variety of nuclear effects. 

\begin{figure}
   \centering
   \includegraphics{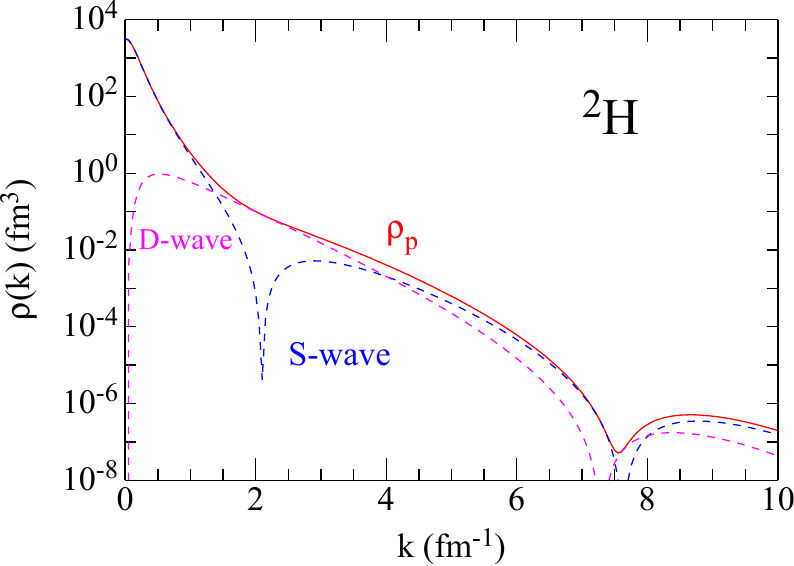}
   \caption{\label{fig:protonDistInDeuteron}The total proton momentum 
      distribution in the deuteron is shown by the red solid line; the 
      contribution from S-wave and D-wave components are shown
   separately by blue and magenta dashed lines.~\cite{Wiringa:2013ala}}
\end{figure}

\begin{figure}
   \centering
   \includegraphics{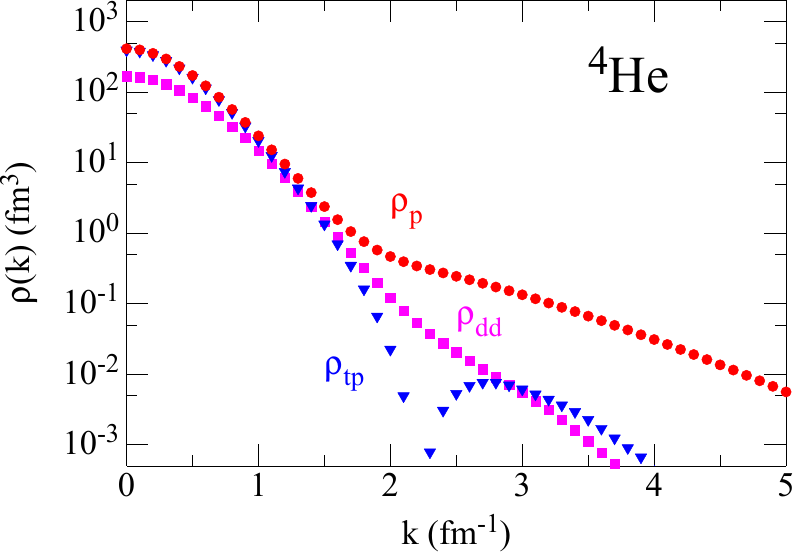}
   \caption{\label{fig:boundDeuteronPDist}The proton momentum distribution in 
      $^4$He is shown by the red circles; the tp cluster distribution is shown 
      by the blue triangles and the dd cluster distribution is shown by the 
   magenta squares.~\cite{Wiringa:2013ala}}
\end{figure}

Neutron DVCS (n-DVCS) is of immediate interest as it is needed to do a flavor 
separation of the GPDs. We propose to measure tagged n-DVCS on $^2$H and $^4$He 
targets starting at $P_{A-1} \simeq$~70~MeV/c  for tagged protons and   
$P_{A-1} \simeq$~120~MeV/c for $^3$He ions. The momentum densities for these 
targets can be seen in Figure~\ref{fig:protonDistInDeuteron} and 
Figure~\ref{fig:boundDeuteronPDist}.
\subsection{n-DVCS with a \texorpdfstring{$^2$H}{Deuteron} Target}
Previous measurements of n-DVCS using a deuteron target required subtracting a 
proton contribution from the total deuteron yields~\cite{Mazouz:2007aa} and 
assumed the validity of the PWIA. The yield for the neutron and coherent 
deuteron can not be separated and the subtraction yields the resulting beam 
spin asymmetry of the combination
\begin{equation}\label{eq:PWIAnDVCSextract}
   D(\vec{e},e^{\prime}\,\gamma)X - H(\vec{e},e^{\prime}\,\gamma)X =
   d(\vec{e},e^{\prime}\,\gamma)d + n(\vec{e},e^{\prime}\,\gamma)n + ...
\end{equation}
which is fit with the CFFs of the neutron and deuteron as free parameters.  
This procedure has a few downsides: it requires a bin by bin equivalent proton 
measurement which is highly prone to systematic effects, the undetected 
spectator system or struck nucleon leaves the center-of-momentum energy, 
$\sqrt{s}$, undetermined, and FSI remain unchecked.   

We propose to measure the recoiling spectator proton, thus, measuring 
$\sqrt{s}$ for every event. Furthermore, the reconstructed missing momentum can 
be used to check for significant final state interactions (see appendix 
\ref{chap:appendixKine}). Comparing $t$ calculated from the virtual and real 
photon momenta to $t$ calculated using the reconstructed missing momentum of 
the neutron (after selection cuts),
\begin{equation}\label{eq:tPhotons}
   t_q = (q_1-q_2)^2
\end{equation}
can provide a measure of the presence of significant final state interactions.  
\subsection{n-DVCS and p-DVCS with a \texorpdfstring{$^4$He}{Helium-4} Target}
A helium target provides the unique opportunity to again measure the neutron 
DVCS beam spin asymmetry, however, now on a bound nucleon with unprecedented 
control over final state interactions. Through the two reactions 
$^4$He$(e,e^{\prime}\gamma\,p\,^3\text{H})$ and 
$^4$He$(e,e^{\prime}\gamma\,^3\text{He})n$ the ratios
\begin{align}\label{eq:BSAratios}
   R_n &= \frac{A_{LU}^{n^{*}}}{A_{LU}^{n}}\\
   R_p &= \frac{A_{LU}^{p^{*}}}{A_{LU}^{p}}
\end{align}
can  provide the leverage needed to definitively make a statement on medium 
modifications.

The proton BSA will be measured by fully detecting the final state; the struck 
proton will be detected in CLAS12 and the recoiling spectator $^3$H will be 
detected in ALERT. The neutron BSA will be measured by tagging a recoil $^3$He 
and without detecting the struck neutron. Exclusivity cuts will ensure the 
n-DVCS event is cleanly selected.
The free proton BSA measurement in Equation\,\ref{eq:BSAratios} will be taken 
from the already approved JLab measurements \cite{E1206119, E1206114}, while 
the neutron BSA will come from the deuteron target measurement discussed above.  
The neutron measurement will have the extra advantage of experimental 
systematics canceling in the ratio because both asymmetries will be measured 
using the same apparatus.
%
\begin{figure}
   \centering
   \includegraphics[width=0.49\textwidth]{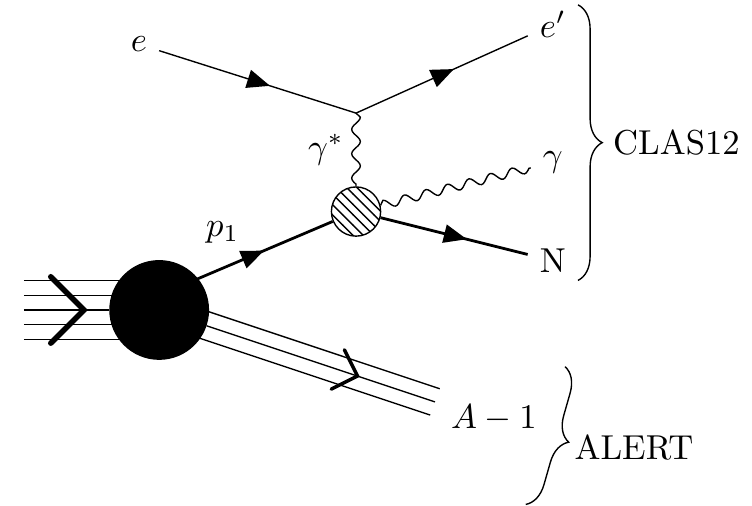}
   \includegraphics[width=0.49\textwidth]{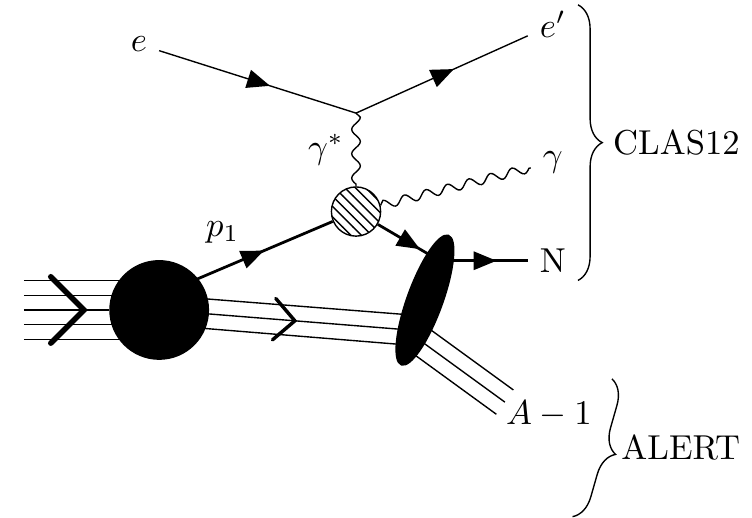}
   \caption{\label{fig:incoherentHe4FSI}Incoherent DVCS on a nuclear target 
   without (left) and with (right) final state interactions.}
\end{figure}

Finally, we consider the fully exclusive proton DVCS reaction where a recoil 
triton is detected as $\bm{P}_{A-1}=-\bm{p}_1$. The fully detected final state 
kinematics present an opportunity to test the PWIA\footnote{Please see appendix 
\ref{chap:appendixKine} for a more detailed explanation.}. One way is to use 
the two momentum transfers, $t_q$ (Equation~\ref{eq:tPhotons}) and
\begin{align}\label{eq:tHadrons}
   t_p &= (p_1-p_2)^2,
\end{align}
which must be the same, i.e., $\delta{t}=t_q-t_p=0$. If a FSI occurs between 
the spectator and the struck nucleon (Figure\,\ref{fig:deuteronFSIdiagram}), such 
as pion exchange, $\delta{t}$ can be non-zero depending which over-determined 
kinematic variables we choose to use (or not use). The reader is referred to 
appendix \ref{chap:appendixKine} for a thorough discussion of this point.  By 
selecting events where $\delta{t}\simeq0$, within the detector resolutions, we 
can be sure that significant final state interactions have not occurred. These 
are events that may contain FSIs that are kinematically indistinguishable from 
the PWIA, but which have an amplitude level influence on the cross section.  
Alternatively, requiring the missing momentum to be back-to-back with the 
recoil spectator provides a cut which is expected to reduce final state 
interactions.

With the final state interactions well under control in the proton DVCS 
channel, charge symmetry suggests that, for the same kinematics, they will be 
similarly understood in the neutron channel. That is the FSIs are assumed to 
follow charge symmetry. Therefore, the proton DVCS BSA measurement on $^4$He is 
crucial for measuring in a model independent way the validity of the PWIA and 
mapping the FSIs for the mirror neutron measurement.

%% file: Detectors.tex
\setlength\parskip{\baselineskip}%
\chapter{Experimental Setup}
\label{chap:setting}
All the different measurements of the ALERT run group require, in addition to 
a good scattered electron measurement, the detection of low energy nuclear 
recoil fragments with a large kinematic coverage. Such measurements have been performed 
in CLAS (BONuS and eg6 runs), where the adequacy of a small additional detector
placed in the center of CLAS right around the target has shown to be the best 
solution. We propose here a similar setup using the CLAS12 spectrometer 
augmented by a low energy recoil detector. 

We summarize in Table~\ref{tab:req} the requirements for the different 
experiments proposed in the run group. By comparison with previous similar 
experiments, the proposed tagged measurements necessitate a 
good particle identification. Also, CLAS12 will be able to handle higher 
luminosity than CLAS so it will be key to exploit this feature in the future 
setting in order to keep our beam time request reasonable.

\begin{table}[ht!]
\centering
\footnotesize
\begin{tabu}{lccc}
\tabucline[2pt]{-}
Measurement  & Particles detected & $p$ range       & $\theta$ range                \rule[-7pt]{0pt}{20pt} \\
\tabucline[1pt]{-}                                                   
Nuclear GPDs & $^4$He             & $230 < p < 400 MeV/c$ & $\pi/4 < \theta < \pi/2$ rad  \rule[-7pt]{0pt}{20pt} \\
Tagged EMC   & p, $^3$H, $^3$He   & As low as possible    & As close to $\pi$ as possible \rule[-7pt]{0pt}{20pt} \\
Tagged DVCS  & p, $^3$H, $^3$He   & As low as possible    & As close to $\pi$ as possible \rule[-7pt]{0pt}{20pt} \\
\tabucline[2pt]{-}
\end{tabu}
\caption{Requirements for the detection of low momentum spectator fragments of the proposed measurements.}
\label{tab:req}
\end{table}

This chapter will begin with a brief description of CLAS12.  
After presenting the existing options for recoil detection and recognize that 
they will not fulfill the needs laid out above, we will describe the design of
the proposed new recoil detector ALERT. We will then present the reconstruction 
scheme of ALERT and show the first prototypes built by our technical 
teams. Finally, we specify the technical contributions of the different 
partners.

%
\section{The CLAS12 Spectrometer}
\begin{figure}
  \begin{center}
    \includegraphics[angle=0, width=0.75\textwidth]{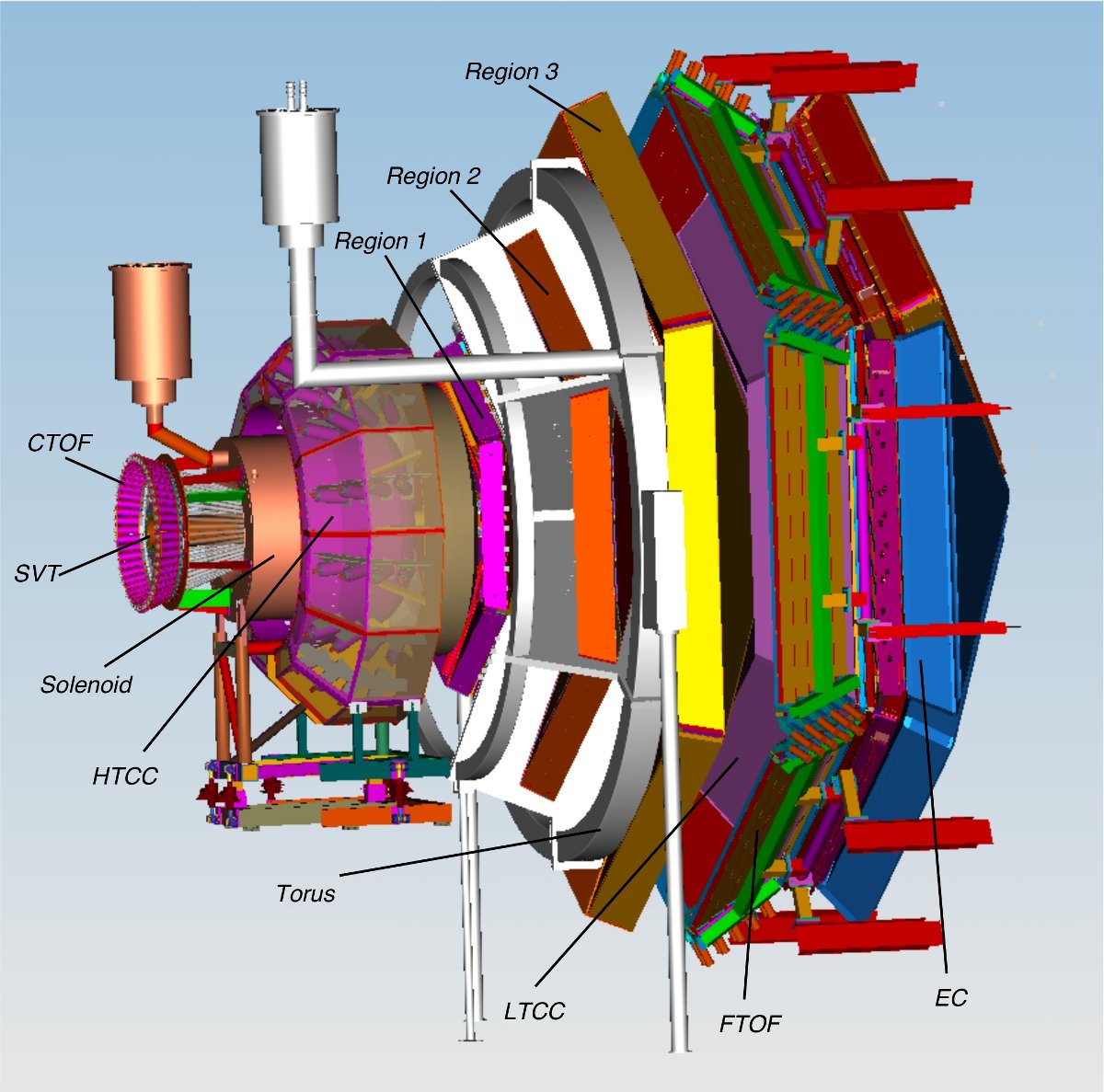}
    \caption{The schematic layout of the CLAS12 baseline design.}
    \label{fig:fd}
  \end{center}
\end{figure}

The CLAS12 detector is designed to operate with 11~GeV beam at an 
electron-nucleon luminosity of $\mathcal{L} = 
1\times10^{35}~$cm$^{-2}$s$^{-1}$. The baseline configuration of the CLAS12 
detector consists of the forward detector and the central detector 
packages~\cite{CD} (see Figure~\ref{fig:fd}). We use the forward detector
for electron detection in all ALERT run group proposals, while DVCS centered
proposals also use it for photon detection. The central
detector's silicon tracker and micromegas will be removed to leave room for
the recoil detector. 

The scattered electrons and photons will be detected in the forward detector which consists 
of the High Threshold Cherenkov Counters (HTCC), Drift Chambers (DC), the Low 
Threshold Cherenkov Counters (LTCC), the Time-of-Flight scintillators (TOF), 
the Forward Calorimeter and the Preshower Calorimeter. The charged particle 
identification in the forward detector is achieved by utilizing the combination 
of the HTCC, LTCC and TOF arrays with the tracking information from the Drift 
Chambers. The HTCC together with the Forward Calorimeter and the Preshower 
Calorimeter will provide a pion rejection factor of more than 2000 up to a 
momentum of 4.9~GeV/c, and a rejection factor of 100 above 4.9 GeV/c. The photons
are detected using the calorimeters.

\section{Available options for a Low Energy Recoil Detector}
We explored available solutions for the low-energy recoil tracker with 
adequate momentum and spatial resolution, and good particle identification for 
recoiling light nuclei (p, $^3$H and $^3$He). After investigating the 
feasibility of the proposed measurements using the CLAS12 Central Detector and 
the BONuS Detector~\cite{bonus6,bonus12}, we concluded that we needed to build 
a dedicated detector. We summarize in the
following the facts that led us to this conclusion.
\subsection{CLAS12 Central Detector}
The CLAS12 Central Detector~\cite{CD} is designed to detect various charged 
particles over a wide momentum and angular range. The main detector package 
includes:
\begin{itemize}
 \item Solenoid Magnet: provides a central longitudinal magnetic field up to 
5~Tesla, which serves to curl emitted low energy M{\o}ller electrons and determine 
particle momenta through tracking in the central detector.
 \item Central Tracker: consists of 3 double layers of silicon strips and 6 
    layers of Micromegas. The thickness of a single silicon layer is  
    \SI{320}{\um}.
 \item Central Time-of-Flight: an array of scintillator paddles with a 
cylindrical geometry of radius 26 cm and length 50 cm; the thickness of the 
detector is 2 cm with designed timing resolution of $\sigma_t = 50$ ps, used 
to separate pions and protons up to 1.2 GeV/$c$.
\end{itemize}

The current design, however, is not optimal for low energy particles 
($p<300$~MeV/$c$) due to the energy loss in the first 2 silicon strip layers. 
The momentum detection threshold is $\sim 200$ MeV/$c$ for protons, $\sim 
350$~MeV/$c$ for deuterons and even higher for $^3$H and $^3$He. These values 
are significantly too large for any of the ALERT run group proposals.

\subsection{BONuS12 Radial Time Projection Chamber}
The original BONuS detector was built for Hall B experiment E03-012 to study 
neutron structure at high $x_B$ by scattering electrons off an almost on-shell 
neutron inside deuteron. The purpose of the detector was to tag the low energy 
recoil protons ($p>60$ MeV/$c$). The key component for detecting the slow 
protons was the Radial Time Projection Chamber (RTPC) based on Gas Electron 
Multipliers (GEM). A later run period (eg6) used a 
newly built RTPC with a new design to detect recoiling $\alpha$ particles in 
coherent DVCS scattering. The major improvements of the eg6 RTPC were full 
cylindrical coverage and a higher data taking rate.

The approved 12~GeV BONuS (BONuS12) experiment is planning to use a similar 
device with some upgrades. The target gas cell length will be doubled, and the 
new RTPC will be longer as well, therefore doubling the luminosity and 
increasing the acceptance. Taking advantage of the larger bore ($\sim 700$ mm) of 
the 5~Tesla solenoid magnet, the maximum radial drift length will be increased 
from the present 3 cm to 4 cm, improving the momentum resolution by 
50\%~\cite{bonus12} and extending the momentum coverage. The main features of 
the proposed BONuS12 detector are summarized in Table~\ref{tab:comp}.

\begin{table}[tbp]
\bgroup
\def\arraystretch{1.1}%
\tabulinesep=1.2mm
\begin{tabu}{lcc}
\tabucline[2pt]{-}                                                   
\textbf{Detector Property}  & \textbf{RTPC}        & \textbf{ALERT}\\
\tabucline[1pt]{-}                                                   
Detection region radius & 4 cm                & 5 cm\\
Longitudinal length & $\sim$ 40 cm         & $\sim$ 30 cm \\
Gas mixture         & 80\% helium/20\% DME & 90\% helium/10\% isobutane \\
Azimuthal coverage  & 360$^{\circ}$               & 340$^{\circ}$\\
Momentum range      & 70-250 MeV/$c$ protons & 70-250 MeV/$c$ protons\\
Transverse mom. resolution & 10\% for 100~MeV/c protons & 10\% for 100~MeV/c protons\\
$z$ resolution & 3~mm & 3~mm \\
Solenoidal field    & $\sim 5$ T           & $\sim 5$ T \\
ID of all light nuclei & No                    & Yes \\
Luminosity      &$3\times10^{33}$ nucleon/cm$^{2}$\!/s & $6\times10^{34}$ nucleon/cm$^{2}$\!/s\\
Trigger             & can not be included  & can be included \\
\tabucline[1pt]{-}                                                   
\end{tabu}
\egroup
\caption{\label{tab:comp}Comparison between the RTPC (left column) and the new tracker (right column).}
\end{table}

\begin{figure}
  \begin{center}
    \includegraphics[angle=0, width=0.5\textwidth]{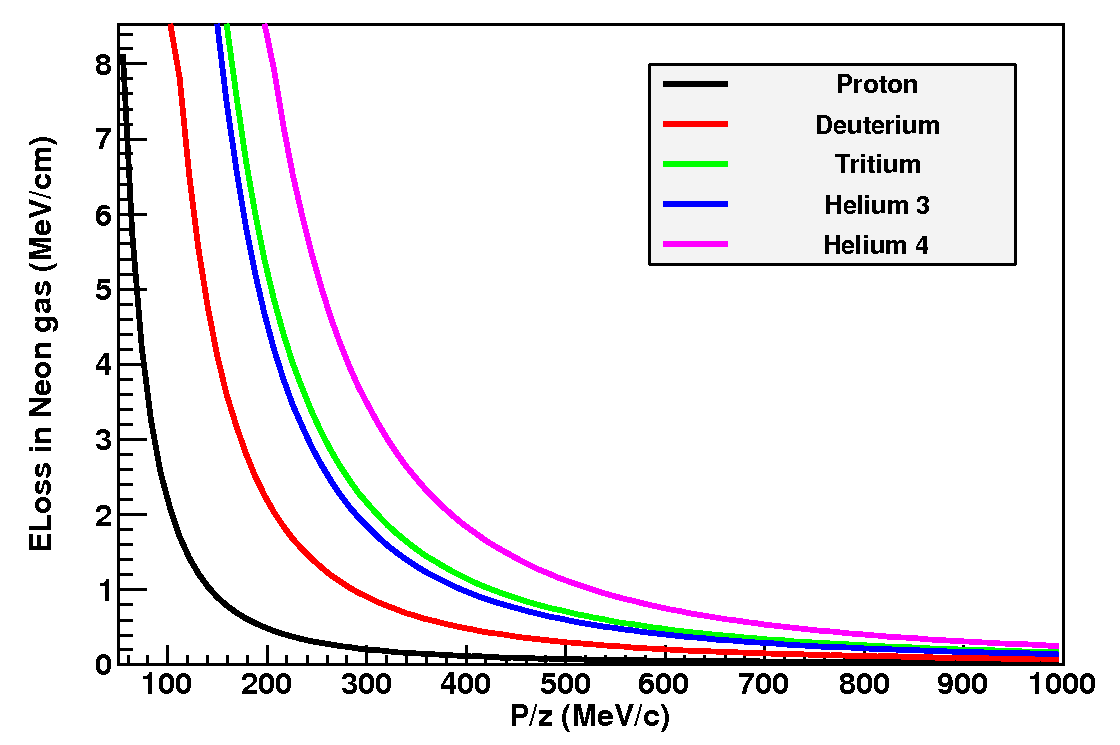}
    \caption{Calculation of energy loss in Neon gas as a function of the particle 
momentum divided by its charge for different nuclei. }
    \label{fig:eloss}
  \end{center}
\end{figure}

In principle, particle identification can be obtained from the RTPC through the 
energy loss $dE/dx$ in the detector as a function of the particle momentum (see 
Figure~\ref{fig:eloss}). However, with such a small difference between $^3$H and 
$^3$He, it is nearly impossible to discriminate between them
on an event by event basis because of the intrinsic width of the $dE/dx$ 
distributions. This feature is not problematic when using deuterium target, 
but makes the RTPC no longer a viable option for our tagged EMC and tagged DVCS 
measurements which require a $^4$He target and the differentiation of $^4$He, 
$^3$He, $^3$H, deuterons and protons.

Another issue with the RTPC is its slow response time due to a long drift 
time ($\sim5~\mu$s). If a fast recoil detector could be included in the trigger 
it would have a significant impact on the background rejection. Indeed, in
about 90\% of DIS events on deuteron or helium, the spectator fragments have too low energy 
or too small angle to get out of the target and be detected. By including
the recoil detector in the trigger, we would not be recording these events anymore.
Since the data acquisition speed was the main limiting factor for 
both BONuS and eg6 runs in CLAS, this 
would be a much needed reduction of the pressure on the DAQ.

\subsection{Summary}
In summary, we found that the threshold of the CLAS12 inner 
tracker is significantly too high to be used for our measurements. On the other hand, the 
recoil detector planned for BONuS12, a RTPC, is not suitable due to its 
inability to distinguish all kind of particles we need to measure.  
Moreover, as the RTPC cannot be efficiently included in the trigger, a lot of  
background events are sent to the readout electronics, which will cause its saturation and 
limit the maximum luminosity the detector can handle. Therefore, we propose 
a new detector design.

\section{Design of the ALERT Detector}
We propose to build a low energy recoil detector consisting of two sub-systems: 
a drift chamber and a scintillator hodoscope.
The drift chamber will be composed of 8 layers of sense wires to provide tracking 
information while the scintillators will provide particle 
identification through time-of-flight and energy measurements. To reduce the 
material budget, thus reducing the threshold to detect recoil particles 
at as low energy as possible, the scintillator 
hodoscope will be placed inside the gas chamber, just outside of the last 
layer of drift wires.

The drift chamber volume will be filled with a light gas mixture (90\% He and 
10\% C$_4$H$_{10}$) at atmospheric pressure. The amplification potential will
be kept low enough in order to not be sensitive to relativistic particles 
such as electrons and pions. Furthermore, a light 
gas mixture will increase the drift speed of the electrons from 
ionization. This will allow the chamber to withstand higher rates and 
experience lower hit occupancy. The fast signals from the chamber and 
the scintillators will be used in coincidence with electron trigger 
from CLAS12 to reduce the overall DAQ trigger rate and 
allow for operation at high luminosity.

\begin{figure}[tbp]
  \begin{center}
    \includegraphics[angle=0, width=0.75\textwidth]%
                    {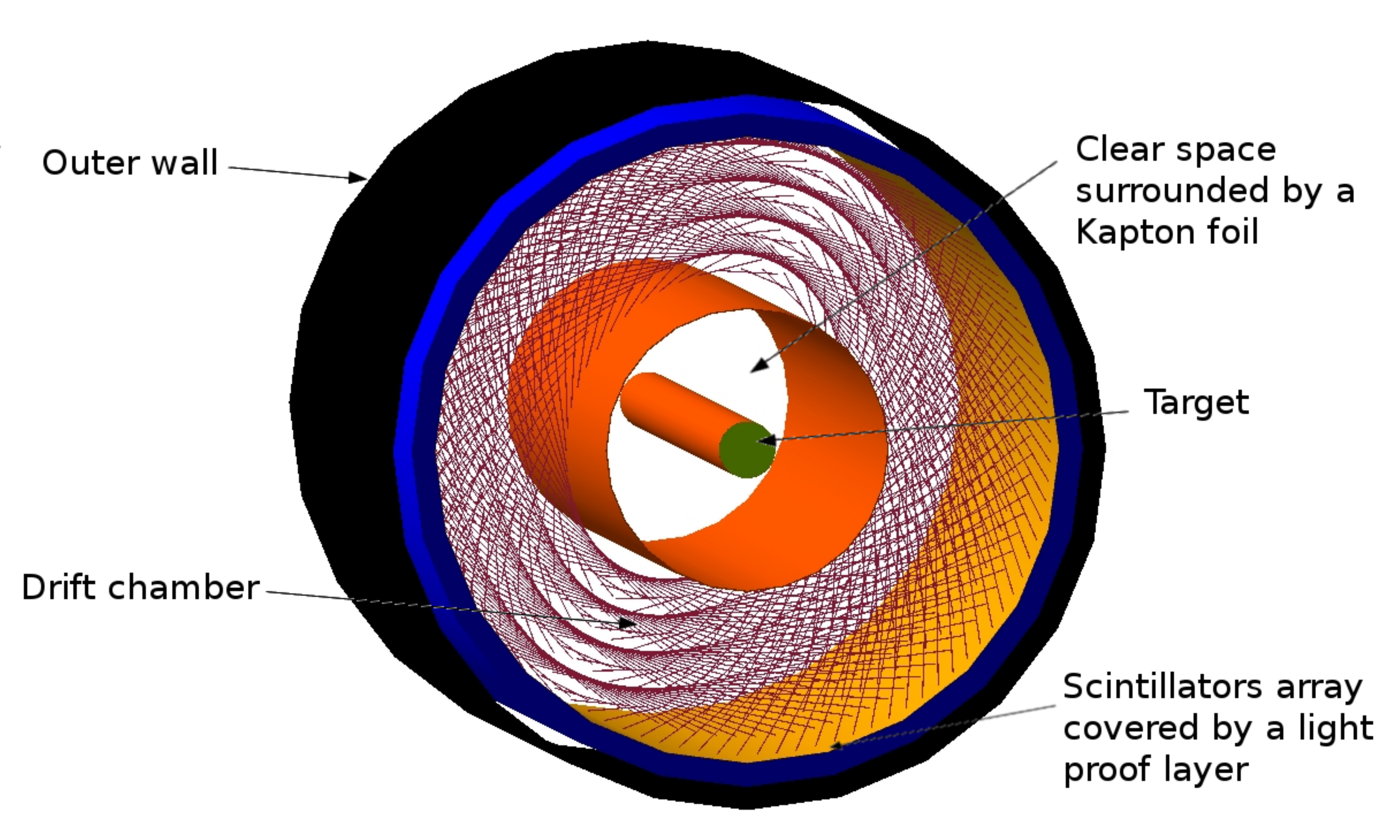}
    \caption{The schematic layout of the ALERT detector design, viewed 
from the beam direction.}
    \label{fig:new_lay}
  \end{center}
\end{figure}

The detector is designed to fit inside the central TOF of CLAS12; the 
silicon vertex tracker and the micromegas vertex tracker (MVT) will be 
removed. The available space has thus an outer radius of slightly more 
than 20~cm. A schematic 
layout of the preliminary design is shown in Figure~\ref{fig:new_lay} and its
characteristics compared to the RTPC design in Table~\ref{tab:comp}. The 
different detection elements are covering about $340^{\circ}$ of the polar 
angle to leave room for mechanics, and are 30~cm long with an effort made to 
reduce the particle energy loss through the materials. From the inside out,
it is composed of:
\begin{itemize}
\item a 30~cm long cylindrical target with an outer radius of 6~mm and target 
   walls \SI{25}{\um} Kapton filled with 3~atm of helium;
\item a clear space filled with helium to reduce secondary scattering from
   the high rate M\o{}ller electrons with an outer radius of 30~mm;
\item the drift chamber, its inner radius is 32~mm and its outer radius is 
85~mm;
\item two rings of plastic scintillators placed inside the gaseous chamber, 
   with total thickness of roughly 20~mm.
\end{itemize}

\subsection{The Drift Chamber}
While drift chambers are very useful to cover large areas at a moderate price, 
huge progress has been made in terms of their ability to withstand higher rates 
using better electronics, shorter distance between wires and optimization of 
the electric field over pressure ratio. Our design is based on other chambers 
developed recently. For example for the dimuon arm of ALICE at CERN, drift 
chambers with cathode planes were built in Orsay~\cite{AliceMuonArmChamber}. 
The gap between sense wires is 2.1~mm and the distance between two cathode 
planes is also 2.1~mm, the wires are stretched over about 1~m. Belle II is 
building a cylindrical drift chamber very similar to what is needed for this 
experiment and for which the space between wires is around 
2.5~mm~\cite{BelleIItdr}. Finally, a drift chamber with wire gaps of 1~mm is 
being built for the small wheel of ATLAS at CERN~\cite{ATLASChamber}. The 
cylindrical drift chamber proposed for our experiment is 300~mm long, and we 
therefore considered that a 2~mm gap between wires is technically a rather 
conservative goal. Optimization is envisioned based on experience with 
prototypes. 

The radial form of the detector does not allow for 90 degrees x-y wires in the 
chamber. Thus, the wires of each layer are at alternating angle of $\pm$ 
10$^{\circ}$, called the stereo-angle, from the axis of the drift chamber.  We 
use stereo-angles between wires to determine the coordinate along the beam 
axis ($z$). This setting makes it possible to use a thin forward end-plate to 
reduce multiple scattering of the outgoing high-energy electrons. A rough 
estimate of the tension due to the $\sim$2600 wires is under 600~kg, 
which appears to be reasonable for a composite end-plate. 

\begin{figure}
  \begin{center}
    \includegraphics[angle=0, width=0.7\textwidth,clip, trim=0mm 0mm 4cm 10cm]{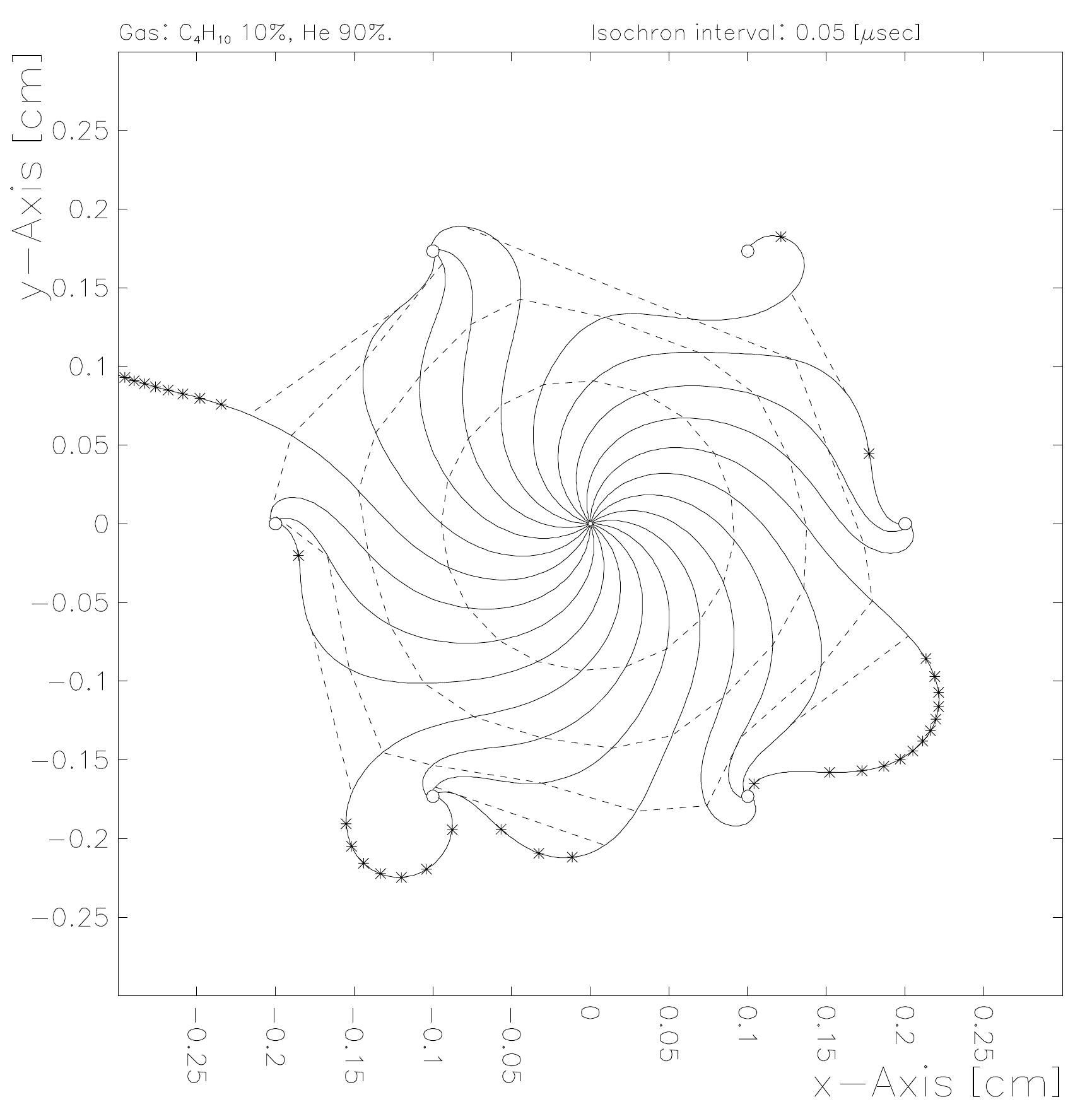}
    \caption{Drift lines simulated using MAGBOLTZ \cite{Magboltz} for one 
sense wire (at the center) surrounded by 6 field wires. The two electric field 
lines leaving the cell disappear when adjusting the voltages on the wires. 
Dashed lines are isochrones spaced by 50~ns. This shows that the maximum drift 
time is about 250~ns.}
    \label{fig:drift_cell}
  \end{center}
\end{figure}

The drift chamber cells are composed of one sense wire made of gold plated 
tungsten surrounded by field wires, however the presence of the 5~T magnetic 
field complicates the field lines. Several cell configurations have been studied with 
MAGBOLTZ~\cite{Magboltz}, we decided to choose a conservative 
configuration as shown in Figure~\ref{fig:drift_cell}. The sense wire is 
surrounded by 6 field wires placed equidistantly from it in a hexagonal 
pattern. The distance between the sense and field wires is constant and equal 
to 2~mm. Two adjacent cells share the field wires placed between them. The 
current design will have 8 layers of cells of similar radius. 
The simulation code MAGBOLTZ is calculating the drift speed and drift paths of 
the electrons (Figure~\ref{fig:drift_cell}). With a moderate electric field, the 
drift speed is around 10~microns/ns, the average drift time expected is thus 
250~ns (over 2~mm). Assuming a conservative 10~ns time resolution, the spatial 
resolution is expected to be around 200~microns due to field distortions and 
spread of the signal.

\begin{figure}
  \begin{center}
    \includegraphics[angle=0, width=0.5\textwidth, trim=5mm 5mm 5mm 15mm, 
    clip]{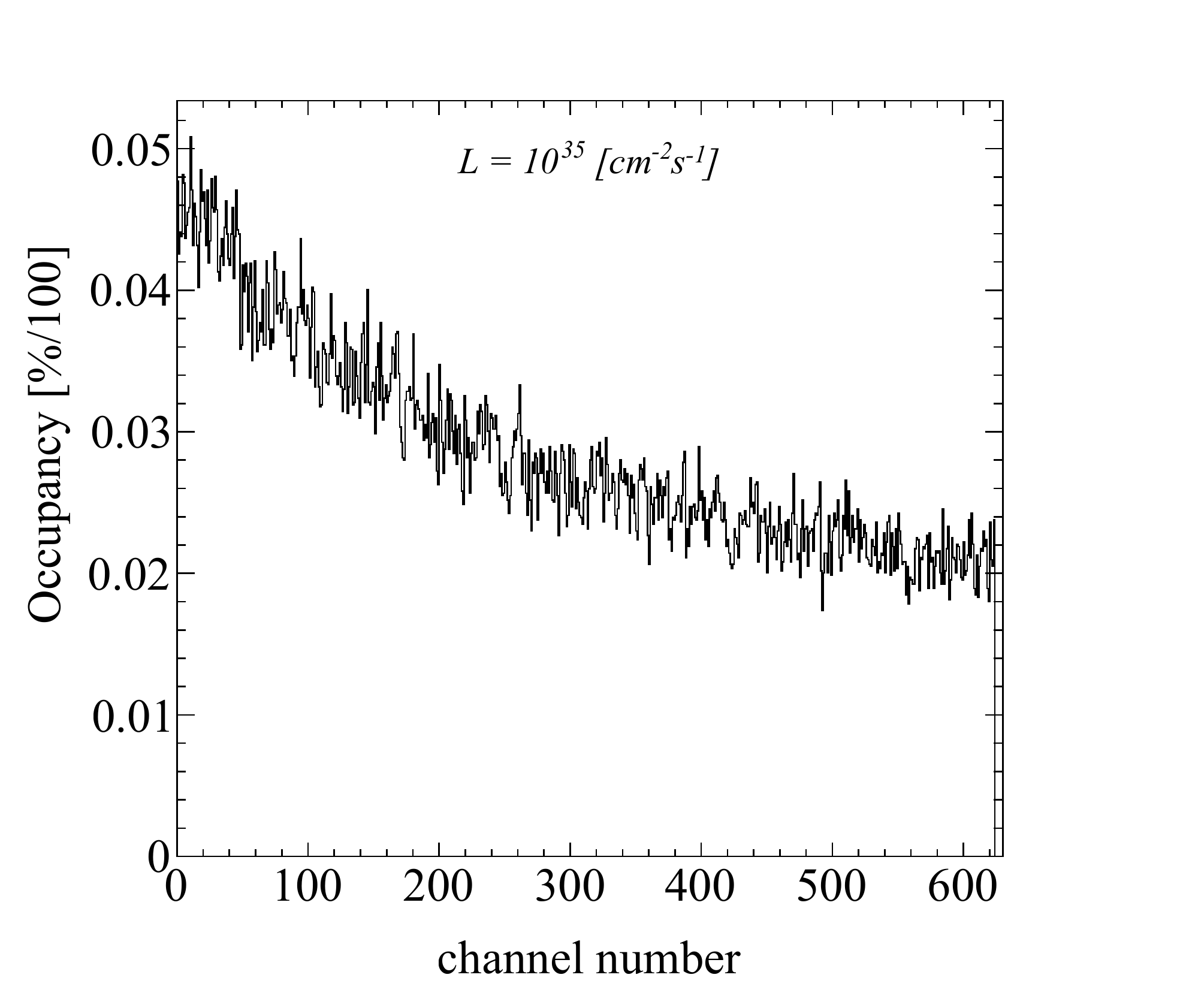}
    \caption{\label{fig:RCoccupancy}A full Geant4 simulation of the ALERT drift 
       chamber hit occupancy
       at a luminosity of $10^{35}$ cm$^{-2}$s$^{-1}$. The channel numbering 
    starts with the inner most wires and works outwards.}
  \end{center}
\end{figure}

\begin{figure}
  \begin{center}
    \includegraphics[angle=0, width=0.7\textwidth, trim=5mm 5mm 5mm 10mm, 
    clip]{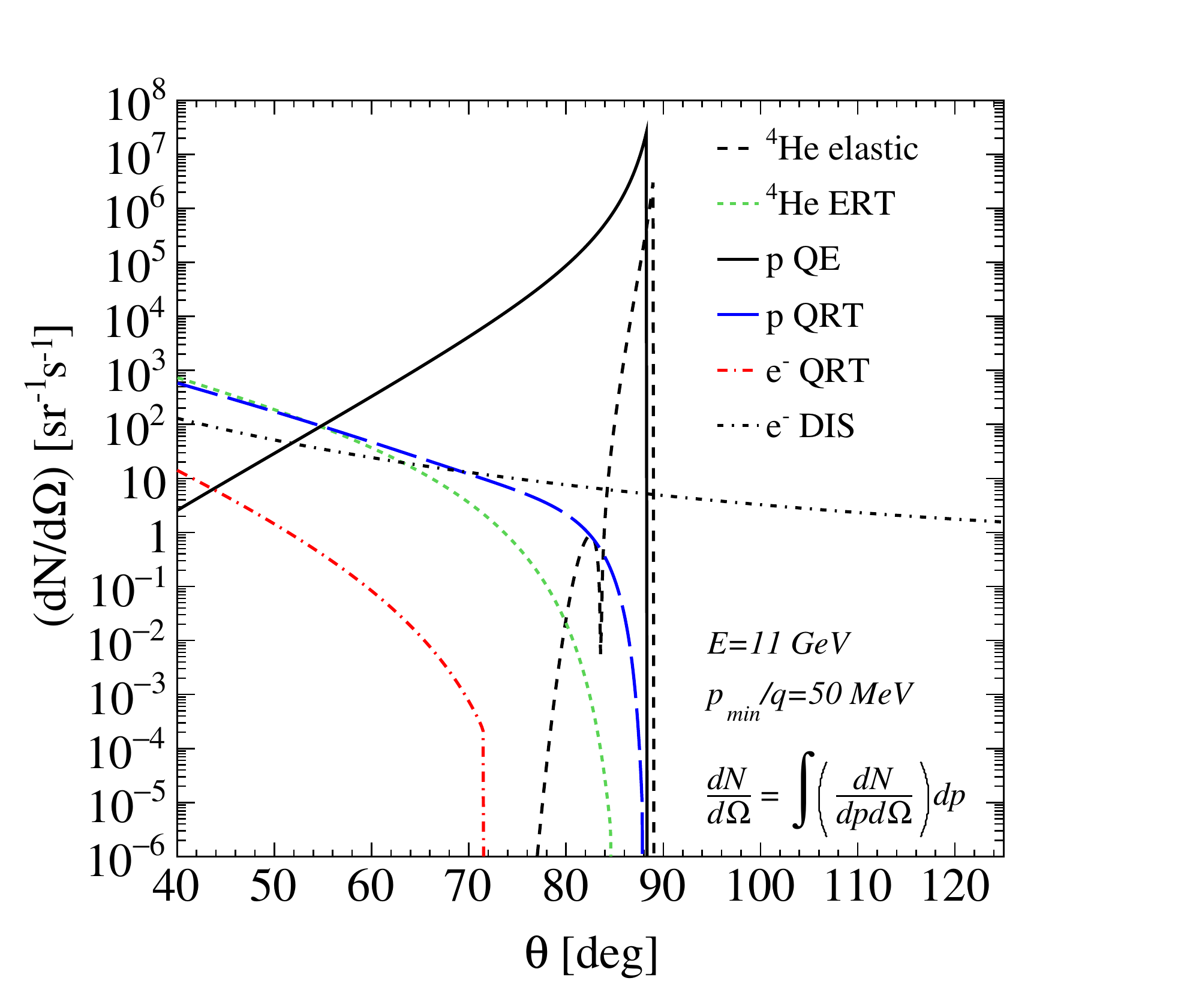}
    \caption{\label{fig:ALERTrates}The rates for different processes as 
    function of angle. The quasi-elastic radiative tails (QRT), $^4$He elastic 
 radiative tail (ERT), and DIS contributions have been integrated over momenta 
 starting at $p/q$ = 50~MeV/c, where $q$ is the electric charge of the particle 
 detected.}
  \end{center}
\end{figure}

The maximum occupancy, shown in Figure~\ref{fig:RCoccupancy},
is expected to be around 5\% for the inner most wires at $10^{35}$~cm$^{-2}$s$^{-1}$
(including the target windows). This is the maximum available luminosity for the 
baseline CLAS12 and is obtained based on the physics channels depicted 
in Figure~\ref{fig:ALERTrates}, assuming an integration time of 200~ns and 
considering a readout wire separation of 4~mm. This amount of accidental hits 
does not appear to be reasonable for a good tracking quality, we therefore 
decided to run only at half this luminosity for our main production runs. This 
will keep occupancy below 3\%, which is a reasonable amount for a drift chamber 
to maintain high tracking efficiency. When running the coherent processes with 
the $^4$He target, it is not necessary to detect the protons\footnote{This 
   running condition is specific to the proposal ``Partonic Structure of Light 
Nuclei'' in the ALERT run group.}, so the rate of accidental hits can then be 
highly reduced by increasing the detection threshold, thus making the chamber 
blind to the protons\footnote{The CLAS {\it eg6} run period was using the RTPC in 
the same fashion.}. In this configuration, considering that our main 
contribution to occupancy are quasi-elastic protons, we are confident that the 
ALERT can work properly at $10^{35}$~cm$^{-2}$s$^{-1}$.

We are currently planning to use the electronics used by the MVT of CLAS12, 
known as the DREAM chip \cite{7097517}. Its dynamic range and time resolution 
correspond to the needs of our drift chamber. To ensure that it is the 
case, tests with a prototype will be performed at the IPN Orsay (see 
section~\ref{sec:proto}).

\subsection{The Scintillator Array} \label{sec:scint}
The scintillator array will serve two main purposes. First, it will provide a 
useful complementary trigger signal because of its very fast response time, 
which will reduce the random background triggers. Second, it will provide 
particle identification, primarily through a time-of-flight measurement, but 
also by a measurement of the particle total energy deposited and path length in 
the scintillator which is important for doubly charged ions.

The length of the scintillators cannot exceed roughly 40~cm to keep the time 
resolution below 150~ps. It must also be segmented to match with tracks 
reconstructed in the drift chamber. Since $^3$He and $^4$He will travel at 
most a few mm in the scintillator for the highest anticipated momenta 
($\sim$~400~MeV/c), a multi-layer scintillator design provides an extra handle on 
particle identification by checking if the range exceeded the thickness of 
the first scintillator layer.

The initial scintillator design consists of a thin (2~mm) inner layer of 60 
bars, 30~cm in length, and 600 segmented outer scintillators (10 segments 
3~cm long for each inner bar) wrapped around the drift chamber. Each of these 
thin inner bars has SiPM\footnote{SiPM: silicon photomultiplier.} detectors 
attached to both ends. A thicker outer layer (18~mm) will be further segmented 
along the beam axis to provide position information and maintain good time 
resolution.

\begin{figure}
  \begin{center}
    \includegraphics[width=0.48\textwidth]{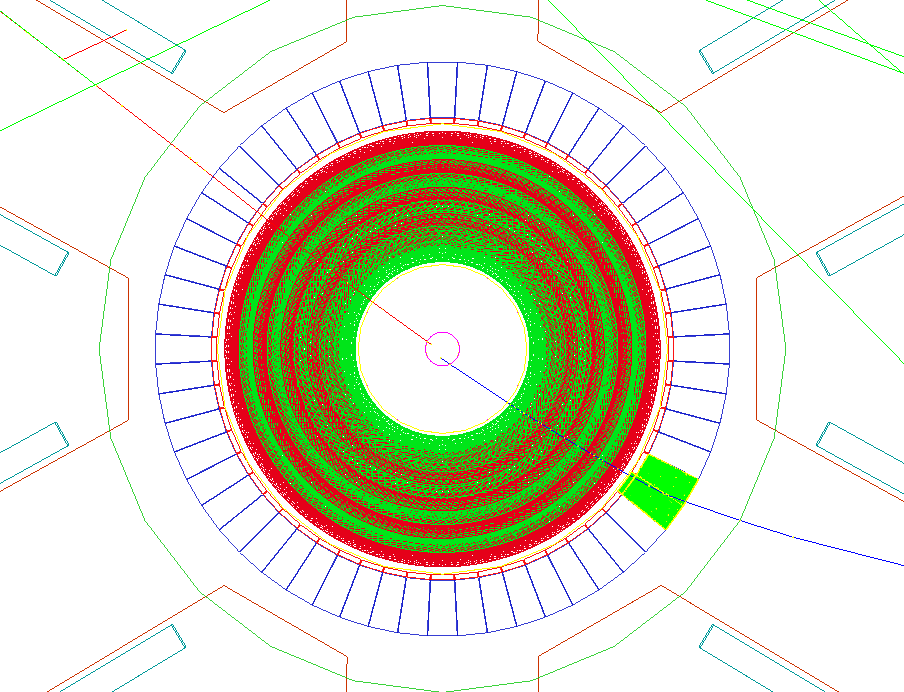}
    \includegraphics[width=0.48\textwidth]{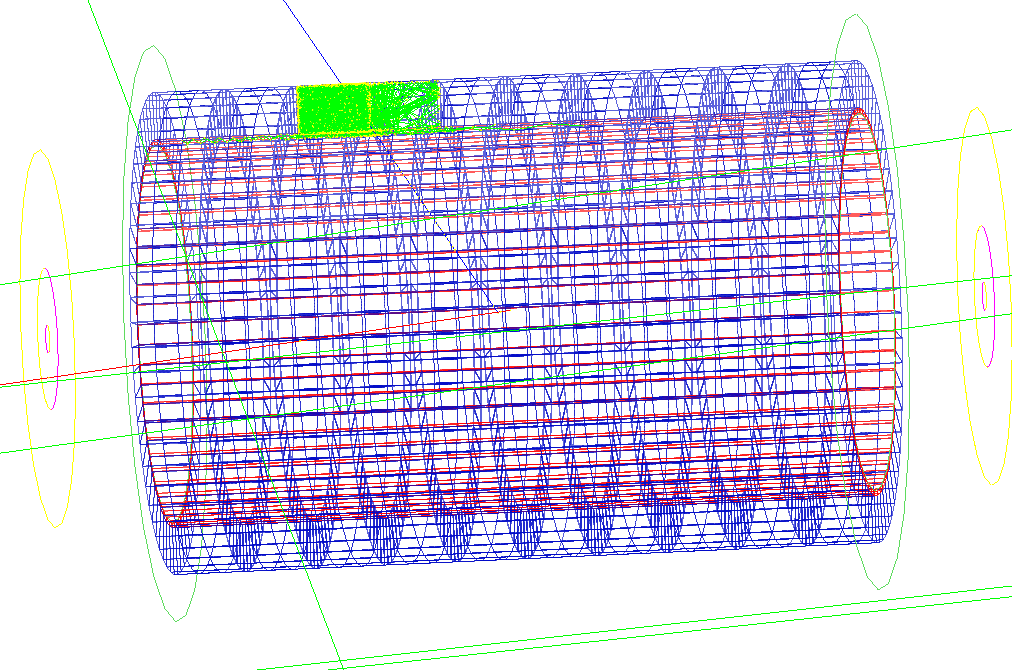}
    \caption{\label{fig:scintHodoscopeDesign}Geant4 simulation of a proton 
    passing through the recoil drift chamber and scintillator hodoscope. The 
 view looking downstream (left) shows the drift chamber's eight alternating 
 layers  of wires (green and red) surrounded by the two layers of scintillator 
 (red and blue). Simulating a proton through the detector, photons (green) are 
 produced in a few scintillators. On the right figure, the dark blue rings are graphical feature showing the contact between the adjacent outer scintillators.}
  \end{center}
\end{figure}

For the outer layer, a dual ended bar design and a tile design with embedded 
wavelength shifting fiber readouts similar to the forward tagger's hodoscope for 
CLAS12~\cite{FThodo} were considered. After simulating these designs, it was 
found that the time resolution was insufficient except only for the smallest 
of tile designs (15$\times$15$\times$7~mm$^3$). Instead of using fibers, a 
SiPM will be mounted directly on the outer layer of a keystone shaped 
scintillator that is 30~mm in length and 18~mm thick. This design can be seen 
in Figure~\ref{fig:scintHodoscopeDesign} which shows a full Geant4 simulation of 
the drift chamber and scintillators. By directly mounting the SiPMs to the 
scintillator we collect the maximum signal in the shortest amount of time.  
With the large number of photons we expect, the time resolution of SiPMs will 
be a few tens of ps, which is well within our target.

\begin{figure}
  \begin{center}
    \includegraphics[angle=0, 
    width=0.48\textwidth]{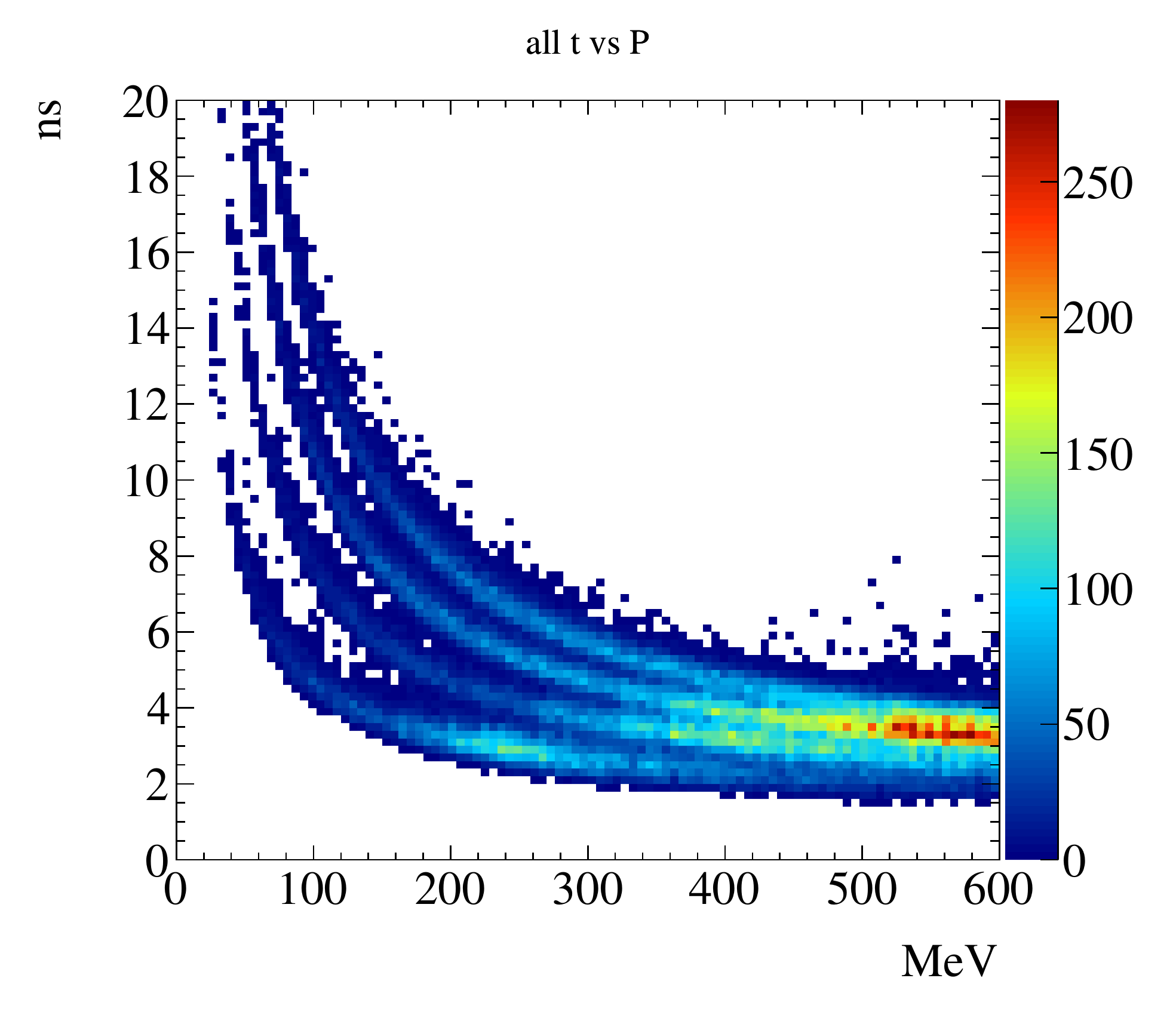}
    \includegraphics[angle=0, 
    width=0.48\textwidth]{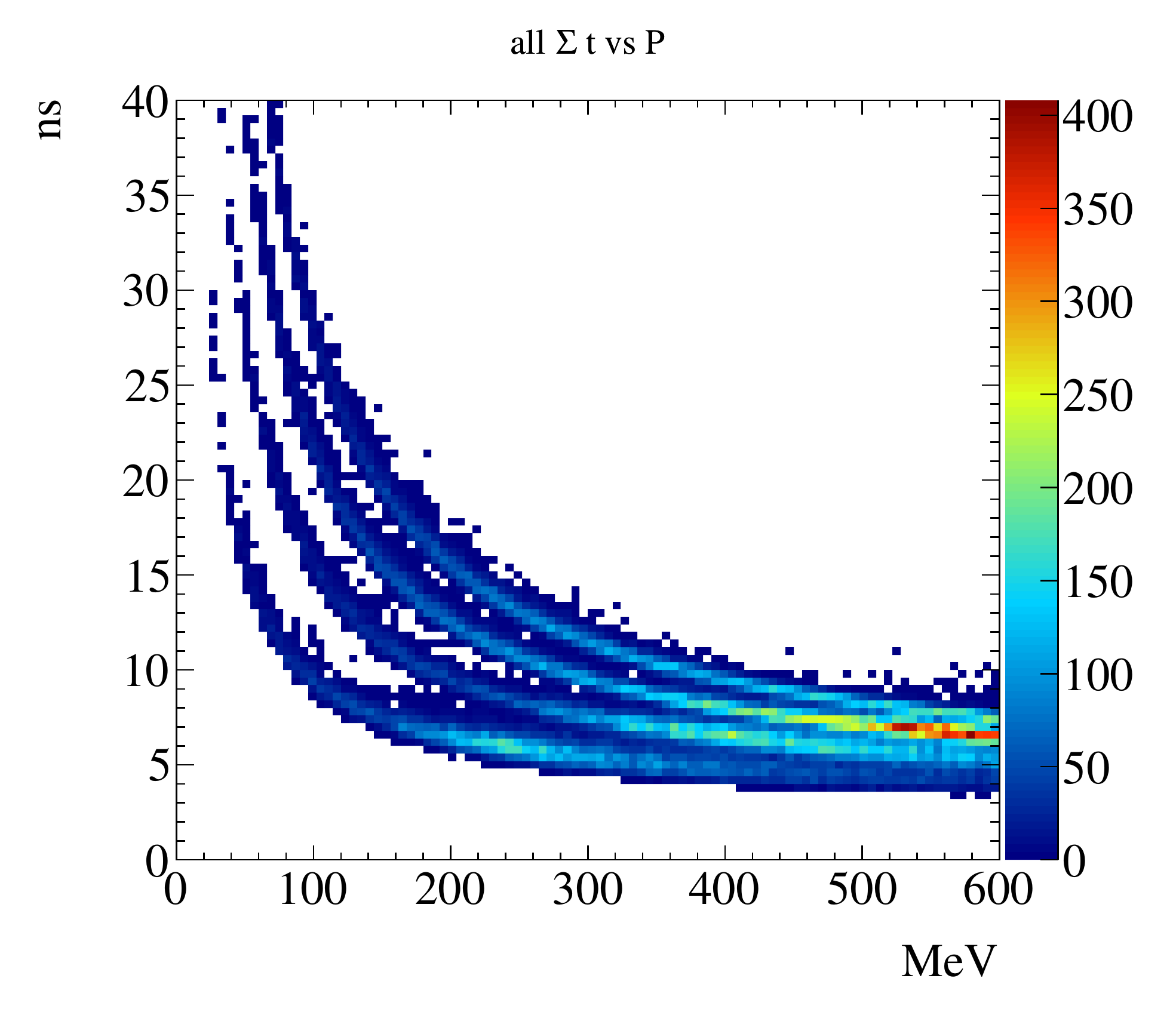}
    \caption{
       \label{fig:scintTimeVsP}Simulated TOF for the various recoil particles 
    vs Momentum. The TOF from just a single readout is shown on the left and 
 the sum of the dual ended readout is shown on the right.   }
  \end{center}
\end{figure}

The advantage of a dual ended readout is that the time sum is proportional to 
the TOF plus a constant. The improved separation of different particles can 
be seen in Figure~\ref{fig:scintTimeVsP}. Reconstructing the position of a hit 
along the length of a bar in the first layer is important for the doubly 
charged ions because they will not penetrate deep enough to reach the second 
layer of segmented scintillator.

The front-end electrons for the SiPMs will include preamplifiers and ASICs\footnote{ASIC: application-specific integrated circuit.}
which provide both TDC and ADC readouts. The PETIROC-2A\cite{PETIROC} ASIC 
provides excellent time resolution ($18$~ps on trigger output with 4 
photoelectrons detected) and a maximum readout rate at about 40k events/s.
Higher readout rates can be handled by using external digitizers by using the 
analog mode of operation and increase this rate by an order of magnitude. The 
ASIC also has the advantage of being able to tune the individual over-bias 
voltages with an 8-bit DAC.

The expected radiation damage to the SiPMs and scintillator material is found 
to be minimal over the length of the proposed experiment. We used the CLAS12 
forward tagger hodoscope technical design report~\cite{FThodo} as a very 
conservative baseline for this 
comparison. We arrived at an estimated dose of 1 krad after about 4.5 months of 
running. The damage to the scintillator at 100 times these radiation levels  
would not be problematic, even for the longest lengths of scintillator 
used~\cite{Zorn:1992ew}.
Accumulated dose on the SiPMs leads to an increased dark current. Similarly
than for scintillators, we do not expect it to be significant over the length of the 
experiment. The interested reader is referred to the work on
SiPMs for the Hall-D detectors~\cite{Qiang:2012zh,Qiang:2013uwa}. A front-end 
electronics prototype will be tested for radiation hardness but we expect  any 
damage to negligible~\cite{commPETIROC}.


\subsection{Target Cell}\label{sec:targetCell}

\begin{figure}
  \begin{center}
    \includegraphics[angle=0, trim={0 0 15cm 0}, clip,
    width=0.99\textwidth]{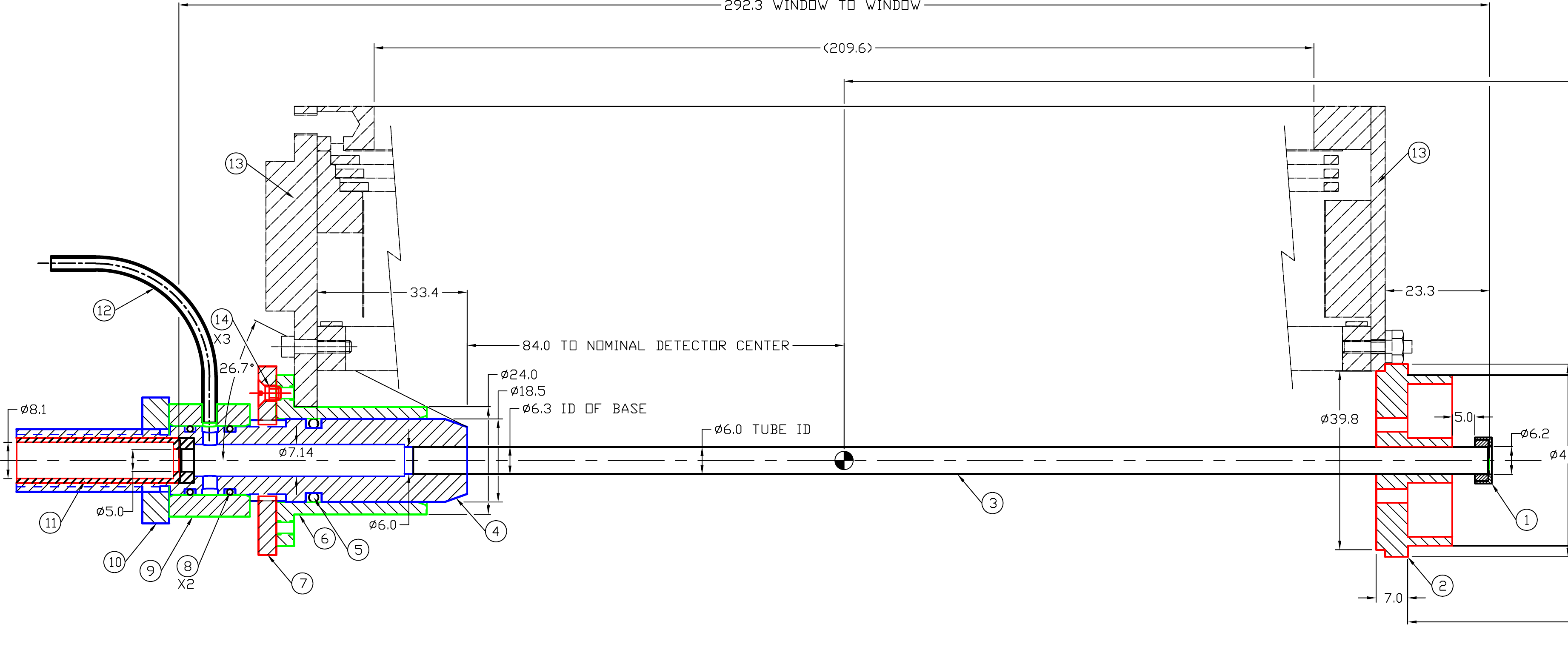}
    \caption{ \label{fig:eg6TargetDrawing}The eg6 target design drawing.}
  \end{center}
\end{figure}

The design of the proposed ALERT target will be very similar to the eg6 target shown 
in Figure~\ref{fig:eg6TargetDrawing}.
The target parameters are shown in Table~\ref{tab:target} 
with the parameters of other existing and PAC approved targets.
Note that, the proposed target has an increased radius of 6~mm compared to all the 
others which have 3~mm radius. This increase compared to the previous CLAS targets has been made
in order to compensate for the expected increase of beam size at 11~GeV. The BONuS12
target is still presently proposed to be 3~mm in radius, if such a target
is operated successfully in JLab, we will definitely consider using a
smaller radius as well, but we prefer to propose here a safer option that we know
will work fine.

\begin{table}
\centering
\caption{Comparison of various straw targets used at JLab.The 
"JLab test targets" correspond
to recent tests performed in JLab for the BONuS12 target, they have
been tested for pressure but have never been tested with beam.
}
\newcolumntype Y{S [ group-four-digits=true,
round-mode=places,
round-precision=1,
round-integer-to-decimal=true,
per-mode=symbol ,
detect-all]}
\tabucolumn Y
\label{tab:target}
\bgroup
\def\arraystretch{1.2}%
\tabulinesep=1mm
\begin{tabu}{l C{1.5cm}C{3.0cm}C{2cm}}
\tabucline[2pt]{-}
\textbf{Experiment} & \textbf{Length} & \centering \textbf{Kapton wall thickness} &
\textbf{Pressure} \\ \tabucline[1pt]{-}
CLAS target (eg6)           & 30~cm & \SI{27}{\um} & 6.0~atm   \\
BONuS12 (E12-06-113) target & 42~cm & \SI{30}{\um} & 7.5~atm \\
JLab test target 1          & 42~cm & \SI{30}{\um} & 3.0~atm   \\
JLab test target 2          & 42~cm & \SI{50}{\um} & 4.5~atm \\
JLab test target 3          & 42~cm & \SI{60}{\um} & 6.0~atm   \\
ALERT proposed target       & 35~cm & \SI{25}{\um} & 3.0~atm   \\
\tabucline[2pt]{-}
\end{tabu}
\egroup
\end{table}

%
\section{Simulation of ALERT and reconstruction} \label{sec:sim}
The general detection and reconstruction scheme for ALERT is as follows. We fit 
the track with the drift chamber and scintillator position information to
obtain the momentum over the charge. Next, using the 
scintillator time-of-flight, the particles are separated and identified by 
their mass-to-charge ratio, therefore leaving a degeneracy for the deuteron and 
$\alpha$ particles.
The degeneracy between deuteron and $\alpha$ particles can be resolved in a few 
ways.  The first and most simple way is to observe that an $\alpha$ will almost 
never make it to the second layer of scintillators and therefore the absence (presence) of a 
signal would indicate the particle is an $\alpha$~(deuteron). Furthermore, as 
will be discussed below, the measured dE/dx will differ for $^4$He and $^2$H, 
therefore, taking into account energy loss in track fitting alone can provide 
separation. Additionally taking further advantage of the measured total energy 
deposited in the scintillators can help separate the $\alpha$s and deuterons.

\subsection{Simulation of ALERT}
\begin{figure}[tbp]
    \begin{center}
        \includegraphics[width=0.45\textwidth]{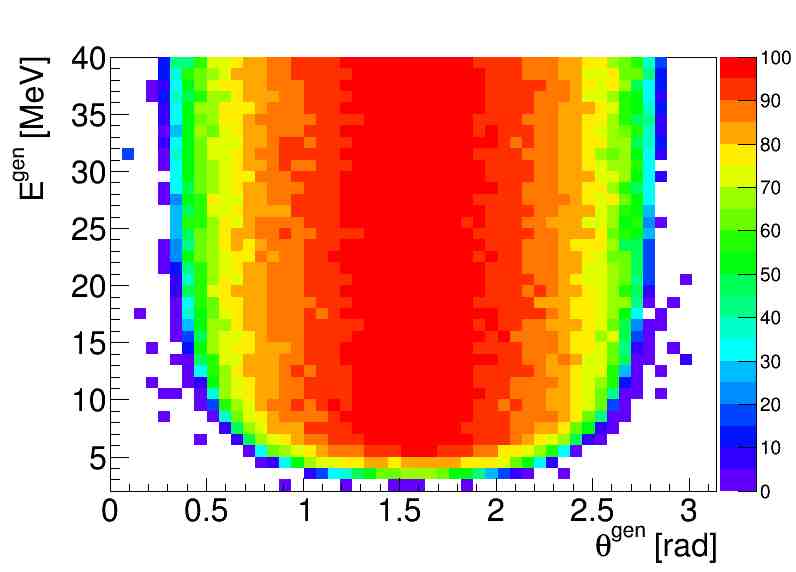}
        \includegraphics[width=0.45\textwidth]{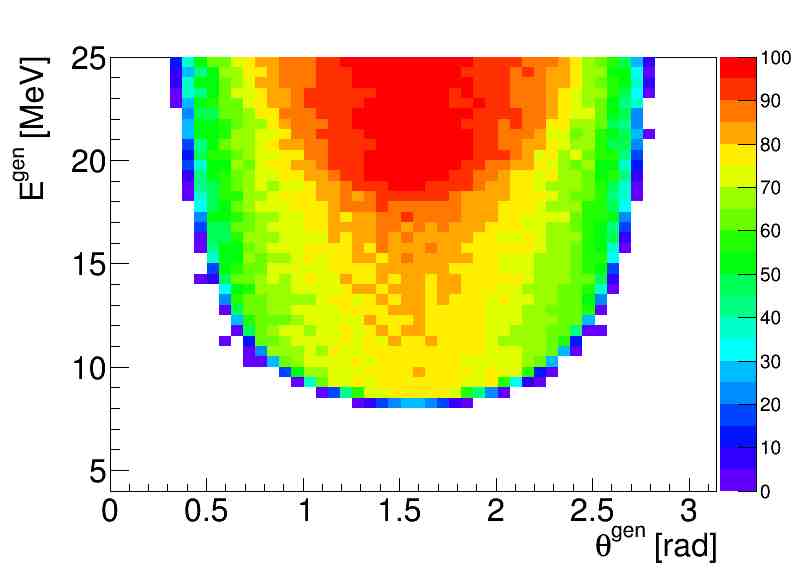}
        \caption{Simulated recoil detector acceptance percentage, for protons (left) and 
$^4$He (right), when requiring energy deposition in the scintillators arrays. 
\label{fig:acceptance}}
    \end{center}
\end{figure}

The simulation of the recoil detector has been implemented with the full 
geometry and material specifications in GEANT4. It includes a 5~Tesla homogeneous 
solenoid field and the entire detector filled with materials as described in the 
previous section. In this study all recoil species are generated with the same 
distributions: flat in momentum from threshold up to 40~MeV 
($\sim$~250~MeV/c) for protons and about 25~MeV for other particles; isotropic 
angular coverage; flat distribution in $z$-vertex; and a radial vertex 
coordinate smeared around the beam line center by a Gaussian distribution of 
sigma equal to the expected beam radius (0.2 mm).
For reconstruction, we require that the particle reaches the scintillator
and obtain the acceptance averaged over the $z$-vertex position shown in 
Figure~\ref{fig:acceptance}.

\subsection{Track Fitting}
\begin{figure}[tbp]
    \begin{center}
        \includegraphics[height=4.5cm, width=0.32\textwidth]{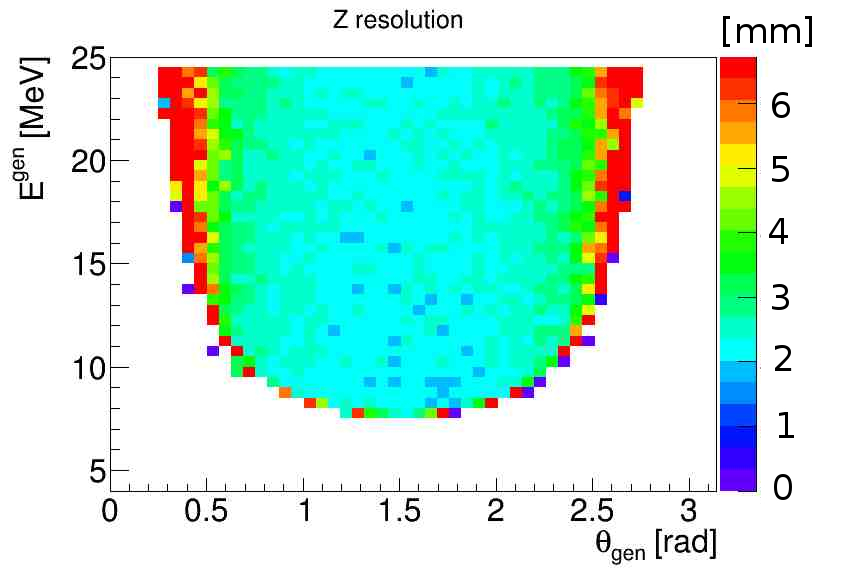}
        \includegraphics[height=4.5cm, width=0.32\textwidth]{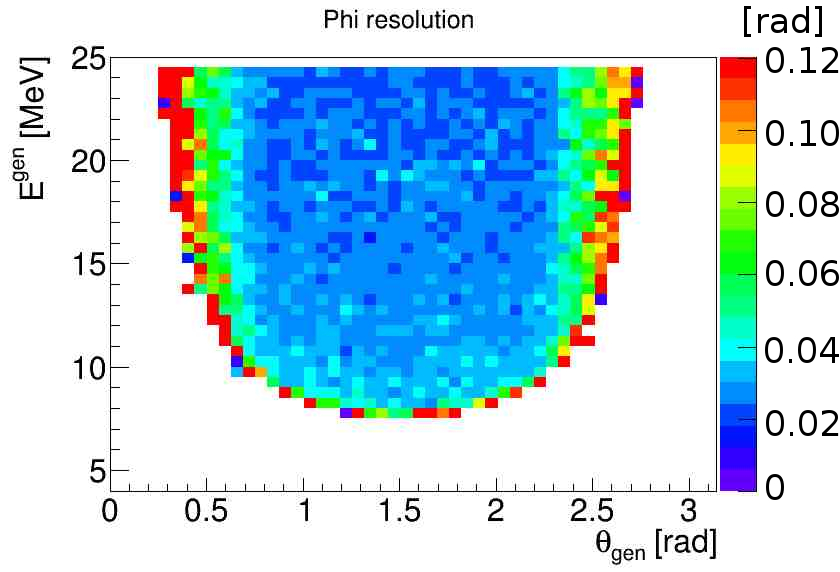}
        \includegraphics[height=4.5cm, width=0.32\textwidth]{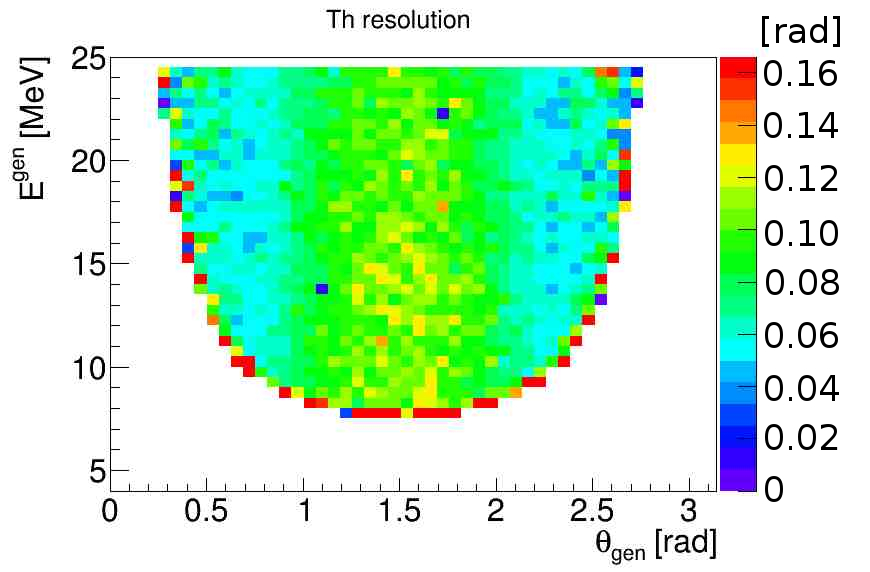}
        \caption{Resolutions for simulated $^4$He:  $z$-vertex resolution in mm (left), azimuthal (center) 
          and polar (right) angle resolutions in radians for the lowest energy
          regime when the recoil track reaches the scintillator.\label{fig:tracking}}
    \end{center}
\end{figure}

\begin{figure}[tbp]
    \begin{center}
        \includegraphics[width=0.45\textwidth]{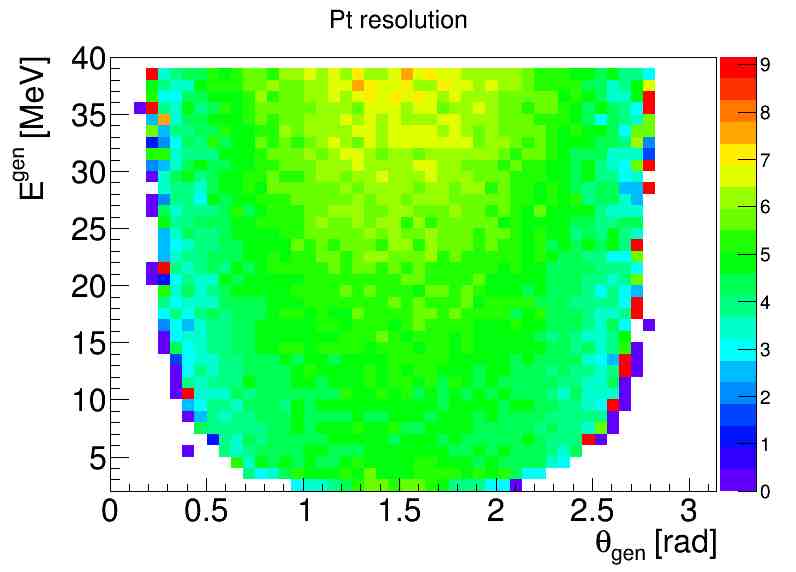}
        \includegraphics[width=0.45\textwidth]{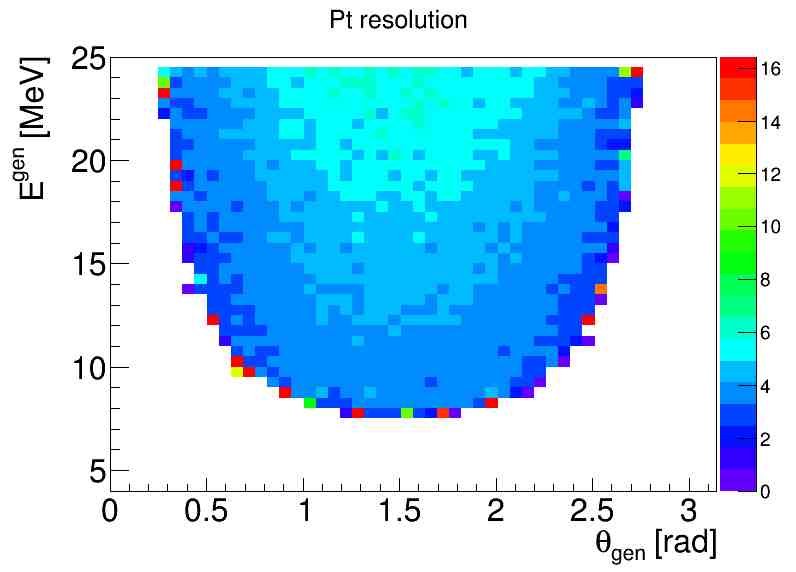}
        \caption{Simulated momentum resolutions (in \%) as a function of energy and 
                 polar angle for protons (left) and $^4$He 
                 (right) integrated over all $z$, when the recoil track reaches the scintillators 
                 array. \label{fig:presolution}}
    \end{center}
\end{figure}

The tracks are obtained using a helix fitter giving the coordinates of 
the vertex and the momentum of the particle. The energy deposited in 
the scintillators could also be used to help determine the kinetic energy of the 
nucleus, but is not implemented in the studies we performed here. 
The tracking capabilities of the recoil detector are investigated 
assuming a spatial resolutions of \SI{200}{\um} for the drift chamber. The wires 
are strung in the $z$-direction with a stereo angle of \ang{10}. The resulting difference between 
generated and reconstructed variables from simulation is shown in 
Figure~\ref{fig:tracking} for $^4$He particles. The momentum resolution for both protons and 
$^4$He is presented in Figure~\ref{fig:presolution}.


\subsection{Particle identification in ALERT}

\begin{figure}[tbp]
    \begin{center}
        \includegraphics[width=0.7\textwidth]{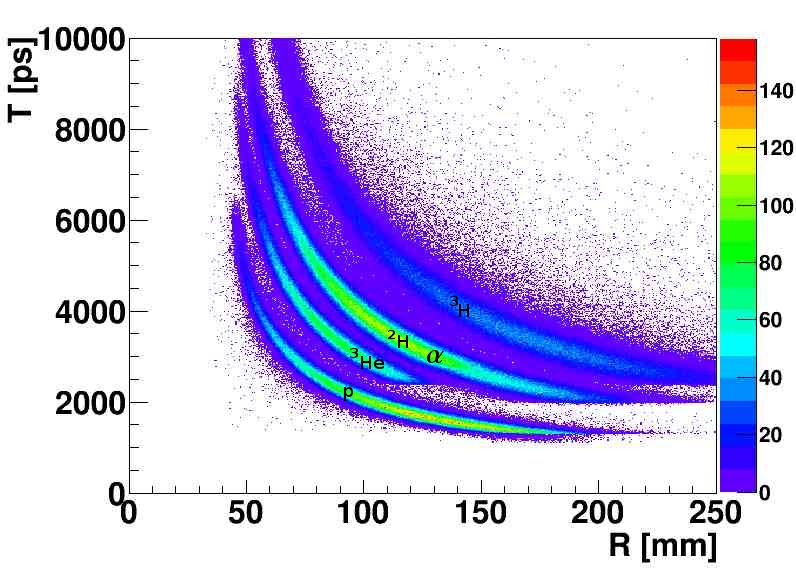}
        \caption{Simulated time of flight at the scintillator versus the 
reconstructed radius in the drift chamber. The bottom band corresponds to 
the proton, next band is the $^3$He nuclei, $^2$H and $\alpha$ are overlapping in 
the third band, the uppermost band is $^3$H\label{fig:SIMtof}. $^2$H and 
$\alpha$ are separated using dE/dx.}
    \end{center}
\end{figure}

The particle identification scheme is investigated using the GEANT4
simulation as well. The scintillators 
have been designed to ensure a 150~ps time resolution. To determine the dE/dx 
resolution, measurements will be necessary for the scintillators and for the 
drift chamber as this depends on the detector layout, gas mixture, 
electronics, voltages... Nevertheless, from \cite{Emi}, one can assume that 
with 8 hits in the drift chamber and the measurements in the 
scintillators, the energy resolution should be at least 10\%.
Under these conditions, a clean separation of three of the five nuclei is shown 
in Figure~\ref{fig:SIMtof} solely based on the time of flight measured by the 
scintillator compared to the reconstructed momentum from the drift chamber. 
We then separate $^2$H and $\alpha$ using dE/dx in the drift chamber and in the 
scintillators.

To quantify the separation power of our device, we simulated an equal quantity 
of each species. We obtained a particle identification efficiency of 99\% for 
protons, 95\% for $^3$He and 98\% for $^3$H and around 90\% for $^2$H and 
$\alpha$ with equally excellent rejection factors. It is important to note that 
for this analysis, only the energy deposited in the scintillators was used, not 
the energy deposited in the
drift chamber nor the path length in the scintillators, thus these numbers
are very likely to be improved when using the full information\footnote{The 
uncertainty remains important about the resolutions that will be achieved 
for these extra information. So we deemed more reasonable to ignore them
for now.}. This analysis indicates that the proposed reconstruction 
and particle identification schemes for this design are quite promising.  
Studies, using both simulation software and prototyping, are ongoing to 
determine the optimal detector parameters to minimize the detection threshold 
while maximizing particle identification efficiency. The resolutions presented 
above have been implemented in a fast Monte-Carlo used to evaluate their impact 
on our measurements.

\section{Drift chamber prototype}
\label{sec:proto}
Since the design of the drift chamber presents several challenges in term of
mechanical assembly, we decided to start prototyping early. The goal is to find a 
design that will be easy to install and to maintain if need be, while keeping the 
amount of material at a minimum. This section presents the work done in Orsay 
to address the main questions concerning the mechanics that needed to be answered:
\begin{itemize}
\item How to build a stereo drift chamber with a 2~mm gap between wires?
\item Can we have frames that can be quickly changed in case of a broken wire?
\item How to minimize the forward structure to reduce the multiple scattering,
while keeping it rigid enough to support the tension due to the wires?
\end{itemize}

\begin{figure}[tbp]
    \begin{center}
        \includegraphics[width=0.4\textwidth]{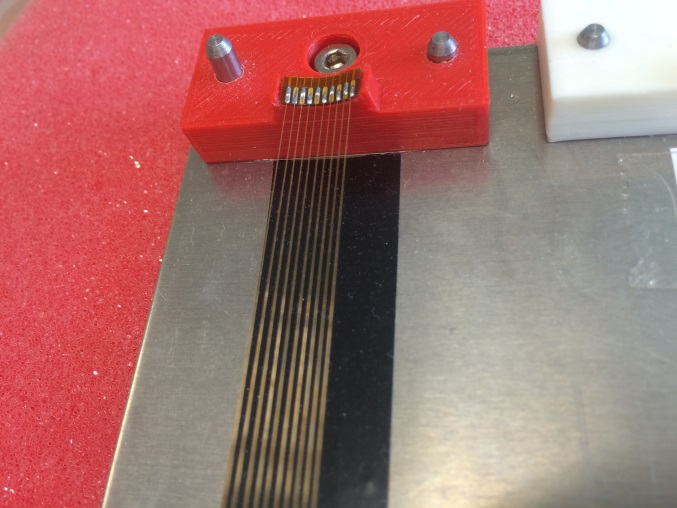}
        \caption{Welded wires on a curved structure with a 2~mm gap between each wire.}
        \label{soldOK}
    \end{center}
\end{figure}

For the first question, small plastic structures realized with a 3D printer 
were tested and wires welded on it, as shown in Figure \ref{soldOK}. This 
demonstrated our ability to weld wires with a 2~mm gap on a curved structure.
 
\begin{figure}[tbp]
    \begin{center}
        \includegraphics[width=0.40\textwidth]{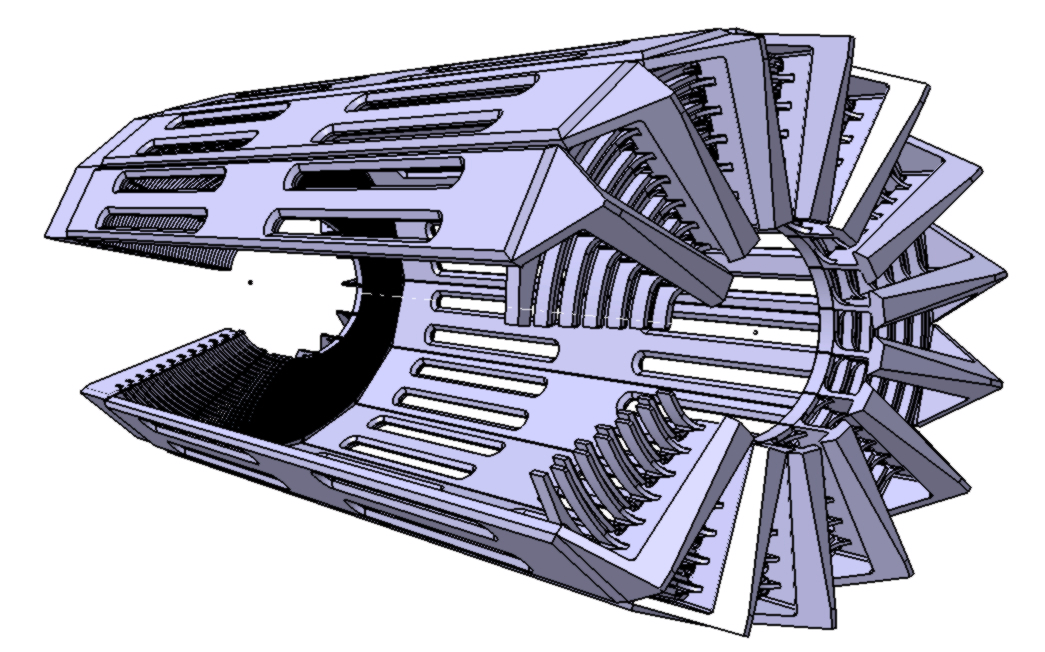}
        \includegraphics[width=0.40\textwidth]{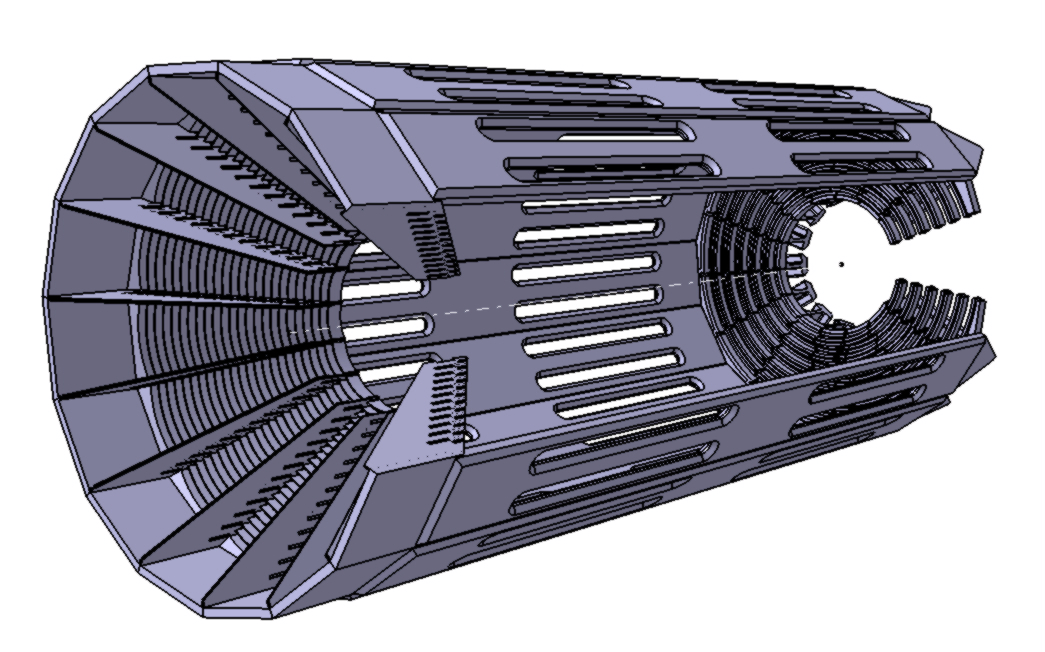}
        \caption{Upstream (left) and downstream (right) ends of the prototype 
        detector in computer assisted design (CAD) with all the sectors included.  \label{wholeView}}
    \end{center}
\end{figure}

\begin{figure}[tbp]
    \begin{center}
        \includegraphics[width=0.4\textwidth]{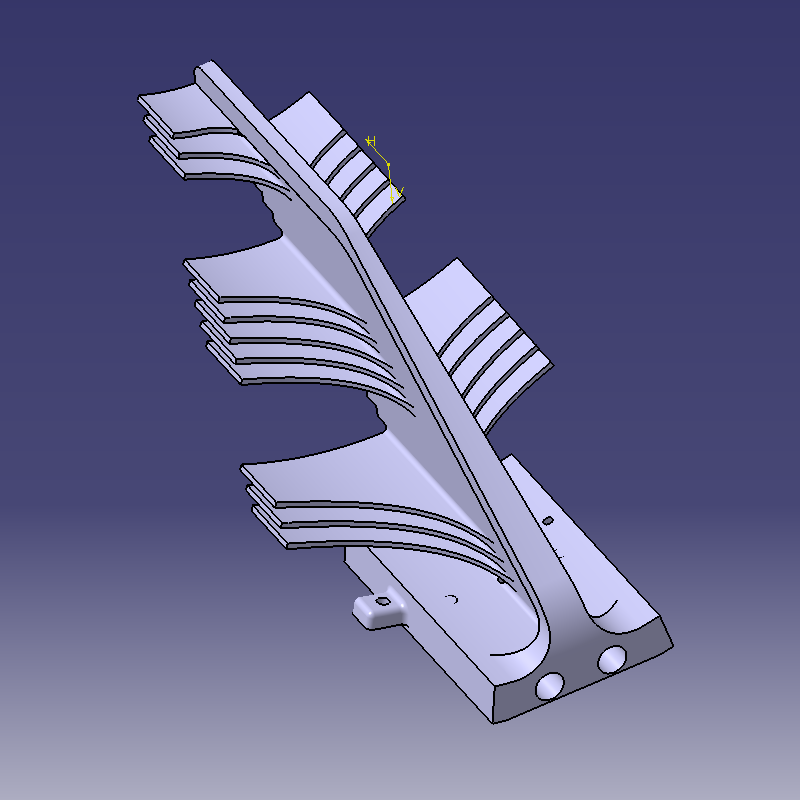}
        \includegraphics[width=0.4\textwidth]{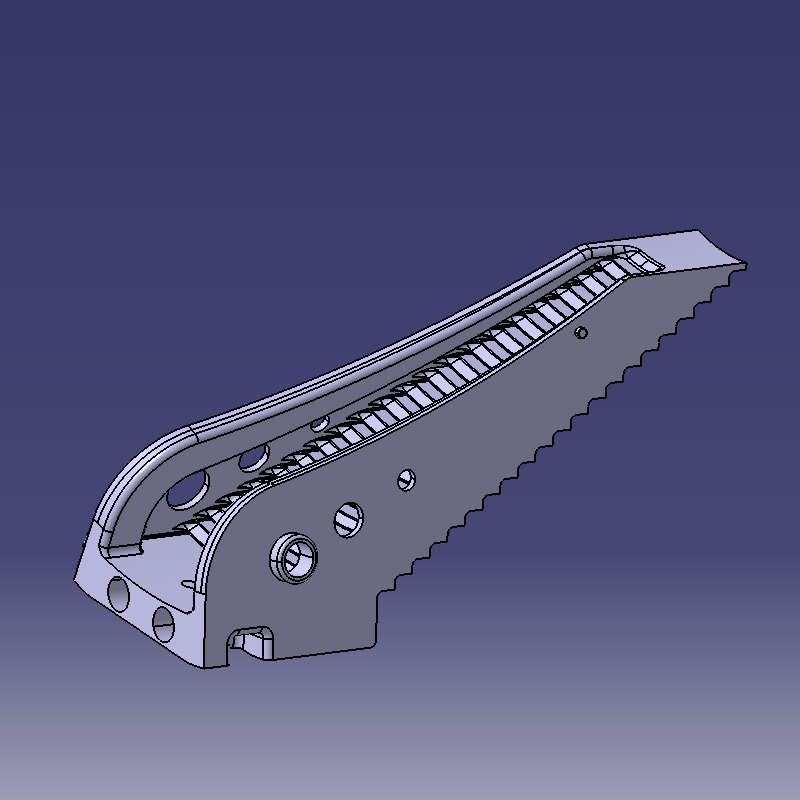}
        \caption{Close up on the CAD of the upstream piece (left) and downstream 
        piece(right) of the drift chamber. Note that the design of the pieces has been
        optimized in comparison of what is shown in Figure~\ref{wholeView}.}
        \label{fig:CAD}
    \end{center}
\end{figure}

To limit issues related to broken wires, we opted for a modular detector made of 
identical sectors. Each sector covers 20$^{\circ}$ of the azimuthal angle 
(Figure~\ref{wholeView}) and can be rotated around the beam axis to be separated 
from the other sectors. This rotation is possible due to the absence of one 
sector, leaving a 20$^{\circ}$ dead angle. Then, if a wire breaks, its sector 
can be removed independently and replaced by a spare. Plastic and metallic 
prototype sectors were made with 3D printers to test the assembling procedure and 
we have started the construction of a full size prototype of one sector.
The shape of each sector is constrained by the position of the wires. It has 
a triangular shape on one side and due to the stereo angle, the other side 
looks like a pine tree with branches alternatively going left and right from 
a central trunk (Figure~\ref{fig:CAD}).

\begin{figure}[tbp]
    \begin{center}
        \includegraphics[width=0.7\textwidth]{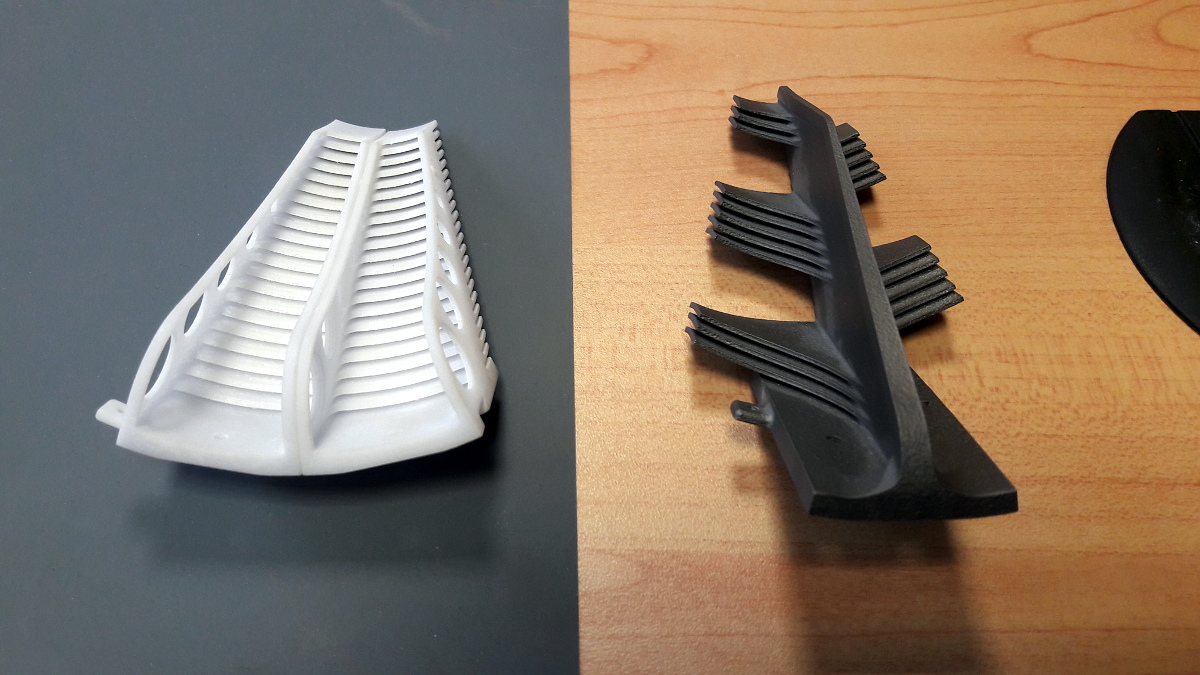}
        \caption{Prototypes for the mechanical parts of the drift chamber made out of plastic
        for the forward part and titanium for the backward.}
        \label{OneSector}
    \end{center}
\end{figure}

Finally, the material used to build the structure will be studied in details with 
future prototypes. Nevertheless, most recent plans are to use high rigidity plastic
in the forward region and metal for the backward structure (as in Figure~\ref{OneSector}). The 
prototypes are not only designed to check the mechanical requirements summarized above 
but also to verify the different cell configurations, and to test the DREAM 
electronics (time resolution, active range, noise). 

\section{Technical contributions from the research groups}
The effort to design, build and integrate the ALERT detector is led by four 
research groups, Argonne National Lab (ANL), 
Institut de Physique Nucl\'eaire d'Orsay (IPNO), Jefferson Lab and Temple 
University (TU). 

Jefferson Lab is the host institution. ANL, IPNO and TU have 
all contributed technically to CLAS12. ANL was involved in the construction of 
the high-threshold Cherenkov counters (HTCC) for CLAS12. ANL has a memorandum 
of understanding (MOU) with JLab on taking responsibility for the HTCC light 
collection system including testing the photomultipliers and the magnetic 
shielding. For the RICH detector for CLAS12, ANL developed full GEANT-4 
simulations in addition to the tracking software. ANL also developed the 
mechanical design of the detector support elements and entrance and exit 
windows in addition to the front-end electronics cooling system. IPNO took 
full responsibility for the design and construction of CLAS12 neutron detector 
(CND). The CND was successfully delivered to Jefferson Lab. TU played an 
important role in the refurbishment of the low threshold Cherenkov counters 
(LTCC), which was completed recently. All 216 photomultipliers have been coated 
with wavelength shifting material (p-Terphenyl) at Temple University, which 
resulted in a significant increase in the number of photoelectrons response.

The three institutions have already shown strong technical commitment to JLab 
12~GeV upgrade, with a focus on CLAS12 and this proposal is a continuation of 
this commitment.

\subsection{Argonne National Laboratory and Temple University}
The ANL medium energy group is responsible for the ALERT scintillator system, 
including scintillation material, light collection device and electronics. 
First results of simulations have led to the design proposed here. This work 
will continue to integrate the scintillator system with the wire chamber. ANL 
will collaborate closely with Temple University to test the light detection 
system. Both institutions will be responsible to assemble and test the detector.

Argonne will provide the electronics and technical support required to
integrate the scintillator detector system into the CLAS12 DAQ. The effort
will minimize the effort required on the part of the Hall B staff.

\subsection{Institut de Physique Nucl\'eaire d'Orsay}
The Institut de Physique Nucl\'eaire d'Orsay is responsible for the wire
chamber and the mechanical structure of the detector design and construction. 
As shown in the proposal, this work
has already started, a first prototype is being built 
to test different cell forms, wire material, wire thickness, 
pressure, etc. This experience will lead to a complete design of the ALERT detector 
integrating the scintillator built at ANL, the gas distribution system and the
electronic connections.

In partnership with {\it CEA Saclay}, IPN Orsay will also test the
use of the DREAM front-end chip for the wire chamber. Preliminary tests were
successful and will continue. The integration of the chip with CLAS12 is
expected to be done by the {\it CEA Saclay}, since they use the same chip to 
readout the CLAS12 MVT. Adaptations to the DAQ necessary when the MVT will be 
replaced by ALERT will be performed by the staff of IPN Orsay.

\subsection{Jefferson Laboratory}\label{sec:jlabContributions}
We expect Jefferson Lab to help with the configuration of the beam line.  
This will include the following items.

\paragraph{Beam Dump Upgrade}
The maximum beam current will be around 1000~nA for the production runs at 
$10^{35}$~cm$^{-2}$s$^{-1}$, which is not common for Hall-B.
To run above 500 nA the ``beam blocker'' will need to be upgraded to handle 
higher power. The beam blocker attenuates the beam seen by the Faraday cup.   
This blocker is constructed of copper and is water cooled. Hall B staff have 
indicated that this is a rather straightforward engineering task and has no 
significant associated costs~\cite{beamBlocker}. 

\paragraph{Straw Target}
We also expect JLab to design and build the target for the experiment as it 
will be a very similar target as the ones build for CLAS BONuS and eg6 runs.
See section \ref{sec:targetCell} for more details. 

\paragraph{Mechanical Integration}
We also expect Jefferson Laboratory to provide assistance in the detector 
installation in the Hall. This will include providing designers at ANL and IPNO 
with the technical drawings required to integrate ALERT with CLAS12. We will 
also need some coordination between designers to validate the mechanical 
integration. 

\paragraph{CLAS12 DAQ Integration}
We also will need assistance in connecting the electronics of ALERT to the 
CLAS12 data acquisition and trigger systems. This will also include help 
integrating the slow controls into the EPICs system.

%% file: Experiment.tex
\setlength\parskip{\baselineskip}%
\chapter{Proposed Measurements}
\label{chap:reach}

We propose to measure the beam spin asymmetry for three DVCS channels using two 
different targets and with tagged spectator systems. The three principal 
reactions are:
\begin{itemize}
   \item $^4$He$(\vec{e},e^{\prime}\,\gamma\,^3\text{H}p)$ -- bound p-DVCS
   \item $^4$H$(\vec{e},e^{\prime}\,\gamma\,^3\text{He})n$ -- bound n-DVCS
   \item $^2$H$(\vec{e},e^{\prime}\,\gamma\,p)n$ -- quasi-free n-DVCS
\end{itemize}
where in the first process the final state is fully detected. Before discussing 
the details of the measurements, we present an overview of the procedure  for 
extracting the $\sin\phi$ harmonic of the BSAs and identify the primary 
deliverables of the experiment.
\section{Asymmetry Extraction Procedure}
\begin{figure}
   \centering
   \includegraphics[width=1.0\textwidth,trim=0mm 0mm 0mm 0mm, 
   clip]{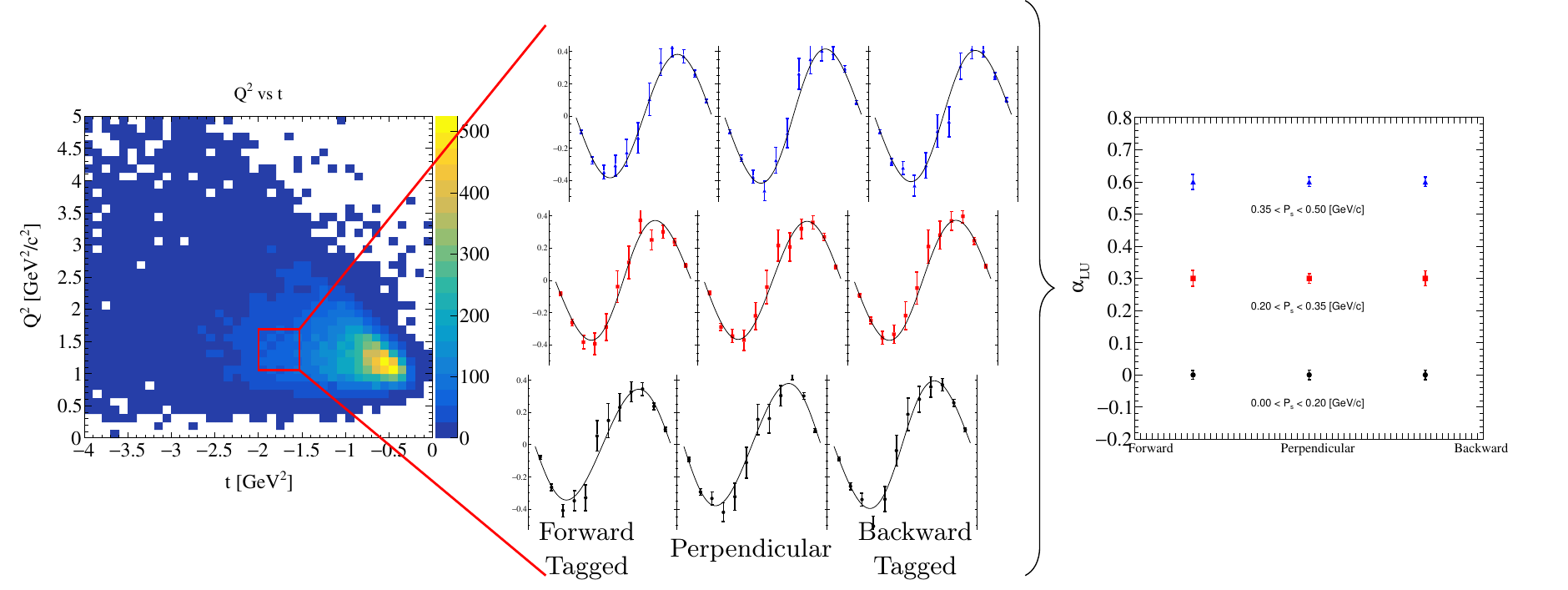}
   \caption{\label{fig:BSAExtraction}A general overview of the BSA extraction 
      procedure for a kinematic bin. For each ($x$, $Q^2$, $t$) bin the BSA is 
      extracted through fitting the asymmetry as function of $\phi$ (see 
      Equation~\ref{eq:ALUfit}) for the various spectator momentum configurations 
      and the $\sin\phi$ harmonic, $\alpha$ is extracted. Note these plots are only 
   showing statistical uncertainties and the values are offset for clarity.}
\end{figure}
Figure~\ref{fig:BSAExtraction} shows how, starting with just one kinematic bin in 
$t$, $Q^2$, and $x$ (which is not explicitly shown), the BSA is extracted for 
three regions of spectator recoil angles relative to the virtual photon direction and three ranges of spectator momenta.  
As indicated, the spectator angles correspond to a forward tagged system, a 
system with perpendicular momenta, and a backward tagged spectator.  In the 
latter angular region FSIs are expected to minimal. Furthermore, three ranges 
of momenta are identified, the lowest corresponding to nucleons moving in the 
mean field  and the highest belonging to nucleons in short range correlated 
pairs.
Fitting the BSA asymmetries yields the $\sin\phi$ harmonic which is shown on 
the right of Figure~\ref{fig:BSAExtraction} for the different spectator kinematic 
regions.

\begin{figure}
   \centering
   \includegraphics[width=1.0\textwidth,trim=0mm 0mm 0mm 0mm, 
   clip]{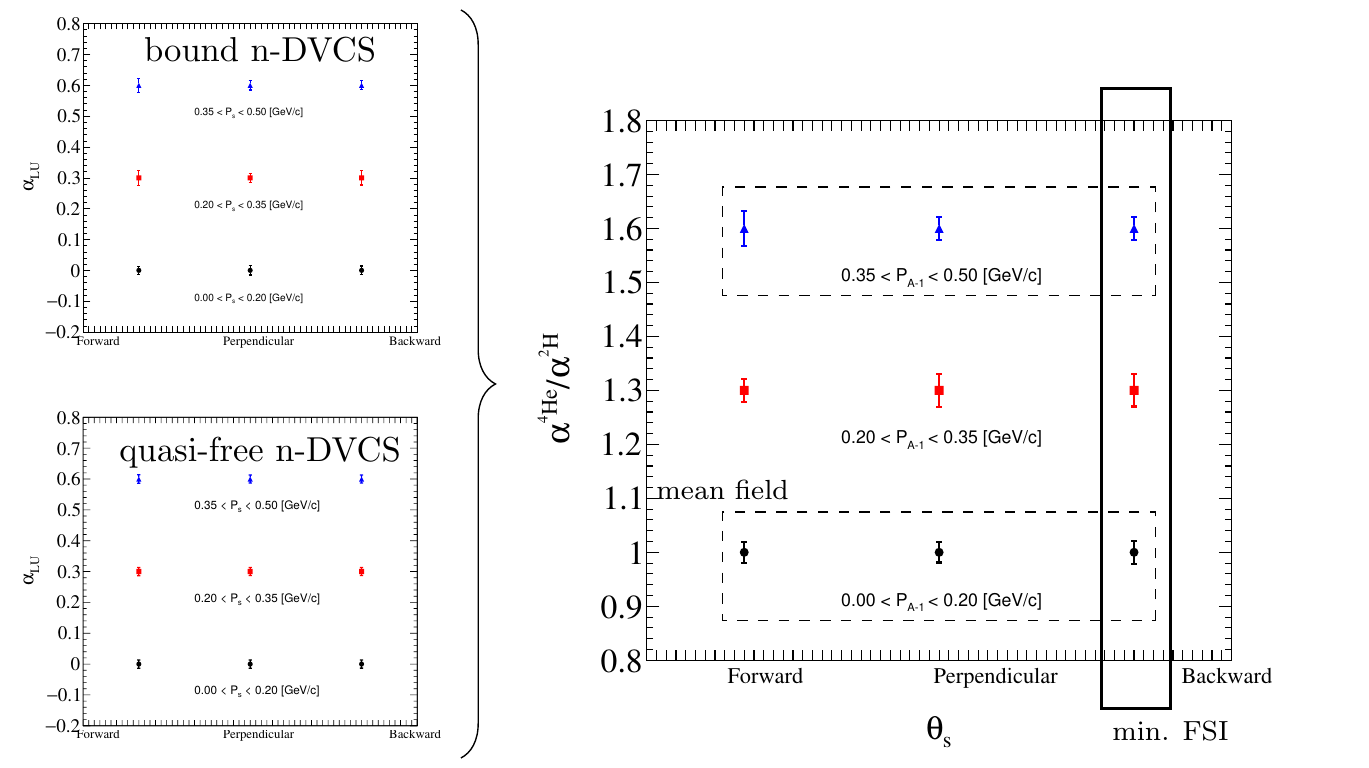}
   \caption{\label{fig:BSARatios}The BSA ratios targeting different nuclear 
      effects with specific spectator kinematics. Note  only statistical 
      uncertainties are shown and the BSA harmonics/ratios are (arbitrarily) 
      offset for clarity.
   }
\end{figure}

The first process above, p-DVCS on $^4$He, provides a model independent way of 
identifying kinematics where final state interactions are minimized (see 
Introduction and Chapter~\ref{chap:physics}). Armed with this information we 
will then measure the n-DVCS beam spin asymmetries on $^4$He and $^2$H knowing 
which kinematics are, or are not, influenced by FSIs.
We will then proceed as shown in Figure~\ref{fig:BSARatios} where the two n-DVCS 
measurements are combined into a ratio of bound neutron to quasi-free neutron.  
These BSA measurements and ratios are the primary deliverables of this 
proposal. They will be measured over a broad range of DVCS kinematics 
accessible to CLAS12 and for the spectator momenta regions noted above using 
the ALERT detector.

\section{Kinematic Coverage}
The kinematic coverage was studied using a newly developed CLAS12 fast 
Monte-Carlo, \texttt{c12sim}, where the CLAS12 detector resolutions were 
replicated based on the Fortran CLAS12 Fast-MC code. Because  \texttt{c12sim} 
is a Geant4 based simulation, the particle transport through the magnetic fields 
was handled by the Geant4 geometry navigation where all other processes were 
turned off. The resolutions for ALERT were obtained through full Geant4 
simulations with all physics processes turned on.
\begin{figure}[htb]
   \centering
   \includegraphics[width=0.49\textwidth,trim=10mm 4mm 10mm 12mm,
   clip]{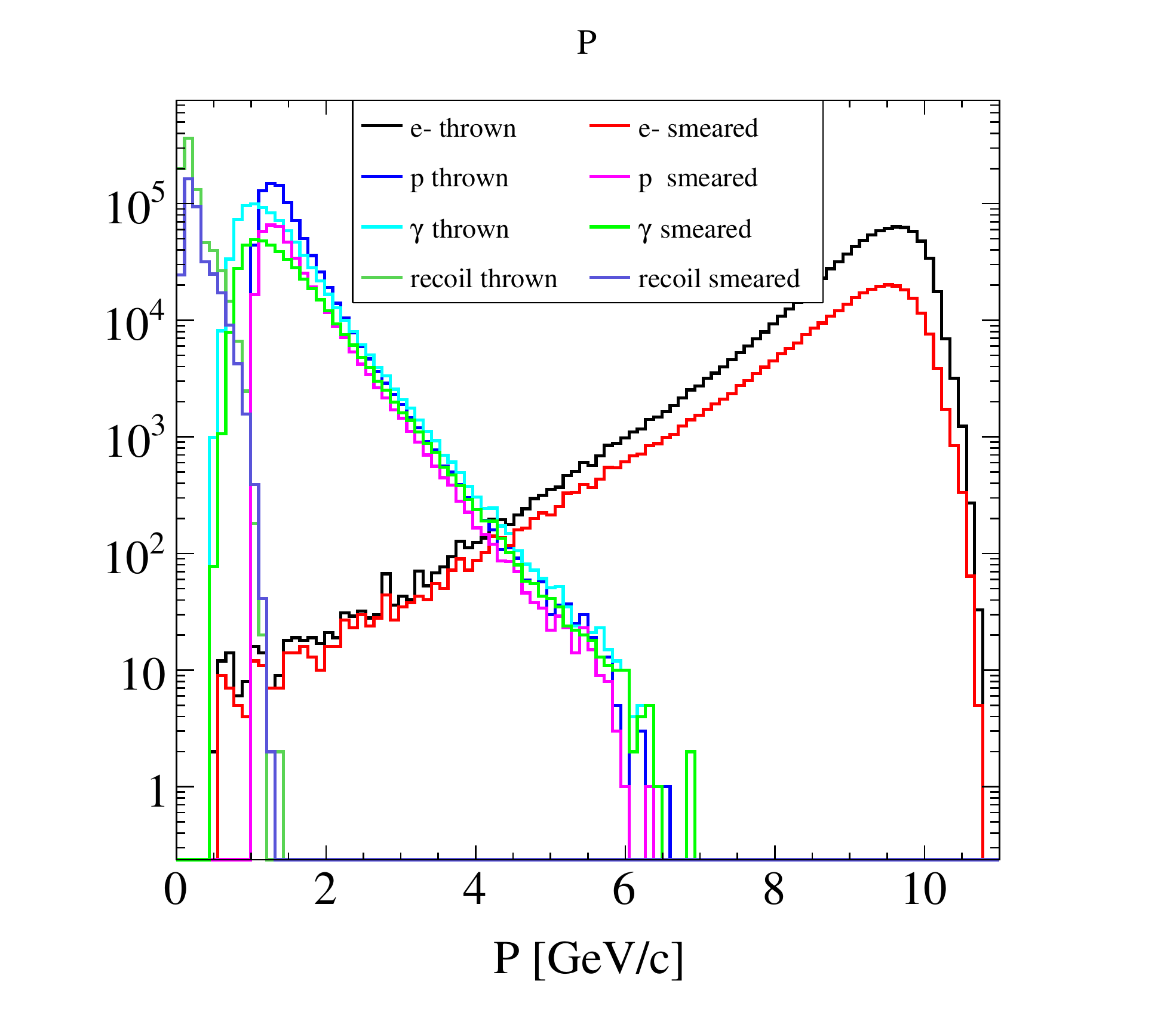}
   \includegraphics[width=0.49\textwidth,trim=10mm 4mm 10mm 12mm,
   clip]{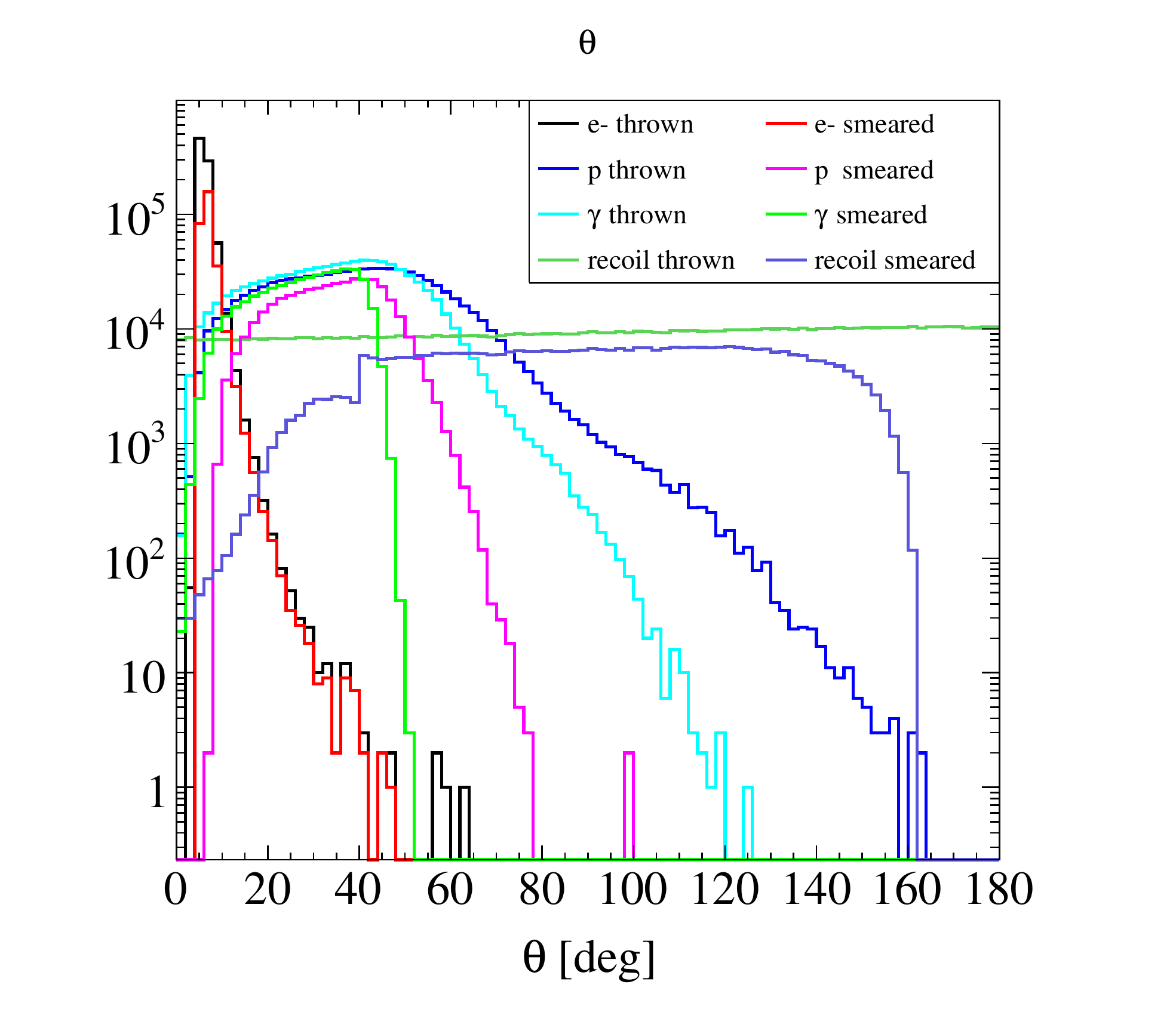}
   \caption{\label{fig:mcPdist}The simulated and detected momentum (left) and 
   angular (right) distributions showing overall detector coverage for the 
 experiment.}
\end{figure}
\begin{figure}[htb]
   \centering
   \includegraphics[width=0.49\textwidth,trim=10mm 4mm 10mm 10mm,
   clip]{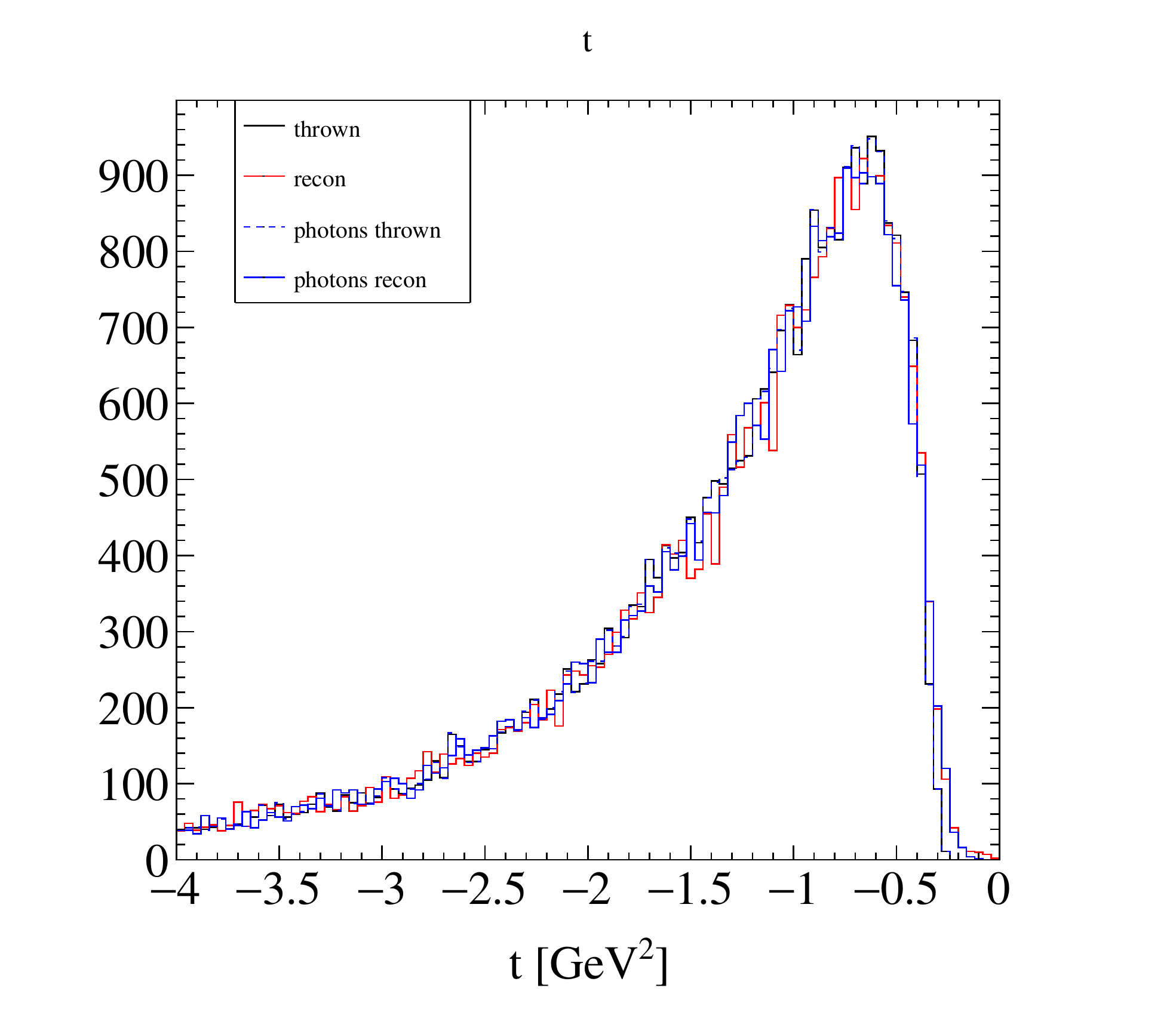}
   \includegraphics[width=0.49\textwidth,trim=10mm 4mm 10mm 10mm,
   clip]{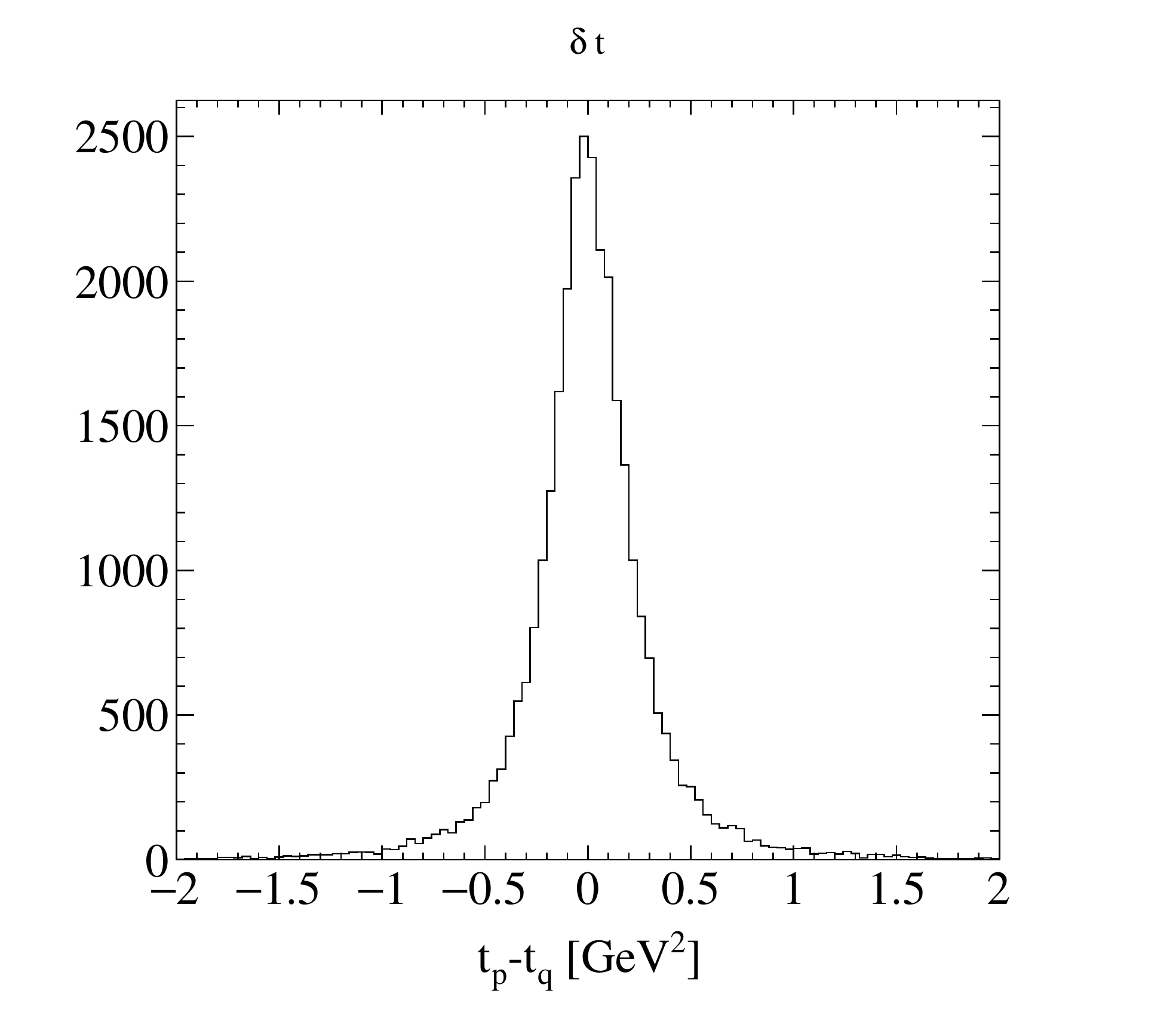}
   \caption{\label{fig:tResolutions}Left: simulated and reconstructed $t$ 
   calculated from the photons ($t_q$) and hadrons ($t_p$). Right: The 
difference between the two momentum transfers, $\delta t= t_p-t_q$.}
\end{figure}

First, we consider the p-DVCS reaction on $^4$He because of its special ability to determine 
the presence of final state interactions through a fully detected final state.
The spectator system, a recoiling $^3$H in the present case, is detected in ALERT 
while the forward electron, photon, and proton are detected in CLAS12. The 
resulting kinematics for the n-DVCS on $^2$H and $^4$He reactions will be quite 
similar, where the key difference is the struck neutron goes undetected. These 
events are then selected via the neutron missing mass cuts.

\begin{figure}
   \centering
   \includegraphics[width=0.49\textwidth,trim=0mm 4mm 2mm 10mm,
   clip]{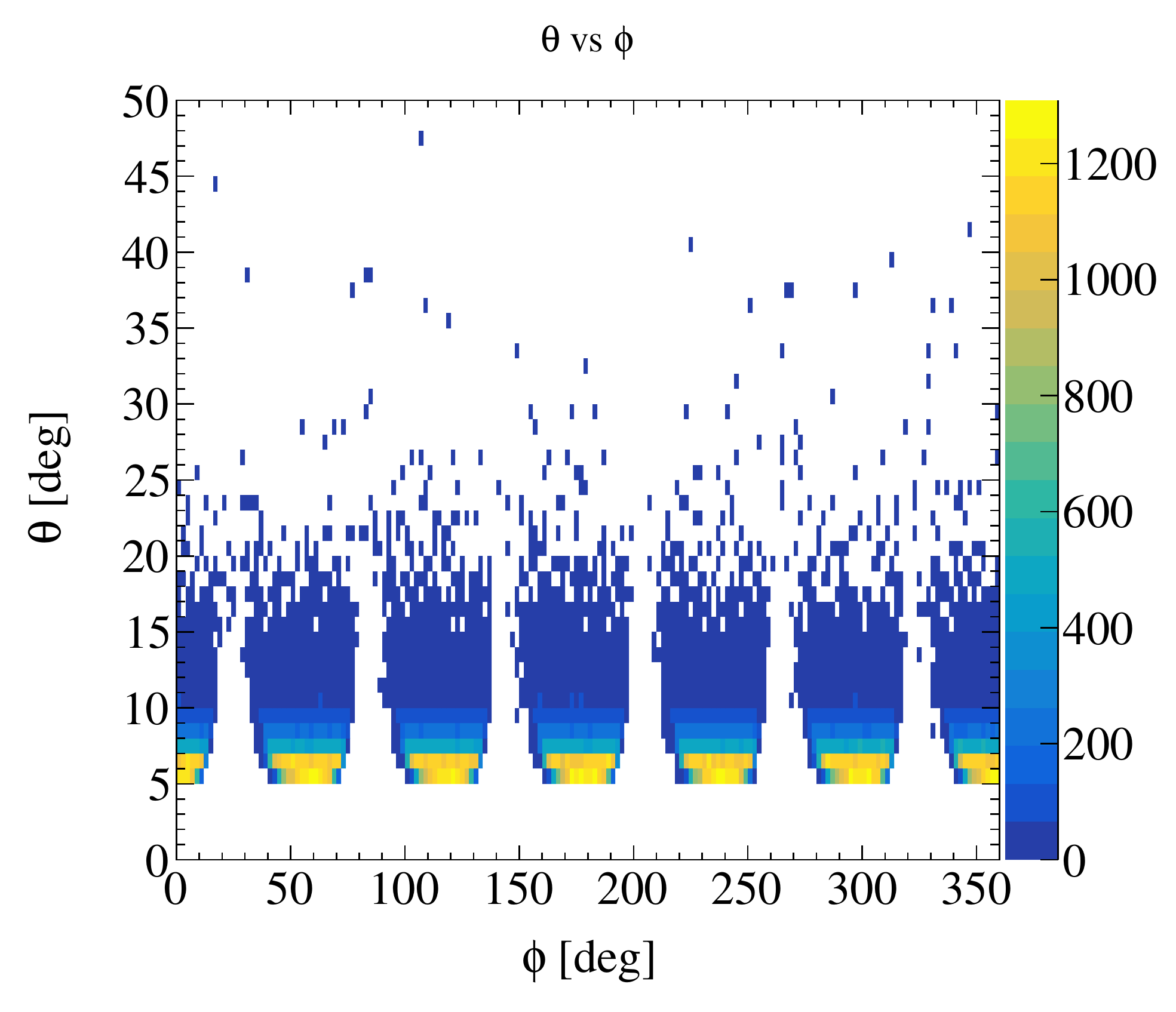}
   \includegraphics[width=0.49\textwidth,trim=0mm 4mm 2mm 10mm,
   clip]{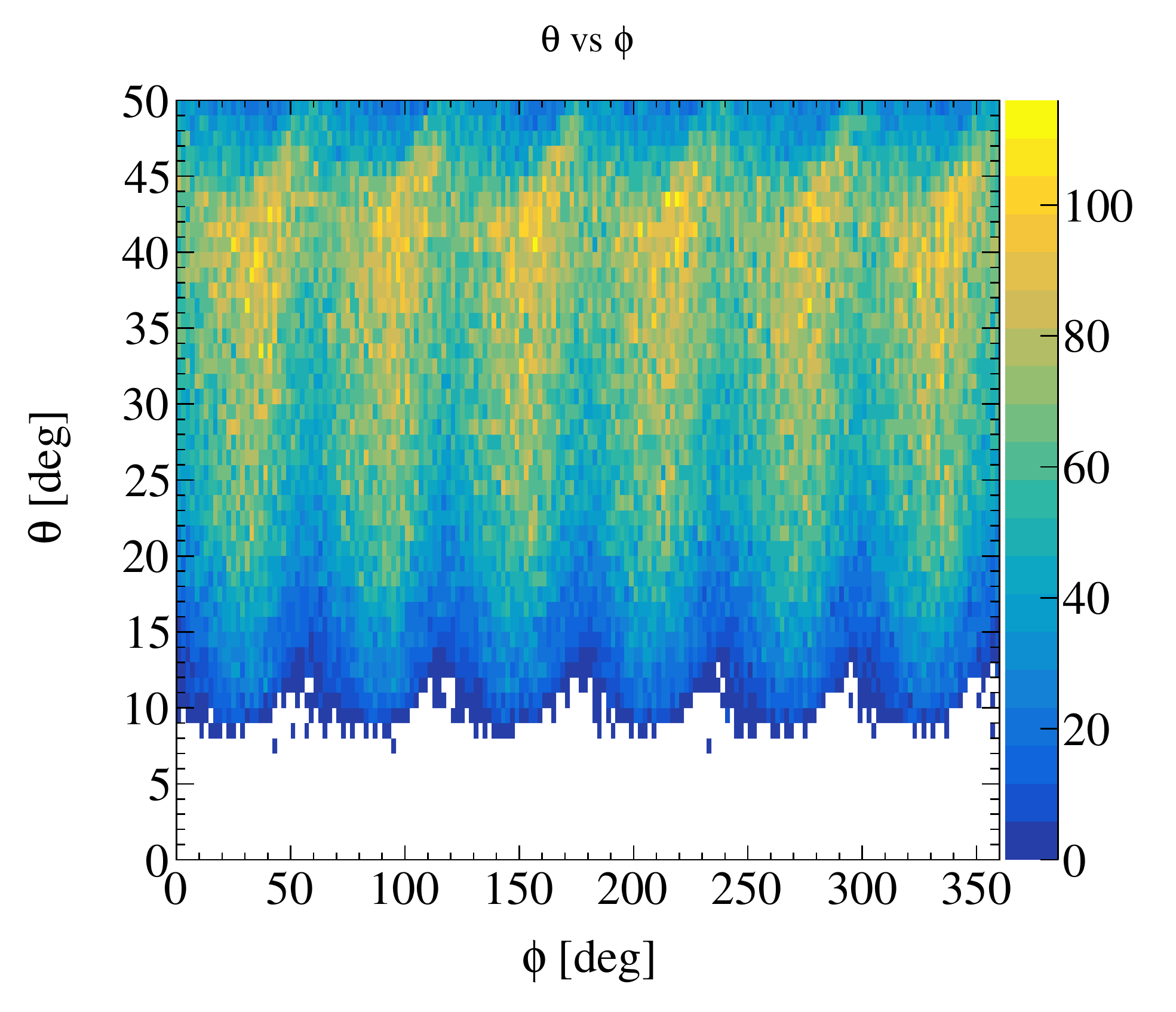}
   \includegraphics[width=0.49\textwidth,trim=0mm 4mm 2mm 10mm,
   clip]{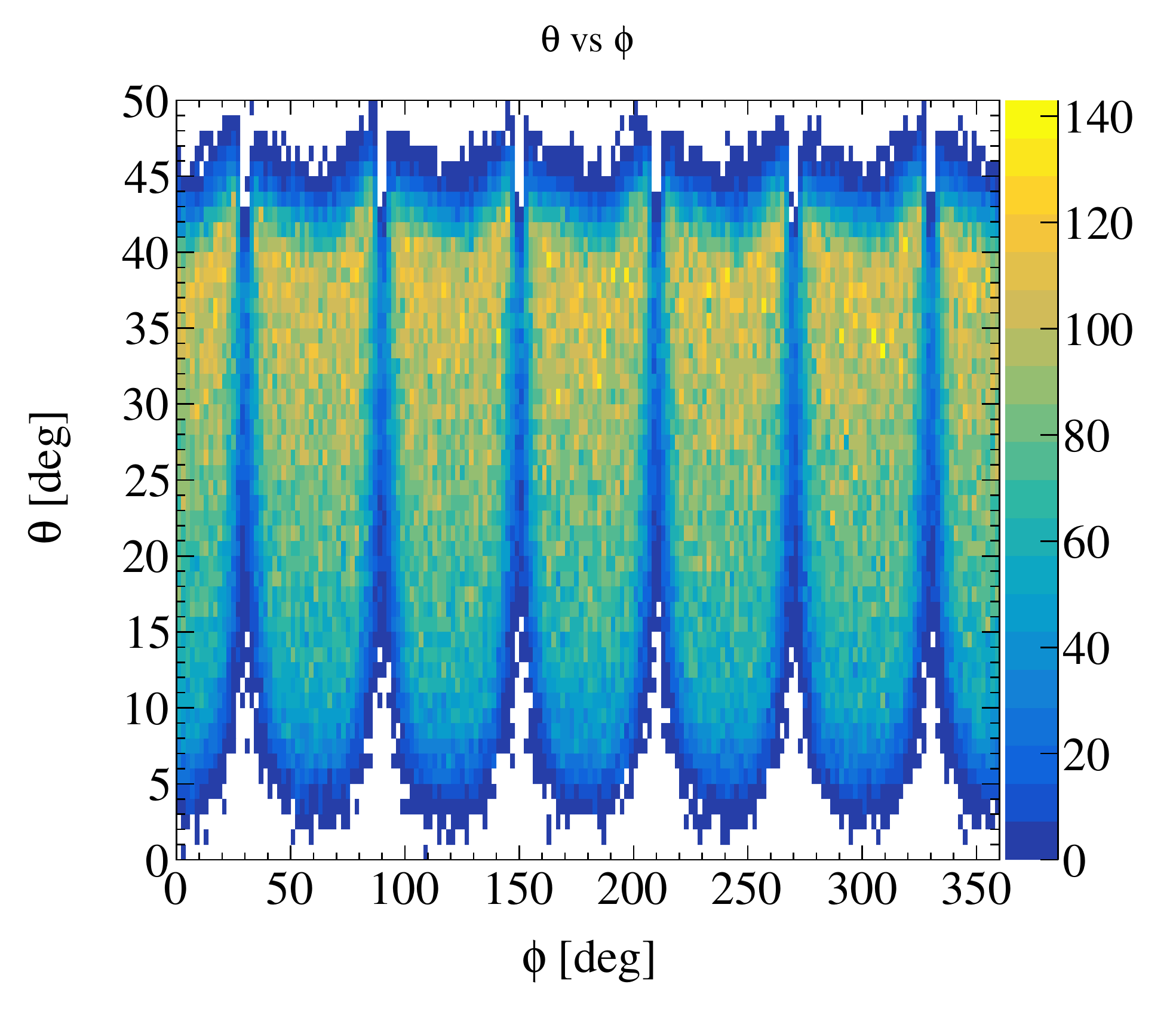}
   \includegraphics[width=0.49\textwidth,trim=0mm 4mm 2mm 10mm,
   clip]{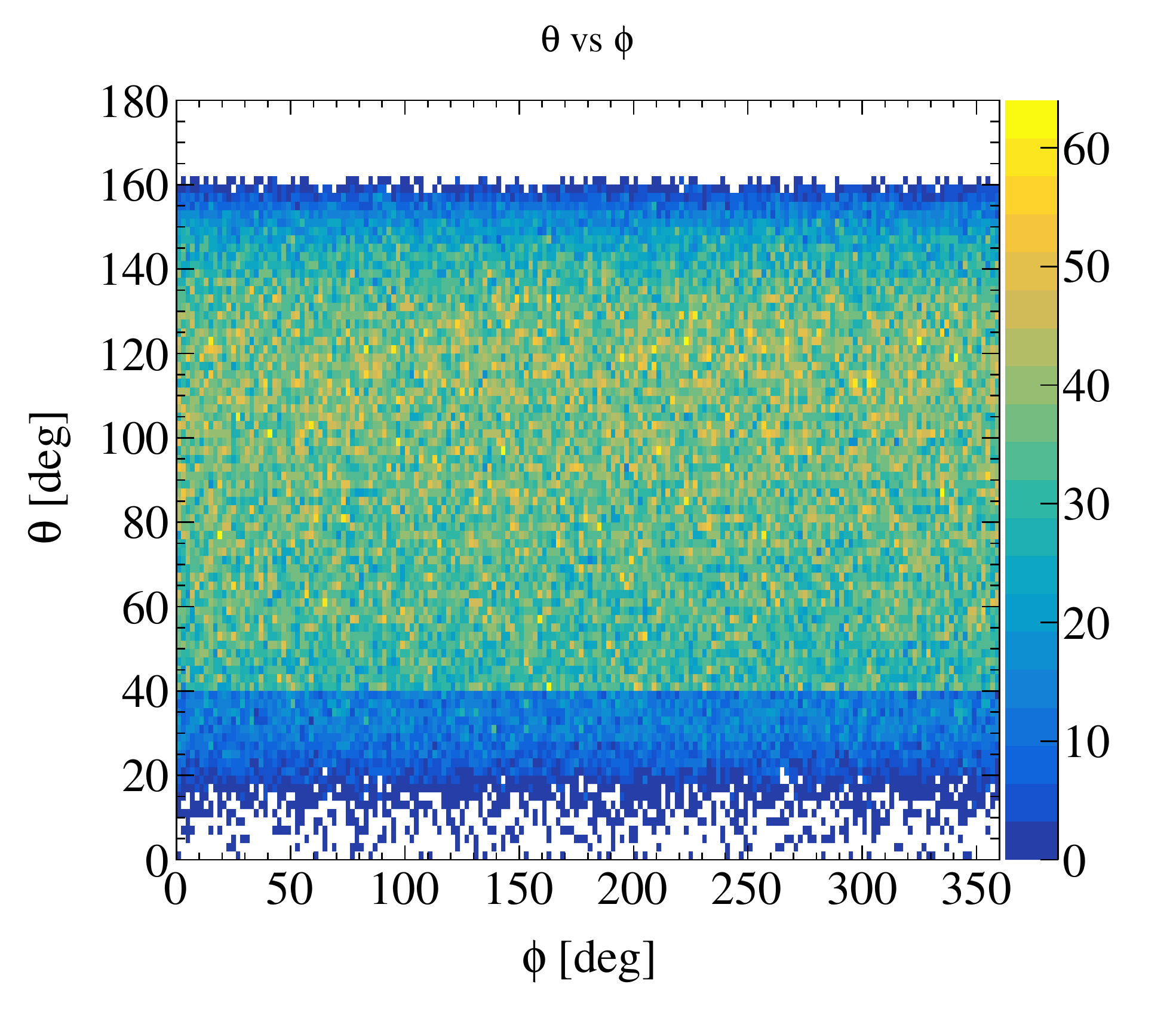}
   \caption{\label{fig:eThetaVsPhi}The angular coverage shown as $\theta$ Vs.  
   $\phi$ for the electron (upper left), proton (upper right), photon (lower 
 left), and recoil spectator (lower right).}
\end{figure}

\begin{figure}
   \centering
   \includegraphics[width=0.49\textwidth,trim=0mm 4mm 2mm 12mm,
   clip]{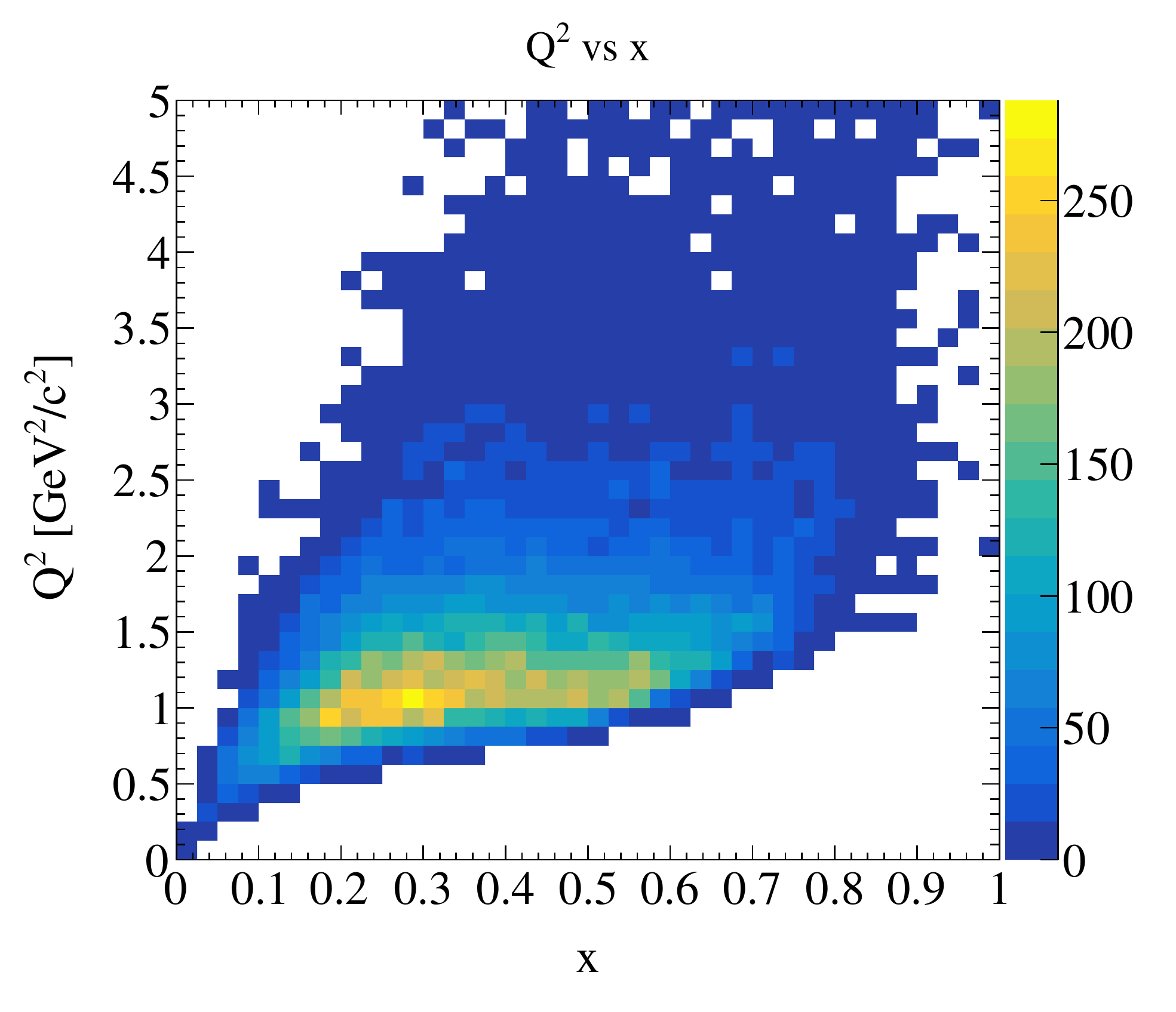}
   \includegraphics[width=0.49\textwidth,trim=0mm 4mm 2mm 12mm,
   clip]{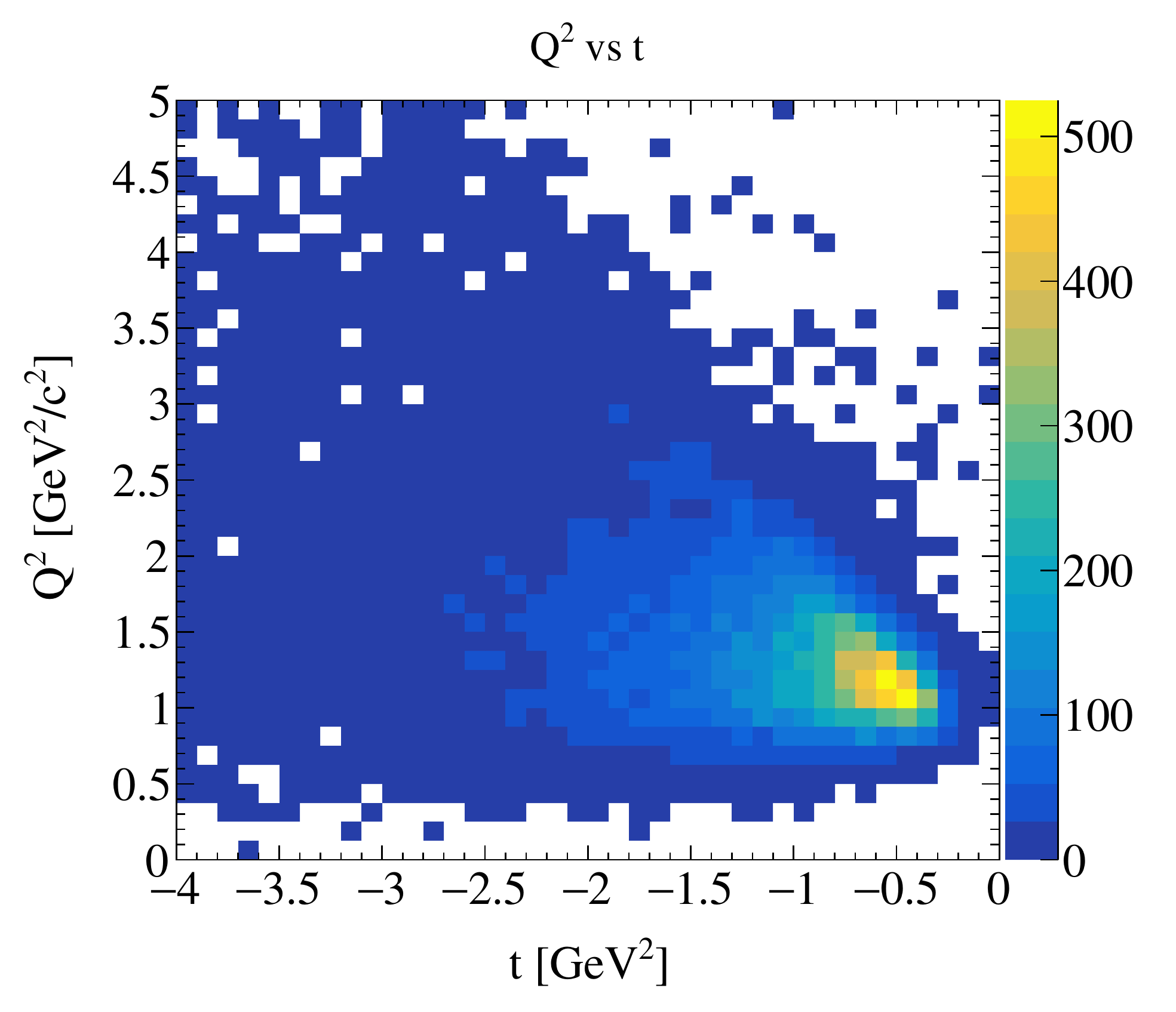}
   \caption{\label{fig:Q2Vsxandt}$Q^2$ plotted against $x$ (left) and $t$ 
   (right).}
\end{figure}

The overall coverage in momentum and scattering angle can be seen in 
Figure~\ref{fig:mcPdist} and angular detector coverage of all the particles can 
be seen in Figure~\ref{fig:eThetaVsPhi}. The bin variables $x$, $Q^2$, and $t$ 
are shown in Figure~\ref{fig:Q2Vsxandt}. See \ref{sec:extraKineProjections} for 
more details on the kinematic coverage.
%

\begin{figure}
   \centering
   \includegraphics[width=0.49\textwidth,trim=0mm 4mm 10mm 10mm,
   clip]{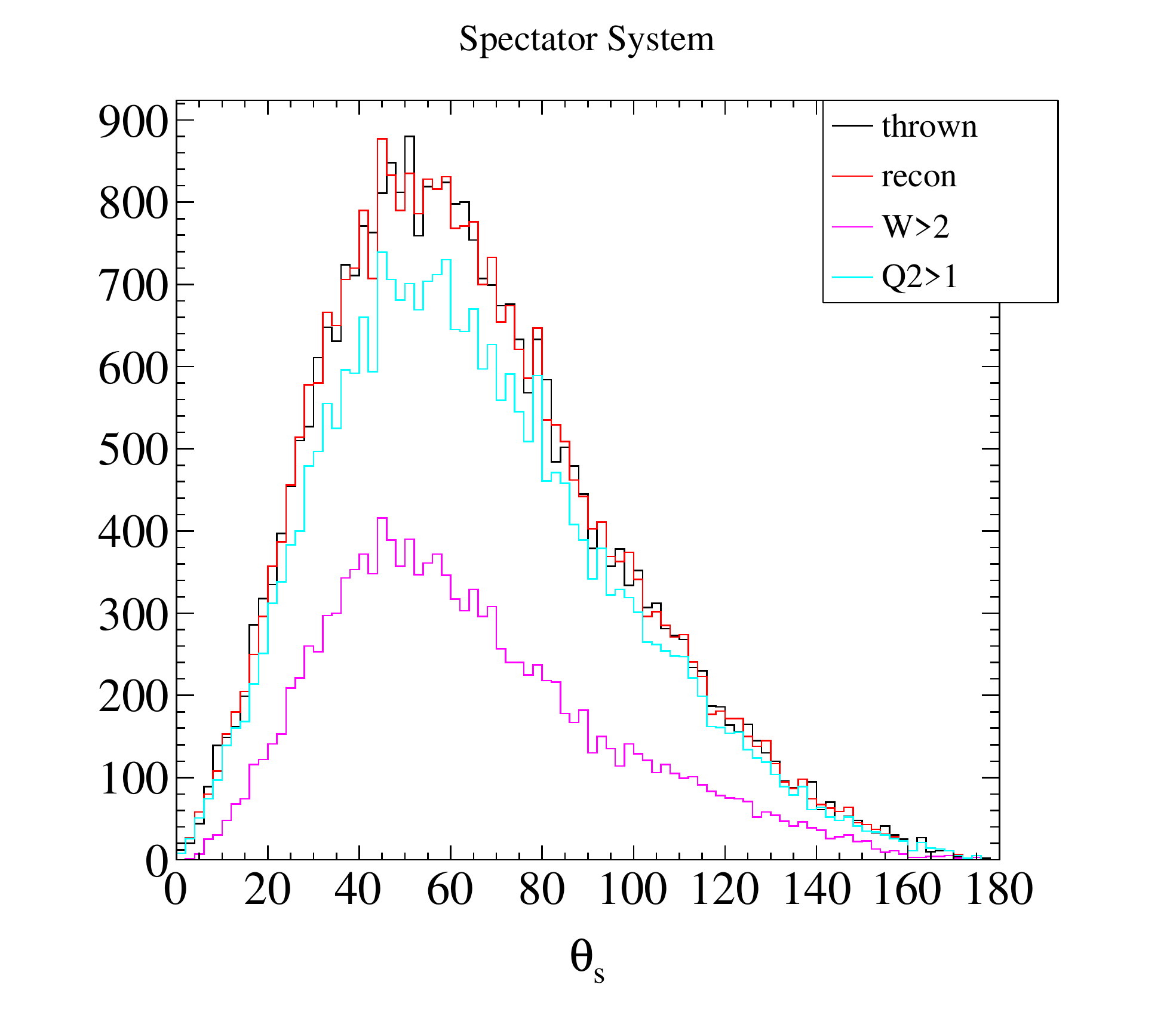}
   \includegraphics[width=0.49\textwidth,trim=0mm 4mm 10mm 10mm,
   clip]{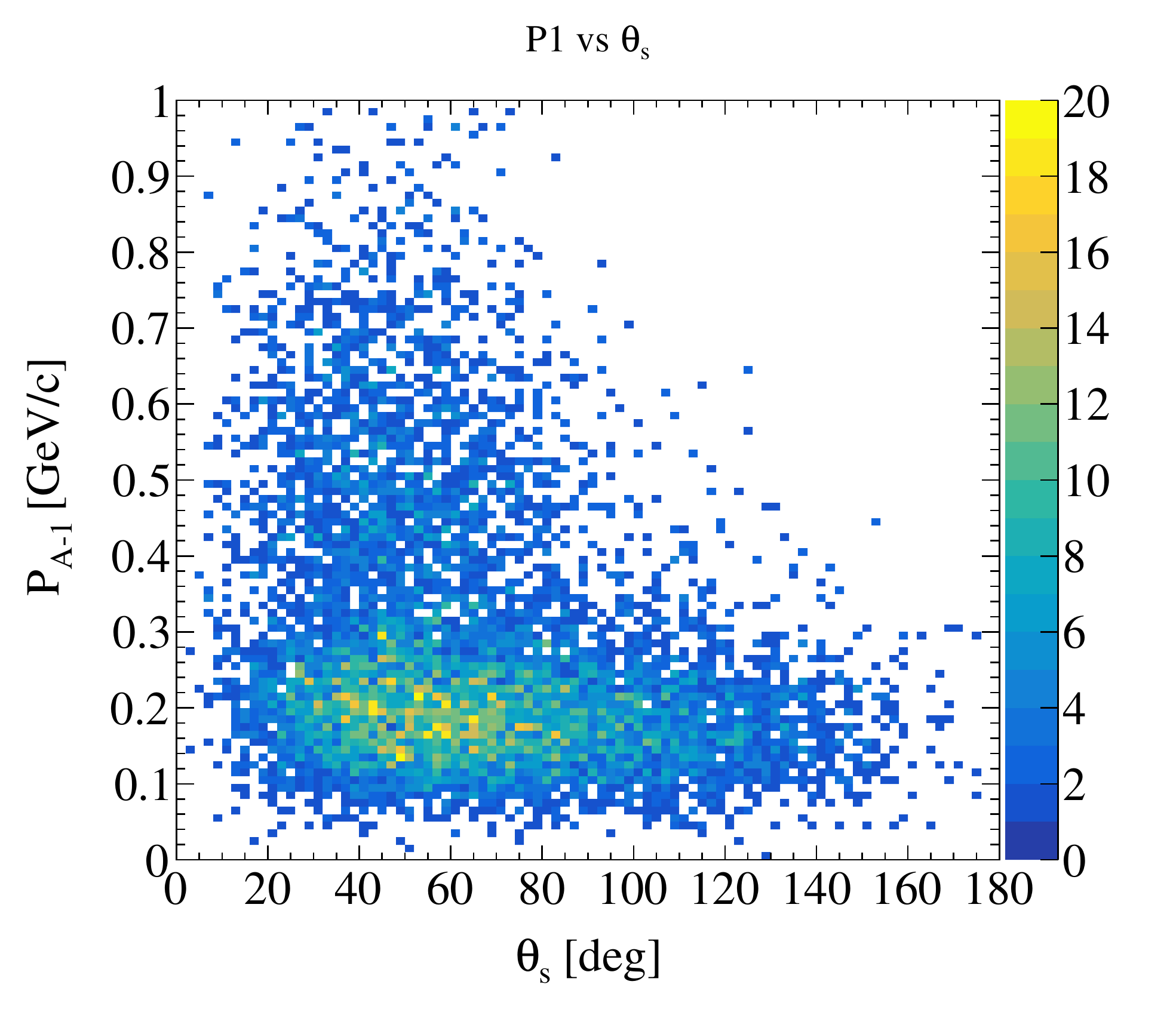}
   \caption{\label{fig:thetasCoverage}Left: Spectator recoil angle, $\theta_s$,
     showing the generated and reconstructed values, also shown is the 
     reconstructed with $Q^2>1$ GeV$^2$ and $W>2$ GeV.  Right: Spectator angle vs 
   reconstructed spectator momentum.}
\end{figure}

As mentioned throughout this proposal and with detail in appendix 
\ref{chap:appendixKine}, the momentum transfer can be reconstructed via using 
the photons, or using the nucleon side of the diagram where we make use use of 
the detected spectator system and the PWIA. Figure\,\ref{fig:tResolutions} shows 
that the resolutions are comparable, thus, allowing for the systematic check of 
FSIs which were not included in the generated events. The spectator angle and 
momentum can be seen in Figure\,\ref{fig:thetasCoverage}, where these results can 
be used along with calculations such as those shown in 
Figure\,\ref{fig:deuteronFSI} to isolate kinematic regions with significant 
FSIs.
%
%

\section{Projections}
\subsection{Beam Spin Asymmetry Extraction}
\begin{figure}
   \centering
   \includegraphics[width=0.79\textwidth,trim=0mm 0mm 0mm 0mm, 
   clip]{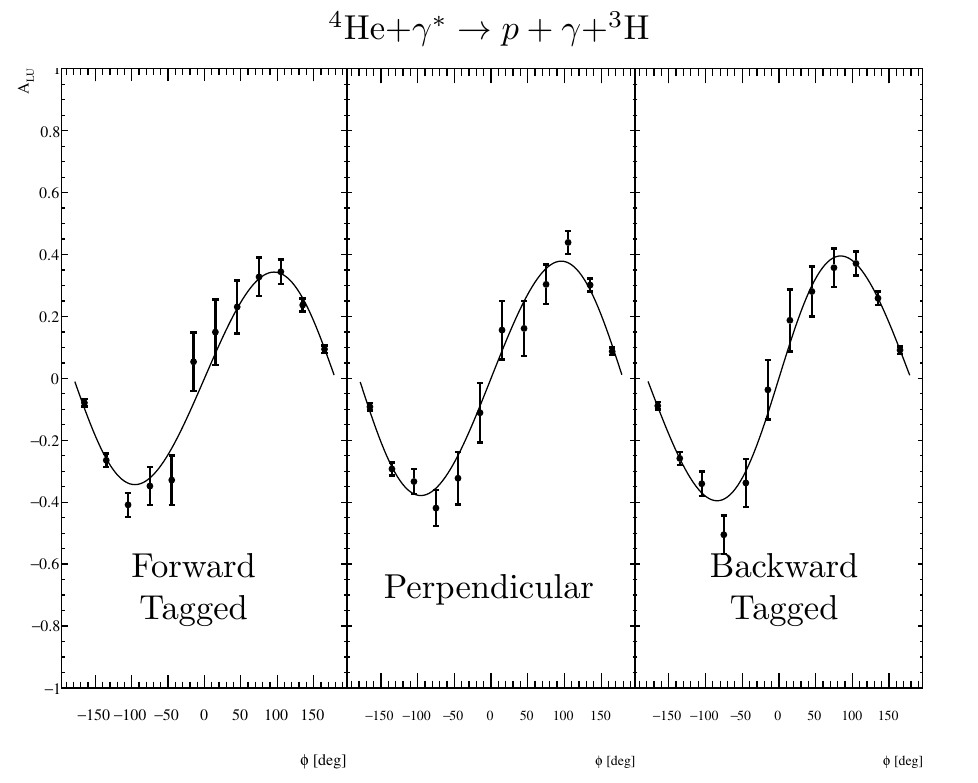}
   \caption{\label{fig:He4ProtonPhiAsyms1}Projections for the statistical 
     uncertainties on $A_{LU}$ for three different bins in spectator angle, all 
     corresponding to the lowest spectator momentum bin. The spectator angles 
     are forward (left), perpendicular (center), and backward (right). Note the 
   low momentum bin corresponds to the mean field nucleons. }
\end{figure}

\begin{figure}
   \centering
   \includegraphics[width=0.49\textwidth,trim=0mm 0mm 0mm 0mm, 
   clip]{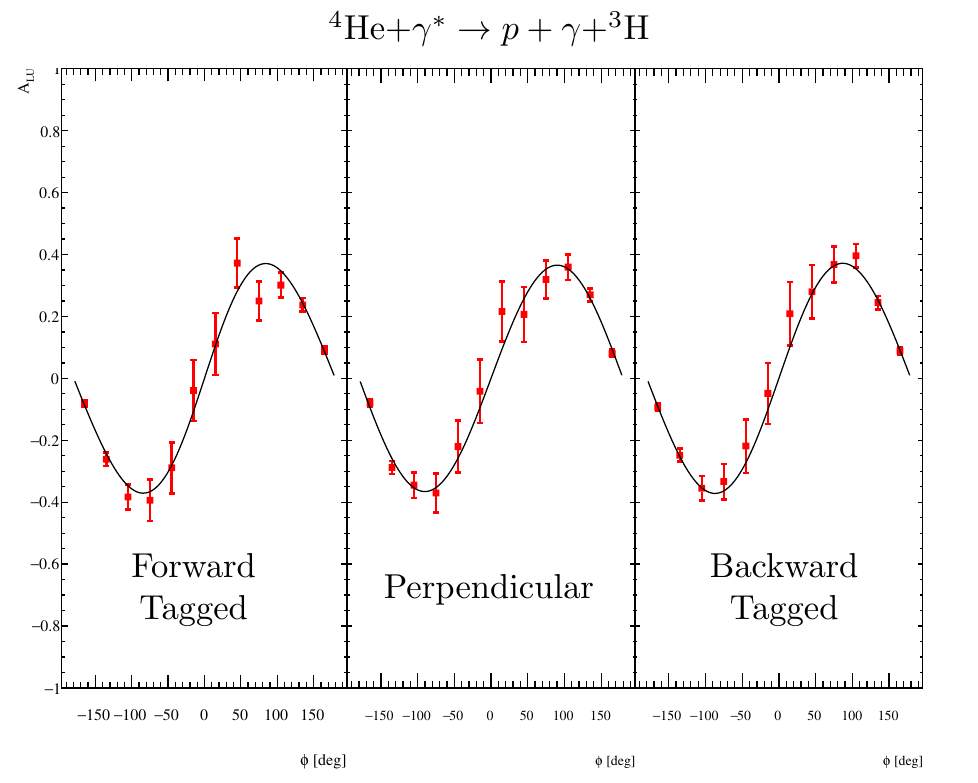}
   \includegraphics[width=0.49\textwidth,trim=0mm 0mm 0mm 0mm, 
   clip]{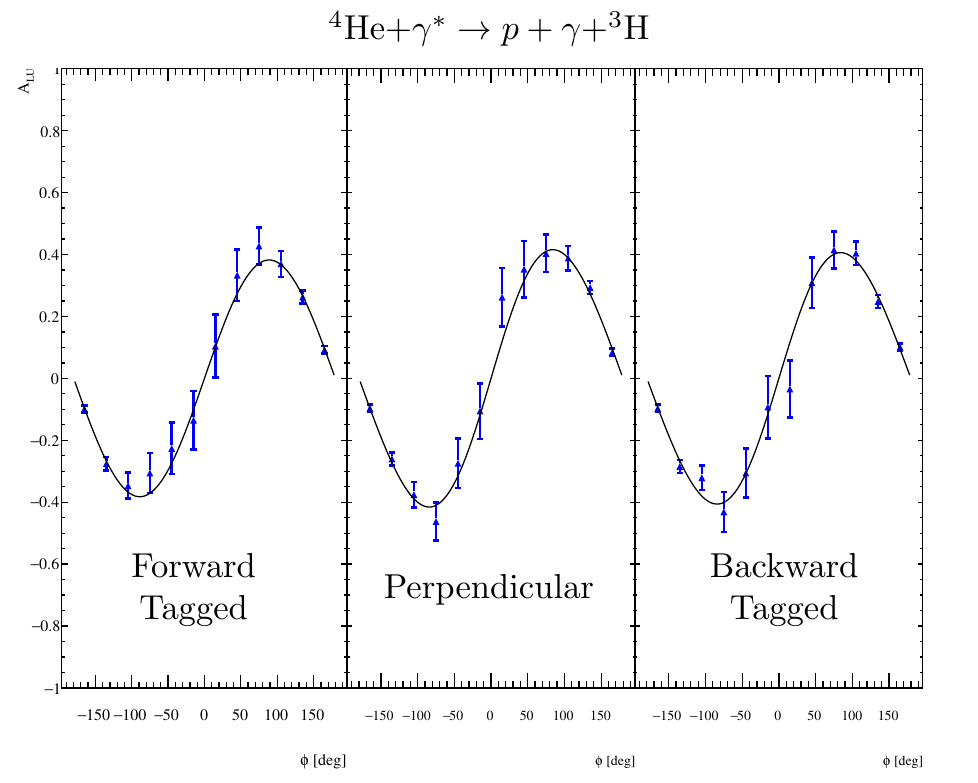}
   \caption{\label{fig:He4ProtonPhiAsyms2}Expected statistical uncertainties of 
      $A_{LU}$ for $\theta_{s}$ bins identical to those  in 
      Figure\,\ref{fig:He4ProtonPhiAsyms1}, but these results show the two higher 
   spectator momentum bins. Note the highest momenta (blue) correspond to SRC 
nucleons.  }
\end{figure}

\begin{figure}
   \centering
   \includegraphics[width=0.49\textwidth,trim=0mm 0mm 0mm 0mm, 
   clip]{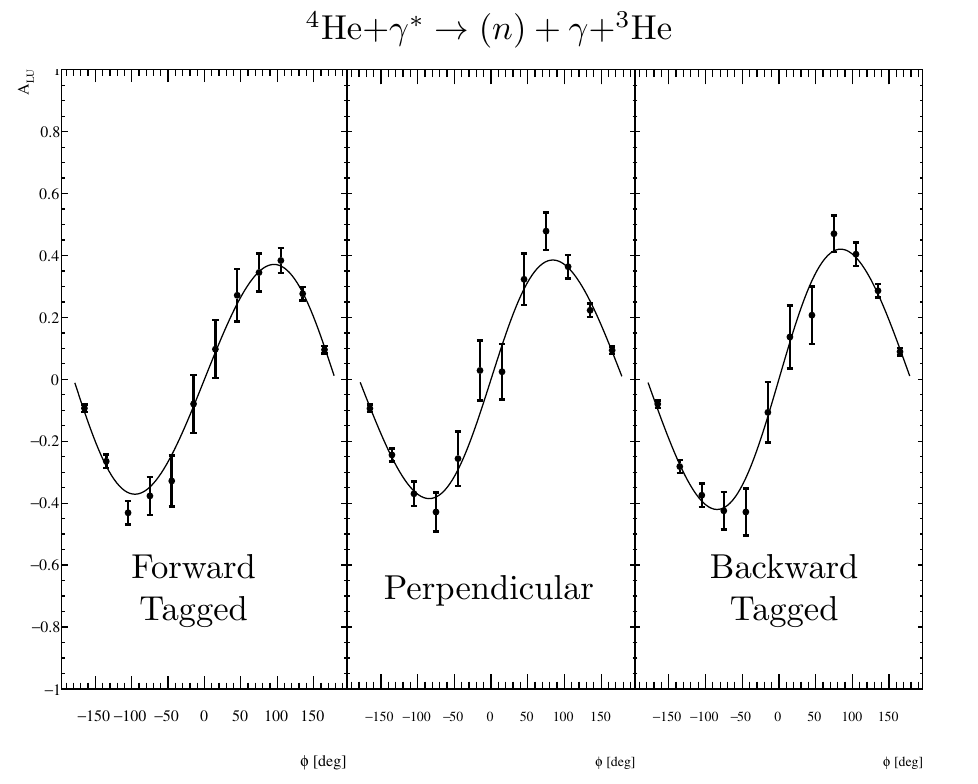}
   \includegraphics[width=0.49\textwidth,trim=0mm 0mm 0mm 0mm, 
   clip]{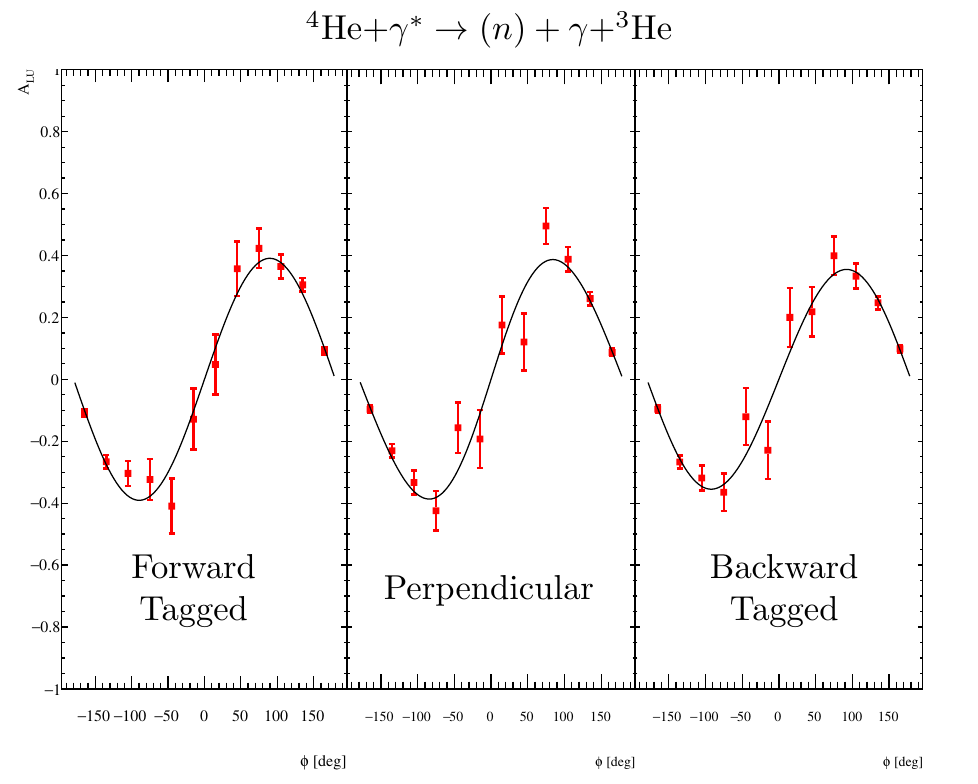}
   \includegraphics[width=0.49\textwidth,trim=0mm 0mm 0mm 0mm, 
   clip]{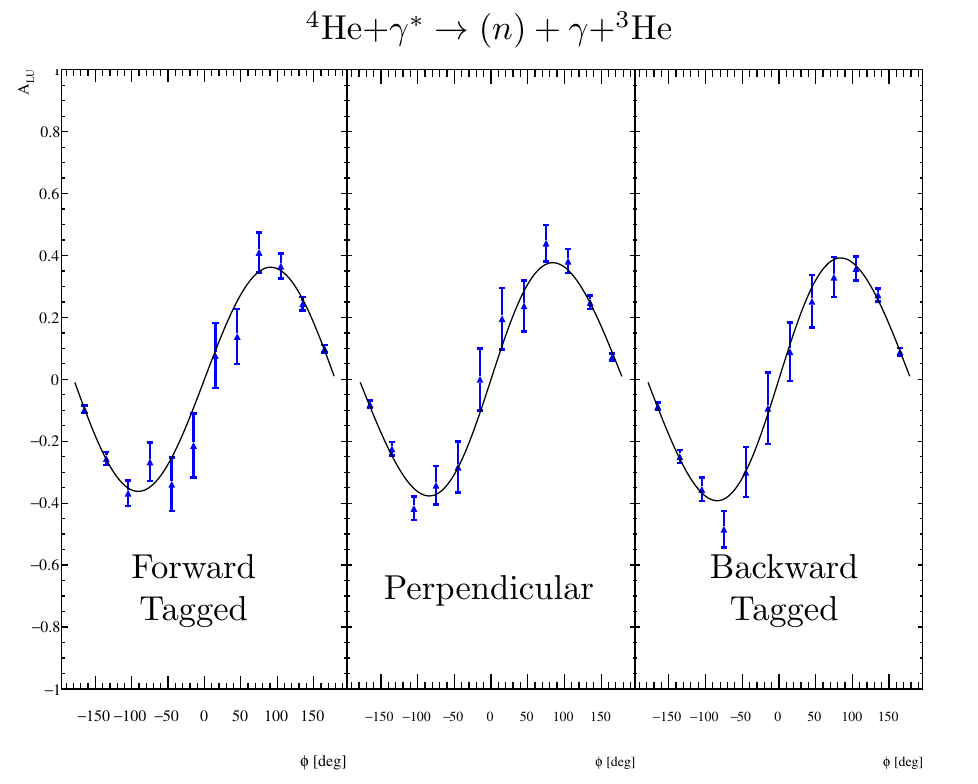}
   \caption{\label{fig:He4NeutronPhiAsyms}Projected statistical uncertainties  
     $A_{LU}$ for $\theta_{s}$ bins identical to those  in 
   Figure\,\ref{fig:He4ProtonPhiAsyms1}, for n-DVCS on $^4$He measurement, in 9 
 different bins of spectator momentum and angle.}
\end{figure}

\begin{figure}
   \centering
   \includegraphics[width=0.49\textwidth,trim=0mm 0mm 0mm 0mm, 
   clip]{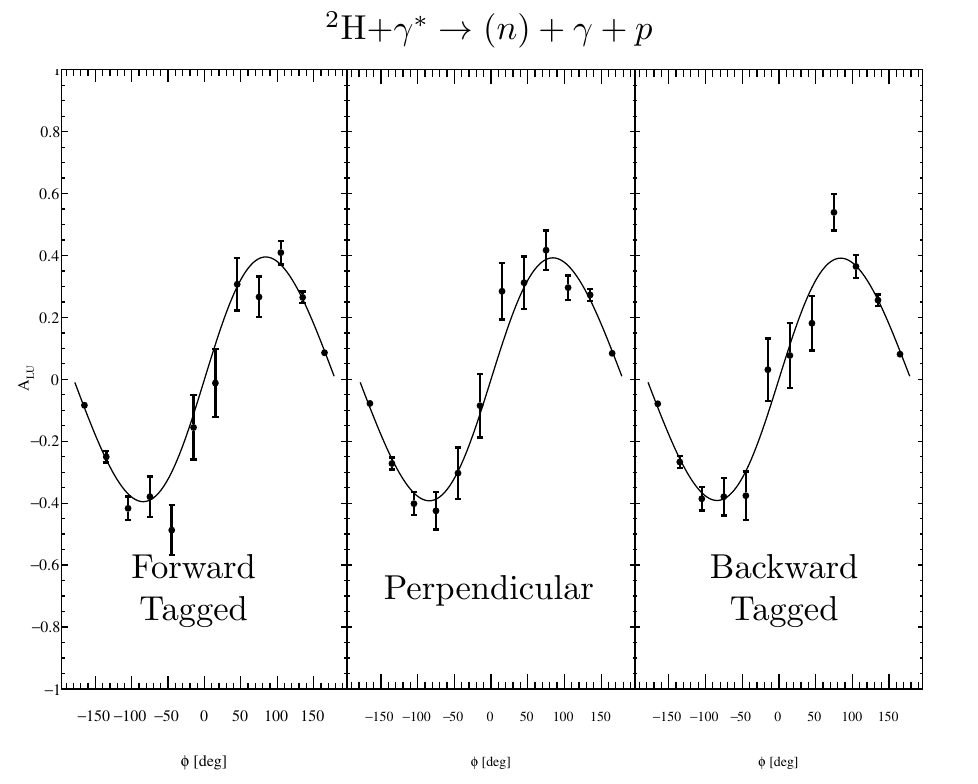}
   \includegraphics[width=0.49\textwidth,trim=0mm 0mm 0mm 0mm, 
   clip]{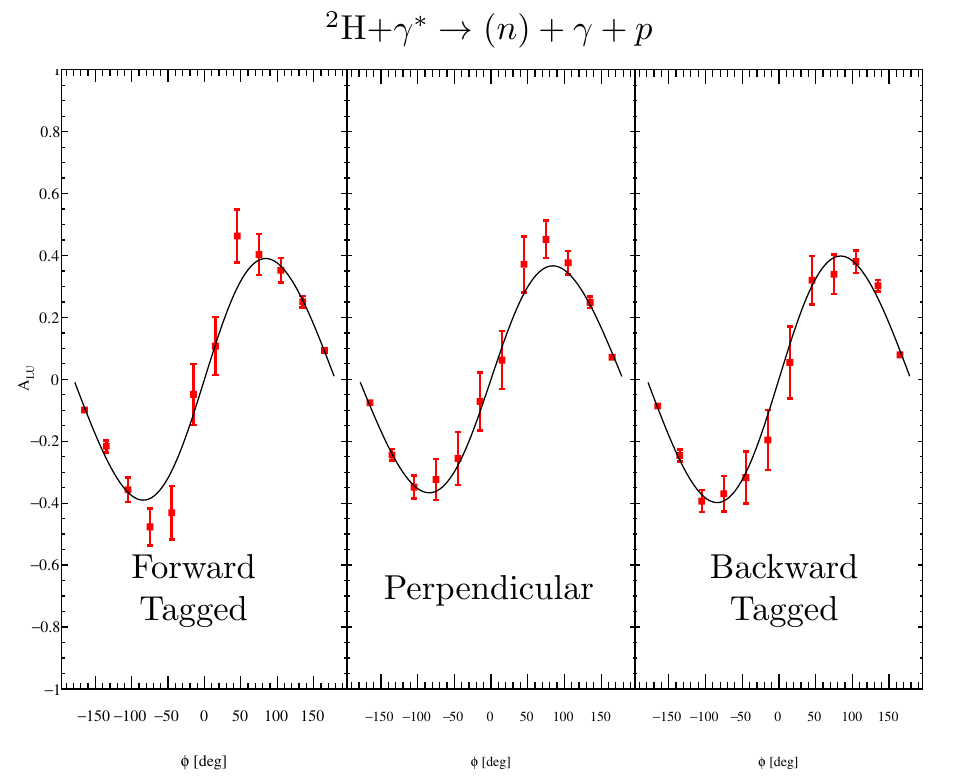}
   \includegraphics[width=0.49\textwidth,trim=0mm 0mm 0mm 0mm, 
   clip]{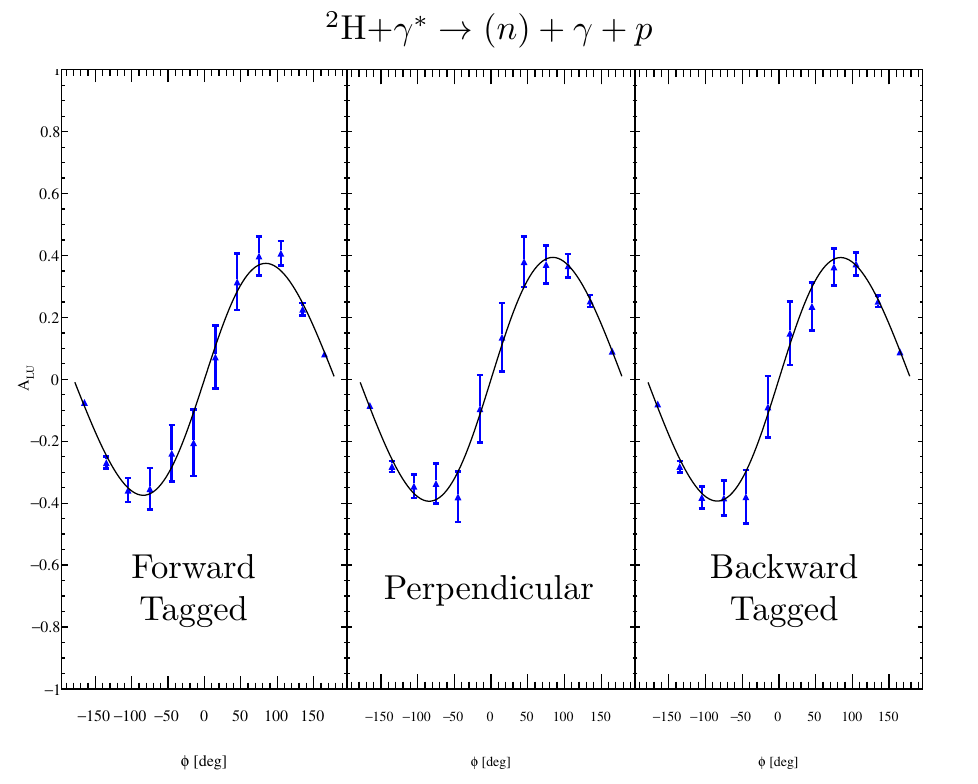}
   \caption{\label{fig:H2NeutronPhiAsyms}Projected statistical uncertainties  
     $A_{LU}$ for $\theta_{s}$ bins identical to those  in 
   Figure\,\ref{fig:He4ProtonPhiAsyms1}, for n-DVCS on $^2$H measurement, in 9 
 different bins of spectator momentum and angle. }
\end{figure}

The measured beam spin asymmetries are binned in 6 variables: $x_B$, $Q^2$, 
$t$, $\phi$, $P_s$, and $\theta_s$. The 6 dimensional data will be reduced to 5 
dimensions by fitting the BSA as a function of $\phi$ to extract harmonic 
content. 
Projections for the statistical uncertainties of these asymmetries and their fits 
are shown for a few bins in Figure~\ref{fig:He4ProtonPhiAsyms1} and Figure~\ref{fig:He4ProtonPhiAsyms2}
for p-DVCS on $^4$He. A few of the $\phi$ binned asymmetries for n-DVCS on $^4$He are shown in Figure~\ref{fig:He4NeutronPhiAsyms}
and similarly in Figure~\ref{fig:H2NeutronPhiAsyms} for n-DVCS on $^2$H.
Note that we are using a simple binning scheme shown in Table\,\ref{tab:bins}.
These bins are likely to change as the cross sections are not well known, 
especially when isolating high momentum spectators.
\begin{table}
   \centering
\vspace*{0.2cm}
   \begin{tabu}{lccccc}
\tabucline[2pt]{-}                                                   
         Bin  & $x$         & $Q^2$       & $t$          & $\theta_s$   & $P_s$ \\ 
        units&             & \si{\GeV^2} & \si{\GeV^2}  & \si{\degree} & \si{\GeV/c} \\
\tabucline[1pt]{-}                                                   
             &0.05         & 1           & 0           & 0.0          & 0.0         \\
             &0.25         & 1.5         & 0.75        & 50           & 0.2         \\
             &0.35         & 2.0         & 1.5         & 100          & 0.35        \\
             &0.5          & 3.0         & 2.5         & 180          & 0.5         \\
             &0.8          & 10          & 6.0 & & \\
\tabucline[2pt]{-}                                                   
\end{tabu}
   \caption{\label{tab:bins}The simple binning scheme used for the proposal.
   Listed here are the bin edges forming each bin.}
\end{table}

\begin{figure}
   \centering
   \includegraphics[width=0.49\textwidth,trim=0mm 4mm 0mm 12mm, 
   clip]{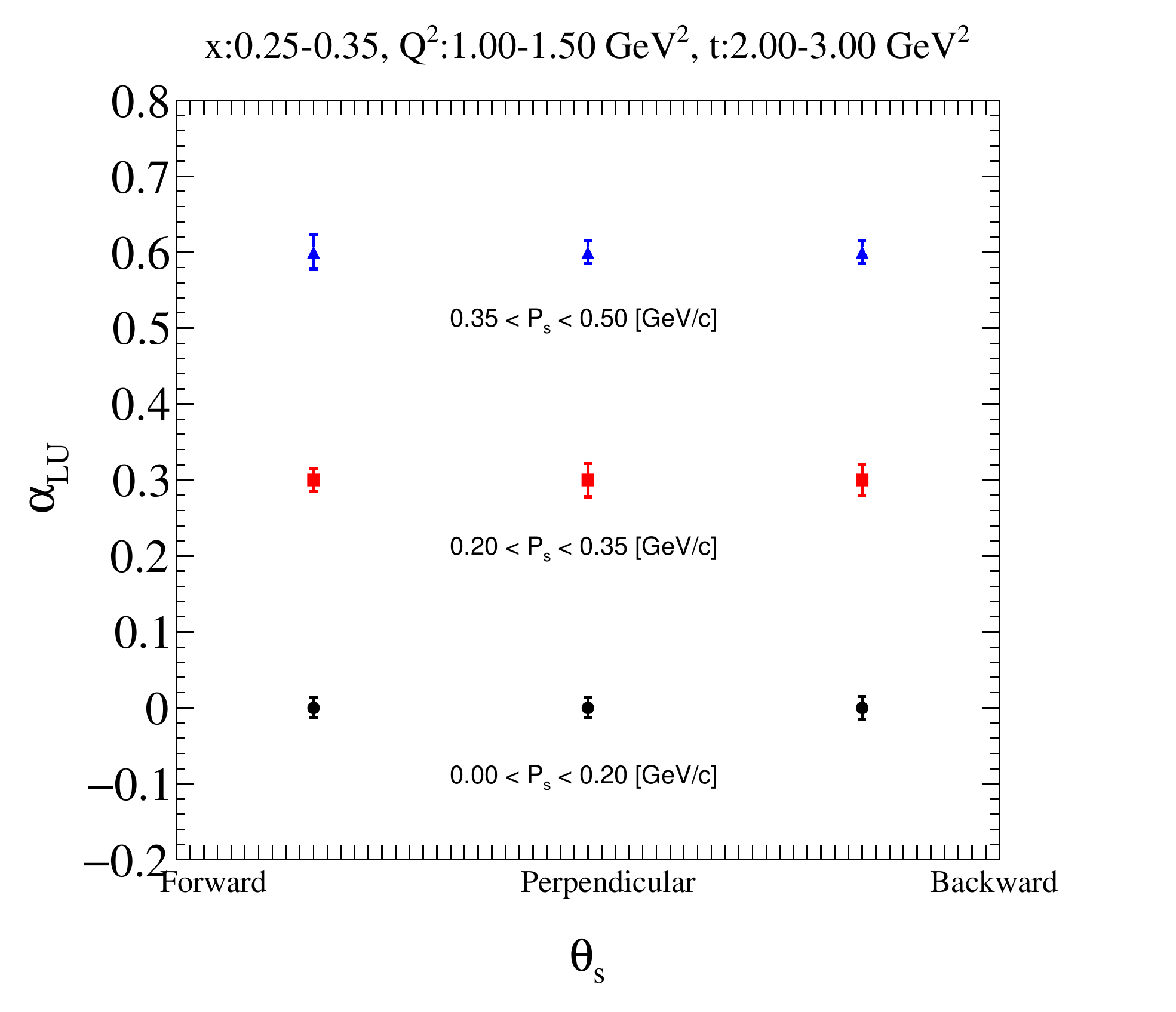}
   \includegraphics[width=0.49\textwidth,trim=0mm 4mm 0mm 12mm, 
   clip]{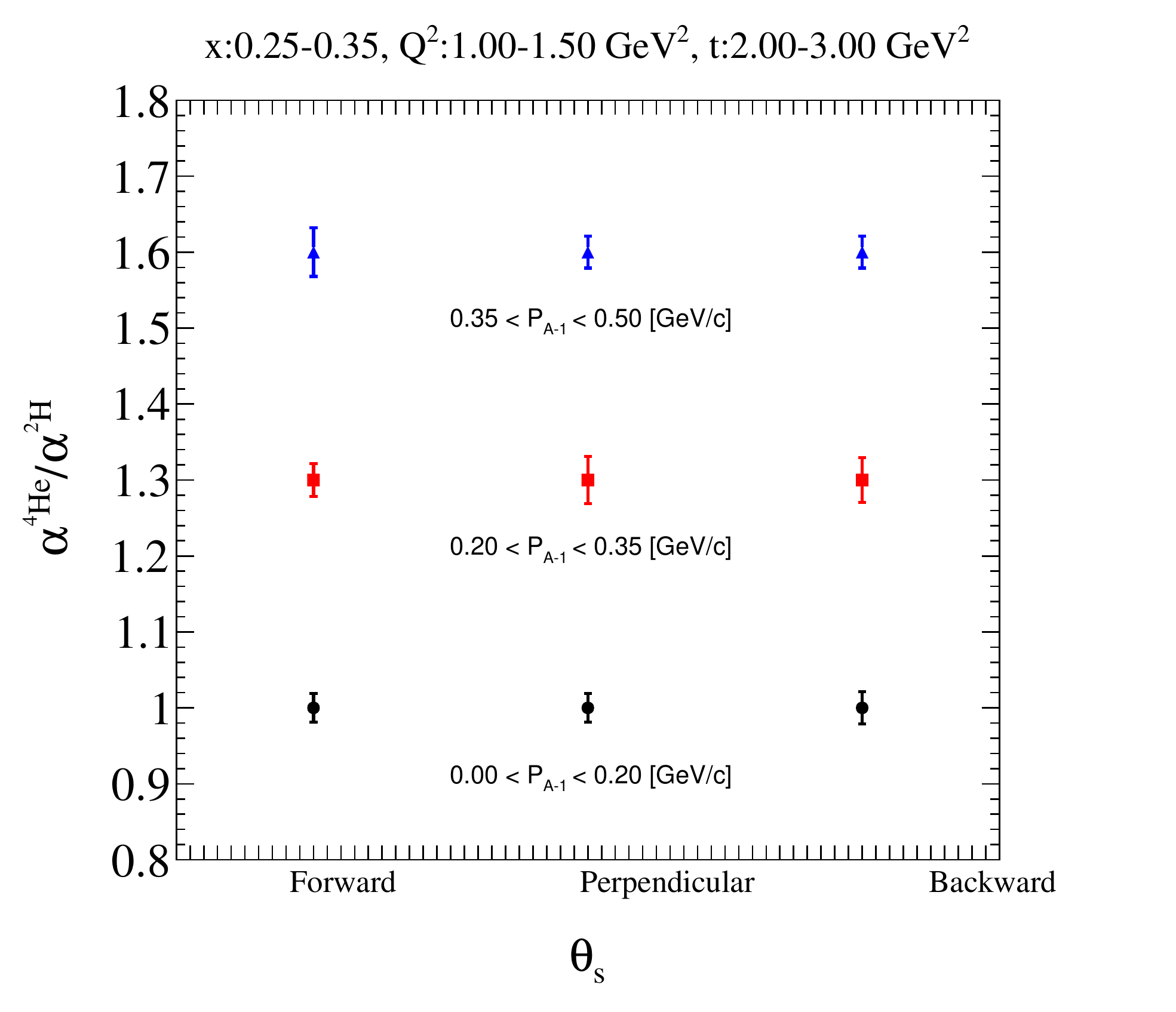}
   \caption{\label{fig:alphaALU}Left: Projected uncertainties after fitting the 
     beam spin asymmetry with equation\,(\ref{eq:ALUfit}) to extract a 
   value of $\alpha$. Each bin in $x$, $Q^2$, and $t$ has 3 bins in $P_{A-1}$ and $\theta_s$, which 
 are offset vertically for clarity. Right: The ratio of $\alpha$s for a bound 
 neutron  and a quasi-free nucleon. }
\end{figure}

The beam spin asymmetries are the primary observables for this experiment and 
will be fit with the following simplified parameterization
\begin{equation}\label{eq:ALUfit}
   A_{LU}(\phi) = \frac{\alpha \sin\phi}{1+\beta\cos\phi}
\end{equation}
where the free parameters $\alpha$ and $\beta$ are related to CFFs and Fourier 
harmonics. As emphasized in section~\ref{sec:polEMCeffect}, the $\sin\phi$ 
harmonic, $\alpha$, is quite sensitive to nuclear effects. 
Therefore, we will extract $\alpha$ for every bin by fitting the 
asymmetry binned in $\phi$ for each kinematic setting. Out of the many 
kinematic settings, Figure~\ref{fig:alphaALU} (left) shows the result of fitting the 
$\phi$ asymmetry for one bin in $x$, $Q^2$, and $t$.

For the n-DVCS measurements, the missing mass cut will select DVCS events. The 
primary assumption is what we will have already observed through the p-DVCS channel
and isolated kinematics where FSIs are minimized. Typically, this corresponds 
to backward low momentum spectators. We will match the kinematics where the FSI 
are observed to be negligible for the proton and look for nuclear effects in 
neutron.  We define the following ratio for the extracted $\alpha$ values from 
DVCS on a quasi-free neutron in $^2$H and from DVCS on a bound neutron in 
$^4$He:
\begin{equation}\label{eq:Ralpha}
   R_{\alpha}^{N} = \frac{\alpha^{(^4\text{He})}_{N^{*}}}{\alpha^{(^2\text{H})}_N}
\end{equation}
where the $N^{*}$ indicates the bound nucleon. We will identify nuclear effects 
by observing  deviations from unity in this ratio and extracting its trend as a 
function of $x$, and for various spectator kinematics limits where we expect mean 
field nucleons or SRC nucleons to dominate.
The projected statistical uncertainty is shown in Figure~\ref{fig:alphaALU} 
(right) and in Fig~\ref{fig:RalphaAllQ2}. See appendix 
\ref{sec:extraRatioProjections} for more BSA ratio projections.

\begin{figure}
   \centering
   \includegraphics[width=0.7\textwidth,trim=0mm 0mm 0mm 0mm, 
   clip]{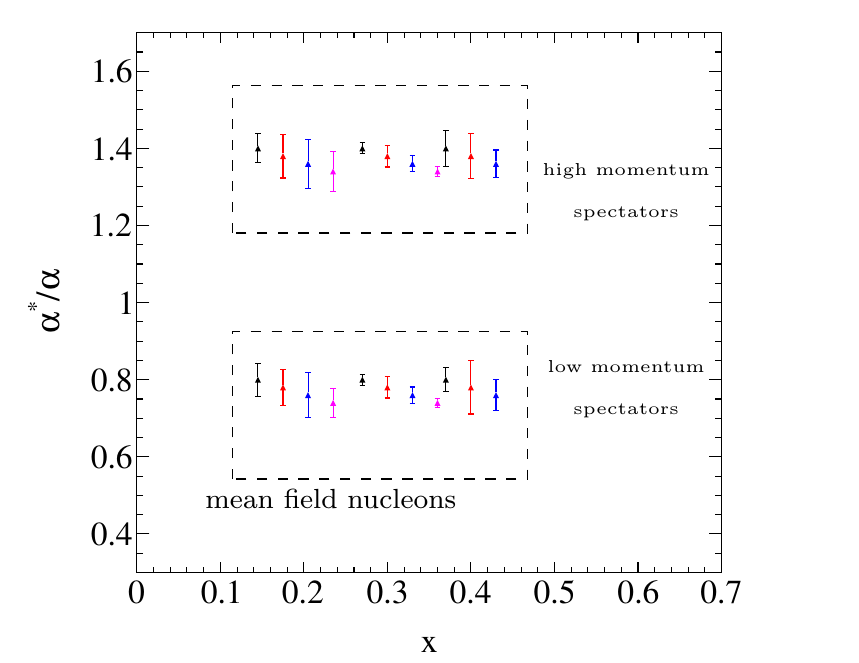}
   \caption{\label{fig:RalphaAllQ2}A subset of the $R_{\alpha}$ ratios  for all $Q^2$ with backward 
      tagged spectators. The highest spectator momenta  bins are offset vertically  above the 
 lowest spectator momenta and the colors indicate the different $t$ bins which are shifted horizontally for clarity.}
\end{figure}

\subsection{Systematic Uncertainties}

We estimate the main sources of systematic uncertainties from those ultimately obtained 
for the CLAS-eg6 experiment's incoherent DVCS measurement~\cite{eg6_note}. 
They are listed in Table~\ref{tab:systematics} along with our estimates for the beam spin
asymmetry systematics. For the BSA the beam polarization will 
dominate our systematic uncertainties followed by the DVCS event selection cuts. 
With the significant improvement of ALERT for detecting the spectator recoils this 
uncertainty is expected to improve by more than a factor of two.

The so-called ``acceptance ratio'' corrects for the $\pi^0$ background 
and is defined for each bin as
\begin{equation}\label{eq:acceptanceRatio}
   R_{\pi^0} = \frac{N_{\pi(\gamma)}}{N_{\pi(\gamma\gamma)}}
\end{equation}
where $N_{\pi(\gamma)}$ and $N_{\pi(\gamma\gamma)}$ are the rates for exclusive 
electro-production of $\pi^0$s where one decay photon is detected and 
where both decay photons are detected, respectively.
The ratio calculated in Equation.~\ref{eq:Ralpha} has the benefit that the acceptance cancels 
in the ratio  under the approximation $R_{\pi^0}(^4\text{He}) \simeq R_{\pi^0}(^2\text{H})$.

External radiative effects on the electron side can be easily understood and studied
using the over-determined kinematics. The exclusivity of the process allows tight cuts 
that remove any initial state radiation. Furthermore, much of the radiative effects will
cancel in the ratio.

\begin{table}
   \centering
   \begin{tabu}{lccc}
\tabucline[2pt]{-}                                                   
      \bf Source & \bf \quad CLAS-eg6 \quad&\quad \bf CLAS12-ALERT \quad & \bf Systematic Type\\
\tabucline[1pt]{-}                                                   
      Beam polarization    & 3.5\%    &    3.5\% & normalization \\
      DVCS event selection & 3.7\%    &    1.0\% & bin-to-bin\\
      Acceptance ratio     & 2.0\%    & $<$1.0\% & bin-to-bin\\
      Radiative Corrections& 2.0\%    & $<$1.0\% & bin-to-bin\\
      Others               & 0.1\%    &    0.1\% & \\
\tabucline[1pt]{-}                                                   
      \textbf{Total}       & 5.5\%    &     4.0\% \\
\tabucline[2pt]{-}                                                   
   \end{tabu}
   \caption{\label{tab:systematics}Estimates of the expected systematic 
   uncertainties compared to CLAS-eg6.}
\end{table}

%% file: Conclusion.tex
\setlength\parskip{\baselineskip}
\chapter{Summary and Answers to PAC44}
\label{chap:conclusion}

In this final chapter we first address issues and concern raised in the PAC44 
report about the ALERT run group. Then, following a brief summary of the 
proposed experiment, we state the beam time request. Then we conclude by 
addressing this proposal's relation to other approved experiments.

\input{PAC_Answer_numbered}

\section{Summary and Beam Time Request}
%
Spectator tagged DVCS on $^2$H and $^4$He is of critical importance for two 
reasons. First and foremost, it identifies the active nucleon in the DVCS 
process. Secondly, spectator tagging provides a handle on the initial nucleon 
momenta, {\it i.e.}, it allows us to separate the mean field nucleons from 
the short range correlated nucleons. Tagged incoherent DVCS uniquely provides  
important leverage for identifying and isolating 
final state interactions while simultaneously probing the struck nucleon
at the parton level. Furthermore, the neutron beam spin asymmetry
is very sensitive to nuclear effects (see section \ref{sec:polEMCeffect}).  
Therefore cleanly extracting the neutron DVCS beam spin 
asymmetry, as we propose to do, in both bound and quasi-free configurations, 
will produce a \emph{high impact result} from which we are able to \emph{unambiguously} conclude
that nucleons are modified in medium at the parton level.
It also allows for the systematic control over FSIs needed to 
definitively observe modified nucleons.

In order to achieve the uncertainties presented in this proposal, we need
20 days of running at 11~GeV with helium target, 20 days at 11~GeV with 
deuterium, both with 80\% longitudinally polarized beam, and 5 days of 
commissioning of the ALERT detector at
2.2~GeV with helium and hydrogen targets.

\section{Relation to other experiments}

This experiment will greatly complement many already approved experiments and
previously conducted experiments.

First, the approved E12-11-003 experiment~\cite{E1211003},
``Deeply Virtual Compton Scattering on the Neutron with
CLAS12 at 11 GeV'' is set to measure the n-DVCS beam spin asymmetry by
directly detecting the struck neutron in the reaction 
$\gamma^{*}+d\rightarrow n+\gamma+(p)$. While we intend to also measure the 
BSA through detection of the spectator proton instead, this is not the main
thrust of this proposal. We aim to observe a medium modified neutron by also 
looking at a similar reaction on the neutron with a helium target where a 
spectator $^3$He is detected.

The approved  E12-06-113 experiment~\cite{bonus12}, ``The Structure of the 
Free Neutron at Large x-Bjorken'' will measure the neutron structure function  
in DIS through a spectator tagging of a recoil proton using the BONuS12 detector.
The reaction $e+d\rightarrow e+p_{s}+(X)$ is aimed at the deuteron's quasi-free
neutron, as is our DVCS BSA with a deuteron target. However, we will also investigate the
bound neutron. Our main result will be the ratio of the BSA $\sin\phi$ harmonics
from bound and quasi-free neutrons, which is a \emph{model independent observable}.

Because we will study the FSIs through the fully detected final state, 
as highlighted throughout this proposal, we will be able to directly test the
validity of the PWIA over a wide range of spectator kinematics.  This information
will directly benefit both experiments mentioned (and many more). The knowledge of the FSIs
can be used to tune the models needed to extract the on-shell neutron structure function.
Furthermore, the neutron DVCS observable will also be sensitive to FSIs which
can be further understood with the results of this experiment.

%
%





%% file: PAC_Answer_numbered.tex
\setlength\parskip{\baselineskip}%
\section{Answers to PAC44 issues}\label{sec:PACAnswers}

\input{PAC_Answer_body}

%% file: PAC_Answer_body.tex
{ \it \textbf{Issues:}}

{ \it 
The Drift Chamber/scintillator technology needs to be demonstrated. We observe 
that a strong program of prototype studies is already underway. }

{\bf Answer:} We feel the technology has no major unknowns, wire chambers and
scintillators have been used for decades as detectors of low energy nuclei
and their properties have been well established. We present in the
proposal a conceptual design demonstrating the feasibility of the detector,
it is common practice to work on the optimization of a certain number of 
parameters after the proposal is approved. In particular, because it is easier to
fund and man a project that has an approved status than a future proposal.
Nevertheless, we remain open to discuss 
the topic in more depth if the committee has any concerns.

{ \it 
The TAC report voiced concerns about the length of the straw cell target and 
the substantial effort needed to integrate the DAQ for this detector into the 
CLAS12 DAQ. }

{\bf Answer:} The TAC and PAC44 raised concerns about the target cell. We have 
added extra discussion  in section \ref{sec:targetCell}, which includes a table 
of existing or planned targets that are similar to the one we proposed. In summary,
our proposed target is twice as wide as the ones used in the 6 
GeV era for the BONuS and eg6 run and should therefore cause no issues. Note 
that the experiment 12-06-113 (BONuS12) is approved with a longer and thinner 
target. Their design will be reviewed by JLab for their experiment readiness 
review (ERR) before the PAC45 meeting. The result of this review should settle 
the question, but in any case, we propose a safer solution based on the 
successful experiments of the 6 GeV era. 

The TAC and PAC44 raised issues regarding integration of ALERT into the CLAS12 
DAQ. First, they raised a concern that the resources necessary for this integration 
are not clearly identified. We have added text in section 
\ref{sec:jlabContributions} outlining the resources provided by each group 
and the technical support they are expected to provide. Secondly, they mentioned 
a concern about the ``substantial effort needed to integrate the DAQ for this 
detector into the CLAS12 DAQ''. We want to emphasize that the read-out systems for 
ALERT are already being used in the CLAS12 DAQ to readout Micromegas detectors. 
Therefore, we will use and build on the experience gained from these systems.

{ \it 
The proposal does not clearly identify the resources (beyond generic 
JLAB/CLAS12 effort) necessary for DAQ integration which may be a substantial 
project. }

{\bf Answer:} As mentioned above, we do not feel this contribution is major,
nevertheless we made this part clearer in the proposal.

{ \it 
During review the collaboration discovered an error in converting the 
luminosity to beam current. This resulted  in a revision that will either 
require doubling the current or the target density. The beam current change 
would require changes to the Hall B beam dump, while raising the target density 
could impact the physics reach of the experiment by raising the minimum 
momentum threshold. }

{\bf Answer:} During the PAC44 proposal submission process the wrong beam 
current was requested. It was a factor of 2 too low. This increased beam 
current brought into contention the issue of possible Hall B beam current 
limits. We chose to use the higher beam current in this new version. 
Based on discussions with the Hall-B and accelerator staff, the only
necessary upgrade necessary to run at \SI{1}{\uA} is with the Hall-B beam blocker.

{ \it 
The precise interplay between final state interactions (FSI) and the tails of 
the initial state momentum distribution in DVCS on 4He was a topic of some 
debate. The collaboration makes an argument that the excellent acceptance of 
the apparatus allows novel constraints that allow selection of kinematic ranges 
where FSI is suppressed. While the originally suggested method to unambiguously 
identify areas of FSI was revised during the review, the committee remains 
unconvinced that the new kinematic selections suggested do not also cut into 
interesting regimes for the initial state kinematics. The committee believes 
that this is model dependent and would like to see more quantitative arguments 
than were provided in this version of the proposal. }

{\bf Answer:} We acknowledge there was an overstatement of the possibilities 
of the Tagged-DVCS proposal on this topic, this has been corrected. We now show 
a reduction, in opposition to the complete suppression previously claimed, in 
events that differ from the PWIA result. This finding is based on a simulation 
using a simple model of FSIs together with a Monte-Carlo event generator.

{ \it 
\textbf{Summary:}

The committee was generally enthusiastic about the diverse science program 
presented in this proposal; in particular the tagged EMC studies and the unique 
study of coherent GPD's on the 4He nucleus. However, the substantial 
modifications made in the proposal during review indicate that it could be 
substantially improved on a reasonably short time scale. We would welcome a new 
proposal that addresses the issues identified by the committee and by the 
collaboration. }

{\bf Answer:} We hope that the new proposals will answer all the questions 
raised by the PAC44 and will make the physics case even more compelling.

{ \it 
We also note that there are multiple experiments, proposed and 
approved, to study the EMC effect, including several with novel methods of 
studying the recoil system. We appreciate the comparisons of recoil 
technologies in this proposal and would welcome a broader physics discussion of 
how the proposed measurements contribute to a lab-wide strategy for exploring 
the EMC effect. }

{\bf Answer:} While no strategy document has been drafted after them, we want to point out
to the PAC that the community of physicist interested by the partonic 
structure of nuclei meets regularly, with often a large focus on what can be done at JLab
(see workshops at Trento\footnote{New Directions in Nuclear Deep Inelastic Scattering \url{http://www.ectstar.eu/node/1221}}, 
Miami\footnote{Next generation nuclear physics with JLab12 and EIC \url{https://www.jlab.org/indico/event/121/}}, 
MIT\footnote{Quantitative challenges in EMC and SRC Research and Data-Mining \url{http://web.mit.edu/schmidta/www/src_workshop/}}, 
and Orsay\footnote{Partons and Nuclei \url{https://indico.in2p3.fr/event/14438/}} for example). 
Nonetheless, we added
in the tagged EMC proposal summary an extension about the 12~GeV approved experiments  
related to the EMC effect. This short annex will hopefully clarify the context 
and the uniqueness of the present experiments.

%% file: appendix.tex
\appendix

\input{tagged_dvcs_kinematics}

\input{DVCS_and_GPD_formalism}

\input{detailed_projections}

%% file: tagged_dvcs_kinematics.tex
\setlength\parskip{\baselineskip}%
-------------------------------------------------------------------------------
\chapter{The Kinematics of Spectator-Tagged DVCS}\label{chap:appendixKine}

This appendix defines and discuses the kinematics of spectator-tagged DVCS. We 
will begin by defining the basic kinematic variables and the plane-wave impulse 
approximation (PWIA). This is followed by an analysis of the fully exclusive 
kinematics where all final-state particles are detected and a discussion of how 
to leverage this extra information for studying FSIs. 

\section{Incoherent DVCS Kinematic Variables}

\subsection{Experimentally Measured Variables}

\begin{figure}
   \centering
   \includegraphics[width=0.60\textwidth]{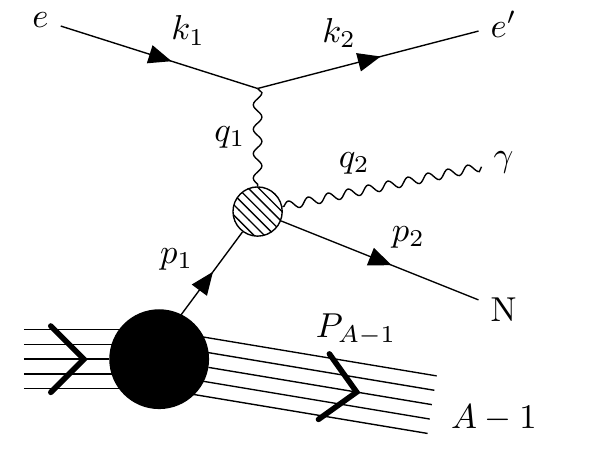}
\caption{\label{fig:appendDVCSkinematics}Tagged incoherent DVCS with labeled momenta.}
\end{figure}

The four-momenta in the tagged incoherent DVCS reaction are defined in 
Figure~\ref{fig:appendDVCSkinematics}.
The momenta are explicitly
\begin{align}
  k_1&=(k_1 ,\bm{k}_1 \simeq k_1^0) &  \label{eq:momenta1}
   k_2&=(k_2 ,\bm{k}_2 \simeq k_2^0) && \text{for $e$ and $e^{\prime}$,}\\
   q_1&=(\nu_1,\bm{q}_1) &
   q_2&=(\nu_2,\bm{q}_2) && \text{for $\gamma^{*}$ and $\gamma$,}\\
   p_1&=(E_1,\bm{p}_1) &
   p_2&=(E_2,\bm{p}_2)&& \text{for initial and struck nucleon,}\\
   p_{A}&=(M_A,\bm{0}) & p_{A-1}&=(E_{A-1},\bm{p}_{A-1})  && \text{for target 
   and spectator nucleus,}\label{eq:momenta2}
\end{align}
and the virtual and real photon momenta are
\begin{align}
   |\bm{q}_1| &= \sqrt{Q^2+\nu_1^2} &\quad& \text{and} & |\bm{q}_2| &= \nu_2\,\text{.} &
\end{align}
The virtual photon energy and four-momentum squared are
\begin{align}
  \nu_1 &= k_1^0 - k_2^0 &\quad& \text{and} & Q^2 &= -q_1^2 = -(k_1-k_2)^2 \simeq 4 k_1^0k_2^0 \sin^2(\theta_{k_1k_2}). &
\end{align}

For the remainder of this chapter we will be considering the process where all 
particles are detected, i.e., $^4$He$(e,e^{\prime}\,\gamma\, p\,+^3$H$)$. The 
incident and scattered electron momenta are experimentally well-determined, thus, 
the virtual photon four-momentum is well defined. The real photon energy and 
direction is measured in the electromagnetic calorimeter and the struck nucleon 
is also detected in the forward CLAS12 detector. The spectator system is 
identified in ALERT, which also measures its momentum. And finally the initial 
nucleus is at rest with mass $M_A$.
Therefore all the momenta in equations \ref{eq:momenta1}--\ref{eq:momenta2} 
(and Figure~\ref{fig:appendDVCSkinematics}) are determined with the exception 
of the initial struck nucleon, $p_1$.  We will return to determining this in 
section~\ref{sec:PWIA}.

\subsection{Momentum Transfer}

\begin{figure}[!htb]
   \centering
   \includegraphics[width=0.40\textwidth]{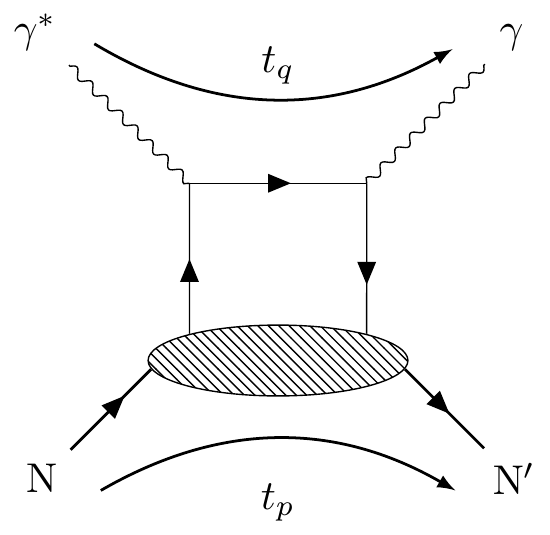}
   \caption{\label{fig:Handbag2}The DVCS handbag diagram showing the two ways 
   the momentum transfer can be calculated. }
\end{figure}

The Mandelstam variable $t$ is the square of the momentum transfer and can 
be calculated on the photon side of the diagram or the hadron side of the diagram 
as illustrated in Figure~\ref{fig:Handbag2}.
We define the former as
\begin{align}
   t_q &= (q_1-q_2)^2 \\
     &= -Q^2 -2\nu_2(\nu_1-q_1\cos\theta_{q_1q_2}) \label{eq:tqDef}
\end{align}
and the latter as
\begin{align}
   t_p &= (p_1-p_2)^2 \\
       &= 2M^2 -2(E_1E_2 - \bm{p}_1\cdot\bm{p}_2)\label{eq:tpDef}.
\end{align}

\subsubsection{Nucleon at Rest}

For DVCS on a fixed proton target, the momentum transfer can be calculated
from the virtual photon momentum, $q_2$, and the \emph{direction} of the 
real photon,
$\hat{\bm{q}}_2=(\cos\phi_{q_2}\sin\theta_{q_2},\sin\phi_{q_2}\sin\theta_{q_2},\cos\theta_{q_2})$.
The momentum transfer squared in this case is calculated as
\begin{equation}\label{eq:tgammagamma}
   \boxed{ t_{\gamma\gamma} = \frac{-Q^2 -2\nu_1( \nu_1 - q_1 \cos\theta_{q_1q_2})}{1+(\nu_1-q_1 \cos\theta_{q_1q_2})/M}}
\end{equation}
where the angle between the virtual and real photons is $\theta_{q_1q_2}$.  

\begin{derivation}[$t_{\gamma\gamma}$ for nucleon at rest.]
  Equation~(\ref{eq:tgammagamma}) is obtained from equation~(\ref{eq:tPhotons}) and 
the real photon's energy determined using the initial nucleon at rest. 
Using the struck nucleon's invariant mass we get
\begin{align*}
   p_2^2 &= M^2\\
         &= M^2 -Q^2 + 2\left(M_1(\nu_1-\nu_2)-\nu_1\nu_2 + \bm{q}_1\cdot\bm{q}_2\right).
\end{align*}
This equation becomes
\begin{align}
   \frac{Q^2}{2} &= \nu_1M_1-\nu_2M_1-\nu_1\nu_2 + \bm{q}_1\cdot\bm{q}_2
\end{align}
which can be solved for $\nu_2$ to yield 
\begin{equation}\label{eq:nu2AtRest}
  \boxed{\nu_2 = \frac{Q^2/2-\nu_1M}{|\bm{q}_1|\cos\theta_{q_1q_2} - M -\nu_1}\,.}
\end{equation}
Putting (\ref{eq:nu2AtRest}) into (\ref{eq:tqDef}) yields the result of (\ref{eq:tgammagamma}).
\end{derivation}

\subsubsection{Bound Nucleon with Fermi Motion}\label{sec:appendxBoundNfermi}

Equation~(\ref{eq:tgammagamma}) is a special case of the more general situation 
where the initial nucleon is not at rest in the lab frame. This is the case 
for a nucleon with non-zero Fermi motion or an electron-proton collider lab 
frame.  Unlike the nucleon at rest case we cannot eliminate \emph{both} $p_1$ 
and $p_2$, instead, we have only the option of eliminating one. This is not a 
problem for an electron-proton collider, where $p_1$ is constant, since we can 
just boost to the frame with $p_1=(M,\bm{0})$ and the analysis can be carried 
out consistently. However, a bound nucleon in a nucleus makes for a lousy 
collider because every scattering event would require a unique analysis frame.

It should be emphasized that the two possible choices above lead to a unique 
opportunity for studying tagged DVCS where the final state is fully detected.  
We will return to this in section \ref{sec:appendixFSIs} only after defining 
the PWIA in section \ref{sec:PWIA}.

\subsection{\texorpdfstring{$t_{\text{min}}$}{t-min} and \texorpdfstring{$t_{\text{max}}$}{t-max}}

\begin{figure}
  \centering
  \includegraphics[width=0.60\textwidth]{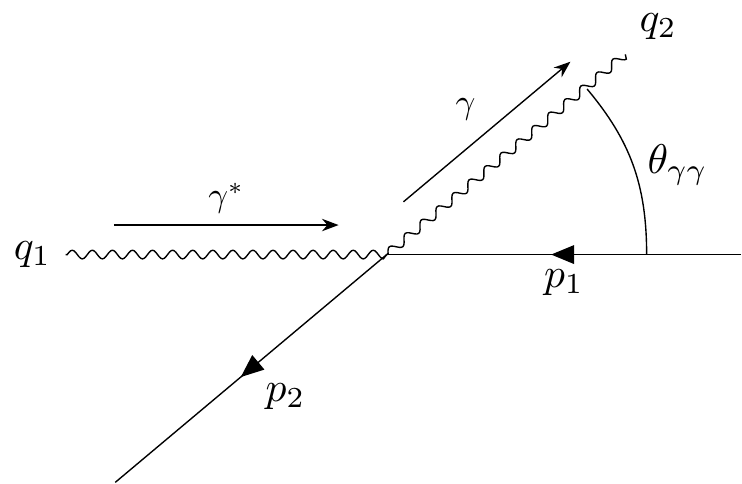}
\caption{\label{fig:NgammaCM}Nucleon-photon center-of-momentum system.}
\end{figure}

The minimum and maximum momentum transfer are easily understood in the virtual 
photon-nucleon center-of-momentum (CM) frame which is shown in 
Figure~\ref{fig:NgammaCM}.
We begin by deriving the real photon's energy in this frame which will be 
useful for deriving further relations between frames.

\begin{derivation}[Real photon energy in the CM frame.]
The center-of-mass energy squared calculated from the final state momenta
is
\begin{align}
   s &= M^2 + 2(E_2\nu_2+|\bm{p}_2||\bm{q}_2|) \\
          &= M^2 + 2\nu_2(\sqrt{\nu_2^2+M^2}+\nu_2) \label{eq:sCMnu2}
\end{align}
where we have used the CM relation $\bm{p}_2=-\bm{q}_2$ and the
fact that the final state nucleon and photon are both on-shell.
Solving (\ref{eq:sCMnu2}) for $\nu_2$ yields
\begin{equation}
   \boxed{ \nu_2^{CM} = \frac{s-M^2}{2\sqrt{s}}}
\end{equation}
where we label the result explicitly as a CM value.
\end{derivation}

The minimum momentum transfer corresponds to the scenario in the CM system
where the virtual photon loses just enough momentum as to become on-shell. That 
is, it transfers only enough momentum to become a real photon and continues to 
propagate in the same direction.  This corresponds to the case where 
$\theta_{\gamma\gamma}=0$.
Using this value in (\ref{eq:tqDef}) gives
\begin{equation}\label{eq:tminCM}
   \boxed{t_{\text{min}} = -Q^2 - \left(\frac{s-M^2}{\sqrt{s}}\right)(\nu_1^{CM}- q_1^{CM})}
\end{equation}
where we have explicitly labeled the frame dependent quantities. Quite similarly, the
maximum momentum transfer corresponds to the case where the particles scatter 
in the opposite direction of their initial momentum. This corresponds to 
$\theta_{\gamma\gamma}=\pi$, yielding the maximum momentum transfer
\begin{equation}\label{eq:tmaxCM}
   \boxed{t_{\text{max}} = -Q^2 - \left(\frac{s-M^2}{\sqrt{s}}\right)(\nu_1^{CM}+q_1^{CM})}.
\end{equation}

As a check, in the case of a real initial photon ($Q^2=0$),
\begin{align}
  t_{\text{min}} &\rightarrow 0  & \text{and} t_{\text{max}}&\rightarrow 
  \left(\frac{M^2-s}{\sqrt{s}}\right)2\nu_1^{CM}.
\end{align}
In the high energy limit were $M^2 \ll s$ or in the massless case  where 
$M$ terms are neglected we find 
\begin{align}
   s &\rightarrow (2 \nu_1^{CM})^2 & \text{and} && 
   t_{\text{max}}&\rightarrow -s
\end{align}
where we find the maximum momentum transfer is simply all the available momentum.

We now need relations for the CM energies between the photon-nucleon CM frame, lab frame, and the frame 
where the initial nucleon is at rest. The CM energy squared in each of these frames is
\begin{align}
   s &= -Q^2 +M^2 +2\left(\nu_1 E_1 - |\bm{q}_1||\bm{p}_1| \cos\theta_{p_1q_1}\right) && \text{any frame} \label{eq:sLabFrame}\\
   s &= -Q^2 +M^2 +2\left(\nu_1^{R} M\right) && \text{$p_1$ rest frame}\label{eq:sRestFrame}\\
   s &= -Q^2 +M^2 +2\left(\nu_1^{CM}\sqrt{|\bm{q}_1^{CM}|^2+M^2} - |\bm{q}_1^{CM}|^2\right) && \text{CM frame}\label{eq:sCMFrame}
\end{align}
where the rest frame and CM frame variables are labeled with $R$ and $CM$, respectively,
while the lab frame variables are not labeled.

Using (\ref{eq:sLabFrame}) and (\ref{eq:sRestFrame}) we find the relation between the 
nucleon rest frame and the lab
\begin{align}
   \nu_1^{R} &= \frac{1}{M}\left(\nu_1 E_1 - |\bm{p}_1||\bm{q}_1|\cos\theta_{p_1q_1} \right) && \text{Lab to nucleon rest frame} \label{eq:nu1R}
\end{align}
and similarly for (\ref{eq:sCMFrame}) and (\ref{eq:sRestFrame})
\begin{align}
\nu_1^{R} &= \frac{1}{M}\left(\nu_1^{CM} 
\sqrt{|\bm{q}_1^{CM}|^2+M^2}+|\bm{q}_1^{CM}|^2 \right) && \text{CM to nucleon 
rest frame.} \label{eq:nu1R2}
\end{align}
Equation \ref{eq:nu1R2} can be turned around and solved for $\nu_1^{CM}$ since 
we need it to calculate the kinematic limits above.
This gives
\begin{equation}
   \boxed{\nu_1^{CM} = 
   \sqrt{\frac{\left(M\nu_1^{R}-Q^2\right)^2}{M^2+2M\nu_1^{R}-Q^2}}}
\end{equation}
which, along with (\ref{eq:nu1R}), can be quite useful for evaluating $t_{min}$ 
and $t_{max}$.

\section{Plane Wave Impulse Approximation}\label{sec:PWIA}

In the following sections we discuss the plane-wave impulse approximation and 
how it \emph{provides a framework for comparison}, even for 
kinematics where it is not expected to apply. We conclude with a detailed 
discussion of the kinematic issues raised around about Fermi motion in 
section~\ref{sec:appendxBoundNfermi}.

\subsection{PWIA Definition}

The plane-wave impulse approximation is a simple model for calculating   an 
incoherent scattering from a bound nucleon.  The PWIA assumes 
\cite{Whelan:1339352}~i) the virtual photon is absorbed by a single nucleon, 
and ii) this nucleon is also the nucleon detected, and iii) this nucleon leaves 
the nucleus without interacting with the A-1 spectator system.  This implies 
the recoiling spectator system has a momentum opposite that of the initial 
struck nucleon,
\begin{equation}\label{eq:p1offshell}
   \bm{p}_1 = -\bm{p}_{A-1}.
\end{equation}
Furthermore, this approximation also implies that the spectator system is 
on-shell, {\it i.e.},
\begin{equation}
   E_{A-1}=\sqrt{|\bm{p}_{A-1}|^2 - M_{A-1}^2}.
\end{equation}
Noting  the initial nucleon can be off-shell, we introduce the following 
definition of the initial nucleon's invariant mass
\begin{equation}
   p_1^2 = \bar{M}^2 \ne  M_N^2.
\end{equation}
The ``off-shellness'' of the struck nucleon is typically characterized by 
$0.7 \lesssim \bar{M}/M_N < 1$.

\subsection{FSI and Off-shellness}\label{sec:appendixFSIs}

From its definition, the PWIA implies all the ``off-shellness'' goes with 
initial nucleon. Where this not the case, the spectator system would be left  
off-shell, and thus, necessitate some final state interaction to put it 
on-shell prior to detection. So here we should emphasize that the PWIA is not 
used throughout this proposal because the authors think it is a correct or even 
a good approximation, but rather, \emph{because it provides a basis for 
comparison}.

\subsection{Measuring Off-shellness in the PWIA}

\begin{figure}
  \centering
   \includegraphics[width=0.31\textwidth]{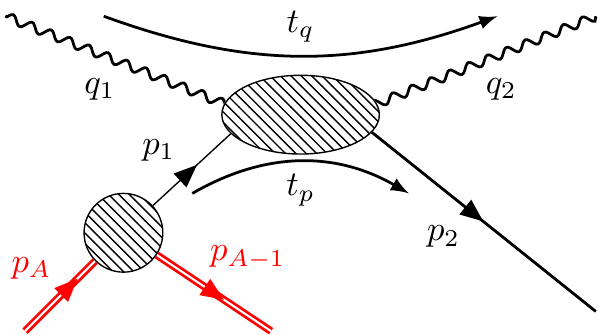}\quad
   \includegraphics[width=0.31\textwidth]{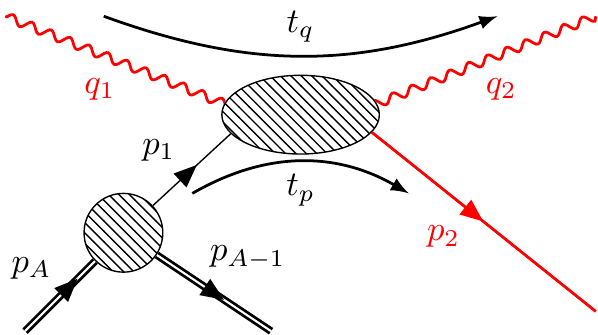}\quad
   \includegraphics[width=0.31\textwidth]{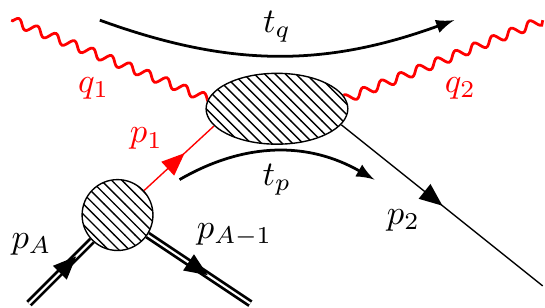}
\caption{\label{fig:offshelldiagram}Highlighted in red are the momenta used to 
calculate the off-shell mass. The diagrams from left to right correspond to  
$\bar{M}_{(0)}$ (\ref{eq:Mbar0}), $\bar{M}_{(1)}$ (\ref{eq:Mbar1}), and 
$\bar{M}_{(2)}$ (\ref{eq:Mbar2}).  }
\end{figure}

The off-shell mass of the nucleon can be determined two different ways with the 
PWIA and a fully detected final state. Starting first with the direct approach 
using the spectator
\begin{align*}
  \bar{M}^2_{(0)} &= (p_A - p_{A-1})^2 \\
                  &= M_A^2 + M_{A-1}^2 - 2 M_A E_{A-1}.\label{eq:Mbar0}\numberthis
\end{align*}
The momenta used in this calculation are highlighted in 
Figure~\ref{fig:offshelldiagram}'s left diagram. The second way to calculate 
the off-shell mass is to use the invariant $p_1^2$ with all the other momenta 
not used in the previous calculation. This yields
\begin{align}
   \begin{split}
     \bar{M}^2_{(1)}(q_1,q_2,p_2) &= M^2  - Q^2 + 2 E_2 (\nu_1 + \nu_2) \\
             &\quad- 2 \left(\nu_1 \nu_2 +  \nu_2 |\bm{p}_2|\cos\theta_{p_2q_2} - \nu_2 |\bm{q}_1| \cos\theta_{q_1q_2} + 
   |\bm{p}_2| |\bm{q}_1|\cos\theta_{p_2q_1}\right)\label{eq:Mbar1}.
   \end{split}
\end{align}
The last way we calculate $\bar{M}$ is to start with the struck nucleon 
invariant mass to eliminate $p_2$ from the expression.  This results in a 
slight more complicated expression:
\begin{align}
  \bar{M}^2_{(2)}(q_1,q_2,p_1) &= \frac{1}{2(\nu_1-\nu_2)}\sqrt{ 
  (a_{\bar{M}}+Q^2+2\bm{q}_1\cdot\bm{p}_1)(b_{\bar{M}}+Q^2+2\bm{q}_1\cdot\bm{p}_1) 
} \label{eq:Mbar2} \\
   \intertext{where}
   a_{\bar{M}} &= 2\nu_1(\nu_2+|\bm{p}_1|)- 
   2\nu_2|\bm{p}_1|(\cos\theta_{p_1q_2}+1+\frac{|\bm{q}_1|}{|\bm{p}_1|}\cos\theta_{q_1q_2})\label{eq:offshellnessA}\\
   b_{\bar{M}} &= 2\nu_1(\nu_2-|\bm{p}_1|)- 
   2\nu_2|\bm{p}_1|(\cos\theta_{p_1q_2}-1+\frac{|\bm{q}_1|}{|\bm{p}_1|}\cos\theta_{q_1q_2}).
\end{align}
The initial nucleon momentum, $\bm{p}_1$, is calculated from the target and 
spectator nuclei using the PWIA.
It is worth noting that $\bar{M}_{(1)}$ does not depend on $p_1$ and  
$\bar{M}_{(2)}$  does not depend on $p_2$. The dependent momenta for each 
calculation are shown in Figure~\ref{fig:offshelldiagram}.

\subsection{Photon Energy as FSI Indicator}\label{sec:nu2FSIindicator}

\paragraph{Calculating $\bm{\nu_2}$.}
Using the over-determined kinematics we can calculate the real photon energy two 
different ways.
\begin{derivation}[$\nu_2(p_2,\bm{\hat{q}}_2, \bar{M}, q_1)$]
  \begin{align*}
    p_2^2 &= M^2\\
          &= -Q^2 +\bar{M}^2 + 2\left(E_1(\nu_1-\nu_2)-\nu_1\nu_2- (\bm{q}_1 - \bm{q}_2)\cdot\bm{p}_1 + \bm{q}_1\cdot\bm{q}_2\right).\numberthis
  \end{align*}
  The last equation becomes
  \begin{align}
    \frac{M^2 - \bar{M}^2 + Q^2}{2} &= \nu_1E_1-\nu_2E_1-\nu_1\nu_2-\bm{q}_1\cdot\bm{p}_1 + \bm{q}_2\cdot\bm{p}_1 + \bm{q}_1\cdot\bm{q}_2
  \end{align}
  which can be solved for $\nu_2$ to yield \begin{equation}\label{eq:nu2A}
    \boxed{
      \nu_2^{(1)} = 
      \frac{(M^2-\bar{M}^2+Q^2)/2-\nu_1E_1+|\bm{q}_1|\,|\bm{p}_1|\cos\theta_{p_1q_1}}
    {|\bm{q}_1|\cos\theta_{q_1q_2} + |\bm{p}_1|\cos\theta_{p_1q_2} - E_1 -\nu_1}}~.
  \end{equation}
  In the case of an on-shell nucleon at rest in the lab 
  ($|\bm{p}_1|\rightarrow0$), (\ref{eq:nu2A}) reduces to (\ref{eq:nu2AtRest}).
\end{derivation}
\begin{derivation}[$\nu_2(p_2,\bm{\hat{q}}_2,\bar{M}, q_1)$]
  Solving for the invariant mass of $p_1$
  \begin{align*}
    p_1^2 &= \bar{M}^2\\
    \begin{split}
      &= -Q^2 +M^2 + 2\Big(\nu_2(q_1\cos\theta_{q_1q_2} -\nu_1 +E_2-|\bm{p}_2| \cos\theta_{p_2q_2})\\
      &\quad\quad -\nu_1E_2 + |\bm{q}_1||\bm{p}_2|\cos\theta_{q_1p_2}\Big)
    \end{split} \numberthis
  \end{align*}
  This becomes
  \begin{equation}\label{eq:nu2B}
    \boxed{
      \nu_2^{(2)} = \frac{(\bar{M}^2 - M^2+Q^2)/2+\nu_1E_2 
      -|\bm{q}_1||\bm{p}_2|\cos\theta_{q_1p_2}}
      {|\bm{q}_1|\cos\theta_{q_1q_2} -|\bm{p}_2| \cos\theta_{p_2q_2}-\nu_1 +E_2}.
    }
  \end{equation}
\end{derivation}

\paragraph{Calculating $\bm{t}$.}
We can now put solutions (\ref{eq:nu2A}) and (\ref{eq:nu2B}) into 
(\ref{eq:tqDef}) to obtain the analogues of $t_{\gamma\gamma}$ in 
(\ref{eq:tgammagamma}) for the case of a bound nucleon with Fermi motion as 
discussed in section \ref{sec:appendxBoundNfermi}. The results are  
\begin{equation}\label{eq:tq1}
  \boxed{t_{q}^{(1)}  = -Q^2 -2(\nu_1-|\bm{q}_1|\cos\theta_{q_1q_2})\frac{(M^2 - \bar{M}^2+Q^2)/2-\nu_1E_1+|\bm{q}_1|\,|\bm{p}_1|\cos\theta_{p_1q_1}}
  {|\bm{q}_1|\cos\theta_{q_1q_2} + |\bm{p}_1|\cos\theta_{p_1q_2} - E_1 -\nu_1}}
\end{equation}
and
\begin{equation}\label{eq:tq2}
  \boxed{ t_{q}^{(2)} = -Q^2 -2(\nu_1-|\bm{q}_1|\cos\theta_{q_1q_2})\bigg[\frac{(\bar{M}^2 - M^2+Q^2)/2+\nu_1E_2 -|\bm{q}_1||\bm{p}_2|\cos\theta_{q_1p_2}}
  {|\bm{q}_1|\cos\theta_{q_1q_2} -|\bm{p}_2| \cos\theta_{p_2q_2}-\nu_1 +E_2}\bigg]}.
\end{equation}



\paragraph{Identifying significant FSIs.}
The equations above for $\nu_2$ and $t$ require the off-shell mass, so we use 
the first PWIA result $\bar{M}_{(0)}$ in (\ref{eq:Mbar0}). To quickly summarize 
the procedure:
\begin{align}
   \big[\text{Eq.~(\ref{eq:Mbar1})}\big]  &\longrightarrow&  
   \bar{M}^{\text{calc}} &= 
   \bar{M}_{(1)}(p_2,\bm{\hat{q}}_2,\nu_2^{\text{exp}})\\
   \big[\text{Eq.~(\ref{eq:nu2B})}\big]            &\longrightarrow& 
   \nu_2^{\text{calc}} &= \nu_2^{(1)}(p_1,\bm{\hat{q}}_2,\bar{M}_{(0)}) \\
   \Big[\nu_2^{\text{calc}} \neq \nu_2^{\text{exp}},  \bar{M}^{\text{calc}} \ne \bar{M}_{(0)} \Big] &\longrightarrow &\text{PWIA} &~\text{modified by FSI.}
\end{align}
In the case where the initial nucleon is on-shell, this reduces to checking 
$\nu_2^{\text{exp}}$ against (\ref{eq:nu2B}). Furthermore, the momentum 
transfer  can be calculated from equation (\ref{eq:tq1}) and compared against 
$t_q$ to verify that we are indeed identifying those events where they differ 
significantly due to FSIs.
This analysis can be turned around, {\it i.e.}, $p_1$ and $p_2$ can be swapped 
in the procedure above:
\begin{align}
  \big[\text{Eq.~(\ref{eq:Mbar2})}\big]   &\longrightarrow& 
  \bar{M}^{\text{calc}} &= 
  \bar{M}_{(2)}(p_1,\bm{\hat{q}}_2,\nu_2^{\text{exp}}) \\
  \big[\text{Eq.~(\ref{eq:nu2A})}\big] &\longrightarrow& \nu_2^{\text{calc}}   
                                       &= 
  \nu_2^{(2)}(p_2,\bm{\hat{q}}_2,\bar{M}_{(0)}) \\
  \Big[\nu_2^{\text{calc}} \neq \nu_2^{\text{exp}},  \bar{M}^{\text{calc}} \ne 
  \bar{M}_{(0)} \Big]   &\longrightarrow  &\text{PWIA} &~\text{modified by FSI.}
\end{align}
Similarly, comparisons of (\ref{eq:tq2}) to $t_q$  can be used to to determine 
the effectiveness of selection cuts. We now will turn our attention to this 
point and try to understand things with a toy model of FSIs.

 
%
%


\section{Toy Model of FSIs}

In this section we discuss a simple toy model of FSIs which was developed in 
order to understand the usefulness of the fully measured final state.
First, we will discuss the toy model and emphasize that more theoretical work 
\emph{or experimental data} is needed before the models can be taken seriously.  
Then we will use the model to test how well the analysis outlined above 
isolates the events with kinematics that are inconsistent with the PWIA due to 
significant FSIs. 

\subsection{Modeling the FSI}

\begin{figure}[!htb]
  \centering
   \includegraphics[width=0.56\textwidth]{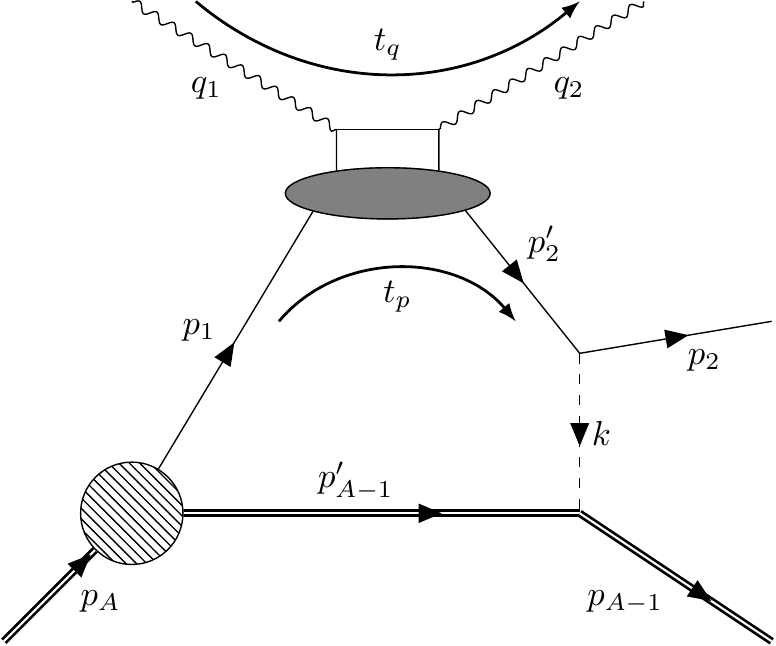}
\caption{\label{fig:toyModelFSI}A toy model of FSIs.}
\end{figure}

The FSIs were modeled as a single momentum exchange as illustrated in 
Figure~\ref{fig:toyModelFSI}. This is obviously far from realistic since any 
rigorous treatment will require amplitude-level calculations. However, as was 
already emphasized, we aim to separate those events which are no longer 
consistent with the PWIA using the momentum measured in the final state.  This 
leaves those events where the FSI exchange produces little kinematic difference 
from the PWIA but may affect the cross section at the amplitude level. With 
these things in mind, we can proceed with the details of the toy model.

To get a feel for the size of the off-shell nucleon in the PWIA, it is worth 
pointing out that the mass difference between $^4$He and $^3$H is
\begin{align}
   M_{^4\text{He}} - M_{^3\text{H}} &= (3.7284 - 2.80943) ~\text{GeV/c}^2 \\
           &= 0.91897~\text{GeV/c}^2\\
   &= 0.97945 M_p
\end{align}
and the mass difference between $^4$He and $^3$He is
\begin{align}
   M_{^4\text{He}} - M_{^3\text{He}} &= (3.7284 - 2.80941) ~\text{GeV/c}^2 \\
           &= 0.91899~\text{GeV/c}^2\\
   &= 0.97943 M_p~.
\end{align}
These differences give a rough estimate of the expected off-shellness in the 
case there are no FSIs present.

A straightforward Monte Carlo was generated, and in order simplify the present 
analysis, the virtual photon kinematics were held fixed at
\begin{align*}
  \nu_1 &= 9~\text{GeV,} &  Q^2 &= 2.65~\text{GeV}^2,
\end{align*}
where for a nucleon at rest this would correspond to  $x=0.157$.
The final state was uniformly sampled from the Lorentz invariant phase space, 
that is, there is no physics in the generated events and therefore all the 
results shown are purely a result of kinematics. However, The initial nucleon 
momentum was sampled from an empirical fit to the nucleon momentum 
distributions and the direction isotropic. The FSI momentum exchanged was 
also isotropic with a value uniformly sampled in the range of $0 < k < 
0.2$~GeV/c.

\begin{figure}[htb]
  \centering
  \includegraphics[width=0.6\textwidth,clip,trim=20mm 2mm 20mm 8mm]{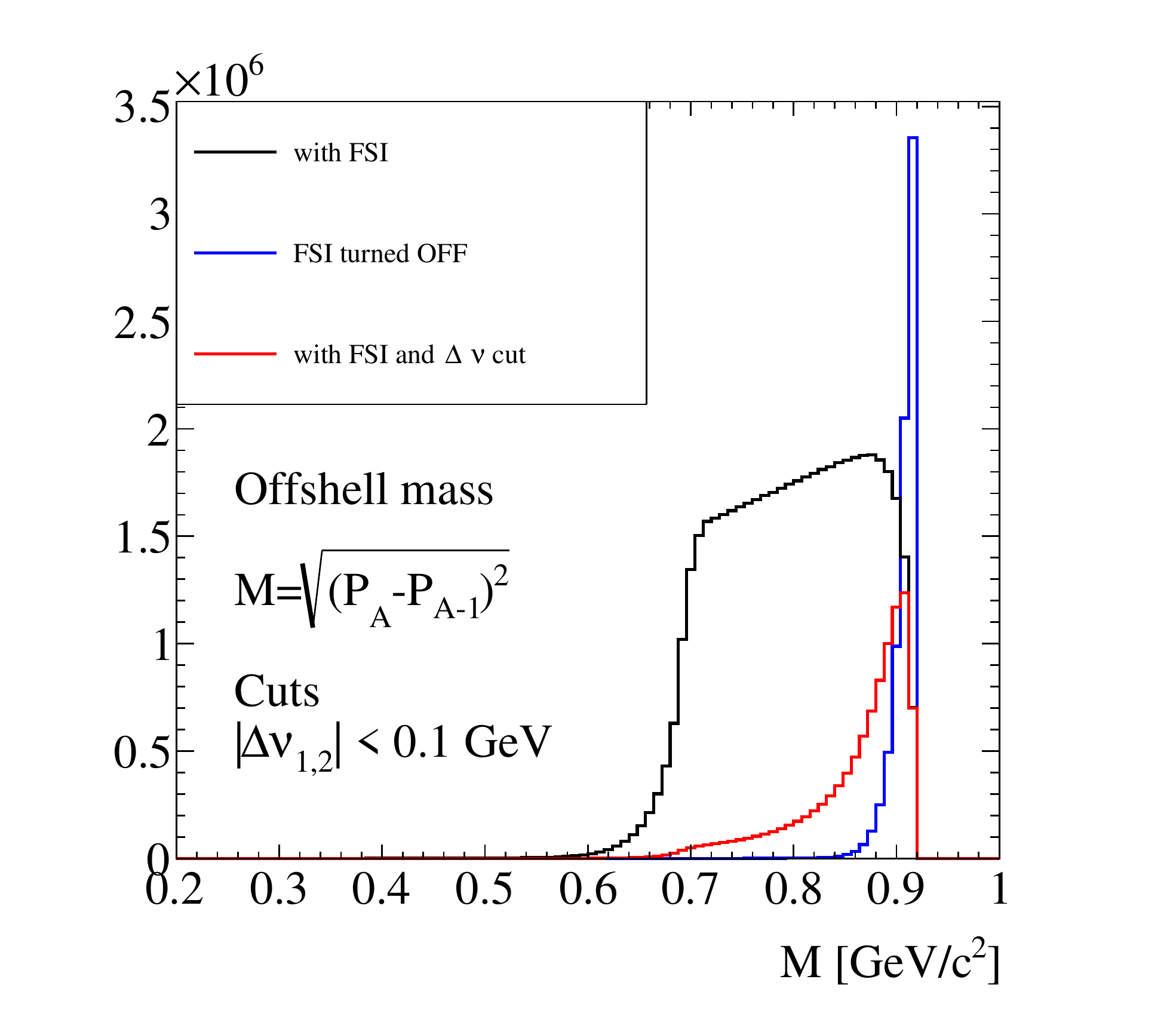}
  \caption{\label{fig:toyModel1}The initial nucleon invariant mass without FSI 
  (blue and reduced by a factor of 5), with FSI turned on (black), and with the 
selection cut (red). See text for more details.}
\end{figure}

\subsection{Toy MC Results}

\begin{figure}[!htb]
  \centering
   \includegraphics[width=0.48\textwidth,clip, trim=17mm 4mm 22mm 8mm]{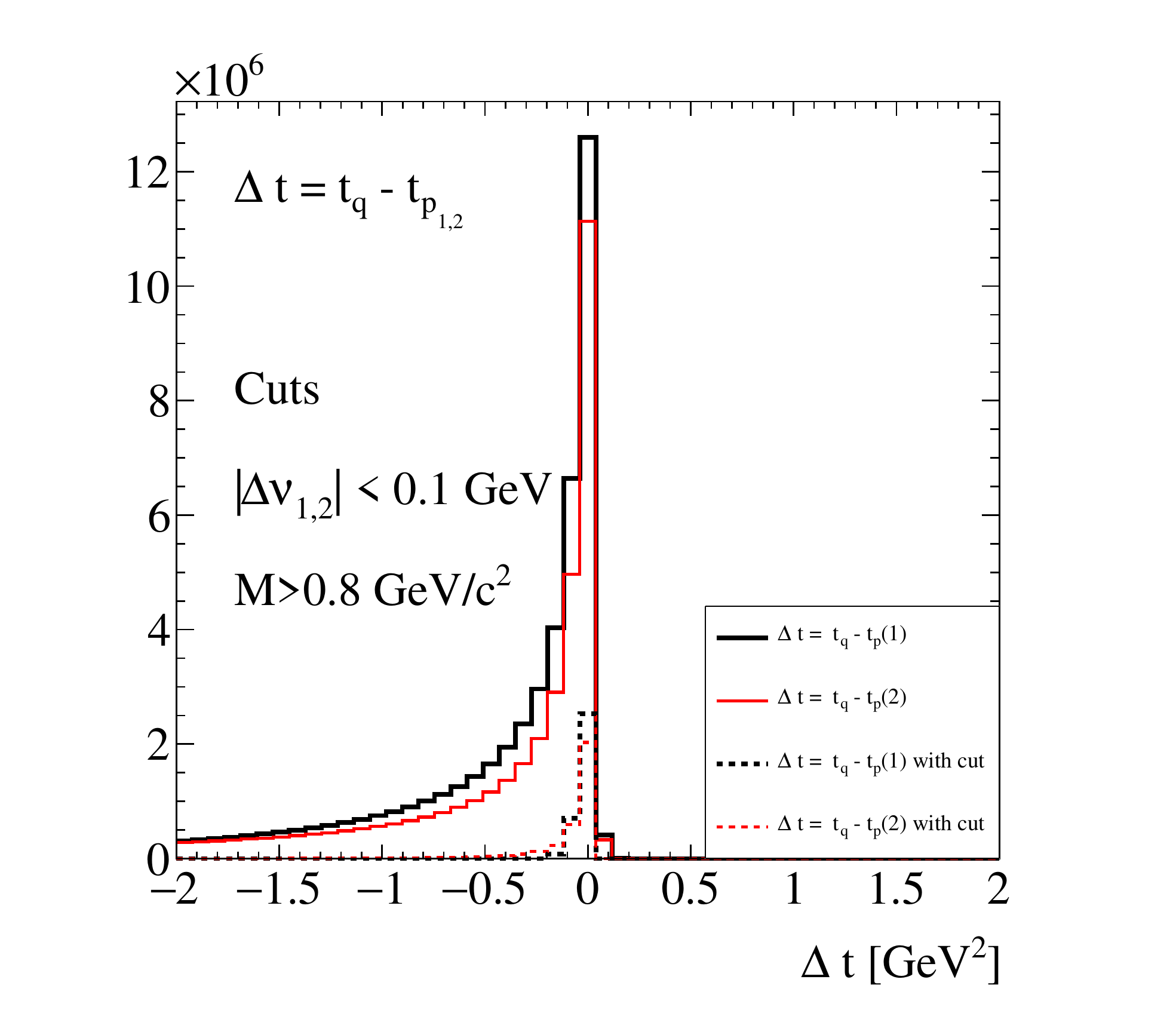}
   \includegraphics[width=0.48\textwidth,clip, trim=17mm 4mm 22mm 8mm]{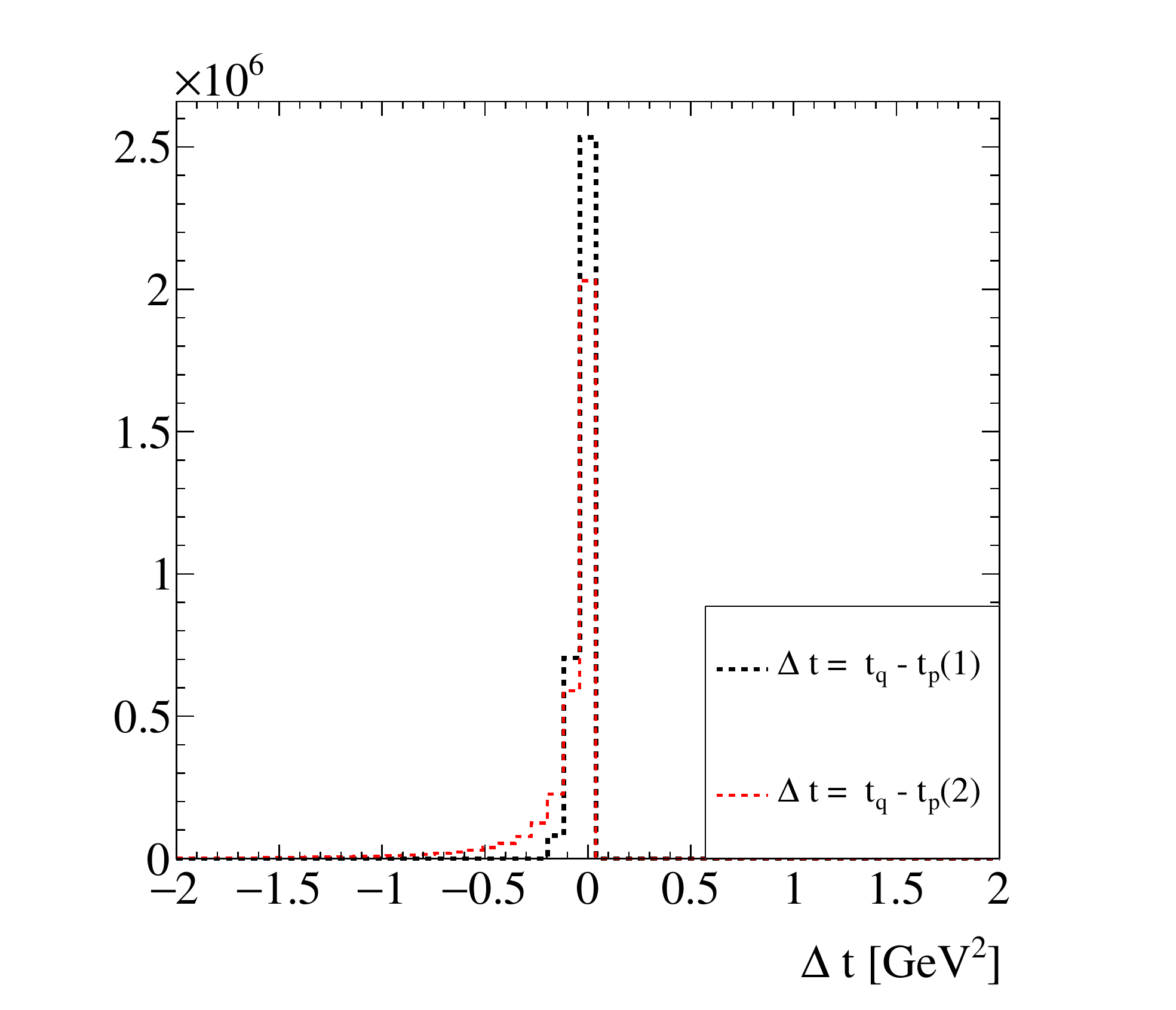}
\caption{\label{fig:toyModel3}The results of equation (\ref{eq:deltattoy}) 
without selection cuts (solid) and with selection cuts (dashed). The right plot 
shows only the results after selection cuts.}
\end{figure}

We now follow the analysis outlined section \ref{sec:nu2FSIindicator}. The 
results for the invariant mass of the initial nucleon calculated with and 
without FSIs are shown in Figure~\ref{fig:toyModel1}. Also shown in red are the  
events that pass a selection cuts:
\begin{align}\label{eq:deltaNu2}
  \Delta\nu_1^{(1)} &> 0.1~\text{GeV,} &\text{and} & &  \Delta\nu_1^{(2)} &> 
  0.1~\text{GeV.}
\end{align}
Furthermore, the results for $\Delta \nu_{(1,2)} = \nu_2^{exp} - \nu_2^{(1,2)}$
are shown in Figure~\ref{fig:toyModel3}.. The dashed histograms have a cut on 
the invariant mass $M_0 > 0.8$~GeV/c$^2$.
\begin{figure}[htb]
  \centering
    \includegraphics[width=0.6\textwidth,clip, trim=17mm 4mm 22mm 8mm]{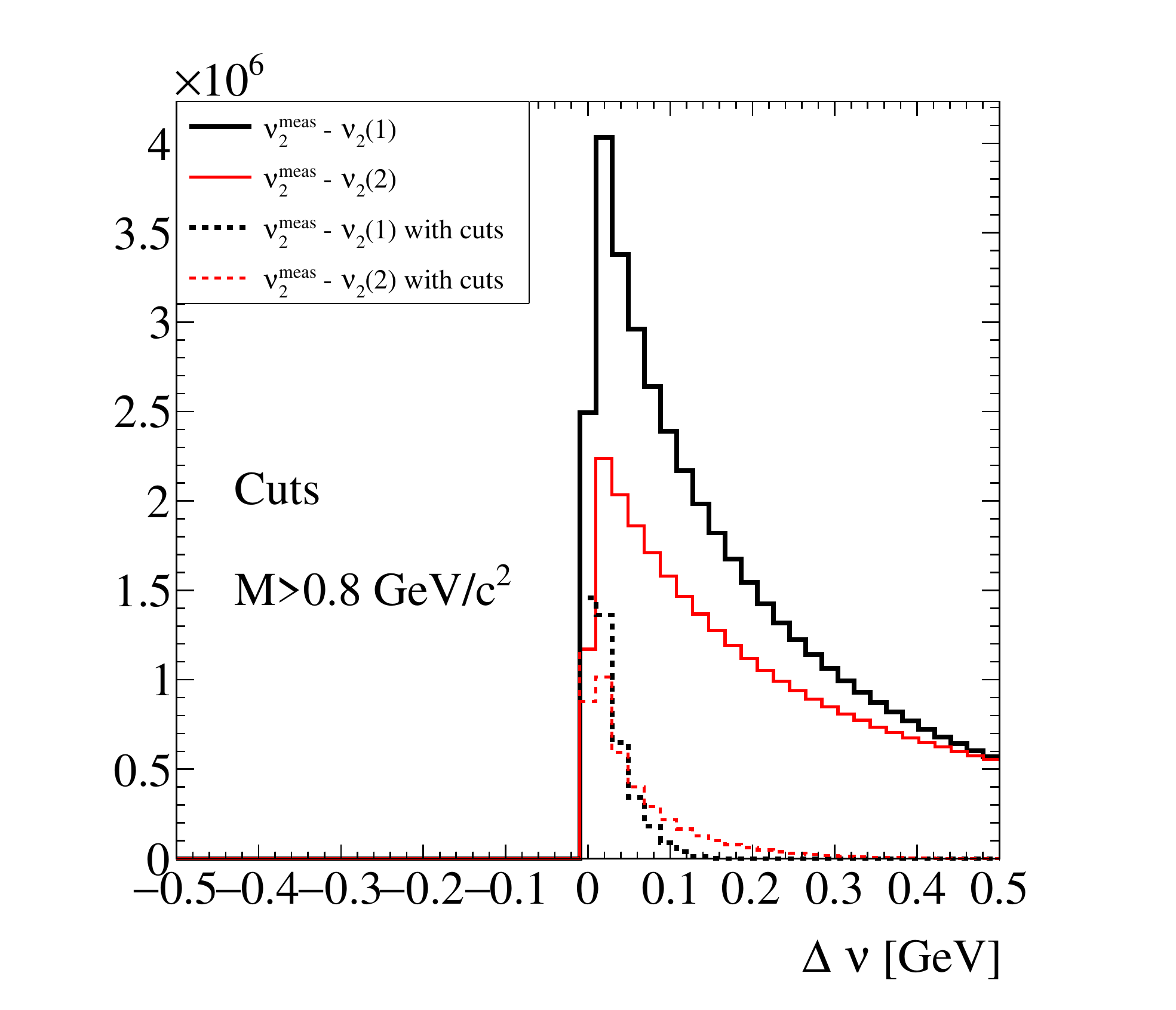}
  \caption{\label{fig:toyModel2}The real photon momentum difference as calculated 
  in equation (\ref{eq:deltaNu2}).  }
\end{figure}

The differences between the experimental momentum transfers are shown in 
Figure~\ref{fig:toyModel3}.
They are defined as
\begin{equation}\label{eq:deltattoy}
\Delta t_{(1,2)} = t_q - t_q^{(1,2)}
\end{equation}
where $t_q^{(1,2)}$ are calculated from (\ref{eq:tq1}) and (\ref{eq:tq2}) 
respectively, and $t_q$ is computed using the directly measured virtual and 
real photons.

\begin{figure}[htb]
  \centering
  \includegraphics[clip, trim=17mm 4mm 22mm 8.5mm,width=0.7\textwidth]{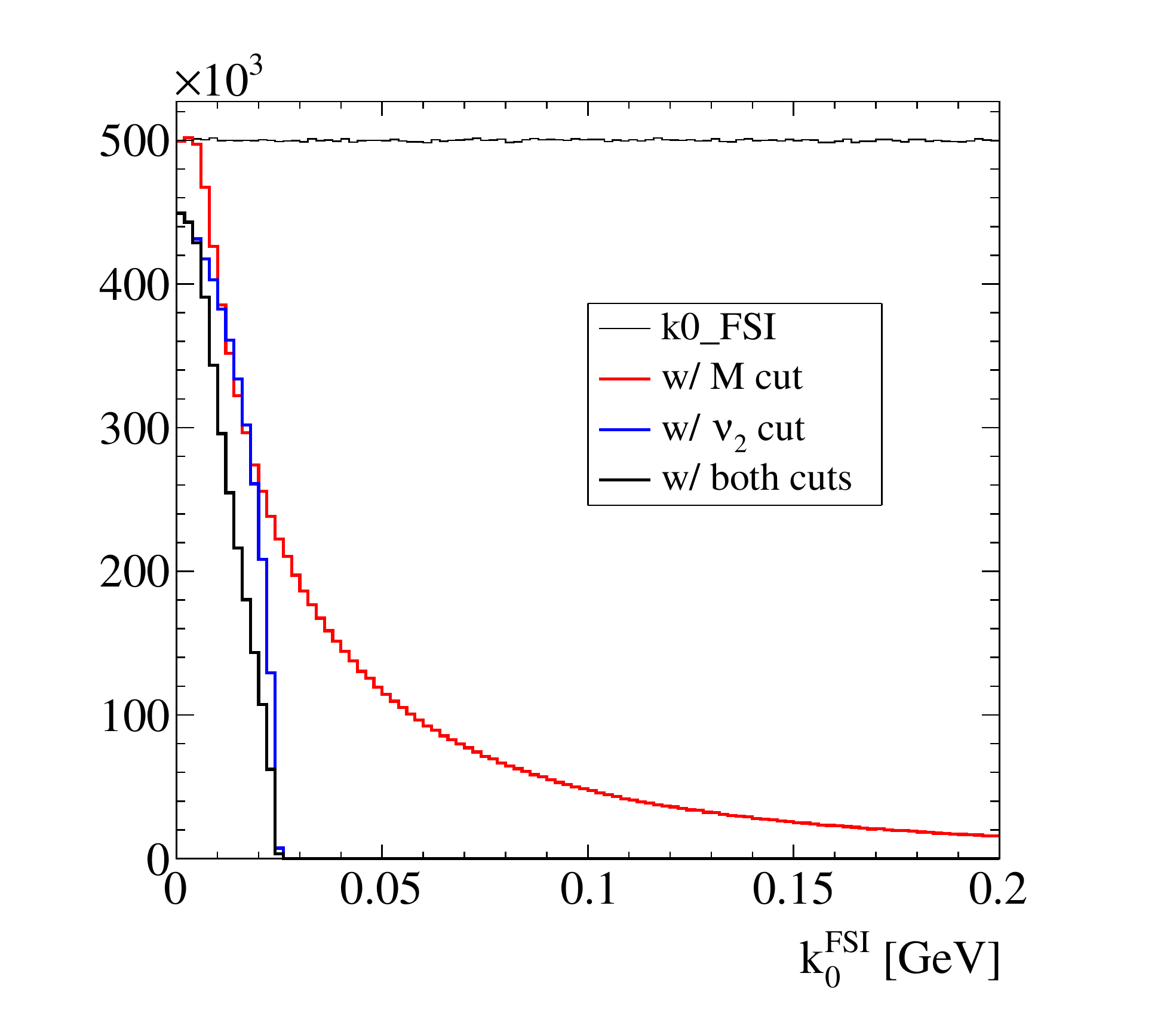}
  \caption{\label{fig:k0FSI}The generated FSI momentum exchange with different  
  cuts applied to select events with small $k_0^{\text{FSI}}$.}
\end{figure}
Figures~\ref{fig:toyModel1}, \ref{fig:toyModel2}, and \ref{fig:toyModel3} 
clearly show that
events with significant FSIs which result in kinematic differences from the 
PWIA can be isolated. The fully exclusive measurement will allow for a unique 
opportunity to study FSIs in this manner.

\begin{figure}[htb]
  \centering
  \includegraphics[clip, trim=17mm 4mm 22mm 15mm,width=0.61\textwidth]{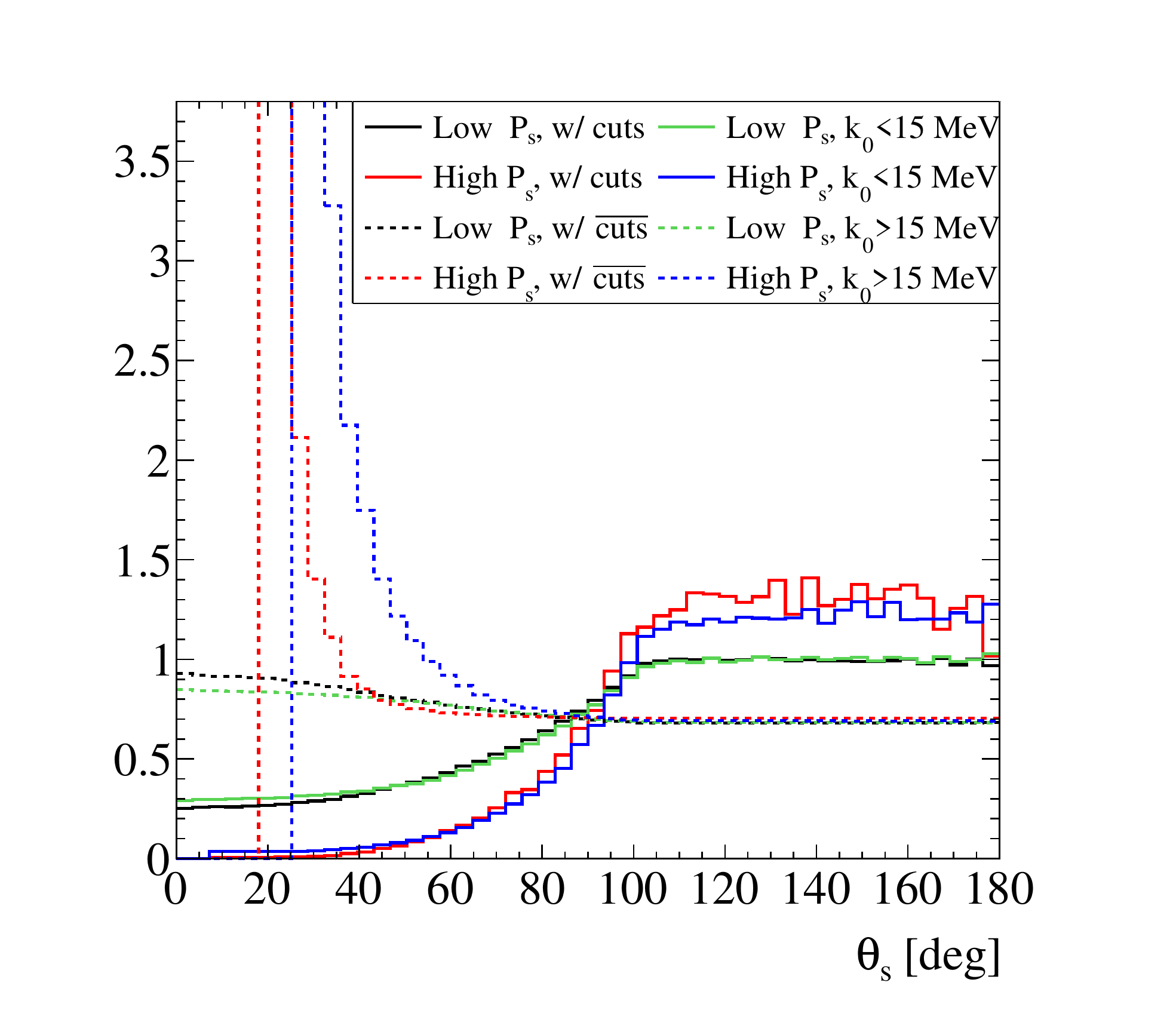}
  \caption{\label{fig:FSIRatios} Ratios (\ref{eq:AppendixFSIRatios})  for the 
    small (solid) and large (dashed) FSI momentum exchange.  The plots labeled 
    with $\overline{\text{cuts}}$ indicate the events are selected with the inverse 
    of the cut discussed above, which in this case will select events with 
  significant FSIs.}
\end{figure}

In order to see the impact of these cuts we take a look at the distribution of 
events versus spectator angle in a fixed bin of ($x$, $Q^2$, $t$, $P_s$). We 
form the ratio of distributions where the denominator is the 
distribution of all events in that bin and numerator includes the specific cuts to 
select a certain type of FSI (small or large momentum exchange). This ratio is 
defined as
\begin{equation}\label{eq:AppendixFSIRatios}
  R=A \frac{N(x,Q^2,t,P_s|\text{FSI cut})}{N(x,Q^2,t,P_s)},
\end{equation}
where $A$ is an arbitrary normalization chosen so that the ratio of backward, 
low-$P_s$, and small $k_0^{\text{FSI}}$ is about 1. The efficacy of such a cut, 
as outlined above, at removing  events with significant FSIs can be seen in 
Figure~\ref{fig:k0FSI}.  For simplicity we limit ourselves to two bins in $P_s$ 
(high and low). We can compare the result against its (unrealistic) counterpart 
by cutting on the FSI momentum exchange as well, i.e., $k_0^{\text{FSI}} < 
15$~MeV or~$k_0^{\text{FSI}} > 15$~MeV which correspond to the solid and dashed 
histograms in Figure~\ref{fig:FSIRatios}.

\subsubsection{Noteworthy Features}

\interfootnotelinepenalty=500
Here we note some observations which are not integral to this proposal but are 
rather interesting. As Figure~\ref{fig:FSIRatios} shows, the experimental cuts 
quite effectively act like a cut on the FSI momentum exchanged. Using this to 
our advantage we could take this one step further and isolate the effects for a 
given momentum exchange by systematically varying the effective 
$k_0^{\text{FSI}}$ cut (e.g. by loosening the values) and subtracting the 
difference\footnote{This is somewhat analogous to using a bremsstrahlung photon 
  beam at different energies in photo-production experiments. Taking the 
  cross-section difference at slightly different beam energies allows the 
  contribution from the high-energy tip of the bremsstrahlung spectrum to be 
isolated.}.  However, we must remember that this is just a crude model and 
reality may be quite different.  Furthermore, even in this simple example we do 
not know how to measure the value of $k_0^{\text{FSI}}$.  We only have cuts 
which are roughly proportional to a range of FSI momentum exchange with an 
unknown proportionality constant.

\subsection{Concluding Remarks}

The FSI problem cannot escape model dependence. The strategy outlined above 
uses the over-determined kinematics to get a handle on FSIs. A PWIA analysis 
permits a kinematic separation \emph{at the event level} yielding roughly two 
event types: i) events with FSI causing significant deviation from PWIA 
kinematics, and ii) events with FSI that produce no discernible difference when 
compared to the PWIA result.

The first type of event can be removed if the goal is finding maximally FSI-free 
events, however, these events are invaluable for studying various models of 
FSIs. The over-determined kinematics will give an extra handle to test various 
models. Using models in agreement with type (i) events, the events of type (ii) 
can be systematically corrected (or even justified to be negligible). More 
theoretical input is needed for accurately modeling but this does not affect 
the impact of this proposals' result.

Perhaps the ultimate extension of the experimental setup would be to measure 
the induced polarization of the struck nucleon. This would be a DVCS version of 
the quasi-elastic scattering experiments where $P_{y}$ gives a measure of FSIs.  
This too would have a model dependence but the combination would be quite 
powerful in understanding the FSIs\footnote{However, this method would require 
a new large recoil polarimeter which does not seem feasible at the moment.}.

%% file: DVCS_and_GPD_formalism.tex
\setlength\parskip{\baselineskip}%
\chapter{DVCS Formalism}

\section{Theory bound nucleon DVCS}\label{sec:GPDs}
\begin{figure}[tbp]
\centering
\includegraphics[scale=0.40]{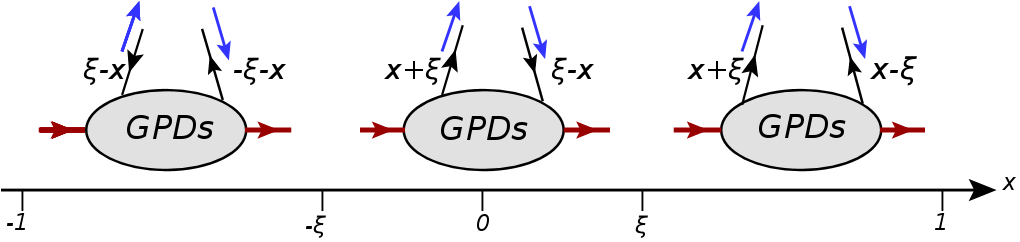}
 \caption{The parton interpretations of the GPDs in three $x$-intervals 
[-1,-$\xi$], [-$\xi$,$\xi$] and [$\xi$,1]. The red arrows indicate the initial 
and the final-state of the proton, while the blue (black) arrows represent 
helicity (momentum) of the struck quark.} \label{fig:GPDpartonicInterp}
\end{figure}

In the infinite-momentum frame, where the initial and the final nucleons go at 
the speed of light along the positive z-axis, the partons have relatively small 
transverse momenta compared to their longitudinal momenta. Referring to figure 
\ref{fig:GPDpartonicInterp}, the struck parton carries a longitudinal momentum 
fraction $x+\xi$ and it goes back into the nucleon with a momentum fraction 
$x-\xi$. The GPDs are defined in the interval where $x$ and $\xi$~$\in$ [-1,1], 
which can be separated into three regions as can be seen in figure 
\ref{fig:GPDpartonicInterp}.  The regions are:
\begin{itemize}
 \item $x\in$ [$\xi$,1]: both momentum fractions $x+\xi$ and $x-\xi$ are 
positive and the process describes the emission and reabsorption of a quark.
 \item $x\in $ [-$\xi$,$\xi$]: $x+\xi$ is positive reflecting the emission of a 
quark, while $x-\xi$ is negative and is interpreted as an antiquark being 
emitted from the initial proton.
 \item $x\in$ [-1,-$\xi$]: both fractions are negative, and $x+\xi$ and $x-\xi$ 
represent the emission and reabsorption of antiquarks.
\end{itemize}

The GPDs in the first and in the third regions represent the probability 
amplitude of finding a quark or an antiquark in the nucleon, while in the 
second region they represent the probability amplitude of finding a 
quark-antiquark pair in the nucleon \cite{Diehl:2001pm}.

\begin{figure}[tbp]
\centering
\includegraphics[scale=0.33]{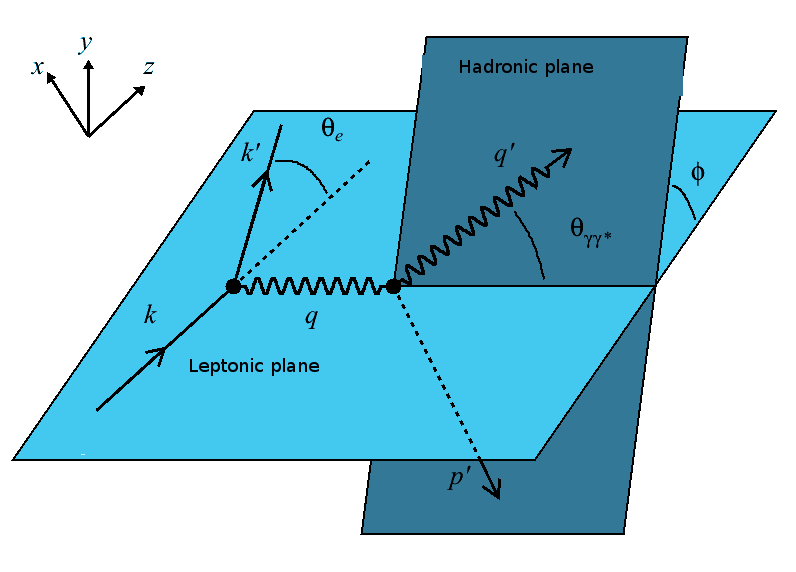}
\caption{The definition of the azimuthal angle $\phi$ between the leptonic and 
the hadronic planes.} \label{fig:phi}
\end{figure}

Following the definition of reference \cite{Ji:1998pc}, the differential DVCS 
cross section is obtained from the DVCS scattering amplitude 
($\mathcal{T}_{DVCS}$) as:
\begin{equation}\label{DVCSCrossSection_tot}
\frac{d^{5}\sigma}{dQ^{2}\, dx_{B}\, dt\, d\phi\, d\phi_{e}} = 
\frac{1}{(2\pi^{4})32}\frac{x_{B}\, 
y^{2}}{Q^{4}}\bigg(1+\frac{4M^{2}x_{B}^{2}}{Q^{2}}\bigg)^{-1/2} 
|\mathcal{T}_{DVCS}|^{2},
\end{equation}
where $\phi_{e}$ is the azimuthal angle of the scattered lepton, 
$y=\frac{E-E'}{E}$ and $Q^{2},x_{B}, t, \phi$ are the four kinematic variables 
that describe the process. The variable $\phi$ is the angle between the 
leptonic and the hadronic planes, as can be seen in figure \ref{fig:phi}.

By neglecting the mass of the quark with respect to the energies of 
$\gamma^{*}$ and $\gamma$, the DVCS scattering amplitude can be parametrized by 
four quark helicity conserving (chiral-even) GPDs: $H$, $E$, $\widetilde{H}$ 
and $\widetilde{E}$ as:
\begin{eqnarray}
\mathcal{T}_{DVCS} =  \sum_{q}(|e|Q_{q})^{2}\epsilon_{\mu}^{\ast} 
\epsilon_{\nu} \Bigg\lbrace \nonumber 
\,\,\,\,\,\,\,\,\,\,\,\,\,\,\,\,\,\,\,\,\,\, 
\,\,\,\,\,\,\,\,\,\,\,\,\,\,\,\,\,\,\,\,\,\, 
\,\,\,\,\,\,\,\,\,\,\,\,\,\,\,\,\,\,\,\,\,\,\,\,\,\,\,\,\,\,\,\,\,\,\,\,\,\,\,\,\,\,\,\, 
\,\,\,\,\,\,\,\,\,\,\,\,\,\,\,\,\,\,\,\,\,\, 
\,\,\,\,\,\,\,\,\,\,\,\,\,\,\,\,\,\,\,\,\,\,\,\,\,\,\,\,\,\,\,\,\,\,\,\,\,\,\,\,\,\,\,\,\,\\
g_{\perp}^{\mu \nu} \int_{-1}^{1} dx \,  \left[ \frac{1}{x-\xi+i\epsilon} 
+\frac{1}{x+\xi-i\epsilon} \right] \times \frac{1}{2} \bar{u}(p') \left[ H^{q} 
\gamma^{+} +  E^{q} i \sigma^{+ \alpha} \frac{\Delta_{\alpha}}{2m_{N}} \right] 
u(p)   \nonumber \\
  + i\epsilon^{\mu \nu + -} \int_{-1}^{1} dx \, \left[ 
\frac{1}{x+\xi-i\epsilon} -\frac{1}{x-\xi+i\epsilon} \right] \times \frac{1}{2} 
\bar{u}(p') \left[ \tilde{H}^{q} \gamma^{+}\gamma_{5} + \tilde{E}^{q} 
\gamma_{5} \frac{\Delta^{+}}{2m_{N}} \right] u(p) \Bigg\rbrace , \nonumber \\
 & &
\label{DVCS_Amplitude_4}
\end{eqnarray}
where $\bar{u}(p')$ and $u(p)$ are the spinors of the nucleon.
 
The GPDs $H$, $E$, $\widetilde{H}$ and $\widetilde{E}$ are defined for each 
quark flavor (q = u, d, s, ... ). Analogous GPDs exist for the gluons, see 
references \cite{Ji:1998pc,Radyushkin:1997ki,Goeke:2001tz} for details. In this 
work, we are mostly concerned by the valence quark region, in which the sea 
quarks and the gluons contributions do not dominate the DVCS scattering 
amplitude.

The GPDs $H$, $E$, $\widetilde{H}$ and $\widetilde{E}$ are called chiral-even 
GPDs because they conserve the helicity of the struck quark. The GPDs $H$ and 
$\widetilde{H}$ conserve the spin of the nucleon, while $E$ and $\widetilde{E}$ 
flip it. The $H$ and $E$ GPDs are called the unpolarized GPDs as they represent 
the sum over the different configurations of the quarks' helicities, whereas 
$\widetilde{H}$ and $\widetilde{E}$ are called the polarized GPDs because they 
are made up of the difference between the orientations of the quarks' 
helicities.

If one keeps the quark mass, another set of GPDs gives contribution to the DVCS 
amplitude. They are called chiral-odd GPDs. They give information about the 
quarks helicity-flip transitions. At leading twist, there are four chiral-odd 
GPDs that parametrize the helicity-flip structure of the partons in a nucleon: 
$H_{T}$, $E_{T}$, $\widetilde{H}_{T}$ and $\widetilde{E}_{T}$ 
\cite{PhysRevD.58.054006}.  Analogous set of chiral-odd GPDs exist for the 
gluon sector (see \cite{PhysRevD.58.054006, Diehl:2003ny}). The chiral-even 
GPDs contribute mostly in the regions where $\xi<x$ and $x<-\xi$, while the 
chiral-odd GPDs have larger contribution in the $x<|\xi|$ region 
\cite{Ji:1998pc}. 

\subsubsection{Basic properties of GPDs} \label{Properties_of_GPDs}

\paragraph{Links to the ordinary FFs and PDFs}

  Links between GPDs and the FFs are constructed by integrating the GPDs over 
the momentum fraction $x$ at given momentum transfer ($t$). Because of Lorentz 
invariance, integrating over $x$ removes all the references to the particular 
light-cone frame, in which $\xi$ is defined. Therefore, the result must be 
$\xi$-independent as can be see in equation \ref{EqGPDsFFlink}:
\begin{eqnarray}
  \int_{-1}^{1}dx \, H^{q}(x,\xi,t)=F^{q}_{1}(t), 	&  & 	\int_{-1}^{1}dx 
\, E^{q}(x,\xi,t)=F^{q}_{2}(t),  \, \, \,\, \, \, \,\, \, \, \,\,  \nonumber \\
  \int_{-1}^{1}dx \, \widetilde{H}^{q}(x,\xi,t)=G^{q}_{A}(t), 	&  & 	
\int_{-1}^{1}dx \, \widetilde{E}^{q}(x,\xi,t)=G^{q}_{P}(t), \, \, \,\, \, \, 
\,\, \, \, \,\, \label{EqGPDsFFlink}
\end{eqnarray}
 where $F^{q}_{1}(t)$ and $F^{q}_{2}(t)$ are the previously introduced Dirac 
and Pauli FFs, $G^{q}_{A}(t)$ and $G^{q}_{P}(t)$ are the axial and pseudoscalar 
electroweak FFs. The latter two can be measured in electroweak interactions; 
see reference \cite{Bernard:1996cc} for more details about the electroweak FFs.
 
\begin{figure}[tbp]
\hspace{-0.1in}
\includegraphics[scale=0.32]{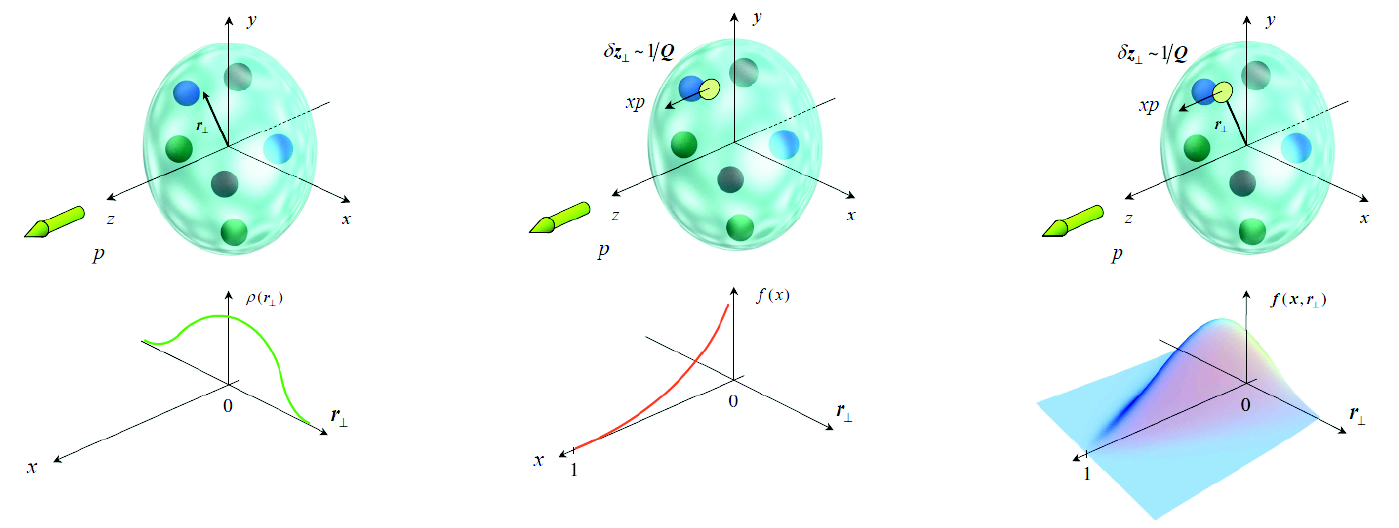}
\caption{The links between the GPDs and the ordinary FFs and PDFs. From left to 
right: the FFs reflect, via a Fourier transform, the two-dimensional spatial 
distributions of the quarks in the transverse plane; the PDFs give information 
about the longitudinal momentum distributions of the partons; finally, the GPDs 
provide a three-dimensional imaging of the partons in terms of both their 
longitudinal momenta and their position in the transverse space plane. The 
figure is from \cite{928a488bce7d45fa933f425b7213a349}. } 
\label{fig:GPDs_FFs_PDFs}
\end{figure}

From the optical theorem, the DIS cross section is proportional to the 
imaginary part of the forward amplitude of the doubly virtual Compton 
scattering (production of a spacelike ($Q^2<0$) virtual photon in the final 
state instead of a real photon) \cite{Guidal:2013rya}. In the limit $\xi 
\rightarrow $0 and t $\rightarrow$0, the GPDs are reduced to the ordinary PDFs, 
such that for the quark sector:
\begin{equation}
H^{q}(x,0,0) = q(x), ~~~~~~~~~~~~~  \widetilde H^{q}(x,0,0) = \Delta q(x),
\end{equation}
where $q(x)$ is the unpolarized PDF, defined for each quark flavor. The 
polarized PDFs $\Delta q(x)$ are accessible from polarized-beam and 
polarized-target DIS experiments. There are no similar relations for the GPDs 
$E$ and $\widetilde{E}$, as in the scattering amplitude, equation 
\ref{DVCS_Amplitude_4}, they are multiplied by factors proportional to $t~(= 
\Delta^{2}$), which vanish in the forward limit. Figure \ref{fig:GPDs_FFs_PDFs} 
summarizes the physics interpretations of the GPDs, the FFs, the PDFs, and the 
links between them.

\paragraph{Polynomiality of GPDs}

The GPDs have a key property which is the polynomiality. This property comes 
from the Lorentz invariance of the nucleon matrix elements. It states that the 
$x^n$ moment of the GPDs must be a polynomial in $\xi$ with a maximum order of 
n+1 \cite{Ji:1998pc}.
\begin{eqnarray}
         \int_{-1}^{1}dx \, x^{n} H^{q}(x,\xi,t) &=& \sum_{(even)i=0}^{n} 
(2\xi)^{i} A_{n+1,i}^{q}(t) + mod(n,2)(2\xi)^{n+1}C_{n+1}^{q}(t), ~~~~~~~~~~ \\
         \int_{-1}^{1}dx \, x^{n} E^{q}(x,\xi,t) &=& \sum_{(even)i=0}^{n} 
(2\xi)^{i} B_{n+1,i}^{q}(t) - mod(n,2)(2\xi)^{n+1}C_{n+1}^{q}(t), ~~~~~~~~~~ \\
        \int_{-1}^{1}dx \, x^{n} \widetilde{H}^{q}(x,\xi,t) &=& 
\sum_{(even)i=0}^{n} (2\xi)^{i} \widetilde{A}_{n+1,i}^{q}(t), ~~~~~~~~~~ \\
        \int_{-1}^{1}dx \, x^{n} \widetilde{E}^{q}(x,\xi,t) &=& 
\sum_{(even)i=0}^{n} (2\xi)^{i} \widetilde{B}_{n+1,i}^{q}(t). ~~~~~~~~~~
\label{Mellinmoment}
\end{eqnarray}
where $mod(n,2)$ is 1 for odd $n$ and 0 for even $n$. Thus, the corresponding 
polynomials contain only even powers of the skewedness parameter $\xi$. This 
follows from time-reversal invariance, i.e. $GPD(x,\xi,t)$ = $GPD(x,-\xi,t)$ 
\cite{Mankiewicz:1997uy}. This implies that the highest power of $\xi$ is $n+1$ 
for odd $n$ (singlet GPDs) and of highest power $n$ in case of even $n$ 
(non-singlet GPDs).  Due to the fact that the nucleon has spin 1/2, the 
coefficients in front of the highest power of $\xi$ for the singlet functions 
$H^q$ and $E^q$ are equal and have opposite signs. This sum rule is the same 
for the gluons \cite{Diehl:2003ny}.

As a consequence of the polynomiality of the GPDs, the first moments of GPDs 
lead to the ordinary form factors, as shown previously in this section. X.~Ji 
derived a sum rule \cite{PhysRevD.55.7114} that links the second moments of the 
quark GPDs $H^q$ and $E^q$, in the forward limit ($t=0$), to the total angular 
momentum ($J_{quarks}~=~\frac{1}{2} \Delta \Sigma + L_{quarks}$), where 
$\Delta\Sigma$ is the contribution of the quark spin to the nucleon spin and 
$L_{quarks}$ is the quarks orbital angular momentum contribution, as:
\begin{equation}
J_{quarks} = \frac{1}{2}\int_{-1}^{1} dx ~ x \left[H^{q}(x, \xi, t=0) + 
E^{q}(x, \xi, t=0) \right]
\label{jis}
\end{equation}
A similar expression exists for the gluons contribution ($J_{gluons}$).

The spin of a nucleon is built from the sum of the quarks' and the gluons' 
total angular momenta, $\frac{1}{2}=J_{quarks}+J_{gluons}$. Regarding the 
experimental measurements, the EMC collaboration \cite{Ashman:1989ig} has 
measured the contribution of the spins of the quarks ($\Delta \Sigma$) to the 
nucleon spin to be around 30$\%$. Therefore, measuring the second moments of 
the GPDs $H$ and $E$ will give access to the quarks orbital momentum 
($L_{quarks}$) which will complete the sector of the quarks in understanding 
the nucleon spin.  For the gluon total angular momentum ($J_{gluons}$), it is 
still an open question how to decompose $J_{gluons}$ into orbital 
($L_{gluons}$) and spin ($\Delta g$) components and to access them 
experimentally, see reference \cite{Lorce:2012rr} for more discussions on this 
subject.

\subsubsection{Compton form factors}

The GPDs are real functions of two experimentally measurable variables, $\xi$ 
and $t$, and one unmeasurable variable, $x$, in the DVCS reaction.  Therefore, 
the GPDs are not directly measurable. In DVCS what we measure are the Compton 
Form Factors (CFFs) that are linked to the GPDs. As shown in equation 
\ref{DVCS_Amplitude_4}, the DVCS scattering amplitude, at leading order in 
$\alpha_{s}$ and leading twist, contains $x$-integrals of the form, 
$\int_{-1}^{+1} dx\frac{GPD^{q}(x,\xi,t)}{x\pm\xi \mp i\epsilon}$, where 
$\frac{1}{x \pm \xi +i\epsilon}$ is the propagator of the quark between the two 
photons. The integrals can be written as:
\begin{equation}
 \int_{-1}^{+1}dx \frac{GPD^{q}(x,\xi,t)}{x \pm \xi \mp i\epsilon} = 
\mathcal{P} \int_{-1}^{1}dx \frac{GPD^{q}(x,\xi,t)}{x \pm\xi} \pm i\pi 
GPD^{q}(x = \mp \xi,\xi,t),
\end{equation}
 where $\mathcal{P}$ stands for the Cauchy principal value integral. The DVCS 
amplitude can be decomposed into four complex CFFs, such that for each GPD 
there is a corresponding CFF. For instance, for the GPD $H^{q}(x,\xi,t)$, the 
real and imaginary parts of its CFF ($\mathcal{H}(\xi, t)$) at leading order in 
$\alpha_{s}$ can be expressed as:
\begin{subequations}
\begin{align}
 \mathcal{H}(\xi, t) =  \Re e(\mathcal{H})(\xi, t) - i\pi \Im 
m(\mathcal{H})(\xi, t)~~~~~~~~~~~~~~~~~~~~~~~~~~~~~~~~~~~~~~~~~~~~\\
\text{with~~~~~~~~~~} \Re e(\mathcal{H})(\xi, t)  =  \mathcal{P} 
\int_{0}^{1}dx[H(x,\xi,t)-H(-x,\xi,t)] \, C^{+}(x,\xi)\\ \text{and ~~~~~~~~~~~} 
\Im m(\mathcal{H})(\xi, t)  =  
H(\xi,\xi,t)-H(-\xi,\xi,t),~~~~~~~~~~~~~~~~~~~~~~~~
\end{align}
\end{subequations}
 where the term corresponding to the real part is weighted by $C^{+}(x,\xi)$ 
($=  \frac{1}{x-\xi} + \frac{1}{x+\xi}$), which appears also in an analogous 
expression for the GPD $E^{q}(x,\xi,t)$. The real parts of the CFFs that are 
associated with the GPDs $\widetilde{H}^{q}(x,\xi,t)$ and 
$\widetilde{E}^{q}(x,\xi,t)$, are weighted by $C^{-}(x,\xi)$ ($=  
\frac{1}{x-\xi} - \frac{1}{x+\xi}$).

\subsubsection{Bethe-Heitler}
\begin{figure}[tbp]
\centering
\includegraphics[scale=0.30]{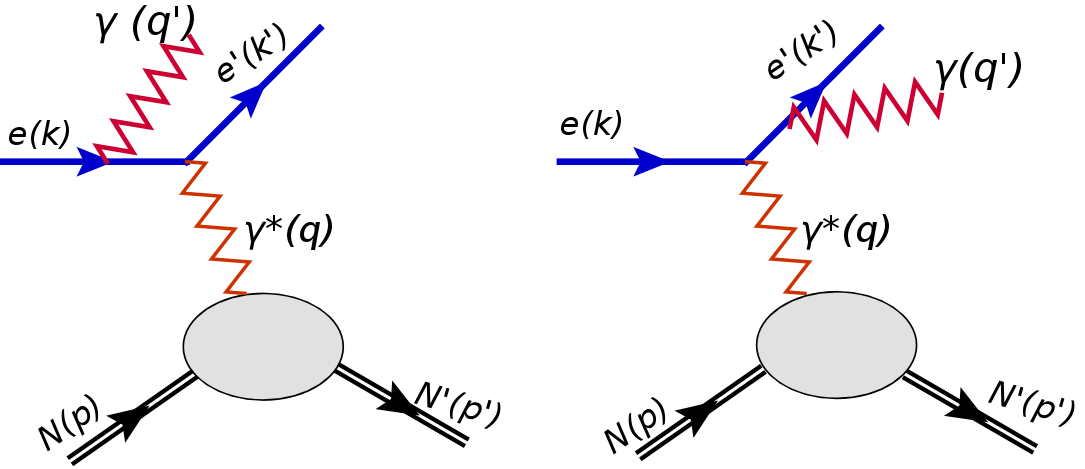}
\caption{Schematic for the Bethe-Heitler process. The final real photon can be  
emitted from the incoming electron (left plot) or from the scattered electron 
(right plot).} \label{fig:BH}
\end{figure}

Experimentally, the DVCS is indistinguishable from the Bethe-Heitler (BH) 
process, which is the reaction where the final photon is emitted either from 
the incoming or the outgoing leptons, as shown in figure \ref{fig:BH}. The BH 
process is not sensitive to GPDs and does not carry information about the 
partonic structure of the hadronic target. The BH cross section is calculable 
from the well-known electromagnetic FFs.
 
The $ep\rightarrow ep\gamma$ differential cross section of a 
longitudinally-polarized electron beam on an unpolarized proton target can be 
written as \cite{PhysRevD.82.074010}:
\begin{equation}
\frac{d^{5}\sigma^{\lambda}}{dQ^{2} dx_{B} dt d\phi d\phi_{e}} = 
\frac{\alpha^{3}}{16 \pi^{2}} \frac{x_{B} \, y^{2}}{Q^{2} \sqrt{1 + (2x_{b}M_{N}/Q)^{2}}} \frac{
|\mathcal{T}_{BH}|^{2} + |{\mathcal{T}}_{DVCS}^{\lambda}|^{2} + {\mathcal{I}}_{BH*DVCS}^{\lambda}}{e^6} 
\label{sigdiff}
\end{equation}
where $\lambda$ is the beam helicity, ${\mathcal{T}}_{DVCS}$ is the pure DVCS 
scattering amplitude, ${\mathcal{T}}_{BH}$ is the pure BH amplitude and 
${\mathcal{I}}^{\lambda}_{BH*DVCS}$ represents the interference amplitude. At 
leading twist, A.~V.~Belitsky, D. Mueller and A. Kirchner have shown that these 
amplitudes can be decomposed into a finite sum of Fourier harmonics, the 
so-called BMK formalism \cite{PhysRevD.82.074010}, as:
\begin{equation}
|\mathcal{T}_{BH}|^{2} =  \frac{e^{6} (1 + \epsilon^{2})^{-2}}{x^{2}_{B} y^{2} 
t \mathcal{P}_{1}(\phi) \mathcal{P}_{2}(\phi)} \left[ c_{0}^{BH} + 
\sum_{n=1}^{2} \Bigg( c^{BH}_{n} \cos(n\phi) + s_{n}^{BH} \sin(\phi) \Bigg) 
\right] \label{TBH}
\end{equation}

\begin{equation}
|\mathcal{T}_{DVCS}|^{2} =  \frac{e^{6}}{y^{2} Q^{2}} \left[ c_{0}^{DVCS} + 
\sum_{n=1}^{2} \Bigg( c_{n}^{DVCS} \cos(n \phi) + \lambda s_{n}^{DVCS} \sin(n 
\phi)\Bigg) \right] \label{TDVCS}
\end{equation}

\begin{equation}
\mathcal{I}_{BH*DVCS} = \frac{\pm e^{6}}{x_B y^{3} t \, \mathcal{P}_{1}(\phi) 
\mathcal{P}_{2}(\phi)} \left[ c_{0}^{I} + \sum_{n=0}^{3} \Bigg( c_{n}^{I} 
\cos(n \phi) + \lambda s_{n}^{I} \sin(n \phi) \Bigg) \right] \label{Tinter} 
\end{equation}
where $\mathcal{P}_{1}(\phi)$ and $\mathcal{P}_{2}(\phi)$ are the BH 
propagators. The leading twist expressions of the DVCS, BH and interference 
Fourier coefficients on a proton target can be found in reference 
\cite{PhysRevD.82.074010}.  The $+$($-$) sign in the interference term stands 
for the negatively (positively) charged lepton beam. In the case of an 
unpolarized proton target, the coefficients of the $\sin(\phi)$ in the BH 
amplitude are zeros.

%% file: detailed_projections.tex
\setlength\parskip{\baselineskip}%
\chapter{Detailed Experimental Projections}\label{sec:extraProjections}

\section{Kinematic Coverage}\label{sec:extraKineProjections}

Here we present many kinematic plots for the tagged DVCS reactions.

\begin{figure}[!htb]
   \centering
   \includegraphics[width=0.49\textwidth,trim=10mm 4mm 10mm 10mm,
   clip]{fastmc2/tagged_dvcs_He4_proton_2/He4_proton_2_5.pdf}
   \includegraphics[width=0.49\textwidth,trim=10mm 4mm 10mm 10mm,
   clip]{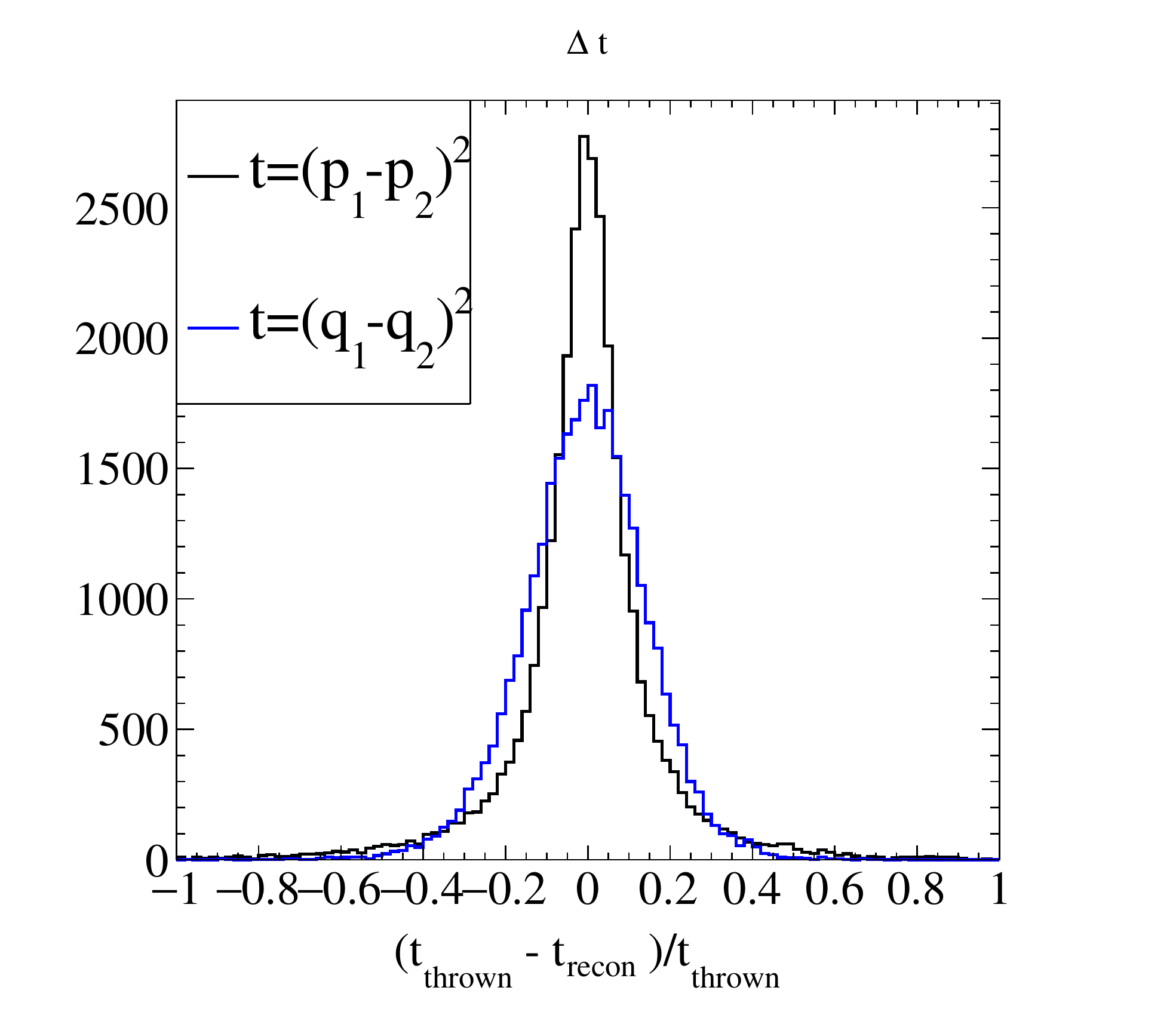}
   \caption{\label{fig:tResolutions2}Left: Thrown and reconstructed $t$ 
   calculated from the photons ($t_q$) and hadrons ($t_p$). Right: The 
   corresponding resolutions for the two methods of determining $t$.  }
\end{figure}
\begin{figure}[!htb]
   \centering
   \includegraphics[width=0.49\textwidth,trim=10mm 4mm 10mm 10mm,
   clip]{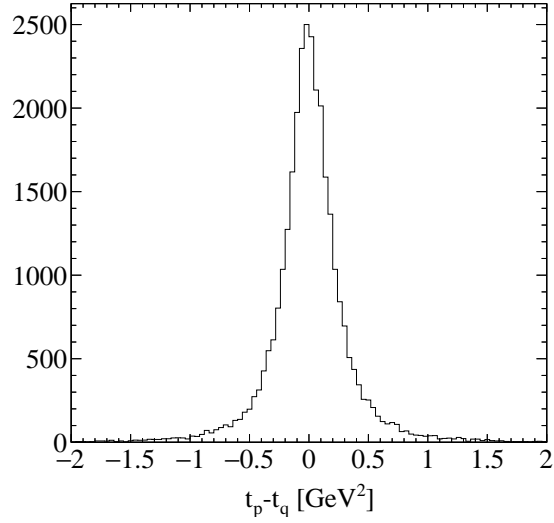}
   \caption{\label{fig:deltat}The difference in the momentum transfers, $\delta 
   t= t_p-t_q$.  }
\end{figure}
\begin{figure}[!htb]
   \centering
   \includegraphics[width=0.49\textwidth,trim=10mm 4mm 10mm 12mm,
   clip]{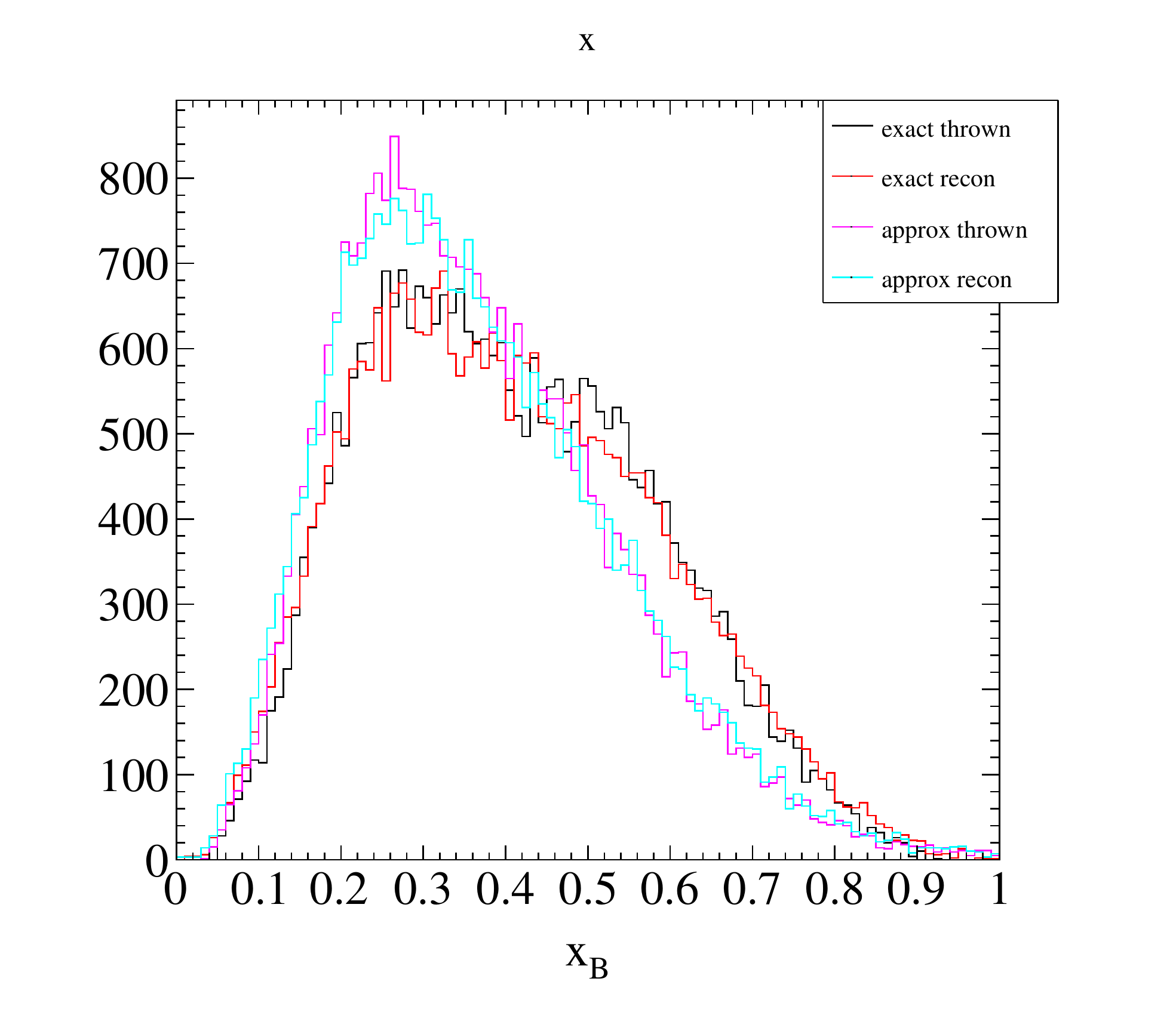}
   \includegraphics[width=0.49\textwidth,trim=10mm 4mm 10mm 12mm,
   clip]{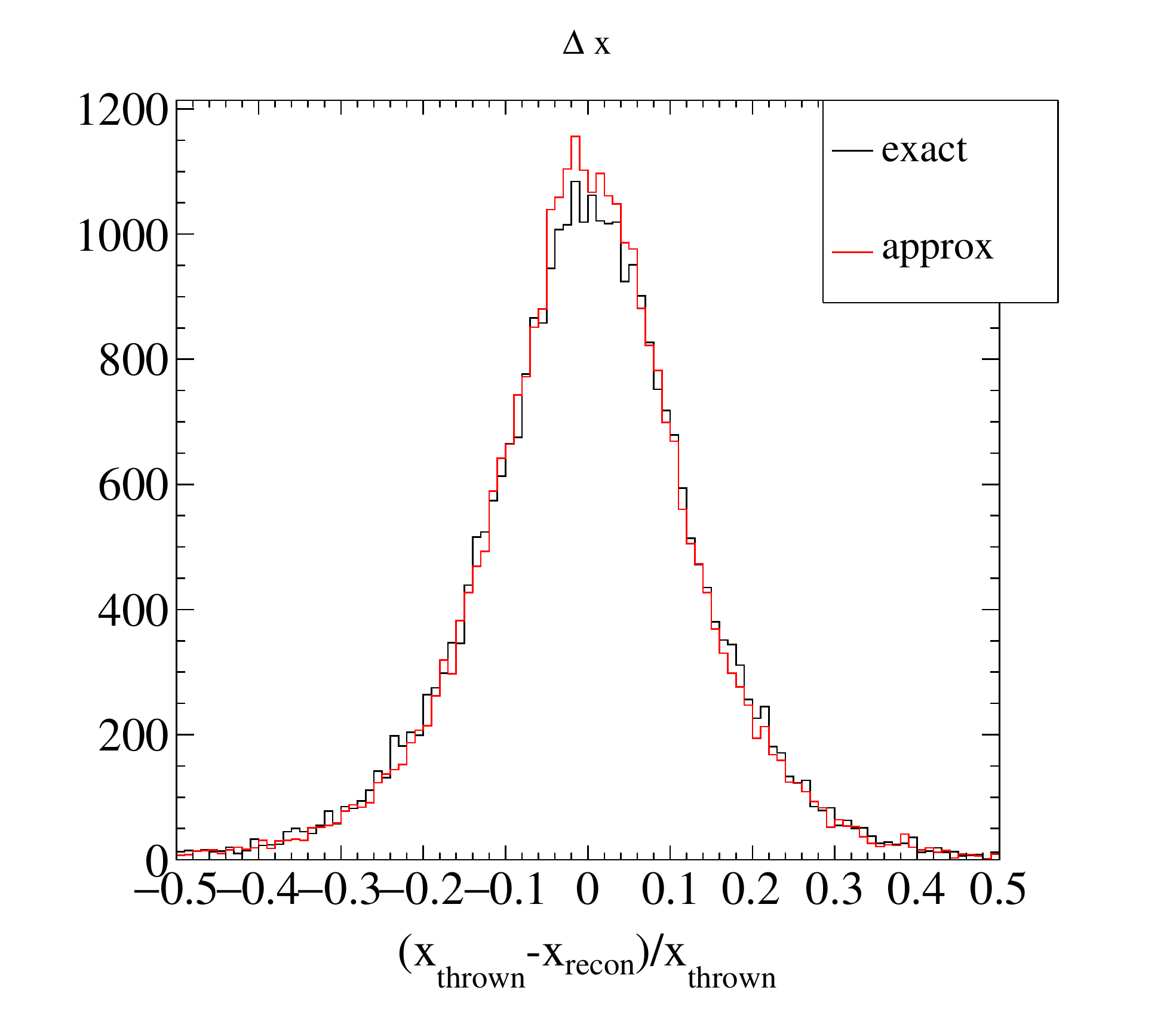}
   \caption{\label{fig:xResolutions}Left: Thrown and reconstructed $x_{B}$ 
   where the approximate calculation assumes the struck nucleon is at rest, 
i.e., $x_{approx}= Q^2/2M\nu$.  Right: Relative resolutions expected.  }
\end{figure}
\begin{figure}[!htb]
   \centering
   \includegraphics[width=0.49\textwidth,trim=10mm 4mm 10mm 12mm,
   clip]{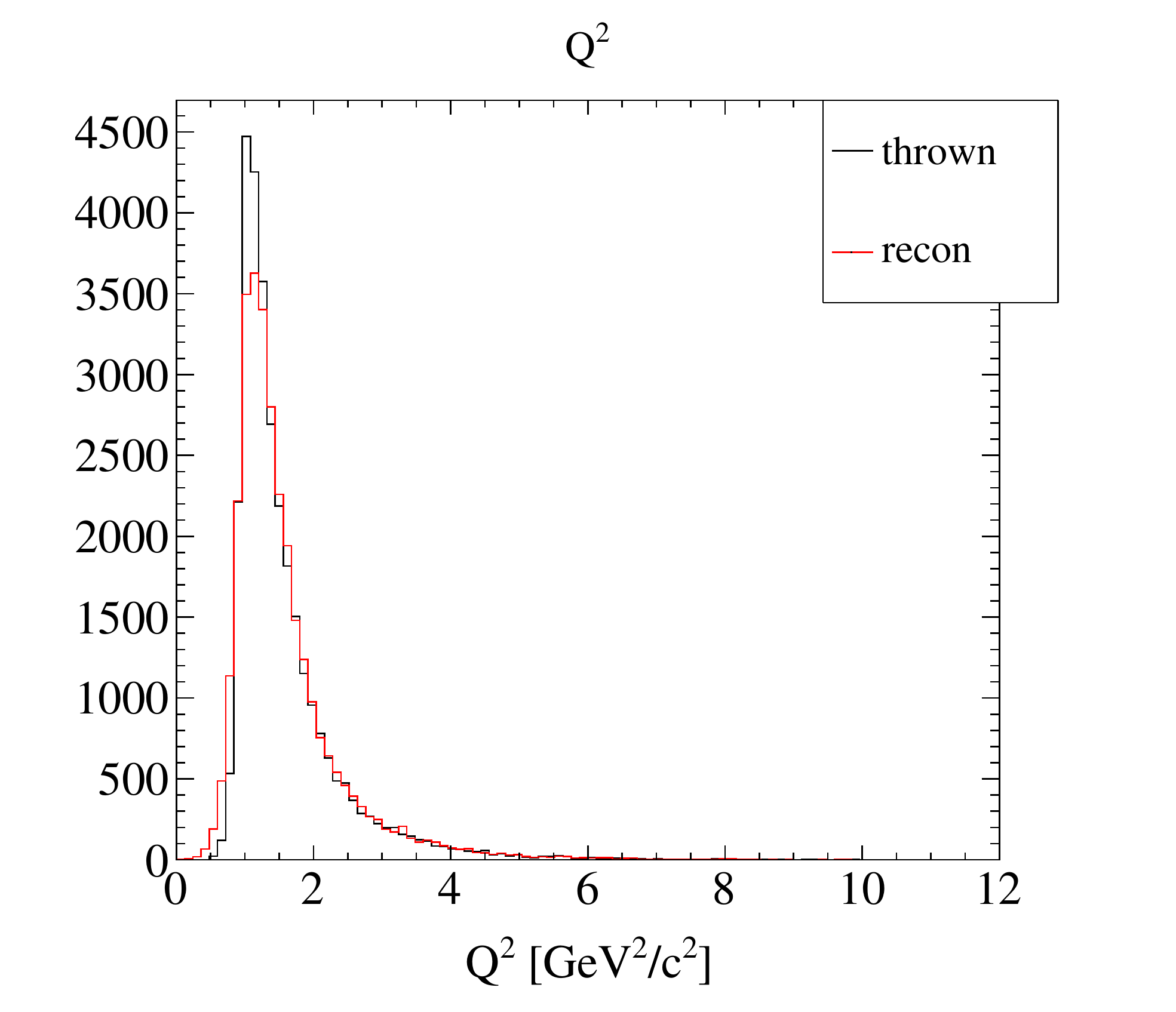}
   \includegraphics[width=0.49\textwidth,trim=10mm 4mm 10mm 12mm,
   clip]{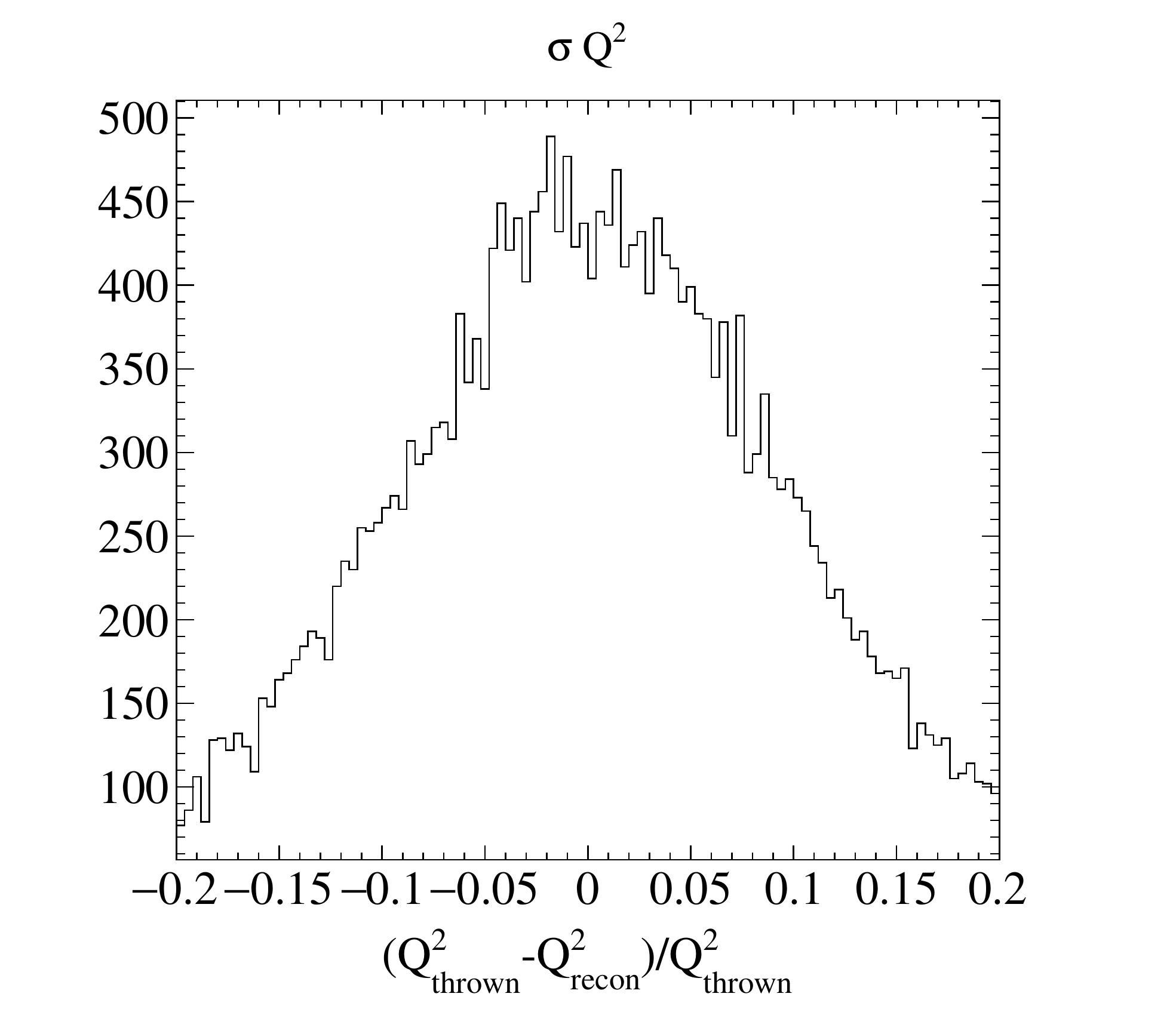}
   \caption{\label{fig:Q2Resolutions}Reconstructed $Q^2$ and its simulated 
   resolution.}
\end{figure}
\begin{figure}[!htb]
   \centering
   \includegraphics[width=0.49\textwidth,trim=10mm 4mm 10mm 12mm,
   clip]{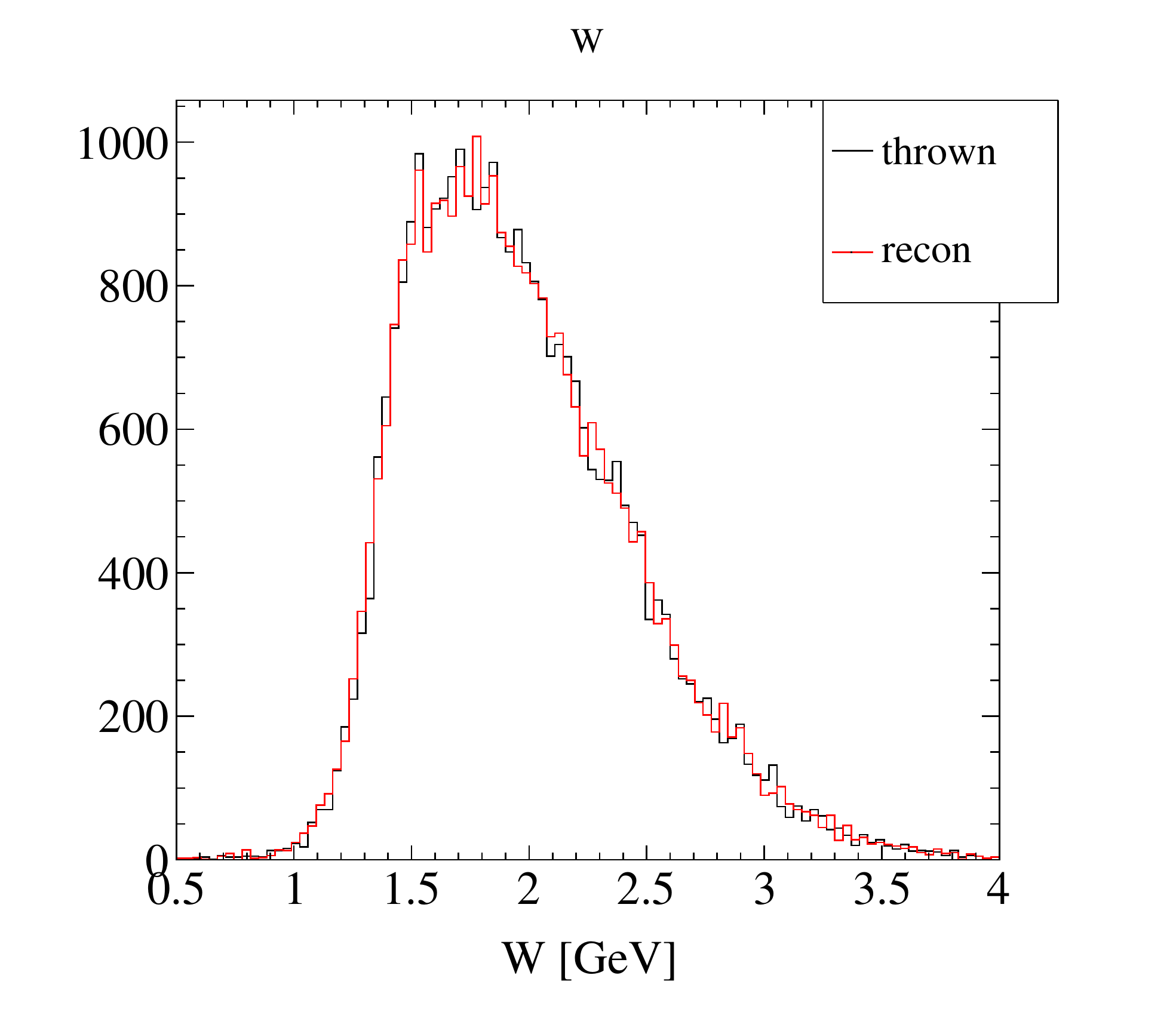}
   \includegraphics[width=0.49\textwidth,trim=10mm 4mm 10mm 12mm,
   clip]{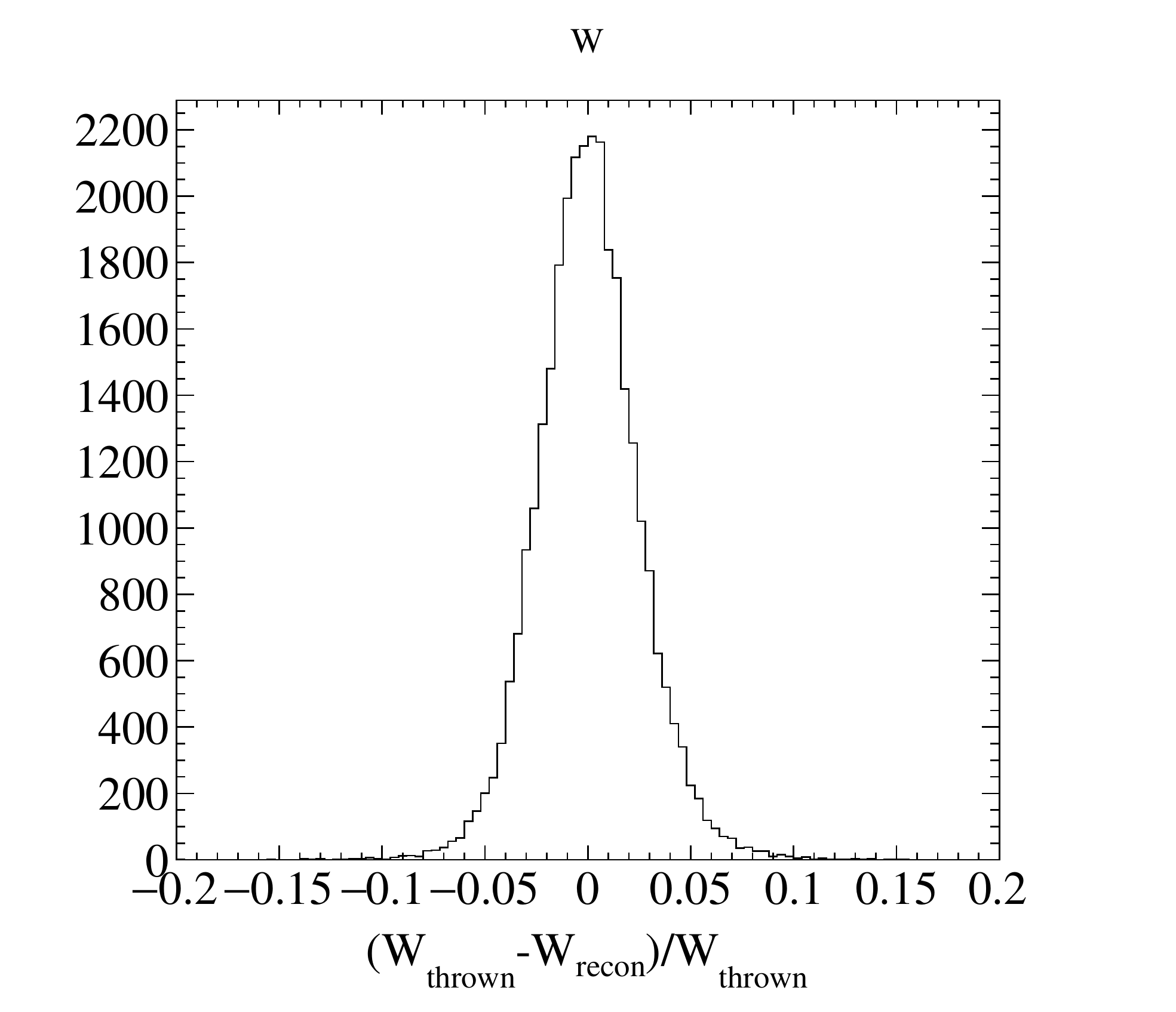}
   \caption{\label{fig:WResolutions}Reconstructed $W$ and its simulated 
   resolution.}
\end{figure}
\begin{figure}[!htb]
   \centering
   \includegraphics[width=0.49\textwidth,trim=0mm 4mm 10mm 10mm,
   clip]{fastmc2/tagged_dvcs_He4_proton_2/He4_proton_2_8.pdf}
   \includegraphics[width=0.49\textwidth,trim=0mm 4mm 10mm 10mm,
   clip]{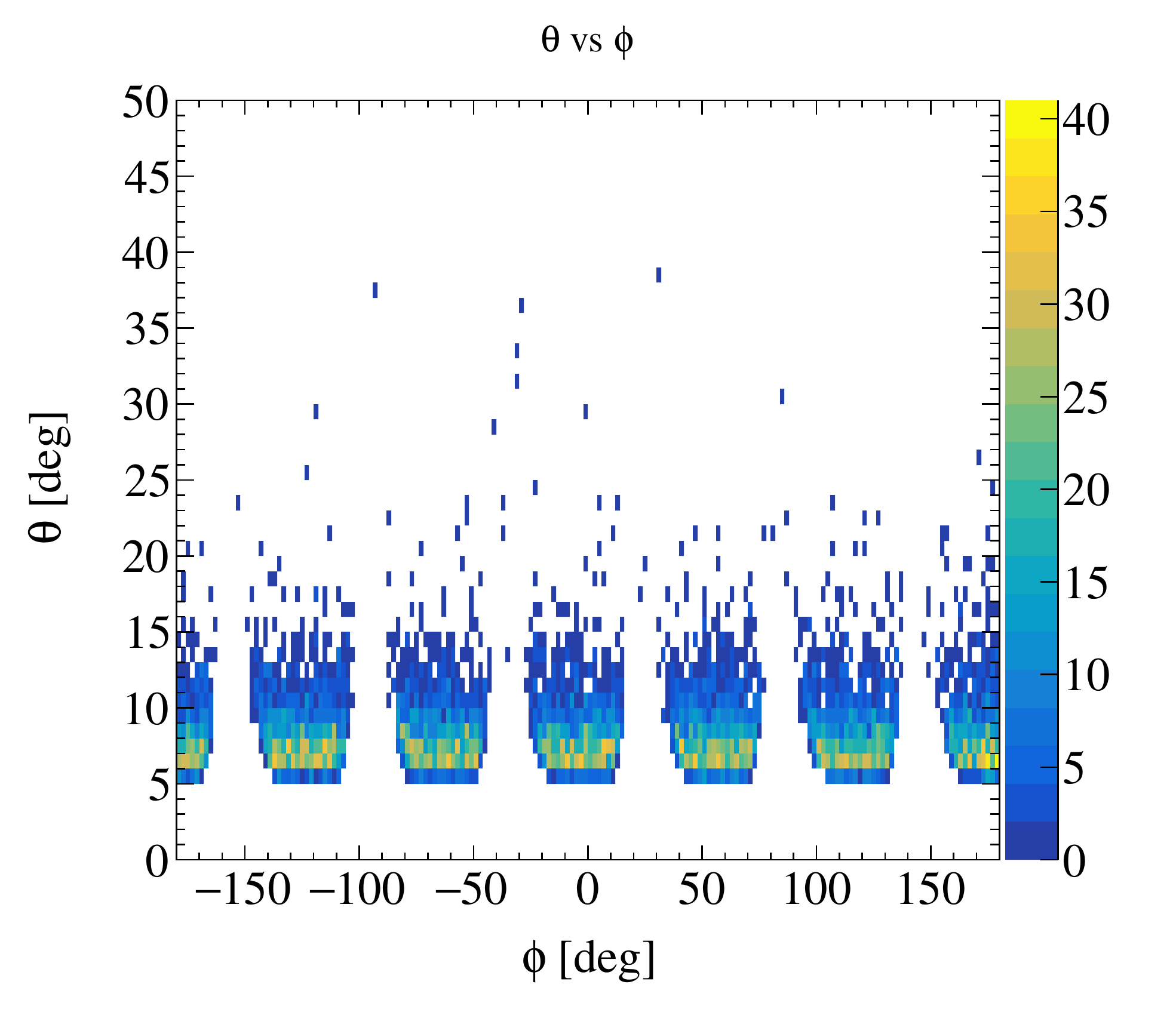}
   \caption{\label{fig:eThetaVsPhi2}Electron $\theta$ Vs. $\phi$, before (left) 
   and after (right) cuts.}
\end{figure}

\section{Projections for \texorpdfstring{$\sin\phi$}{sin(phi)} harmonic of the BSA}\label{sec:extraRatioProjections}

\begin{figure}[!htb]
   \centering
   \includegraphics[width=0.4\textwidth,trim=4mm 5mm 10mm 12mm, 
   clip]{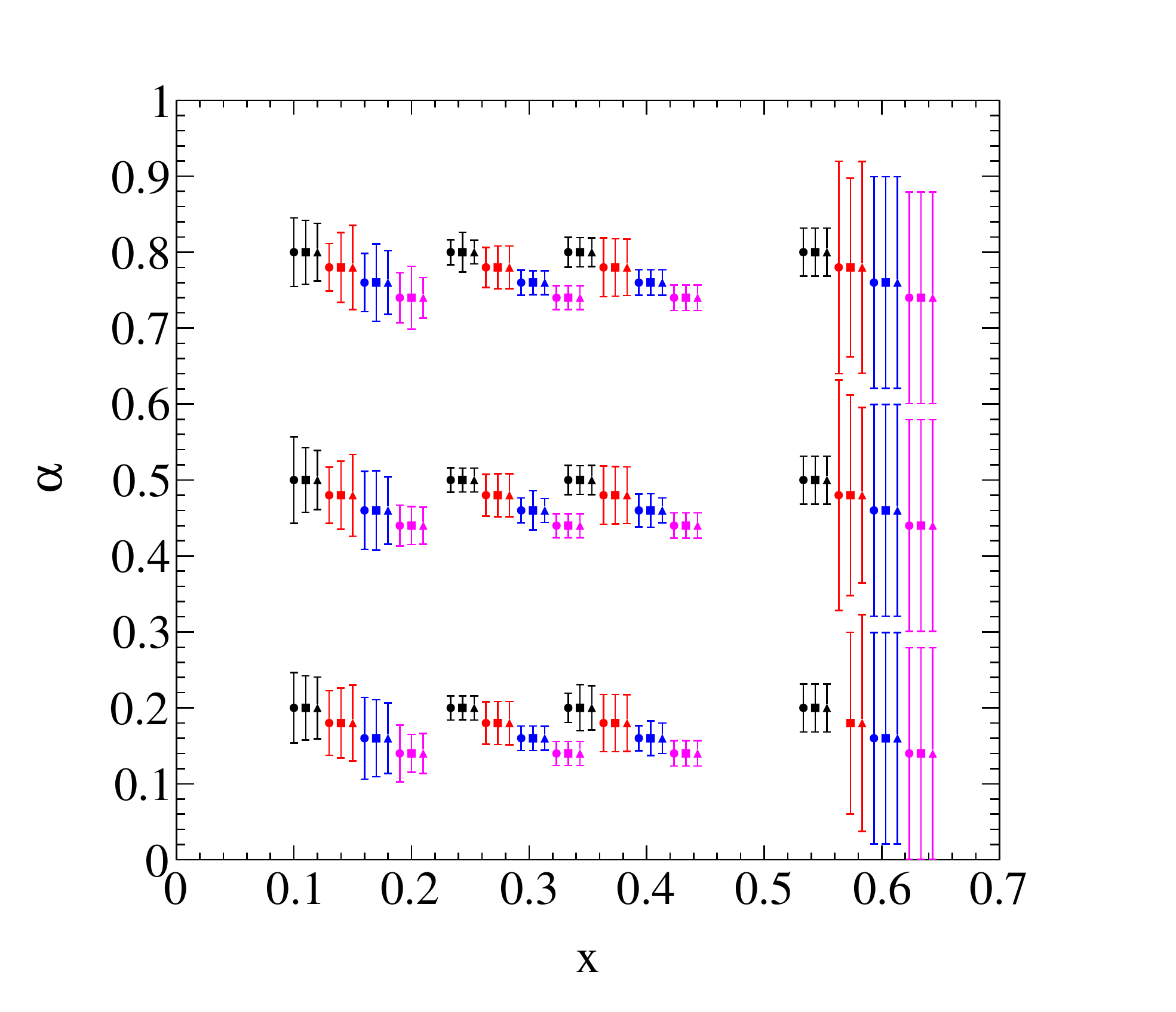}
   \includegraphics[width=0.4\textwidth,trim=4mm 5mm 10mm 12mm, 
   clip]{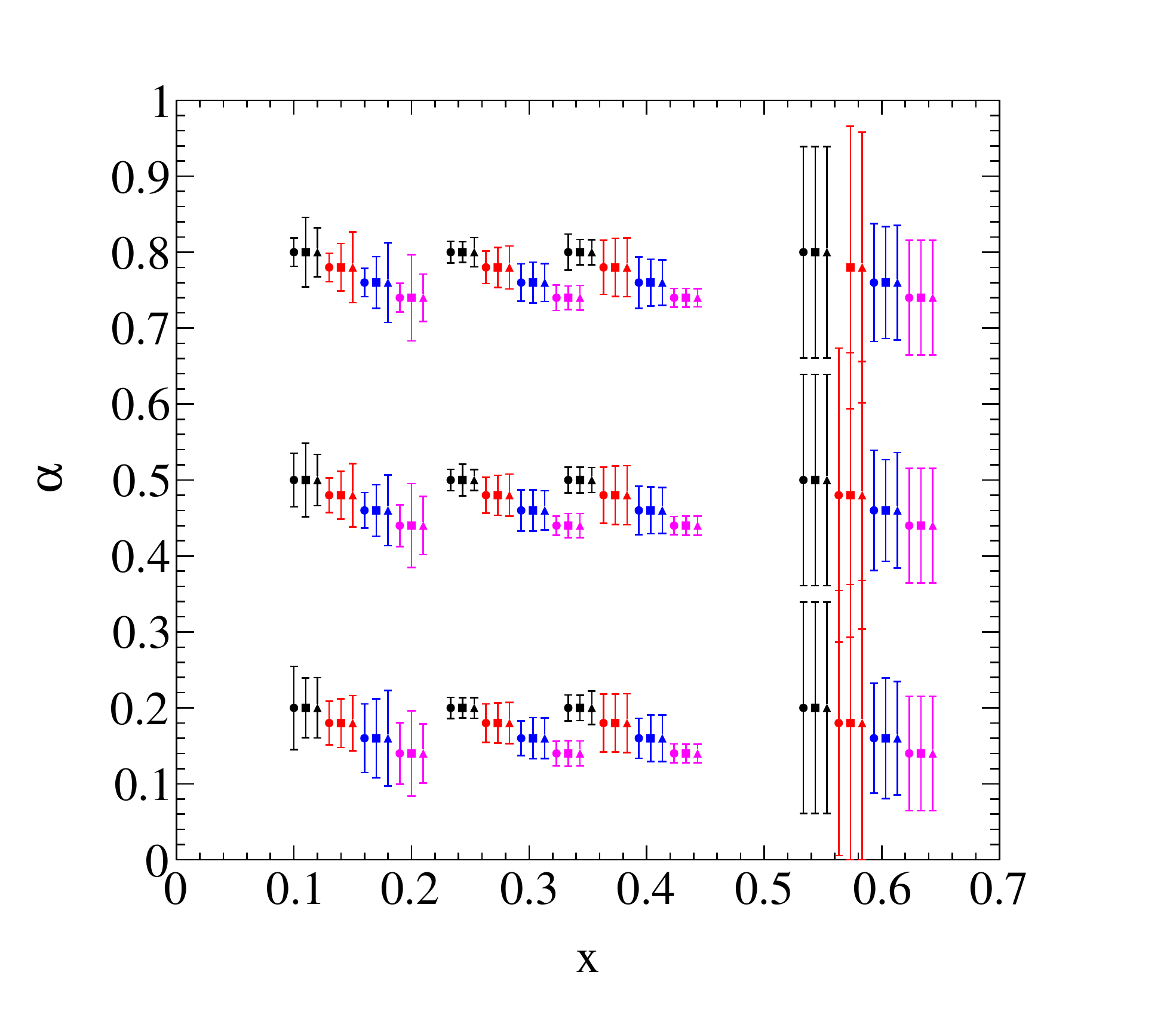}
   \includegraphics[width=0.4\textwidth,trim=4mm 5mm 10mm 12mm, 
   clip]{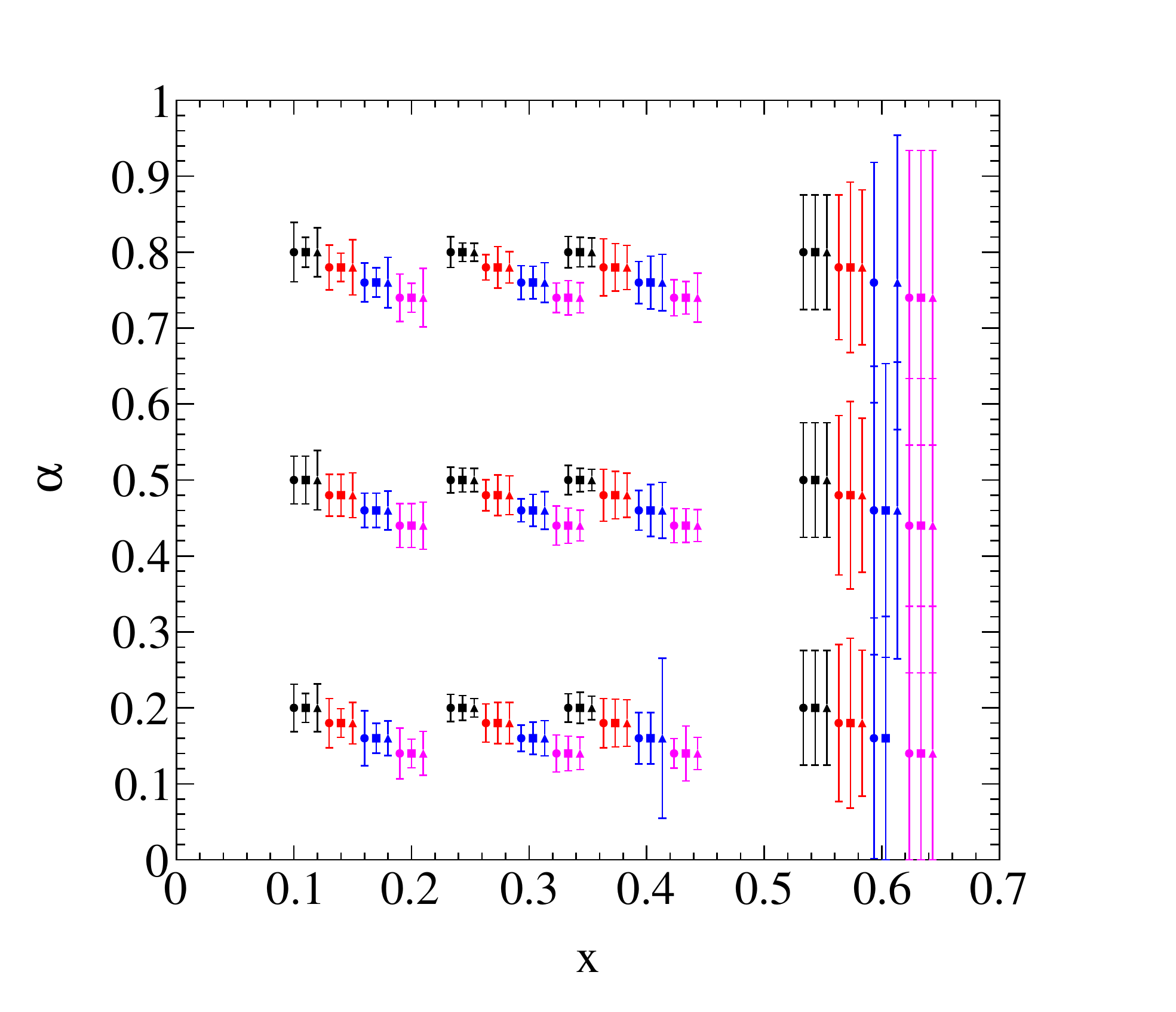}
   \includegraphics[width=0.4\textwidth,trim=4mm 5mm 10mm 12mm, 
   clip]{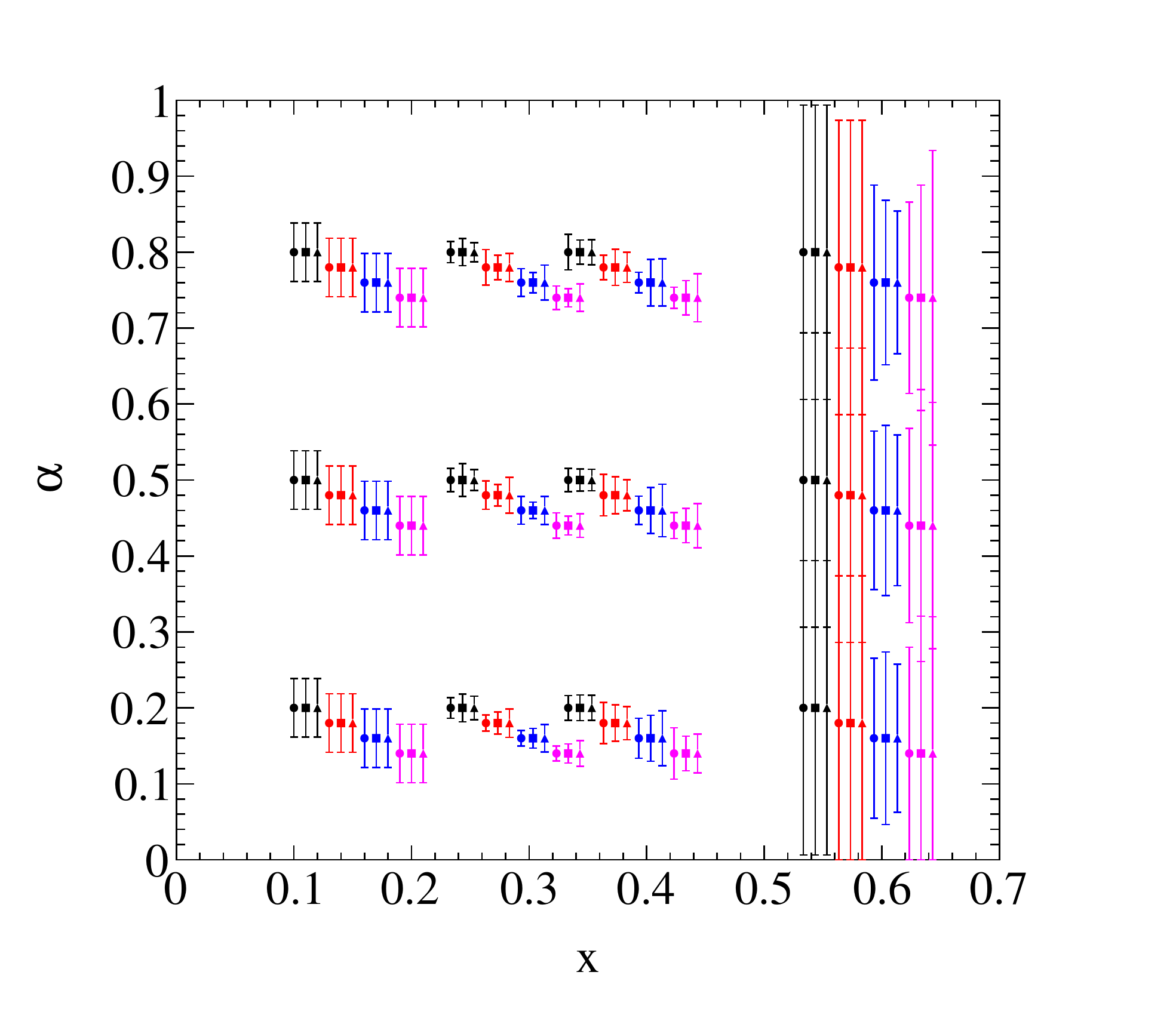}
   \caption{\label{fig:nDVCS4HeAlpha}Projected statistical uncertainty for
      the n-DVCS $\alpha$ from a $^4$He target. The points are offset for clarity.
   Each plot shows the results for different $Q^2$ bins starting with the
lowest in the upper left and the highest in the lower right. The horizontal 
bands of points starting from low to high are for the three spectator momentum 
bins (like Figure~\ref{fig:alphaALU}) and the different symbols indicate the 
spectator angle bins. The points color along with points grouped with a 
slight negative slope are for different $t$ bins starting with black for the lows $|t|$ bin.
}
\end{figure}
\begin{figure}[!htb]
   \centering
   \includegraphics[width=0.4\textwidth,trim=4mm 5mm 10mm 12mm, 
   clip]{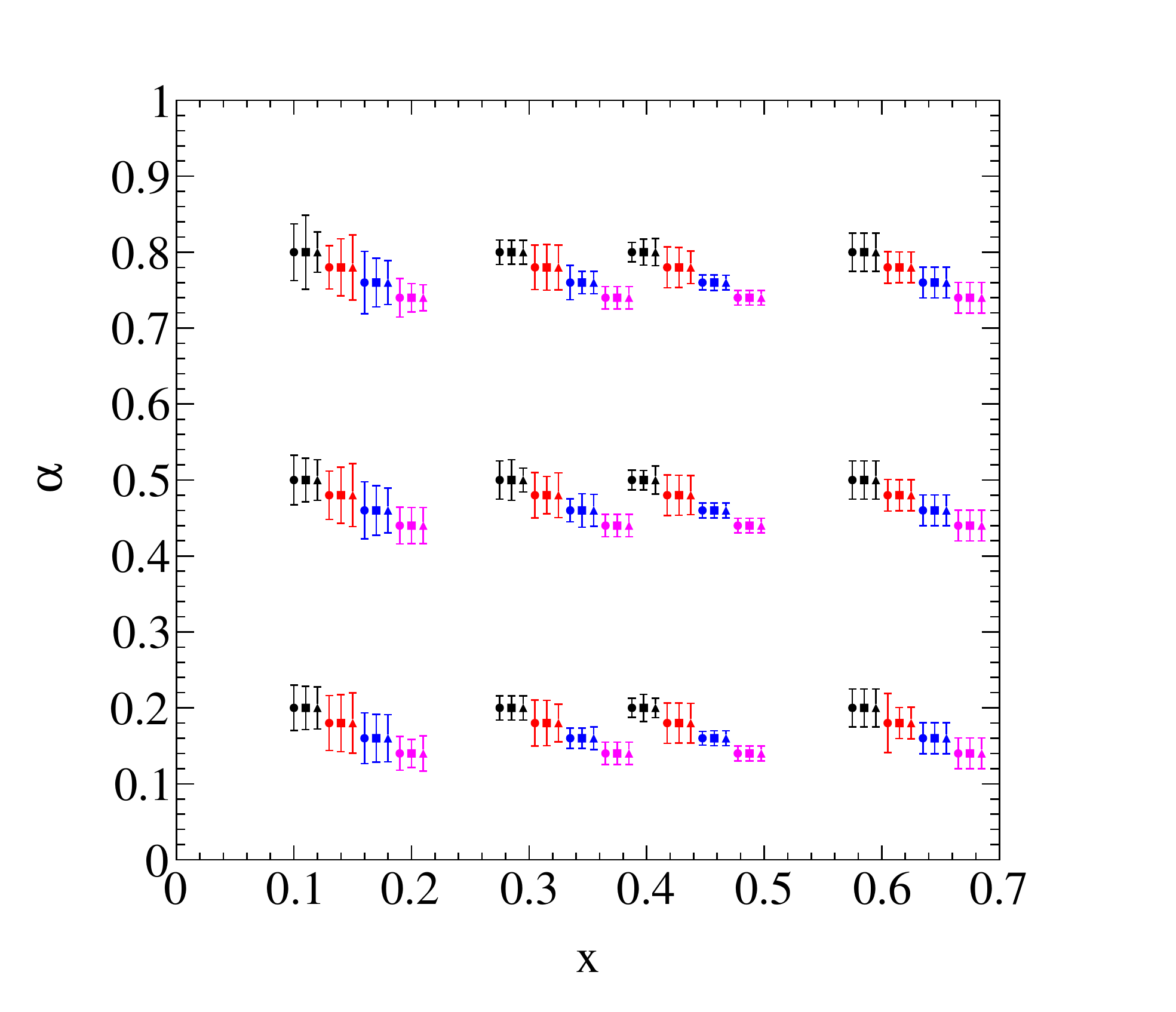}
   \includegraphics[width=0.4\textwidth,trim=4mm 5mm 10mm 12mm, 
   clip]{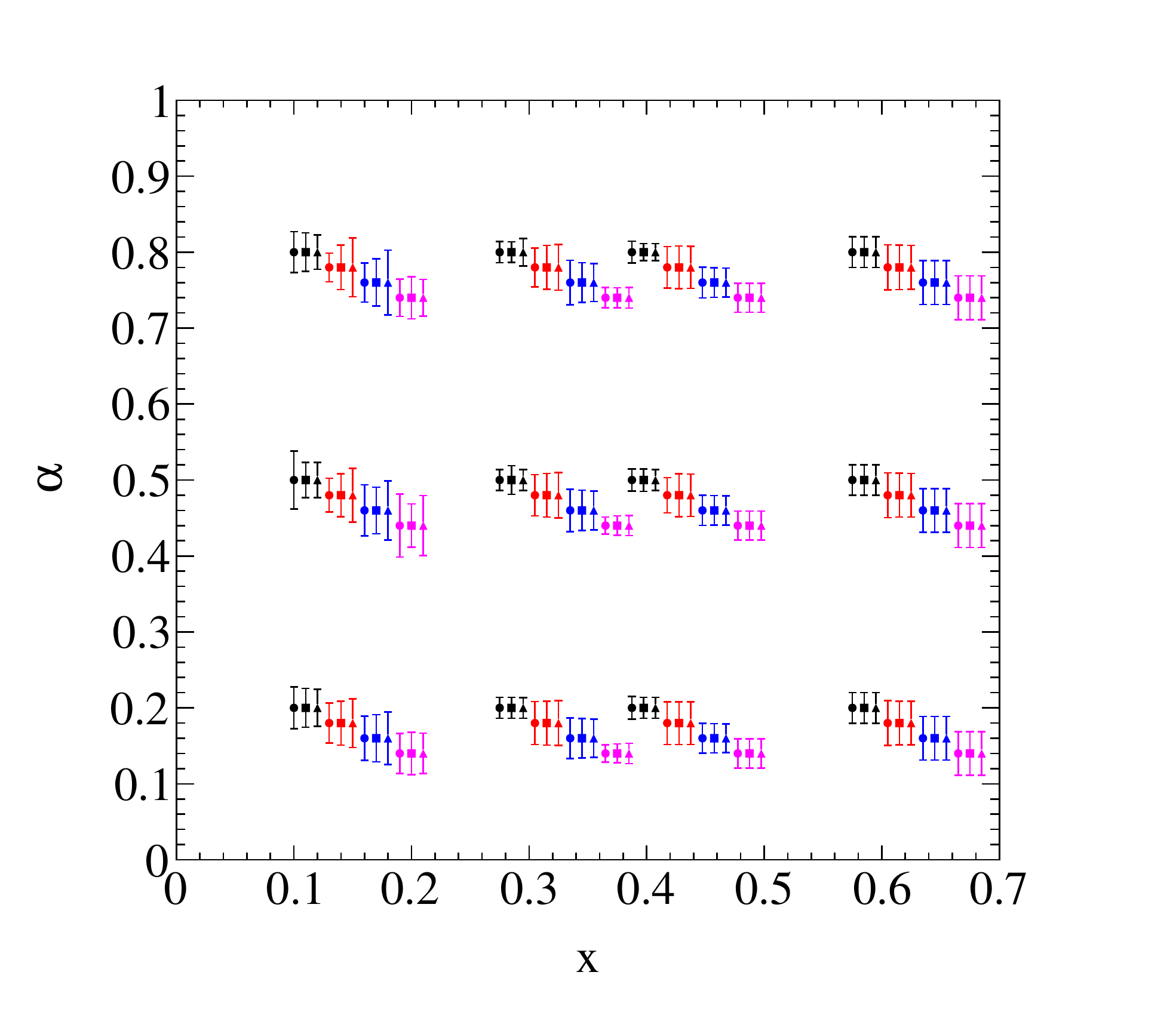}
   \includegraphics[width=0.4\textwidth,trim=4mm 5mm 10mm 12mm, 
   clip]{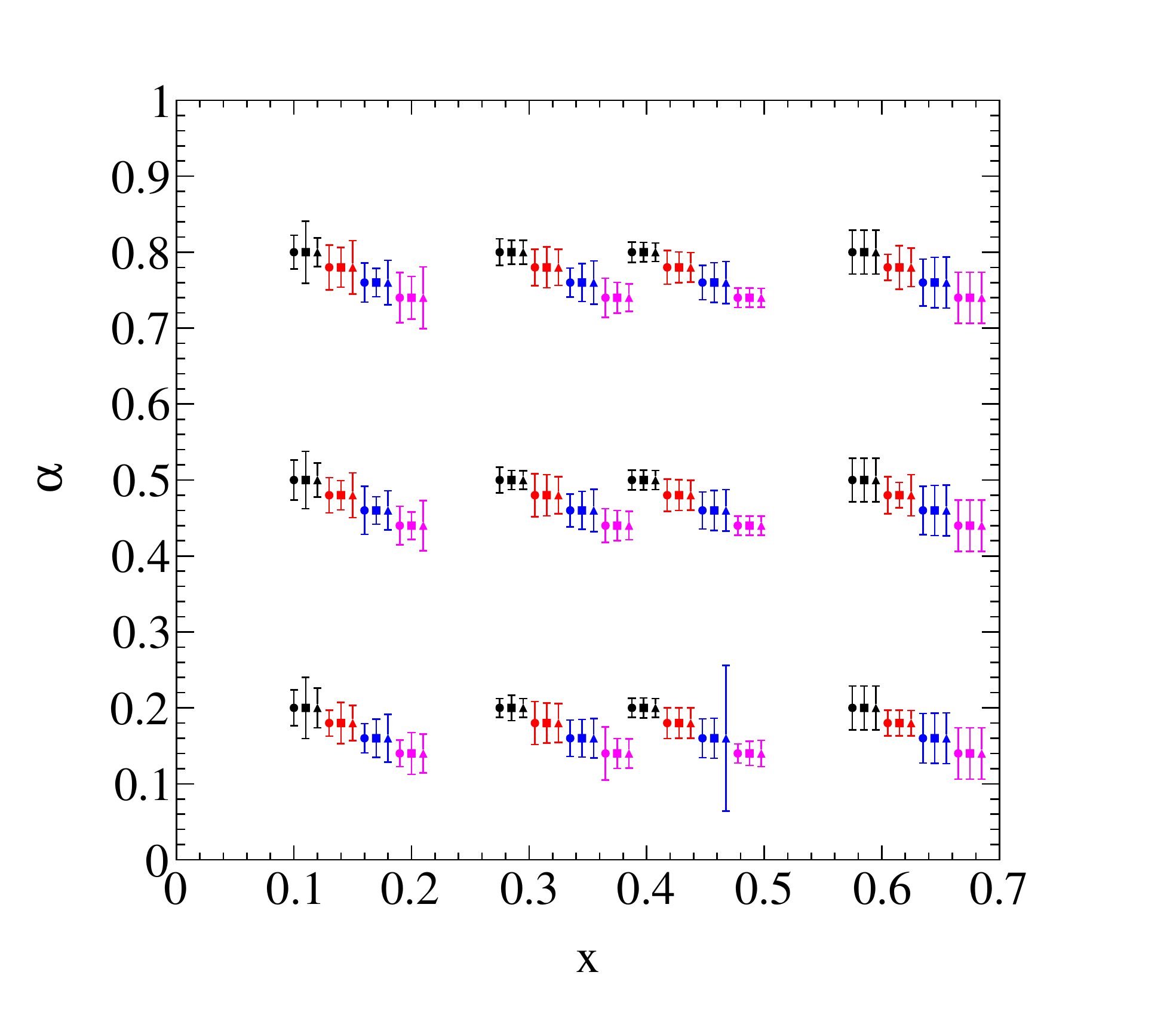}
   \includegraphics[width=0.4\textwidth,trim=4mm 5mm 10mm 12mm, 
   clip]{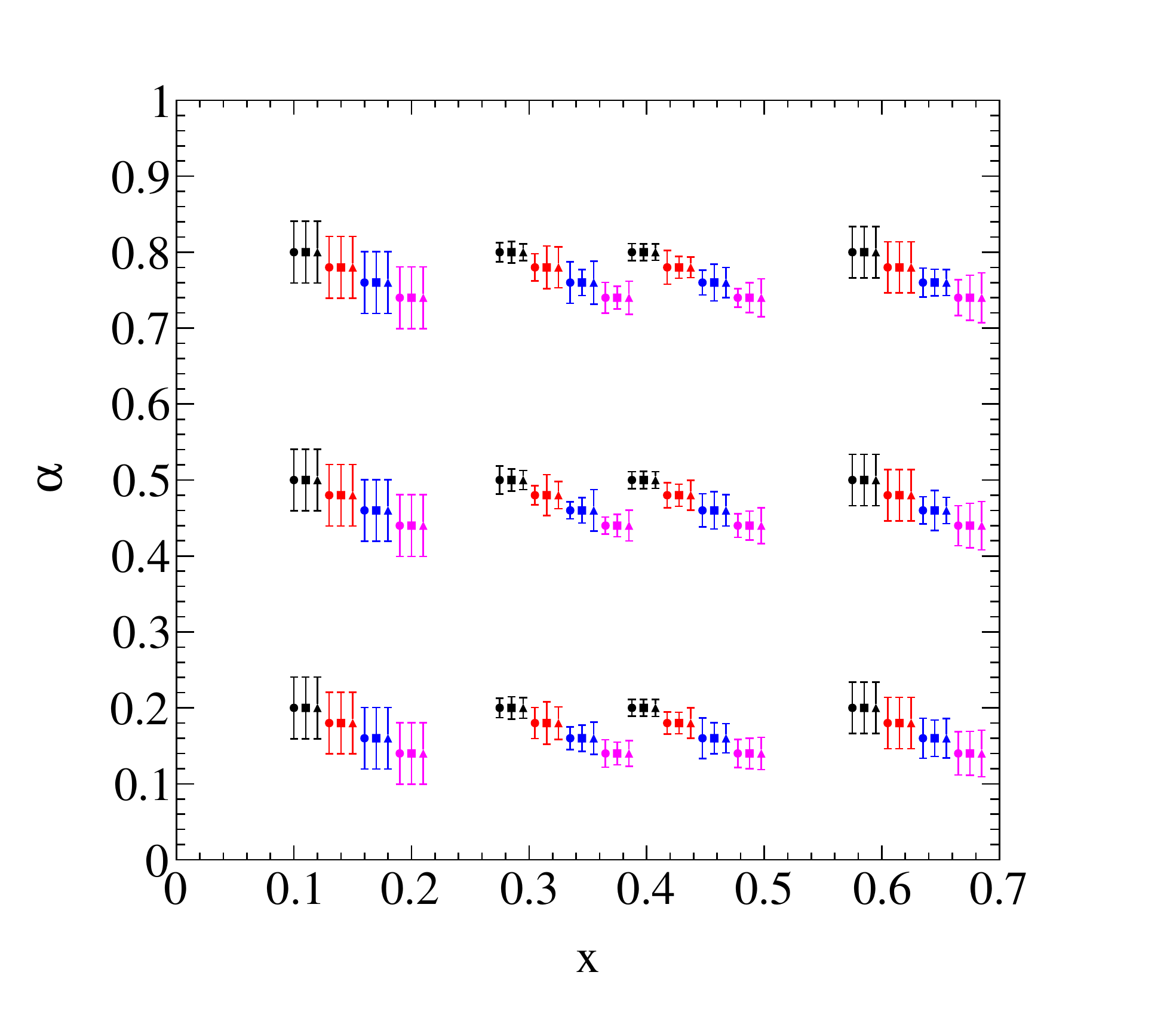}
   \caption{\label{fig:pDVCS4HeAlpha}Same as Figure~\ref{fig:nDVCS4HeAlpha} except 
   for p-DVCS on a $^4$He target.}
\end{figure}

\begin{figure}[!htb]
   \centering
   \includegraphics[width=0.4\textwidth,trim=4mm 5mm 10mm 12mm, 
   clip]{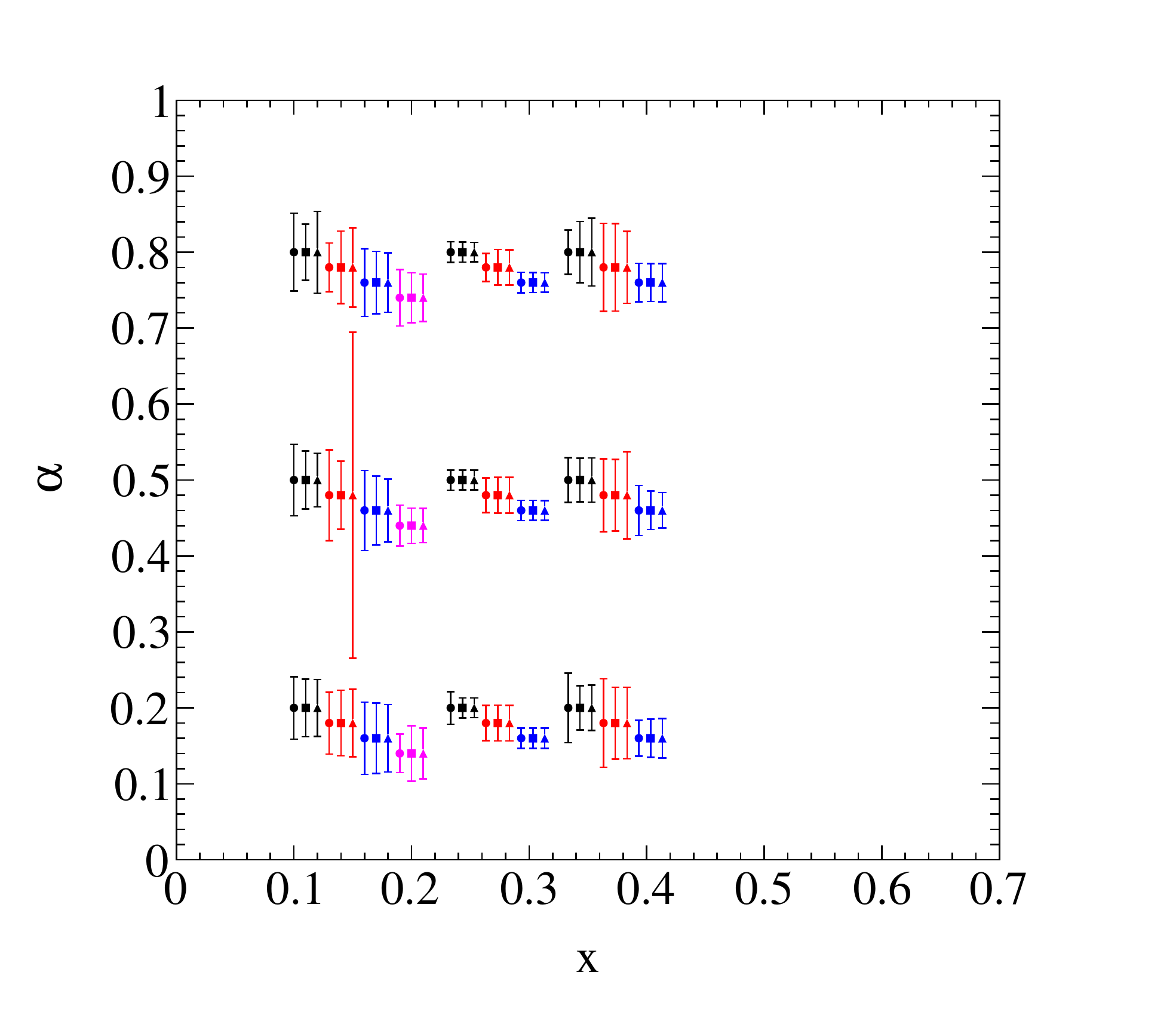}
   \includegraphics[width=0.4\textwidth,trim=4mm 5mm 10mm 12mm, 
   clip]{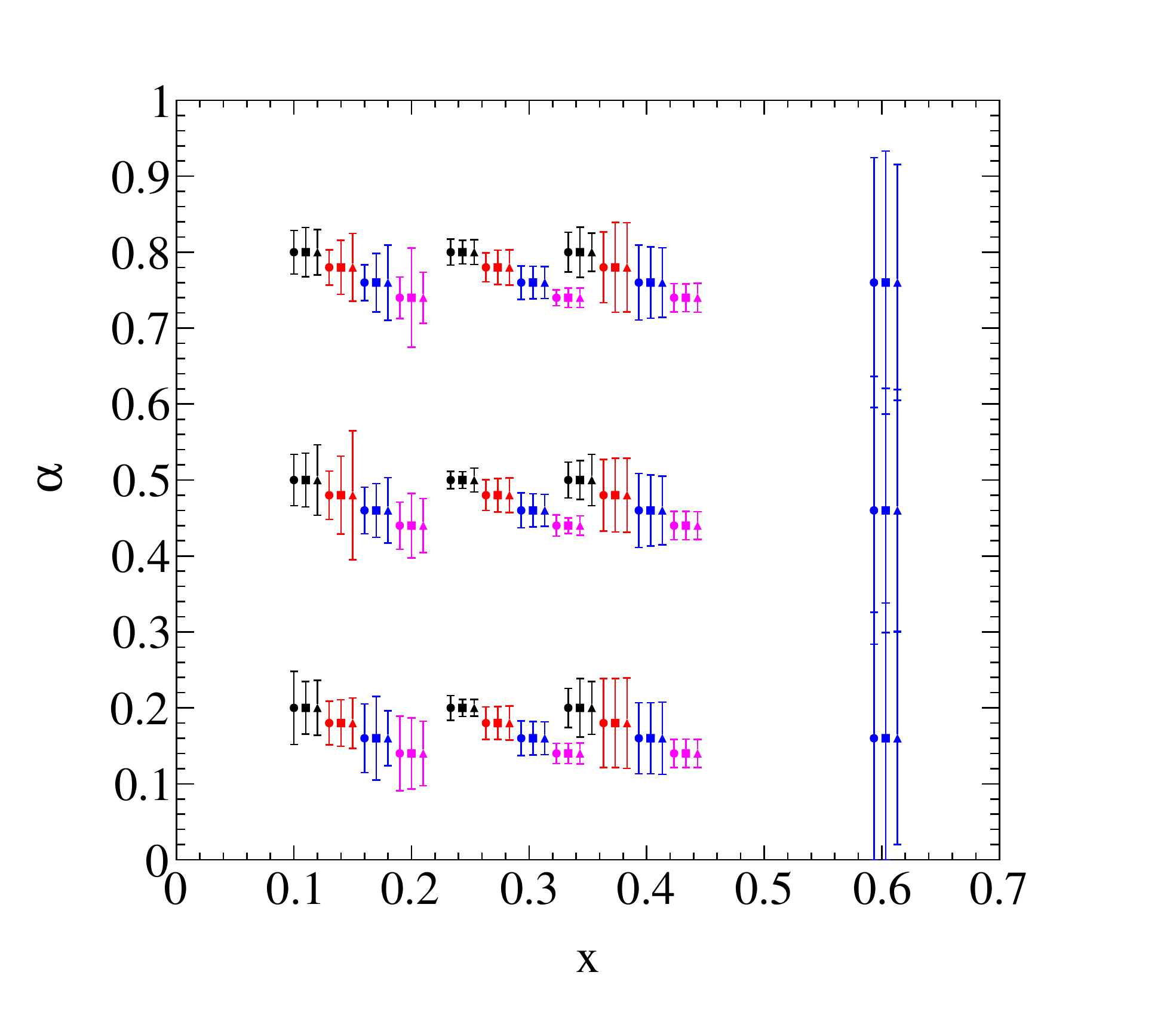}
   \includegraphics[width=0.4\textwidth,trim=4mm 5mm 10mm 12mm, 
   clip]{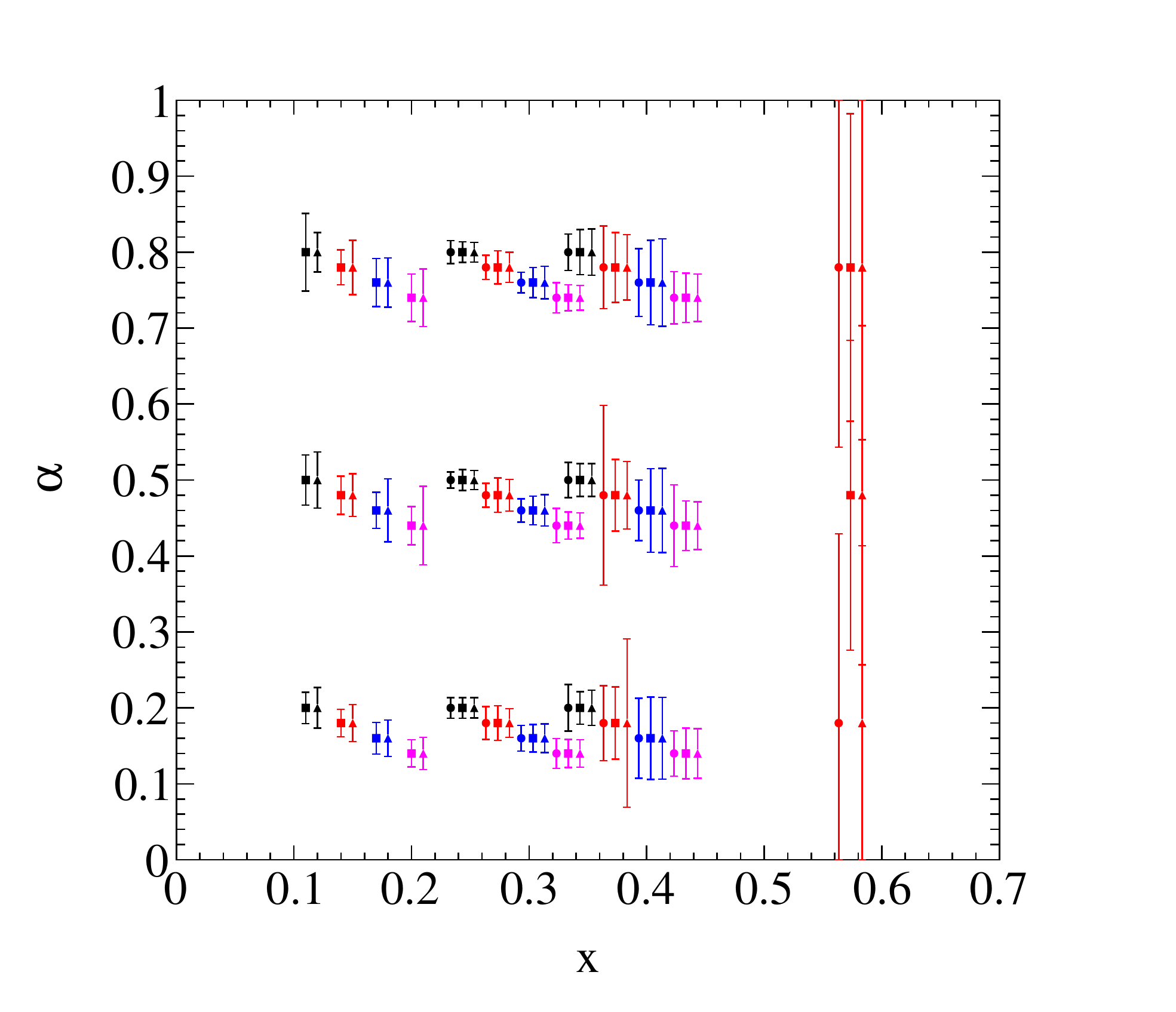}
   \includegraphics[width=0.4\textwidth,trim=4mm 5mm 10mm 12mm, 
   clip]{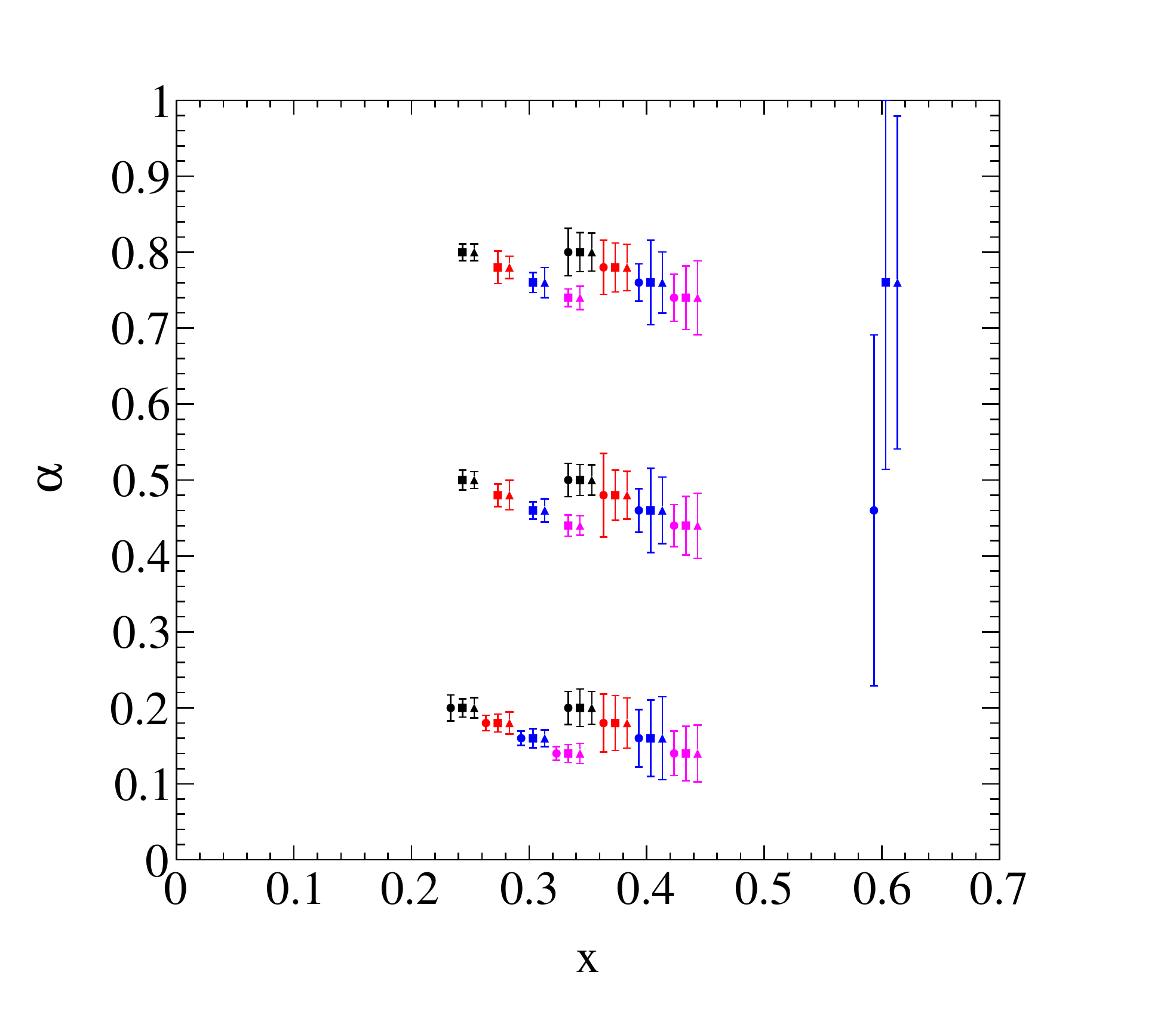}
   \caption{\label{fig:nDVCS2HAlpha}Same as Figure~\ref{fig:nDVCS4HeAlpha} except 
   for n-DVCS on a $^2$H target.}
\end{figure}

\newpage

\subsection{Off-forward EMC Effect Ratio}

\begin{figure}[!htb]
   \centering
   \includegraphics[width=0.4\textwidth,trim=4mm 5mm 10mm 12mm, 
   clip]{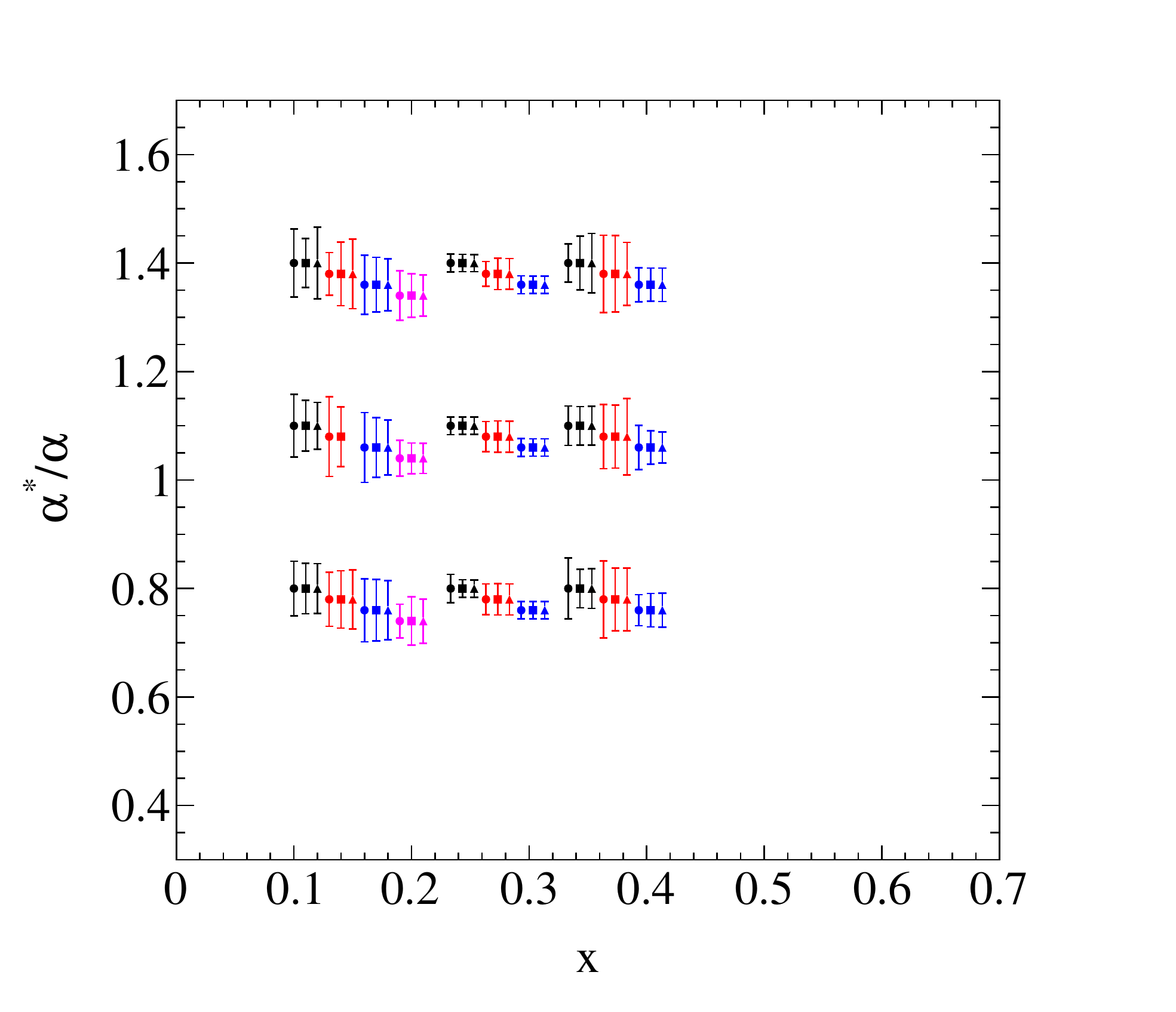}
   \includegraphics[width=0.4\textwidth,trim=4mm 5mm 10mm 12mm, 
   clip]{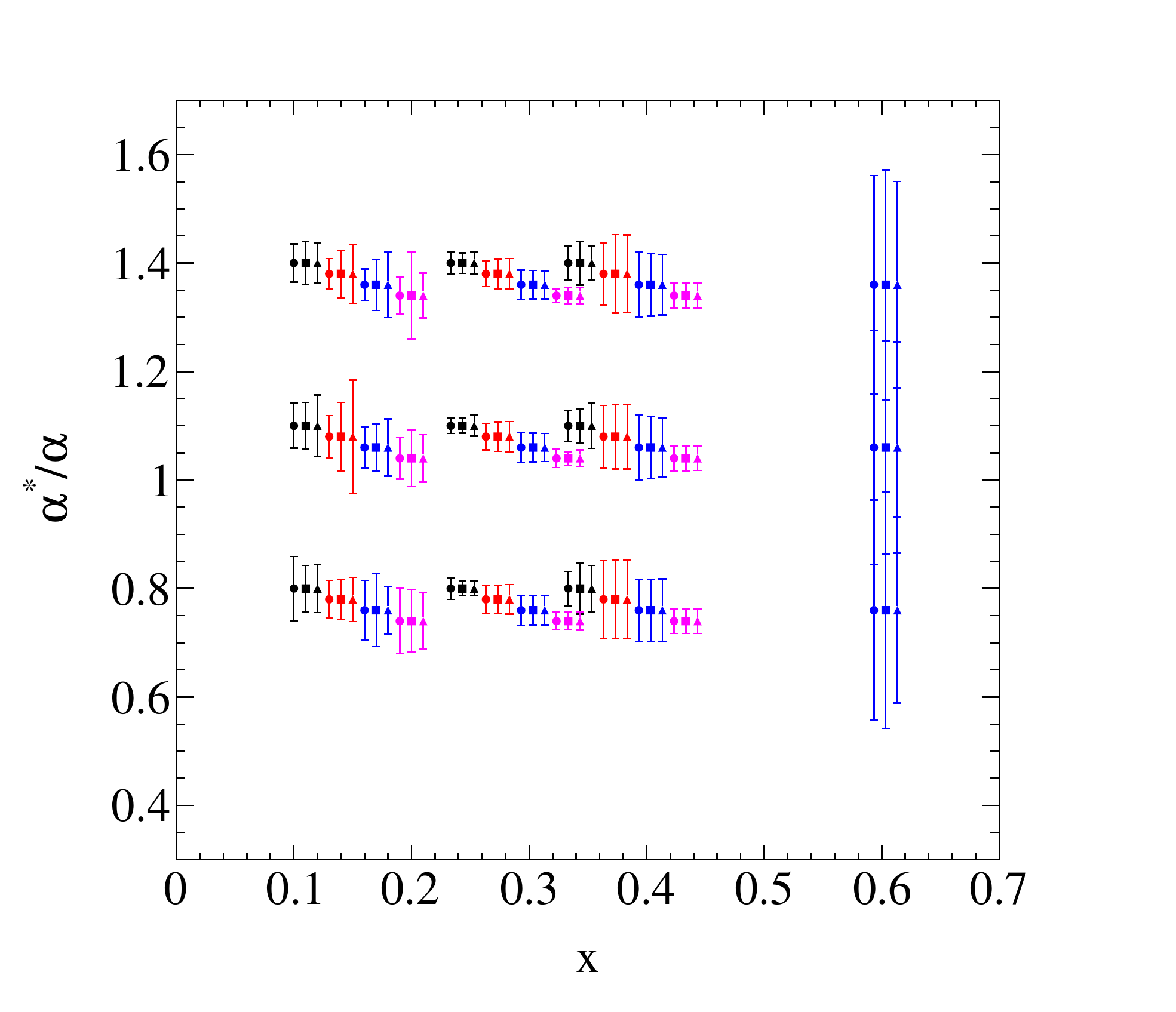}
   \includegraphics[width=0.4\textwidth,trim=4mm 5mm 10mm 12mm, 
   clip]{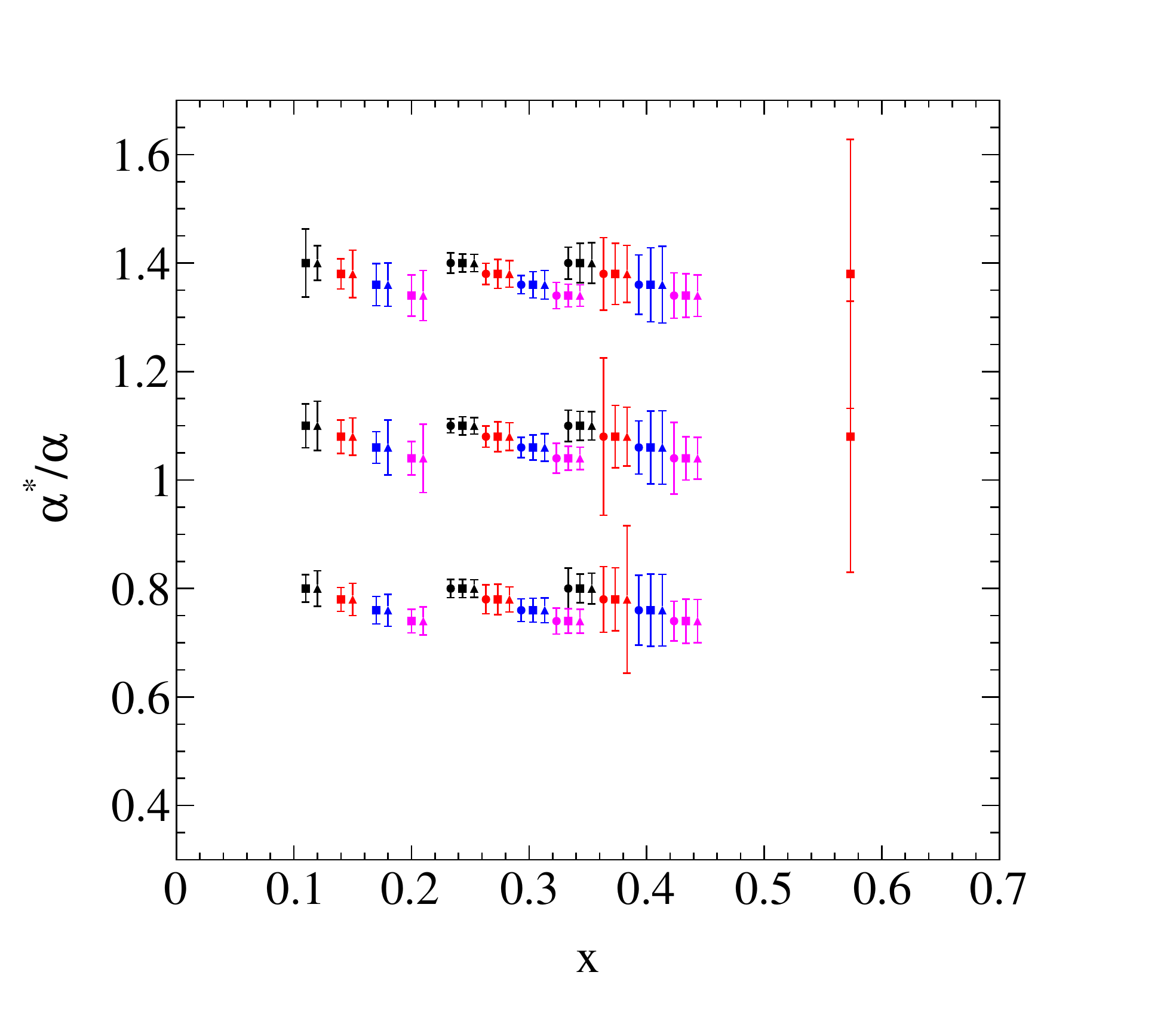}
   \includegraphics[width=0.4\textwidth,trim=4mm 5mm 10mm 12mm, 
   clip]{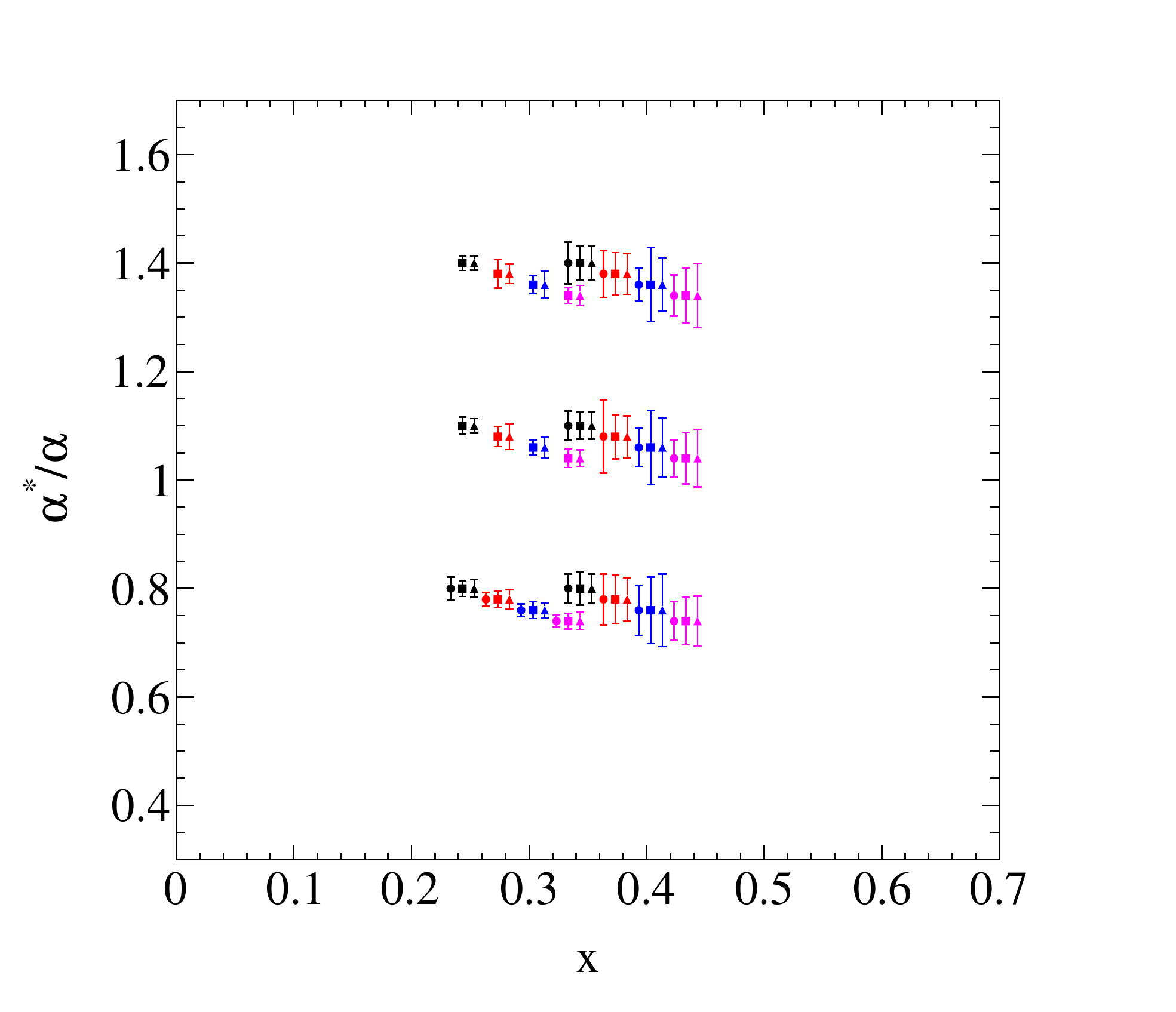}
   \caption{\label{fig:nDVCSAlphaRatio}The off-forward EMC ratio for a 
      bound neutron to a quasi-free neutron (see Equation~\ref{eq:Ralpha})
      for the same kinematics and binning described
      in Figure~\ref{fig:nDVCS2HAlpha}}
\end{figure}

The projected statistical uncertainties for the off-forward 
EMC ratio, defined in Equation~\ref{eq:Ralpha}, is shown in Figure~\ref{fig:nDVCSAlphaRatio}.
